\pdfoutput=1 
\documentclass[11pt,a4paper,twoside,openright]{report}
\usepackage[applemac]{inputenc}
\usepackage[T1]{fontenc}
\usepackage{graphicx}
\usepackage[francais]{babel}
\usepackage{indentfirst}
\usepackage{amsmath}
\usepackage[psamsfonts]{euscript}
\usepackage[psamsfonts]{amssymb}
\usepackage{natbib}
\usepackage{epsfig}
\usepackage{deriv}
\usepackage{fancyhdr}
\usepackage{minitoc}
\usepackage{shorttoc}
\usepackage{pdfpages}
\usepackage{aeguill}
\usepackage{appendix}

\newcommand{\Rsolar}{\mbox{$R_{\odot}\,$}}

\newcommand{\kms}{\mbox{$\mbox{km\,s}^{-1}$}\,}

\usepackage{natbib}
\usepackage[colorlinks=true,linkcolor=blue,allcolors=blue]{hyperref}%
\usepackage{url} 
\def\kms{km s$^{-1}$}

\newcommand*\getsto{\aa@centerstack{\gets}{\to}}
\newcommand*\lid{\aa@centerstackskip{<}{=}{1.2\p@}{1.2\p@}{1\p@}{0.9\p@}}
\newcommand*\gid{\aa@centerstackskip{>}{=}{1.2\p@}{1.2\p@}{1\p@}{0.9\p@}}

\includepdfset{pages=-,pagecommand={\thispagestyle{fancy}}}

\usepackage{caption}

\definecolor{astral}{RGB}{46,116,181}
\definecolor{orange}{RGB}{181,80,80}
\usepackage{mathptmx}% Times Roman font
\usepackage[scaled=.90]{helvet}% Helvetica, served as a model for arial
\usepackage{xcolor}
\usepackage{titlesec}
\titleformat{\chapter}[display]
  {\normalfont\sffamily\huge\bfseries\color{astral}}
  {\chaptertitlename\ \thechapter}{20pt}{\Huge}
\titleformat{\section}
  {\normalfont\sffamily\Large\bfseries\color{orange}}
  {\thesection}{1em}{}
\titleformat{\subsection}
  {\normalfont\sffamily\Large\bfseries\color{orange}}
  {\thesubsection}{1em}{}

\begin{document}
\begin{titlepage}
\begin{center}

{\Large Université Côte d'Azur - UFR Sciences}\\
Ecole Doctorale Sciences Fondamentales et Appliquées\\

\vspace*{1.2truecm}

 {\huge Thèse d'Habilitation à Diriger des Recherches}  \\ [0.5cm]

{\bf {\huge  Les étoiles pulsantes et les binaires à éclipses comme indicateurs de distances dans l'univers}}\\ [1.5cm]

par \\

{\bfseries \Large{Nicolas  NARDETTO}} \\ [1cm]

soutenue le mercredi 20 Décembre 2017
\end{center}

\begin{flushleft}
{\huge Jury} : \\
\end{flushleft}

\begin{center}
\begin{tabular}{lll}

M. Xavier DELFOSSE    & Université Grenoble-Alpes     & \emph{Rapporteur}\\
\vspace{0.10cm}
M. Pascal FOUQUE  & Université Paul Sabatier - Toulouse 3                  & \emph{Rapporteur}\\
\vspace{0.10cm}
Mme Agnès LEBRE  & Université de Montpellier                   & \emph{Rapportrice} \\
\vspace{0.10cm}
Mme Yveline LEBRETON    & Observatoire de Paris-Site de Meudon                & \emph{Examinatrice}\\
\vspace{0.10cm}
Mme Karine PERRAUT      & Université Grenoble-Alpes          & \emph{Examinatrice} \\
\vspace{0.10cm}
M. Ennio PORETTI   &  Osservatorio Astronomico di Brera        & \emph{Examinateur}\\
\vspace{0.10cm}

\end{tabular}
\end{center}

\vspace{0.5cm}
\begin{center}

Université Côte d'Azur, Observatoire de la Côte d'Azur, CNRS, Laboratoire Lagrange, France Nicolas.Nardetto@oca.eu

\end{center}
\end{titlepage}

%\chapter*{Résumé}

 \strut \thispagestyle{empty} \vfill \pagebreak
 \dominitoc

\tableofcontents
\newpage \thispagestyle{empty}

 %\listoffigures
%\newpage \thispagestyle{empty}

 %\listoftables
%\newpage \thispagestyle{empty}

%\newpage
%\chapter*{Résumé}

 %\clearemptydoublepage

%\newpage
%\chapter*{Abstract}

 %\clearemptydoublepage

%\chapter*{{\it Introduction}}
%\addstarredchapter{Introduction}
%\markboth{\uppercase{Introduction}}{\uppercase{Introduction}}

 %\clearemptydoublepage

%%%%%%%%%%%%%%%%%%%
%%%%%%%CHAPTER I
%%%%%%%%%%%%%%%%%%%

\chapter{Introduction}

Depuis l'aube de l'humanité, nous cherchons à comprendre et à caractériser notre univers. La question de la détermination des distances dans l'univers a toujours été centrale, et a plusieurs fois amené à des changements de paradigmes. 

\section{Un peu de cosmogonie: Notre vision de l'univers au XXIe siècle}\label{Sect_cosmogonie}

La vision de l'univers à l'époque de la Grèce antique se résumait à une Terre plate, placée au centre, un monde supra-lunaire parfait, appartenant au domaine des dieux, comprenant le Soleil, les 5 planètes (Mercure, Vénus, Mars, Jupiter, Saturne) et, au delà, la ‘‘sphère des fixes'' sur laquelle se trouvaient les étoiles. En 300 avant JC, Aristote observe les éclipses de Lune et remarque la courbure de l'ombre de la Terre sur la Lune, ce qui donne la première preuve que la Terre est ronde. Magellan l'expérimentera presque 2 millénaires plus tard. A partir de la taille de l'ombre, Aristarque de Samos établira que la Terre a un diamètre trois fois plus important que la Lune et que sa distance à la lune est de 64 rayons terrestres. A la même époque, Eratosthène utilise les positions relatives des ombres à Syène (aujourd'hui Assouan) et à Alexandrie pour en déduire la circonférence de la Terre. Alors qu'il n'y pas d'ombre dans les puits de Syène, les gnomons indiquent un angle de 7.2° à Alexandrie, soit $1/50$ d'un tour complet (360°). La distance entre Syène et Alexandrie étant de 5000 stades, la circonférence de la Terre est 50 fois plus importante, soit environ 250000 stades ($\simeq$ 40000 km).  Cette mesure permet à Aristarque de Samos d'en déduire la distance de la Terre à la Lune. Ce dernier, par une mesure d'angle liée au premier quartier de Lune, estime alors pour la première fois la distance entre la Terre et le Soleil. Malgré Héraclite qui propose que la Terre tourne sur elle-même, et Pythéas, le Marseillais, qui mesure son inclinaison, notre planète semble indiscutablement immobile sous nos pieds et finalement le modèle géocentrique de Ptolémée (IIe siècle avant JC) s'impose avec la Terre immobile et ronde au centre de l'univers et le Soleil et les planètes autour. Ce modèle, complexe, qui nécessite des épicycles pour reproduire le mouvement des planètes, va perdurer pendant 1700 ans. 

En 1543, Copernic revient à un modèle héliocentrique (car Aristarque de Samos l'avait déjà proposé bien avant lui), mais celui ci ne reproduit toujours pas les observations et des épicycles restent nécessaires. Le changement de paradigme va finalement trouver son  origine dans les observations minutieuses (à 1 minute de précision) de Tycho Brahé à Uraniborg sur l'île de Ven, dans le détroit du Sund (cette île aujourd'hui suédoise appartenait à l'époque au Danemark). Tycho Brahé est contre le modèle héliocentrique et ce n'est qu'à sa mort que son disciple, Kepler, va étudier en détail les données pour en déduire le mouvement elliptique des planètes. Grâce à ces ellipses, le modèle héliocentrique reproduit alors bien mieux les observations. Ainsi,  la troisième loi de Kepler permet de déduire la distance au Soleil de toutes les planètes (à partir de la distance Terre-Soleil). En 1610, Galilée observe les phases de Vénus, ce qui valide indubitablement le modèle héliocentrique. Il montre également que le monde supra-lunaire n'est pas parfait; il observe effectivement les cratères sur la Lune, les taches du Soleil, les anneaux de Saturne (même s'il ne sait pas encore qu'il s'agit d'anneaux). Il découvre également les 4 plus gros satellites de Jupiter, illustrant le fait qu'un astre peut tourner autour d'une autre planète que la Terre. Mais une question demeure: pourquoi ne sent-on pas la Terre bouger sous nos pieds ? Galilée pose alors les principes de ce qui deviendra plus tard le "référentiel Galiléen": un individu enfermé dans la cale d'un bateau naviguant sur une mer calme, ne peut faire aucune expérience pour prouver que le bateau est en mouvement rectiligne uniforme. Notre vision du système solaire est alors complétée par la découverte d'Uranus (Herschel; 1778) et, par le calcul d'abord, de Neptune (Le Verrier; 1846). Mais une chose demeure: la sphère des fixes. 

Il y a 3000 ans, les Egyptiens avaient tout de même repéré une étoile variable sur la sphère des fixes, Algol. Un objet dont l'éclat varie dans le ciel parfait supra-lunaire ne pouvait être pour les Grecs que l'oeil de la Méduse, qui deviendra ensuite chez les Arabes {\it Ras El Ghul}, c'est-à-dire la tête du démon. Bien plus tard, en 1596, une autre variable est découverte, Mira Ceti, par Fabricius, dont le nom même signifiant "La merveilleuse" montre bien le changement de vision par rapport aux étoiles et à la sphère des fixes. En 1784, Goodrick découvre la première Céphéide, $\delta$ Cep. C'est également lui qui proposera le concept de "binaire à éclipses" pour expliquer la variabilité d'Algol. A son époque, Tycho Brahé tenta de mesurer la parallaxe d'une nova dans la constellation de Cassiopée en 1572, mais la tentative fut infructueuse. C'est bien \cite{bessel1838} qui mesure pour la première fois la parallaxe annuelle d'une étoile, 61 Cyg ($\pi = 0.314$) et en déduit une distance correspondant à 3.18 parsecs. La sphère des fixes est donc bien plus éloignée que ce que l'on imaginait alors. 
%En 1910, Hertzsprung comprend que la couleur des étoiles est reliée à leur température (loi de Wien, Prix Nobel de Physique de 1911), ce qui conduira  au diagramme Hertzsprung-Russell. 
Un nouveau changement de paradigme se prépare alors avec les théories de la relativité restreinte \citep{Einstein1905} et générale \citep{Einstein1916}. Ces théories permettent d'abord de comprendre l'avance du périhélie de Mercure (43 seconde d'arc par siècle de plus que la valeur prédite par la théorie Newtonienne). Mais elles seront également très utiles pour  décrire l'univers dans son ensemble, comme nous le verrons plus loin. En 1908, Henrietta Leavitt \citep{leavitt08} dénombre 1777 Céphéides dans les nuages de Magellan,  des étoiles supergéantes jaunes très brillantes. Puis en 1912, \citet{leavitt1912} étudient 25 d'entre elles dans le petit nuage de Magellan et découvrent alors une propriété très particulière: leur période de pulsation (typiquement de quelques jours à quelques mois) est reliée à la moyenne de leur luminosité intrinsèque (relation \emph{PL}), ce qui en fait des indicateurs de distance uniques. Nous reviendrons sur les propriétés qui font que les Céphéides pulsent. En déterminant la distance de ces étoiles dans les amas globulaires, \citet{shapley1918}  montre que le Soleil ne se trouve pas au centre de la Voie Lactée mais sur le bord (du fait de la distribution non isotrope des amas). A cette époque, il y a donc environ un siècle seulement, un débat fait rage entre les partisans d'un univers qui se réduit à la Voie Lactée  \citep{shapley1919} et ceux qui pensent qu'il est bien plus grand et constitué d'"univers îles" (i.e. de galaxies) comme l'avait imaginé le philosophe Kant un peu plus tôt \citep{curtis1920}. De 1926 à 1929, Hubble détermine ainsi la distance de la galaxie du triangle \citep{Hubble26} et de la galaxie d'Andromède \citep{Hubble29a}, que l'on pense être alors simplement des nébuleuses spirales (classées M33 et M31 par Messier), et trouve une distance bien au delà des limites de la Voie Lactée. Hubble détermine également la distance de la galaxie de Barnard (NGC 6822) \citep{Hubble25}. On comprend alors que ces ‘‘nébuleuses spirales'' sont des "univers îles", ce qui met fin au débat Shapley-Curtis. Hubble trouvera néanmoins une distance pour ces galaxies deux fois trop petite, du fait d'une confusion entre les Céphéides de type I (qu'il a effectivement observées dans M31 et M33) et les Céphéides de type II (W Virginis, moins métalliques et moins brillantes de 1.5 magnitude environ), et sur lesquelles repose effectivement l'étalonnage de la relation \emph{PL} effectué par Shapley.  Hubble mesure ainsi la distance de 46 galaxies. Un peu plus tôt, dès 1914, Vesto Slipher avait observé ces "nébuleuses spirales" à l'aide d'un spectroscope et avait trouvé un décalage vers le rouge des raies spectrales pour 11 nébuleuses parmi les 15 observées \citep{Slipher1914, Slipher1917}. Par définition, si une raie spectrale de longueur d'onde d'émission $\lambda_\mathrm{0}$ est observée à une longueur d'onde de $\lambda$, alors le décalage vers le rouge (ou {\it redshift}) est: 

\begin{equation}
z = \frac{\lambda}{\lambda_\mathrm{0}}-1
\end{equation}

Pour les objets proches (< 2 milliards d'années-lumière, < 600 Mpc ou encore $z<0.15$), le décalage vers le rouge peut être assimilé à une vitesse de récession dont la formule est simplement donnée par l'effet Doppler $v=cz$. Survient alors une découverte incroyable dont la paternité fait toujours débat \footnote{Lire http://www.nature.com/news/2011/110627/full/news.2011.385.html et l'introduction de "L'invention du big bang", Jean-Pierre Luminet (Le Seuil, 1997).}. En 1927, Georges Lemaître rédige un article en Français dans la revue {\it Annales de la Société scientifique de Bruxelles}, établissant que l'univers est en expansion \citep{lemaitre27}. Il montre en effet que plus une galaxie est  éloignée de nous, plus elle s'éloigne rapidement. \cite{hubble29b} établit alors à son tour une valeur de la pente de cette relation, ou {\it constante de Hubble}, $H_\mathrm{0}$, où l'indice 0 indique que c'est l'expansion actuelle de l'univers.  Si l'univers accélère ou décélère  avec le temps, $H$ peut varier selon l'époque de l'univers considérée. Commence alors l'ère de la cosmologie.

\section{Un peu de cosmologie}

L'expansion de l'univers découle naturellement de la relativité générale couplée avec l'hypothèse d'un univers homogène (c'est-à-dire ayant la même densité de matière en tout point, à un instant donné) et isotrope (identique dans toutes les directions). Pour décrire les équations de la dynamique de l'univers, nous devons également supposer que le contenu de l'univers se comporte comme un fluide parfait, c'est-à-dire dont toutes les propriétés intéressantes (du point de vue de la dynamique cosmique) sont décrites par sa densité de masse-énergie $\rho$ et sa pression moyenne $p$. 
Dans ce cas, le tenseur d'énergie-impulsion prend une forme simple et les équations d'Einstein se réduisent à un ensemble de deux équations différentielles, les équations de Friedmann:

\begin{equation}\label{Eq_Fri1}
\dot{a}^2 - \frac{1}{3} ( 8 \pi G \rho + \Lambda ) a^2 = - k
\end{equation}

\begin{equation}\label{Eq_Fri2}
\frac{\ddot{a}}{a} = - \frac{4}{3} \pi G (\rho + 3 p/c^2) + \frac{1}{3} \Lambda 
\end{equation}

Dans ces équations, $a=a(t)$ est le facteur d'échelle de l'univers. Celui-ci est fondamentalement relié au décalage vers le rouge (mesuré par exemple par la position d'une raie en émission dans une galaxie) du fait que la quantité $z+1$ est égale au rapport entre la taille de l'univers maintenant ($a_\mathrm{0}$) et la taille de l'univers au moment de l'émission de la lumière ($a$). $\Lambda$ est la constante cosmologique. Elle correspond à une répulsion universelle et a été introduite à l'origine par Einstein pour forcer l'univers à être statique, en accord avec la vision de Newton qui imaginait un univers homogène, statique et infini. A la découverte de l'expansion de l'univers par Hubble, la constante a été enlevée, pour ensuite être ré-introduite avec la découverte de l'expansion accélérée. $k$ est le terme de courbure de l'univers et vaut -1, 0 ou 1 selon que l'univers est courbé négativement, spatialement plat, ou courbé positivement \footnote{$k$ est ici le terme de courbure de l'univers à ne pas confondre avec le $k$-facteur des Céphéides discuté plus loin}. Afin de résoudre ces équations, nous devons introduire une relation entre la pression et la densité: 

\begin{equation}\label{Eq_Etat}
p=\omega \rho
\end{equation}

pour chacun des composants de l'univers, c'est-à-dire la matière et le rayonnement essentiellement. Ces composantes de densité varient avec le facteur d'échelle $a$ lorsque l'univers est en expansion et donc varient avec le temps. Au temps t, on peut définir le paramètre de Hubble:

\begin{equation}\label{Eq_H}
H(t) = \frac{\dot{a}}{a}
\end{equation}

et la constante de Hubbleqw $H_\mathrm{0}$ est la valeur du paramètre de Hubble prise à l'instant $t=0$. Si $\Lambda = 0$, on peut décrire la cinématique de l'univers simplement en utilisant la première équation de Friedman (Eq.~\ref{Eq_Fri1}). Si on ajoute l'hypothèse que l'univers est plat ($k=0$), alors on obtient:

\begin{equation}\label{Eq_rhoc}
\rho = \rho_\mathrm{c} = \frac{3 H^2}{8\pi G} 
\end{equation}
	
où $\rho_\mathrm{c}$ est la densité critique. Ainsi, plusieurs cas peuvent se présenter. Si l'univers a une densité $\rho$ < $\rho_\mathrm{c}$, alors $ k< 0$ et l'univers est courbé négativement, ce qui implique $\dot{a}>0$ et donc une expansion éternelle. Si à l'inverse $\rho$ > $\rho_\mathrm{c}$, alors $k~>~0$ et l'univers est courbé positivement, ce qui implique qu'il existe un instant $t$ pour lequel $\dot{a}=0$. A ce point, l'expansion va s'arrêter et l'univers va commencer à se contracter. Si la constante cosmologique $\Lambda$ est positive, alors l'univers est quasiment contraint de s'expandre indéfiniment, à moins que la densité de matière soit bien plus grande que $\Lambda$ et dans ce cas l'univers peut se recontracter avant que l'expansion domine. 

On peut introduire des quantités sans dimensions pour décrire la densité d'énergie associée à la constante cosmologique $\Omega_\mathrm{\Lambda}=\Lambda /3H_\mathrm{0}^2$ et la courbure de l'espace temps,  $\Omega_\mathrm{k}= k /H_\mathrm{0}^2$. En réarrangeant l'Eq.~\ref{Eq_Fri1}, on obtient alors:

\begin{equation}\label{Eq_Friv2}
\frac{H^2}{H_\mathrm{0}^2} = \frac{\rho}{\rho_\mathrm{c}} - \Omega_\mathrm{k} a^{-2} + \Omega_\Lambda 
\end{equation}

La densité pour une composante particulière de l'univers (matière ou rayonnement) X peut être exprimée comme une fraction de la densité critique sous la forme:

\begin{equation}\label{Eq_Friv2}
\frac{\rho_\mathrm{X}}{\rho_\mathrm{c}} = \Omega_\mathrm{X} a^{\alpha} 
\end{equation}

où l'exposant $\alpha$ représente la dilution de la composante avec l'expansion de l'univers et est reliée au paramètre $\omega$ (Eq.~\ref{Eq_Etat}) par l'équation $\alpha = -3 (1+ \omega)$. L'équation \ref{Eq_Friv2} est valable seulement si $\omega$ est constant. Ainsi, pour la matière ordinaire, $\alpha=-3$: la densité diminue comme $\frac{1}{R^3}$ lorsque l'univers se dilate, ce qui est intuitif. Pour le rayonnement, on a $\alpha=-4$, car en plus de la dilatation géométrique, il y a une perte d'énergie liée au décalage vers le rouge au fur et à mesure que l'univers se dilate. Ce décalage vers le rouge n'est pas lié à l'effet Doppler (mais assimilable à un effet Doppler en deçà de $z=0.15$) mais bien à un effet relativiste gravitationnel. Enfin, la densité d'énergie associée à la constante cosmologique reste la même quelle que soit la taille de l'univers, ainsi on  $\alpha=0$ et $\omega=-1$. Une solution avec $\omega < -\frac{1}{3}$ (appelée quintessence)  peut également produire de l'expansion. Par ailleurs, il n'y a aucune raison pour que $\omega$ reste constant avec $z$, et les observations futures pourront probablement contraindre des modèles du type $\omega=\omega_\mathrm{0} + \omega_\mathrm{1} z$. Le terme d'énergie noire regroupe tous ces types de modèles. 

Finalement, on obtient la variation du {\it paramètre} de Hubble en fonction de la {\it constante} de Hubble:

\begin{equation}\label{Eq_Ht}
H^2 = H_\mathrm{0}^2 (\Omega_\Lambda + \Omega_\mathrm{m} a^{-3} + \Omega_\mathrm{r}a^{-4} - \Omega_\mathrm{k}a^{-2})
\end{equation}

où $\Omega_\mathrm{r}$ est la densité d'énergie de radiation et $\Omega_\mathrm{m}$ est la densité d'énergie de matière. Cette équation indique que le rayonnement dominait en terme d'énergie lorsque l'univers était de petite taille (a petit). De son côté, $\Omega_\Lambda$ implique une force répulsive responsable de l'expansion de l'univers comme nous l'avons vu, {\it mais elle correspond également à une densité d'énergie qui affecte la courbure de l'univers}. Ainsi, on a:

\begin{equation}\label{Eq_somme}
\Omega_\Lambda + \Omega_\mathrm{m} + \Omega_\mathrm{r} - \Omega_\mathrm{k} = 1
\end{equation}

Dans cette équation, si $\Omega_\mathrm{k}=0$, l'univers est plat, et la somme totale de l'énergie de densité de la matière et du rayonnement ajoutée à la densité d'énergie liée à la constante cosmologique doit correspondre à la densité critique dont la formulation est alors plus complexe que l'Eq.~\ref{Eq_rhoc}. Il est à noter que les univers proches de $\Omega_\mathrm{k}=0$ ont tendance à s'éloigner de ce point d'équilibre, ce qui pose le problème de la ''platitude de l'univers''. La période d'inflation dans l'histoire de l'univers permet de résoudre ou disons contourner ce problème. L'univers observable, plat ou quasi-plat, n'est qu'une partie d'un univers plus grand courbé ou pas, fini sans bord ou infini, un peu comme l'horizon sur Terre. Ces quelques rappels de cosmologie nous ont permis de replacer la constante de Hubble dans son contexte, et nous allons voir maintenant comment cette quantité peut être mesurée concrètement de nos jours. 

%%%%%%%SECTION 
\chapter{Les différents moyens de déterminer la constante de Hubble (Ho)}

En moins d'un siècle, nous sommes donc passés d'un univers statique de la taille d'une galaxie à la vision d'un univers observable en expansion, d'un âge d'environ 13.6 milliards d'années et contenant 100 milliards de galaxies. Pour mesurer l'expansion de l'univers (i.e. $H_\mathrm{0}$), il faut une vitesse d'éloignement et une distance. Mesurer des vitesses d'éloignement est relativement aisé: il suffit d'observer avec un spectrographe un objet présentant une raie d'émission. En revanche, la question des distances est extrêmement délicate. Il faut soit une chandelle standard (i.e. un objet dont la luminosité est connue), soit une règle standard (i.e. un objet dont la longueur est connue), et ensuite nous pouvons utiliser la magnitude apparente ou le diamètre angulaire pour estimer la distance. Ce qui veut dire que comprendre la physique de ces objets est incontournable si l'on cherche à estimer sans équivoque leur luminosité intrinsèque ou leur taille réelle. Hubble trouva une valeur autour de 500 km/s/Mpc, soit environ dix fois plus grande que les estimations actuelles. Cette valeur initialement trop élevée et les révisions progressives vers de plus basses valeurs ({\it cf} \citet{trimble96, tammann06})  s'expliquent d'abord par un biais dans les échantillons de départ \citep{behr51}, puis par une confusion entre des étoiles brillantes et certaines régions HII \citep{humason56, sandage58} et enfin, entre les Céphéides de type I et II \citep{baade56}. 

Par ailleurs, pour mesurer $H_\mathrm{0}$, il faut que les objets en question soient suffisamment loins pour que leur vitesse soit essentiellement due à l'expansion générale de l'univers (ce que l'on appelle dans la littérature, ‘‘The Hubble flow''). On sait par exemple que le système solaire se déplace à 390 \kms vers la constellation du Lion, tandis que la Voie Lactée dans son ensemble, se déplace à 600 \kms dans la direction de l'amas du Centaure (lieu connu sous le nom de grand attracteur ou concentration de Shapley). Ainsi, il faut sortir du Superamas de l'Hydre et du Centaure pour s'affranchir des vitesses liées au potentiel gravitationnel local et donc atteindre quelques dizaines de Mpc, soit autour de $z \simeq 0.01$. Pour les distances autour de $z=0.15$, la relation entre le décalage vers le rouge $z$ et la luminosité (ou le diamètre angulaire) n'est plus linéaire et dépend de la densité de matière $\Omega_\mathrm{m}$ et d'énergie noire $\Omega_\mathrm{\Lambda}$, et aussi de la constante de Hubble elle-même. D'ailleurs, on définit en cosmologie des distances particulières. Les deux plus importantes sont la distance de diamètre angulaire $D_\mathrm{A}$ qui fait le lien entre le diamètre angulaire apparent et la taille réelle d'un objet (comme une galaxie), et la distance de luminosité $D_\mathrm{L}=(1+z)^2D_\mathrm{A}$, qui relie le flux observé d'un objet à sa luminosité intrinsèque. 

\begin{figure}[htbp]
\begin{center}
%\resizebox{1.0\hsize}{!}{\includegraphics[clip=true]{f0.eps}}
\resizebox{1.0\hsize}{!}{\includegraphics[clip=true]{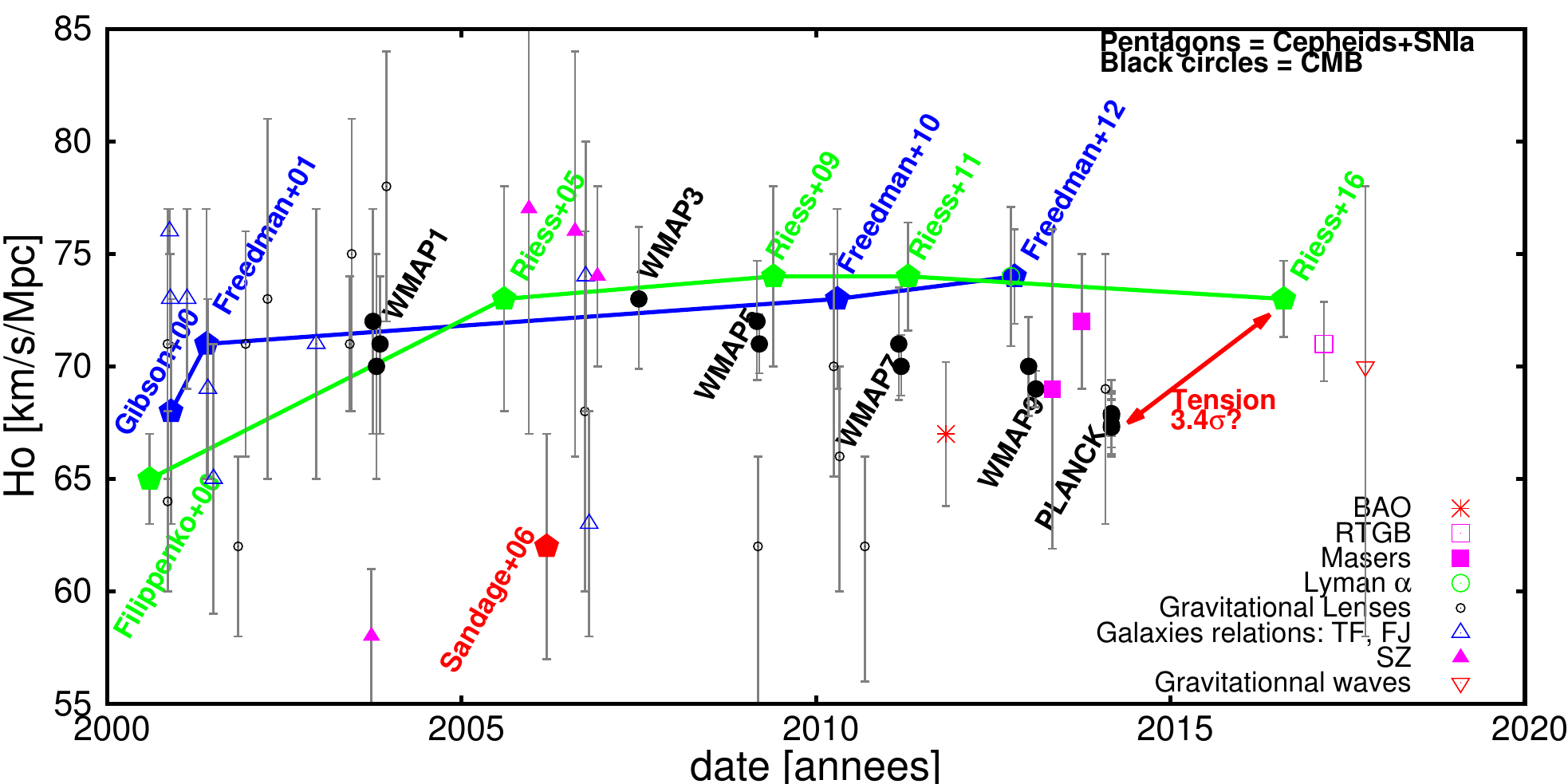}}
\end{center}
\caption{Evolution de la détermination de la constante de Hubble depuis l'an 2000. Les différentes méthodes sont indiquées. Les deux groupes qui utilisent les Céphéides et les supernova de type Ia ont des résultats compatibles: d'une part le  {\it Carnegie Hubble Project} de Wendy Freedman (en bleu) et d'autre part, le projet SHOES d'Adam Riess (en vert). Seule la valeur de \cite{sandage06}, autour de 61,km/s/Mpc, est significativement plus faible. Cette incohérence est discutée dans la revue de \cite{jackson15}, et est attribuée à la façon de gérer l'impact de la métallicité sur la relation \emph{PL} des Céphéides (voir Sect.~\ref{sect_plz}). Les valeurs issues des observations du rayonnement de fond cosmologique (WMAP 1, 3, 5, 7, 9 et Planck) diminuent progressivement au cours des études et des analyses pour finalement se situer autour de 3.4$\sigma$ en dessous des valeurs basées sur les SN Ia. La littérature scientifique se réfère à cette incohérence sous le terme de {\it tension}. Les autres méthodes qui permettent d'étalonner la relation entre z et la luminosité des SNIa à savoir les masers et les relations liées aux Galaxies, Tully-Fisher (TF) et Faber-Jackson (FB), ne sont pas encore assez précises et exactes pour espérer lever la tension. Seule la valeur récente liée au {\it Red Tip of Giants Branch} RTGB donne une précision de 2\% compatible avec les projets CHP et SHOES. On note également que les méthodes de détermination de distance à 1 étage: {\it Baryonic Acoutic Oscillatio} (BAO), Lyman $\alpha$, les lentilles gravitationnelles et la méthode de Sunyaev-Zeldovitch (SZ) présentent des incertitudes qui ne permettent pas actuellement de résoudre la tension et il en est de même de la méthode récente basée sur les ondes gravitationnelles.}
%Positive velocities correspond to a redshift or receding motion.} 
\label{Fig_Ho}
\end{figure}

Malheureusement, il n'existe pas d'objet dont on connaisse la luminosité (ou la taille) en une seule étape, sans ambiguité, et pouvant être observé jusqu'à quelques dizaines de Mpc. L'approche utilisée est donc de déterminer la distance d'objets proches pour ensuite étalonner la luminosité des objets plus lointains. Le processus peut se répéter plusieurs fois et on obtient finalement ce que l'on pourrait appeler un véritable échafaudage des échelles de distances dans l'univers, tel que celui représenté sur la Figure~\ref{Fig_distances}. Chaque avancée dans ce domaine a permis de comprendre la physique des objets utilisés. Le contenu astrophysique des méthodes utilisées est clairement un désavantage pour l'estimation de la constante de Hubble, mais il est incontournable. Le nombre d'étapes utilisées l'est également car chacune d'entre elle apporte son lot d'incertitudes statistiques et systématiques. Ainsi, si l'on regarde les valeurs de la constante de Hubble obtenues depuis l'an 2000 par les différentes méthodes que nous allons maintenant décrire, on obtient un intervalle de valeur entre typiquement 60 et 75 km/s/Mpc (Fig.~\ref{Fig_Ho}).

%%%%%%%SECTION 
\section{L'étalonnage des échelles de distance dans l'univers}\label{Sect_d}

Si on considère d'abord l'approche liée à l'échafaudage des distances, on constate que le chemin le plus court pour atteindre la constante de Hubble nécessite au minimum deux objets astrophysiques: les Céphéides et les supernova de type Ia. En effet, comme nous l'avons déjà mentionné dans la Sect.~\ref{Sect_cosmogonie}, les Céphéides possèdent la propriété particulière que leur période de pulsation est reliée à leur luminosité moyenne intrinsèque. Cependant, pour utiliser la relation période-luminosité (\emph{PL}), il faut l'étalonner, ce qui constitue véritablement la base de l'``échafaudage'' des échelles de distance dans l'Univers. La relation a été étalonnée pour la première fois par \citet{hertzsprung13} en utilisant des parallaxes statistiques, mais le moyen le plus direct est d'utiliser la méthode de la parallaxe trigonométrique: du fait du mouvement de la Terre autour du Soleil, les Céphéides proches se déplacent de manière apparente par rapport aux étoiles lointaines, ce qui donne leur distance. La combinaison de la distance et de la magnitude apparente fournit la magnitude absolue (module de distance) et donc la luminosité intrinsèque. Il est alors aisé de combiner la période et la luminosité pour en déduire la relation \emph{PL}. Cependant, actuellement (i.e. fin 2017), il n'existe que 7 Céphéides pour lesquelles nous avons une mesure de parallaxe à mieux que 10\% \citep{benedict07}. Aussi, une autre approche consiste à mesurer la distance du LMC (ou SMC) par une méthode indépendante (par exemple les binaires à éclipses) pour en déduire la relation \emph{PL}. Le LMC se trouve à environ 50 kpc, soit 2 à 3 ordres de grandeur plus proche que les galaxies intéressantes pour mesurer la constante de Hubble, qui sont effectivement situées entre 5 et 40 Mpc. Les Céphéides, avec les géantes rouges, sont les seuls objets observables à la fois dans le LMC et dans ces galaxies. L'arrivée du {\it Hubble Space Telescope} (HST) a été très importante pour que ceci soit possible. Les instruments futurs comme l'E-ELT le seront également. On ne saurait donc trop souligner l'importance des Céphéides pour l'étalonnage des échelles de distance, car sans ces étoiles, il serait très difficile de faire le lien entre la Voie Lactée, le LMC et les galaxies externes. Mais même l'observation des Céphéides avec le HST ne permet pas d'atteindre des distances suffisantes pour établir la constante de Hubble directement (i.e. hors du ‘‘Hubble flow''). La dernière étape consiste donc à utiliser des objets astrophysiques encore plus brillants, c'est-à-dire les supernovae de type Ia (SNIa). Il s'agit de binaires dont l'une des composantes (une géante rouge) transfère de la masse vers une naine blanche, une étoile dont la stabilité provient de la dégénérescence des électrons. Lorsque la masse de la naine blanche dépasse la masse limite de Chandrasekhar, celle-ci explose. Bien que la magnitude absolue de l'explosion ne soit pas tout à fait constante, les SNIa ont des courbes de lumière similaires \citep{pskovskii67,barbon73, Tammann82}, et il existe en particulier une très bonne corrélation entre le pic de brillance et la chute de luminosité 15 jours après le pic, une quantité que l'on note $\Delta m_\mathrm{15}$ \citep{phillips93,hamuy96}. Le point clef pour contraindre la constante de Hubble consiste donc à observer jusqu'à 40 Mpc (i.e z ~ 0.01) avec le Hubble Space Telescope des galaxies qui contiennent à la fois des Céphéides et des SNIa. On dispose à ce jour d'une vingtaine de galaxies qui répondent à ce critère. Ces SNIa permettent alors de fixer le point zéro de la relation entre z et la distance de luminosité des SNIa, ces dernières étant observables jusqu'à environ $z=6$. 

Cette approche est ainsi utilisée par deux groupes indépendants, le groupe du ‘‘Canergie Hubble Program'' (CHP\footnote{\url{http://chp.obs.carnegiescience.edu/wiki/Main_Page}}) mené par Wendy Freedman, dont les valeurs de $H_\mathrm{0}$ sont indiquées en bleu sur la Fig.~\ref{Fig_Ho} et le groupe ‘‘Supernova $H_\mathrm{0}$ for the Equation of State'' (SHOES) dirigé par Adam Riess, le co-récipiendaire du Prix Nobel de physique 2011 (en vert sur la Fig.~\ref{Fig_Ho}).  Les deux groupes s'accordent (dans les barres d'erreur) sur la constante de Hubble. 
%La valeur de \cite{sandage06}, probablement due à des effets de métallicité, sera discuté plus tard. 
Les méthodes sont assez similaires et pas totalement déconnectées: 

\begin{enumerate}
\item \cite{freedman12} ont utilisé des observations uniques du satellite {\it Spitzer} à 3.6 $\mu m$ et les distances de 10 Céphéides Galactiques obtenues par des mesures de parallaxes trigonométriques avec le {\it Hubble Space Telescope} \citep{benedict07} pour contraindre le point-zéro de la relation \emph{PL}. La pente de la relation  \emph{PL} fut quant à elle déduite de l'observation  {\it Spitzer}  de 90 Céphéides dans le LMC. Cette relation \emph{PL} fut ensuite utilisée pour étalonner la relation des supernovae Ia obtenue par \citet{riess11}. 
\item \cite{riess16} ont utilisé 4 indicateurs de distances géométriques pour étalonner la relation PL des céphéides: (1) les mégamasers de NGC~4258, (2) 8 binaires à éclipses du Grand Nuage de Magellan (LMC), (3) 15 parallaxes HST de Céphéides Galactiques\footnote{Ces 15 parallaxes correspondent aux dix de \citet{benedict07} dont la précision s'échelonne entre 5\% et 12\%, trois d'Hipparcos \citep{esa97,vanleeuwen07} et deux autres observées par l'équipe SHOES en utilisant un nouveau mode d'observation sur le HST \citep{casertano16}.} et enfin (4) 2 binaires à éclipses situées dans la galaxie M31 \footnote{Il est intéressant de noter que ces estimations de distance n'utilisent pas la version purement observationnelle de la méthode de détermination de distance des binaires à éclipses (comme c'est le cas pour le LMC), mais sont basées sur des modèles stellaires. Nous reviendrons sur ce point dans la Sect.~\ref{Chap_EBs}}. Leur meilleure estimation de la constante de Hubble est ainsi de $H_0=73.24\pm 1.74$ km s$^{-1}$ Mpc$^{-1}$, ce qui donne une incertitude de 2.4\% (en incluant les incertitudes statistiques et systématiques). 
\end{enumerate}

\begin{figure}[htbp]
\begin{center}
\includegraphics[width=14cm]{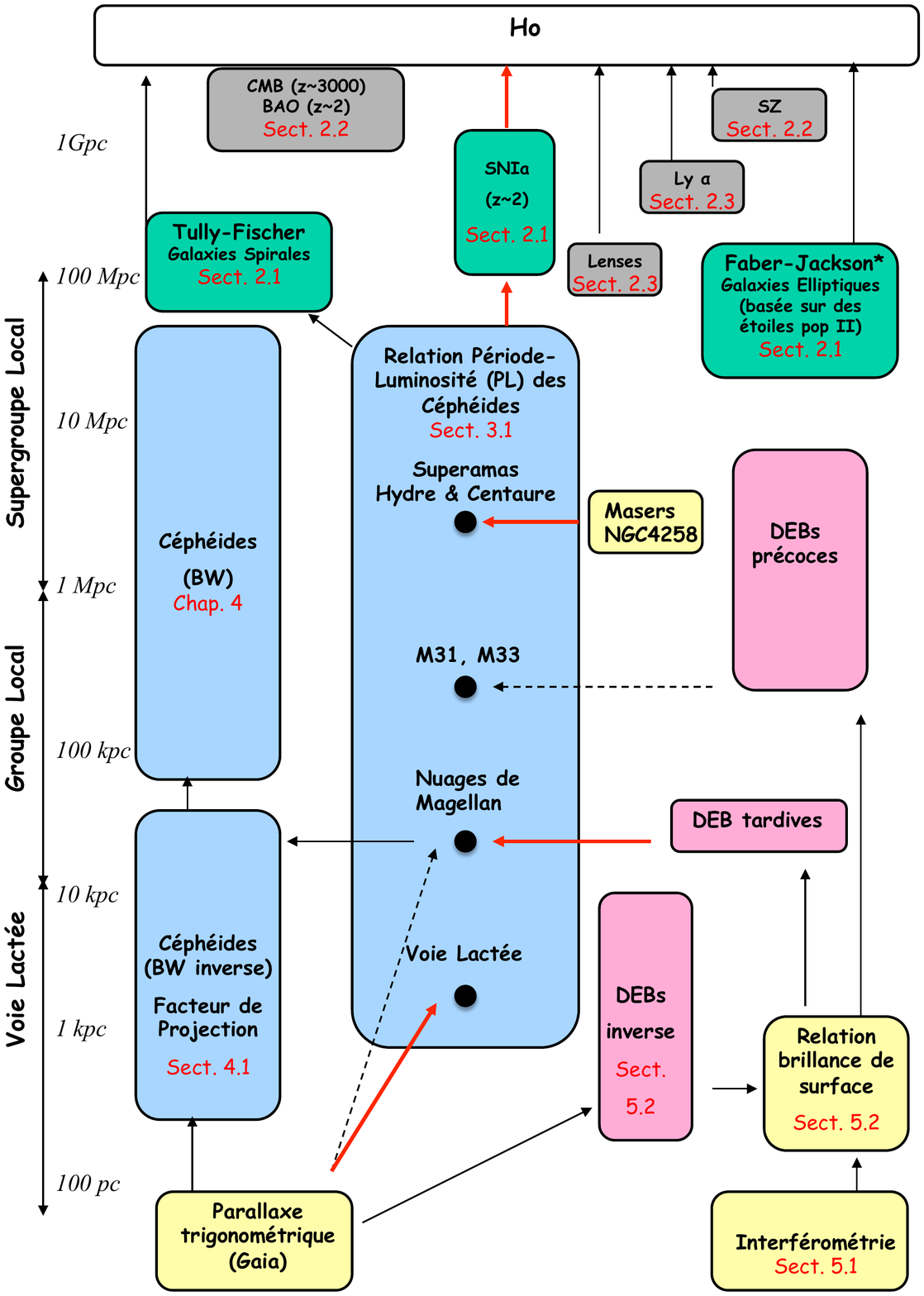}
\end{center}
\vspace*{-10mm} \caption{ \footnotesize   Organigramme des méthodes de détermination de distances dans l'univers. Les flèches rouges indiquent le ‘‘chemin'' suivi par \citet{riess16} pour contraindre $H_\mathrm{0}$: les parallaxes trigonométriques des céphéides Galactiques, la distance du LMC déduite des binaires à éclipses détachées tardives (DEB) et la distance au maser de NGC 4258. Ces distances de référence permettent de contraindre la relation $PL$ des Céphéides et d'étalonner la relation des supernova de type Ia. Les différentes parties de ce diagramme sont explicitées dans les sections indiquées.}\label{Fig_Organigramme}
\end{figure}

Dans cette démarche, \cite{riess16} utilisent un mégamaser dans  NGC~4258 pour étalonner la relation \emph{PL}, la seule alternative au LMC \citep{chaussen84}. Cette galaxie présente une coquille de masers qui sont orientés quasiment selon la tranche (‘‘edge-on'') \citep{miyoshi95,greenhill95}  et apparemment en rotation circulaire Képlerienne. La mesure des vitesses des masers individuels et de leur accélération permet de calculer la taille réelle de la coquille de masers, et donc la distance de la galaxie à 3\%, soit une précision légèrement moins bonne que la précision actuellement obtenue sur le LMC (environ 2\%). La méthode a subi quelques améliorations avant d'atteindre cette précision \citep{Herrnstein99,Humphreys05,argon07}. Néanmoins, des études indiquent que la distance à NGC~4258 pourrait être biaisée par un problème de dégénérescence dans la procédure d'ajustement des paramètres du maser \citep{reid13}. Par ailleurs, les valeurs de $H_\mathrm{0}$ issues uniquement de cette méthode sont moins précises que celles obtenues dans le cadre des projets CHP ou de SHOES (carrés magenta sur la Fig.~\ref{Fig_Ho}). Ainsi, le LMC (150 fois plus proche que NGC~4258) reste une référence pour construire l'échafaudage des échelles de distances dans l'univers. L'approche de \cite{riess16} basée sur ces 4 objets astrophysiques est représentée sur la Fig.~\ref{Fig_Organigramme} par des flèches rouges. Dans la suite, nous aborderons différentes façons d'améliorer ce ‘‘chemin'' vers la constante de Hubble. 

Il existe néanmoins une alternative aux SNIa pour la détermination de $H_\mathrm{0}$. Toutes ces méthodes reposent sur une corrélation entre une propriété facilement observable de ces galaxies et leur luminosité.  Ainsi, la vitesse de rotation des galaxies spirales vues par la tranche $v$ est proportionnelle à la luminosité de la galaxie ($L \propto v^4$). En effet, le nombre d'étoiles contenues dans la galaxie détermine à la fois sa masse et donc sa vitesse de rotation, mais aussi son éclat. Il s'agit de la relation de Tully-Fisher \citep{tully77}. Cette relation nécessite néanmoins d'être étalonnée, notamment par les Céphéides. Le même type de relation peut être établie pour les galaxies elliptiques; il s'agit alors de la relation de Faber-Jackson \citep{faber76}. Les galaxies elliptiques étant plus âgées dans l'histoire de l'univers, elles sont aussi moins métalliques et ne contiennent en général pas ou peu de Céphéides. Il faut alors utiliser des étoiles pulsantes de type II pour étalonner leur relation (RR Lyrae, W Vir), mais ces étoiles étant moins brillantes que les Céphéides, la méthode reste moins favorable. Enfin, il est possible de trouver des combinaisons plus complexes, comme le paramètre $D_\mathrm{n}$ \citep{dressler87,lynden88} que l'on peut lier à la luminosité, ou la méthode type ‘‘fundamental plane'' \citep{dressler87,Djorgovski87} qui relie quant à elle trois propriétés: la luminosité à l'intérieur d'un rayon effectif (en deçà duquel la moitié de la luminosité de la galaxie est émise), le rayon effectif et la dispersion de vitesse des étoiles. Enfin, le degré de résolution des étoiles dans une galaxie dépend de sa distance. Une dernière méthode repose donc sur la distribution de brillance de la galaxie \citep{tonry88}. Les valeurs de $H_\mathrm{0}$ déduites des méthodes Tully-Fischer et Faber Jackson sont indiquées sur la Figure~\ref{Fig_Ho} par les triangles bleus. La précision et la dispersion des mesures laissent penser que ces approches ne sont pas au même point de maturité que la méthode SNIa. La conclusion est la même si l'on considère les autres méthodes liées aux propriétés des galaxies.

%\vspace*{-10mm}

\section{Les méthodes liées au rayonnement de fond cosmologique}\label{Sect_cmb}

Les indicateurs de distance sont des objets astrophysiques complexes, résultats d'une longue évolution de l'univers vers la complexité.  Loin dans le passé n'existaient ni galaxies, ni étoiles. Par ailleurs, nous avons vu avec l'Eq.~\ref{Eq_Ht} qu'à une époque reculée, la densité de rayonnement dominait sur la densité de matière. Le rayonnement présent était en étroite interaction avec la matière. L'univers primordial était opaque. Ce n'est qu'à la fin de l'ère primordiale, après environ 300000 ans, que la matière de l'univers est devenue transparente: les électrons alors libres dans l'univers se sont combinés avec les protons pour former de l'hydrogène neutre. A partir de ce moment s'est également initié un processus de contraction locale (comparée à l'expansion globale) qui aboutira, des centaines de millions d'années plus tard, à l'apparition des premières étoiles et galaxies. Le rayonnement qui prédominait à l'époque de la (re)combinaison était à une température d'environ 3000K, donc essentiellement composé de photons du domaine de la lumière visible ou ultraviolette. Aujourd'hui, ce rayonnement, du fait de l'expansion de l'univers, est beaucoup moins énergétique, avec un pic d'émission dans le domaine des micro-ondes (0.001mm). Le caractère thermique du rayonnement se traduit par la forme du spectre: une loi de corps noir. La satellite Planck a mesuré une température de 2.728K, avec une incertitude extrêmement faible. Ce rayonnement fossile ("Cosmic Background Microwave", CMB) correspondant à $z_\mathrm{rec}=3000/2.7 \simeq 1100$, représente la trace la plus ancienne que l'on ait jamais enregistrée. L'univers était alors environ 1100 fois plus petit que maintenant.

Au début de l'univers, des structures existaient sous la forme de légères fluctuations de densité ($\delta \rho \simeq 0.01$) dans un fluide constitué de photons et de baryons. Les gradients de pression associés à la force de rappel gravitationnelle ont conduit à des oscillations, très similaires à des ondes acoustiques, que l'on appelle "Baryonic Acoustic Oscillations (BAO)". Ce motif s'est alors figé au moment de la recombinaison alors même que l'univers était encore en expansion. Le pic le plus significatif que l'on obtient dans le CMB correspond ainsi à ce pic acoustique. La taille des structures angulaires dans le CMB est donc relié à la taille de ‘‘l'horizon acoustique'' au moment de la recombinaison et à la distance angulaire entre nous et $z_\mathrm{rec}$, qui dépend justement des paramètres cosmologiques et de la courbure de l'univers. Plus l'univers est courbé négativement, plus la distance angulaire est grande et plus le pic acoustique dans le CMB se déplace vers une taille caractéristique plus petite. Ainsi, le satellite Planck et les missions précédentes ont montré que l'univers est spatialement plat ($k=0$ ou $\Omega_\mathrm{k}=0$ dans l'Eq.~\ref{Eq_somme}), c'est à dire que l'on a $ \Omega_\mathrm{r} + \Omega_\mathrm{m} + \Omega_\mathrm{\Lambda} \simeq  1$. Il faut noter que l'on ne pourra jamais montrer que l'univers est absolument plat, seulement qu'il est plat dans un certain domaine de confiance.   

Mais la structure géométrique globale de l'univers n'est pas la seule propriété que l'on peut déduire du spectre de fluctuation du CMB. Le second pic permet de contraindre $\Omega_\mathrm{b} H^2$, c'est la densité de matière des baryons, alors que le troisième pic est sensible à la densité de matière totale (incluant la matière noire) $\Omega_{m} H^2$. Ainsi, $H_\mathrm{0}$ peut être déterminée à partir du spectre du CMB, bien que cette quantité reste dégénérée essentiellement avec $w$, le paramètre de l'Eq.\ref{Eq_Etat} qui définit l'équation d'état de l'énergie noire. En effet, le CMB nous indique que l'univers est quasi-plat et nous donne la densité d'énergie liée à la matière, ce qui permet d'en déduire la densité d'énergie liée à l'énergie noire. Il faut alors faire une hypothèse sur l'équation d'état de l'énergie noire ($w$) pour déterminer la pression répulsive qui sera justement à l'origine de l'expansion de l'univers et donc $H_\mathrm{0}$. Sur la Figure~\ref{Fig_Ho}, j'ai indiqué les valeurs de $H_\mathrm{0}$ déduites du rayonnement de fond cosmologique avec WMAP puis Planck. Les valeurs de $H_\mathrm{0}$ ainsi que les références utilisées dans ce graphique se trouvent dans le tableau de l'annexe~\ref{Tab.Ho}. WMAP a publié des résultats de manière régulière après 1, 3, 5, 7 et 9 observations respectivement. On constate une diminution progressive et assez curieuse des valeurs de $H_\mathrm{0}$ dans le cas de WMAP au fil des années. Mais, on peut finalement noter que les résultats WMAP-9 et Planck sont compatibles.

La valeur de \citet{riess16} est 3.4$\sigma$ plus élevée que celle déduite du rayonnement de fond cosmologique établie par le satellite Planck, $H_0=66.93\pm 0.62$ km s$^{-1}$ Mpc$^{-1}$ \citep{planck16_XIII}. La "tension" ainsi obtenue est aujourd'hui extrêmement débattue: soit elle provient d'erreurs systématiques, soit elle implique une nouvelle physique. Il faut voir effectivement que la valeur de H$_0$ déduite à partir de l'échafaudage des échelles de distance (Céphéides, supernova de type Ia) est déterminée à z$\simeq < 0.15$, alors que la valeur déduite à partir du rayonnement de fond cosmologique correspond à $z=1100$. 
%Nous reviendrons sur ce résultat dans une section suivante. 

Il existe néanmoins un moyen remonter les échelles de distance depuis le CMB dans ce qu'on pourrait appeler un étalonnage des échelles de distance {\it inverse}. En effet, les BAO (du fait de la vitesse du son qui est connue dans le plasma) fournissent une règle (ou distance) de référence caractéristique à l'époque du CMB, qui va conditionner la structure et l'évolution des amas de galaxies. Ainsi, si l'on mesure les corrélations dans la distribution des galaxies à grande échelle avec par exemple le  {\it ‘‘Sloan Digital Sky Survey''} (SDSS), on trouve des distances typiques de l'ordre de 100 Mpc. En comparant cette échelle de distance obtenue autour de $z=0.35$ avec celle du CMB correspondant à $z=1100$, on obtient un moyen de contraindre les paramètres cosmologiques et donc $H_\mathrm{0}$. Une valeur récente obtenue par cette méthode (BAO) est indiquée sur la Fig.~\ref{Fig_Ho} et est compatible avec les résultats du CMB. Il est à noter que les analyses données par WMAP ou Planck prennent cette contraindre additionnelle en compte dans leur analyse. 

La dernière méthode liée au rayonnement de fond cosmologique est la méthode Sunyaev-Zel'dovich (S-Z). Le principe de la méthode \citep{sunyaev72}, ainsi que la façon dont elle est utilisée pour déduire la constante de Hubble \citep{silk78} ont fait l'objet de quelques revues \citep{Birkinshaw99,carlstrom02}. La méthode repose sur la physique de gaz chaud ($10^8$K)  dans des amas, qui émettent du rayonnement X en émission par l'effet bremsstrahlung, générant une brillance de surface dont la formule dépend de la densité d'électrons et de leur température. Par ailleurs, les électrons du gaz chaud interagissent avec les photons du CMB (diffusion par effet Compton) et les décalent en fréquence. Apparaissent ainsi des "trous" dans l'émission radio du CMB (car les photons ont été retirés de ces fréquences pour être décalés vers les hautes fréquences). La combinaison des rayonnements X et radio permet de déterminer la distance du nuage. Cette estimation reposent sur quelques hypothèses concernant le nuage mais ne dépend d'aucune chandelle stellaire. En mesurant le décalage vers le rouge du nuage $z$, on obtient une contrainte sur $H_\mathrm{0}$. Cette méthode s'applique autour de $z=0.15$ à $1$. \citet{reese02} a néanmoins identifié un budget d'erreur de l'ordre de 20 à 30\% pour la détermination de distance d'un amas, ce qui explique la grande dispersion des mesures de $H_\mathrm{0}$ par cette méthode indiquées sur la Fig.~\ref{Fig_Ho}.

\section{Les méthodes liées aux quasars}\label{Sect_quasars}

La théorie de la relativité générale exprime l'effet de la gravitation comme une courbure de l'espace-temps. Ainsi, le rayonnement issu d'une source lointaine, comme un quasar, est dévié par le champ gravitationnel d'un objet massif (une galaxie ou un amas de galaxies) situé entre la source et l'observateur. Ceci peut générer plusieurs sortes d'effets: déplacement de l'image, amplification, déformation ou même la création d'images multiples. \citet{Refsdal64} a montré que si la source d'arrière-plan est variable, alors il est possible de déterminer une distance absolue et donc la constante de Hubble. En effet, si le quasar devient subitement plus brillant, l'observateur, du fait de la lentille gravitationnelle, le percevra à des instants différents pour les différentes images du quasar. Le décalage temporel ainsi mesuré dépend essentiellement de la distance des objets, i.e. de la lentille et de la source, ainsi que du potentiel gravitationnel. Si le décalage vers le rouge $z$ peut être mesuré pour ces objets, alors la constante de Hubble peut être déterminée directement. La première lentille gravitationnelle fut découverte par \citet{walsh79} et depuis lors, la méthode a été appliquée à de nombreuses occasions (voir Fig.~\ref{Fig_Ho}). La principale difficulté provient de la dégénérescence entre le potentiel gravitationnel considéré et la constante de Hubble. 

Une autre façon de procéder est d'utiliser le rayonnement issu de quasars, dont les raies en absorption (forêt de Lyman $\alpha$) sont le résultat d'un long processus d'absorption de la matière le long de la ligne de visée. La distribution de ces raies donne une information sur la distribution de matière et donc contiennent des informations cosmologiques intéressantes dont $H_\mathrm{0}$ peut être déduite \citep{tytler04,mcdo05,fontribera13,delubac14}. Cette méthode a donné récemment \citep{chavez12} une valeur de $H_\mathrm{0}$ très proche du résultat de \citep{freedman12} (voir figure~\ref{Fig_Ho}).

Enfin, depuis deux jours\footnote{Ces lignes ont été écrites le 18 octobre 2017} nous savons qu'il existe une autre méthode de détermination de la constante de Hubble basée sur la détection des ondes gravitationnelles ! On ne parle plus de chandelles stellaires pour déterminer la constante de Hubble mais de sirènes car la fréquence des ondes gravitationnelles est audible. La détection des ondes gravitationnelles générées par GW170817 (la collision de deux étoiles à neutrons) par Virgo et Ligo le 17 Août 2017  \citep{virgo17b} ainsi que sa contrepartie électromagnétique ont permis de définir la galaxie hôte du phénomène, il s'agit de NGC 4993. La distance de la galaxie est déduite directement à partir du signal gravitationnel (42.9 $\pm$ 3.2 Mpc), tandis que le décalage vers le rouge est mesuré de la façon habituelle à partir des ondes électromagnétiques. La valeur obtenue est $H_\mathrm{0}=70.0_\mathrm{-8.0}^{+12.0}$ \kms Mpc$^{-1}$ \citep{virgo17a} \footnote{Cette valeur est fortement corrélée à l'inclinaison du système binaire. \citet{guidorzi17} ont affiné ce résultat à partir d'observations X et ont obtenu  $H_\mathrm{0}=75.5_\mathrm{-7.3}^{+14.0}$ \kms Mpc$^{-1}$.}. La précision obtenue (environ 15\%) n'est pas suffisante pour résoudre la tension sur la constante de Hubble, mais ouvre une nouvelle voie fort intéressante. 

\begin{figure}[htbp]
\begin{center}
%\resizebox{1.0\hsize}{!}{\includegraphics[clip=true]{f0.eps}}
\resizebox{1.0\hsize}{!}{\includegraphics[clip=true]{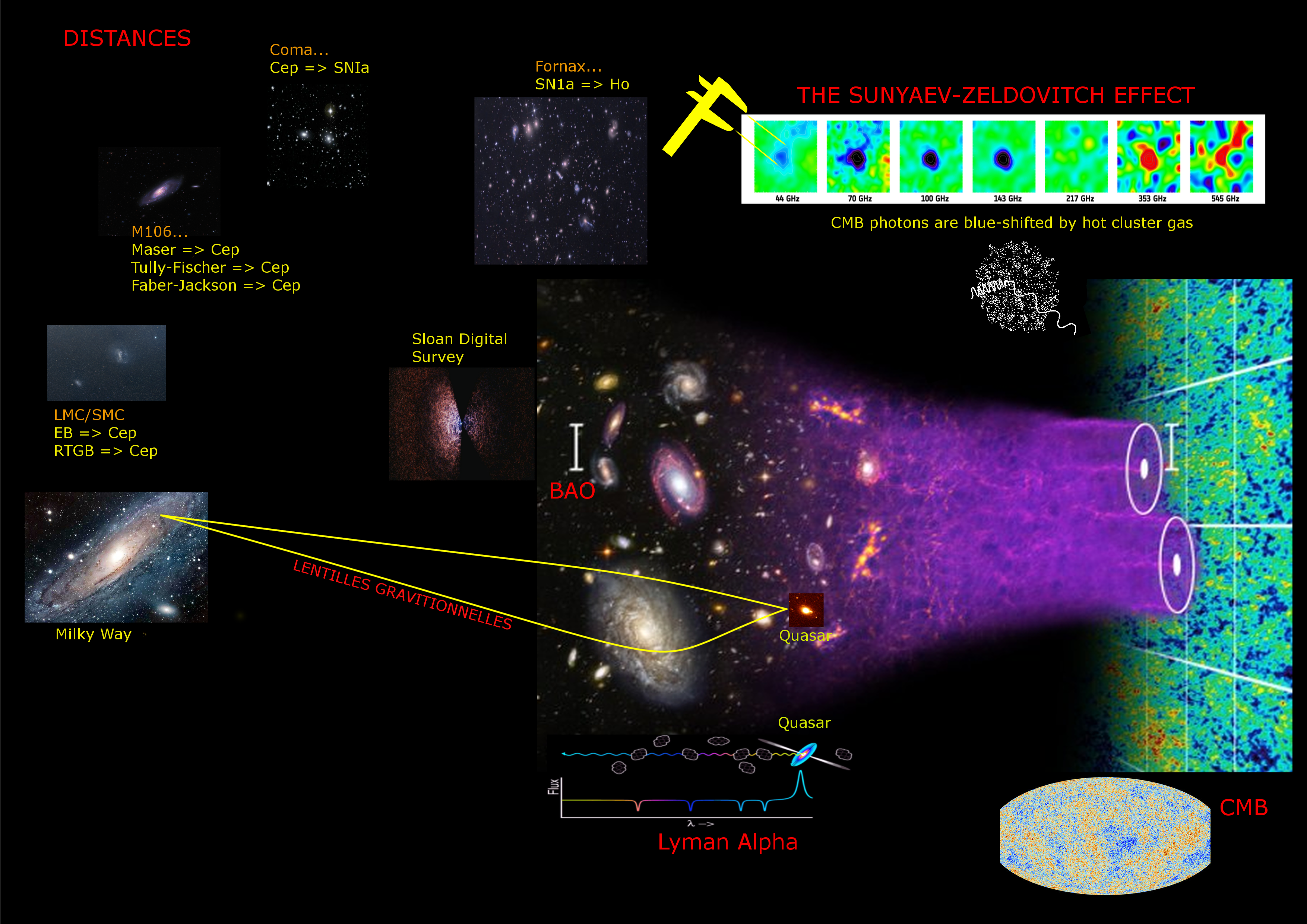}}
\end{center}
\caption{Montage représentant les différentes méthodes de détermination de la constante de Hubble discutées dans ce manuscrit. Outre la méthode liée à l'étalonnage des échelles de distances dans l'univers basée essentiellement sur les Céphéides et supernova de type Ia (Sect.~\ref{Sect_d}), on peut étudier le rayonnement de fond cosmologique (CMB), ou l'effet Sunyaev-Zeldovitch qui repose lui aussi sur les propriétés des photons du CMB (Sect.~\ref{Sect_cmb}). D'autres méthodes reposent essentiellement sur les quasars, telles les lentilles gravitationnelles et les forêts Lyman $\alpha$  (Sect.~\ref{Sect_quasars}). Enfin, les oscillations baryoniques acoustiques (BAO) dans le CMB fournissent une règle de référence à partir de laquelle les structures, et en particulier les amas de galaxies, vont se former à grande échelle. Ainsi, l'étude de la position des galaxies par les grands {\it surveys} photométriques et spectroscopiques (par exemple le ''Sloan Digital Sky Survey'' SDSS), permettent de contraindre H(z) \citep{reid10, reid12}.} 
%Positive velocities correspond to a redshift or receding motion.} 
\label{Fig_DessinHo}
\end{figure}

L'ensemble des méthodes que nous avons vues sont décrites schématiquement dans la Figure~\ref{Fig_DessinHo}. 

\newpage
\section{Comment dépasser la tension sur $H_\mathrm{0}$ ?}

Parmi les différentes approches que nous avons vues pour contraindre la constante de Hubble $H_\mathrm{0}$, il apparaît clairement sur la Fig.~\ref{Fig_Ho} que les plus précises reposent sur l'étalonnage des échelles de distance dans l'univers et sur le rayonnement de fond cosmologique (incluant la contrainte supplémentaire des oscillations acoustiques baryoniques, BAO). Les méthodes totalement indépendantes telles que S-Z, les lentilles gravitationnelles et les forêts de Lyman~$\alpha$ restent relativement imprécises, et ne permettent pas, actuellement, de résoudre la tension obtenue entre les méthodes reposant sur les SNIa \citep{riess16} et le satellite Planck \citep{planck16_XIII}. Par ailleurs, concernant les distances, le chemin le plus robuste reste celui des Céphéides et des SNIa. Les mégamasers sont prometteurs, tandis que les méthodes liées aux propriétés des galaxies (Tully-Fisher et Faber-Jackson) donnent des résultats plutôt imprécis par rapport aux SNIa. Ainsi que peut-on dire à propos de cette tension ? 

D'abord, le problème de la tension dépasse le ‘‘simple'' problème de la mesure de la constante de Hubble $H_\mathrm{0}$. Le modèle $\Lambda$CDM qui décrit l'univers dont la dynamique est décrite par les Equations~\ref{Eq_Ht} et \ref{Eq_somme} fait intervenir 6 paramètres. Mais finalement, lorsque l'on regarde les différentes façons de contraindre $H_\mathrm{0}$, on s'aperçoit que le jeu de paramètres clefs peut se résumer, de manière simpliste, à $\Omega_\mathrm{k}$, $\Omega_{m}$,  $w$ et $H_\mathrm{0}$. L'approche CMB contraint {\it effectivement} ces 4 paramètres, et ce d'autant mieux si les BAO sont observées à bas décalage vers le rouge. Cependant, la contrainte est bien plus robuste pour les 2 premiers que pour les 2 derniers. \citet{planck16_XIII} ont bien précisé dans leur papier: {\it ‘‘CMB experiments provide indirect and highly model-dependent estimates of the Hubble constant. It is therefore important to compare CMB estimates with direct estimates of H0, since any significant evidence of a tension could indicate the need for new physics.''}. Cette dégénérescence explique-t-elle pourquoi les valeurs de $H_\mathrm{0}$ décroissent progressivement lorsque l'on considère successivement WMAP-1 à 9, puis les résultats du satellite Planck ? Ces études donnent effectivement des tables de résultats avec de l'ordre de 10 à 20 paramètres ajustés, ce qui implique un travail complexe au niveau de la gestion de la dégénérescence entre les paramètres. 

De l'autre côté, l'échafaudage des échelles basé sur les Céphéides et les SNIa permet avant tout de mesurer le taux d'expansion de l'univers maintenant, c'est à dire à $z=0$, $H_\mathrm{0}$, mais aussi en fonction de $z$ (i.e. $H(z)$, et donc $w$) si des SNIa sont détectées suffisamment loin (autour de $z=2$). Cette approche a elle aussi ses difficultés:
\begin{enumerate}
\item Les périodes des Céphéides de la Voie Lactée sont courtes ($P<10$ jours) en comparaison des périodes des Céphéides dans les galaxies contenant des SNIa, et le recouvrement est finalement faible. Ainsi, la solution qui consiste à utiliser uniquement les parallaxes HST des Céphéides Galactiques pour contraindre la relation des SNIa paraît instable à elle seule (cf. l'annexe A de \citet{Efstathiou14}). La situation changera sensiblement avec l'arrivée de {\it Gaia} et la détermination de distances de Céphéides Galactiques à longues périodes. 
\item Ensuite, la solution qui consiste à utiliser le LMC pour étalonner la relation \emph{PL} est sensible à la métallicité, étant donné que le LMC a une métallicité un peu plus faible que les galaxies lointaines. Ainsi, la relation \emph{PL} est-elle dépendante de la métallicité ? Si oui, comment cette correction doit-elle être prise en compte ? Sur la Figure~\ref{Fig_Ho} la valeur de $H_\mathrm{0}$ déduite par \cite{sandage06}, autour de 61 km/s/Mpc est significativement plus faible que celle de \citet{riess16}. Cette incohérence est discutée dans la revue de \cite{jackson15} et est attribuée à la façon de gérer l'impact de la métallicité sur la relation \emph{PL} des Céphéides. Nous y reviendrons dans la Sect.~\ref{sect_plz}.
\item Enfin, il a été noté récemment que le pic des SNIa peut être corrélé au taux de formation des étoiles \citep{rigault15}. Les auteurs avancent qu'un biais de +1.8\kms Mpc$^{-1}$ est possible dans la valeur de $H_\mathrm{0}$ déduite à partir des Céphéides et des SNIa. 
\end{enumerate}

Un dernier point à considérer est que l'approche CMB utilise d'abord des données (i.e. le rayonnement de fond cosmologique) dont les propriétés sont liées à l'époque où l'univers était beaucoup plus petit (i.e. à $z=1 100$), pour ensuite ‘‘remonter'' en quelque sorte l'échafaudage des échelles de distances {\it en sens inverse}, et utiliser les BAO, et même les SNIa, ce qui fournit finalement une valeur de $H_\mathrm{0}$. De manière opposée, la méthode liée à l'échafaudage des échelles de distances dans l'univers, basée sur les Céphéides et les SNIa, reposent sur des données à $z<0.15$ et la mesure de $H_\mathrm{0}$ est dans ce cas plus directe. Ainsi, comparer des valeurs de $H_\mathrm{0}$ découlant de données obtenues d'une part à $z=1100$ et, d'autre part, à $z<0.15$ n'est pas forcément évident, et la tension pourrait tout aussi bien révéler la présence d'une ‘‘nouvelle physique'' à prendre en compte. 

Nous en arrivons donc à la conclusion que déterminer les distances dans l'univers à l'aide des Céphéides et des SNIa reste indispensable pour contraindre $H_\mathrm{0}$ et, de manière indirecte, l'ensemble des paramètres du modèle $\Lambda$CDM. Les méthodes liées au CMB seul, bien que très puissantes, ne peuvent pas contraindre tous les paramètres du modèle sans introduire des dégénérescences. Aussi comment peut-on avancer et tenter de résoudre la ‘‘tension'' liée à la détermination de $H_\mathrm{0}$ ? Du côté de l'échafaudage des échelles de distances qui nous intéresse dans ce manuscrit, un des points clefs (outre les problématiques liées aux SNIa que nous n'aborderons pas ici) est d'étudier tout ce qui pourrait biaiser la relation \emph{PL} des Céphéides, c'est-à-dire s'intéresser à toutes les propriétés physiques des Céphéides qui pourraient, à période égale, changer leur luminosité. Ainsi, \citet{riess16} indiquent que les plus grandes incertitudes sur $H_\mathrm{0}$ proviennent pour 1.3\% des distances des galaxies de référence (LMC, NGC 4258, M31), et pour 1.1\% de notre connaissance de la relation PL: métallicité, dispersion, rougissement. En revanche, \citet{riess16} ne prennent pas en compte l'impact potentiel de l'environnement des Céphéides et de la binarité, qui peuvent, s'ils ne sont pas considérés, générer une dispersion dans la relation \emph{PL}.  Ainsi, les quatre principales questions sont en définitive :  

\begin{enumerate}
\item La pente de la relation \emph{PL} est-elle sensible à la métallicité et donc à la galaxie considérée ? 
\item Quelle est la distance du LMC, de NGC 4258, de M31, et d'autres galaxies contenant des Céphéides, et donc le point-zéro de la relation \emph{PL} ? Ce point-zéro est-il sensible à la métallicité ? 
\item La dispersion de la relation PL peut-elle être améliorée dans l'infrarouge en considérant l'impact de l'environnement et de la binarité ? 
\item Un autre point mérite l'attention. Serait-il possible de court-circuiter la relation PL des Céphéides et d'utiliser directement les méthodes dites de Baade-Wesselink pour déterminer la distance des Céphéides extragalactiques, de manière individuelle, sans passer par une relation statistique ? 
\end{enumerate}

Les points 1/ et 2/ sont des objectifs majeurs du projet international Araucaria\footnote{\url{https://araucaria.camk.edu.pl/}; \url{https://en.wikipedia.org/wiki/Araucaria_Project}} de détermination de distance dans le groupe local. Les points 3/ et 4/ sont étudiés dans le cadre de l'ANR "UnlockCepheids". Nous aborderons successivement ces quatres points dans les sections du chapitre suivant. 

\chapter{La relation \emph{PL} des Céphéides et les distances dans le Groupe Local}\label{Chap_Araucaria}

\begin{figure}[htbp]
\begin{center}
\includegraphics[width=15cm]{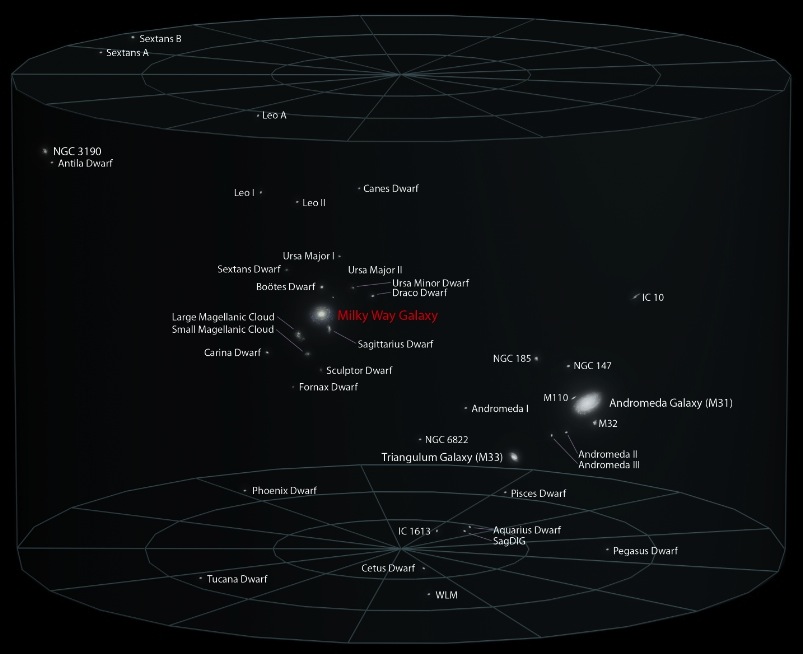}
\includegraphics[width=15cm]{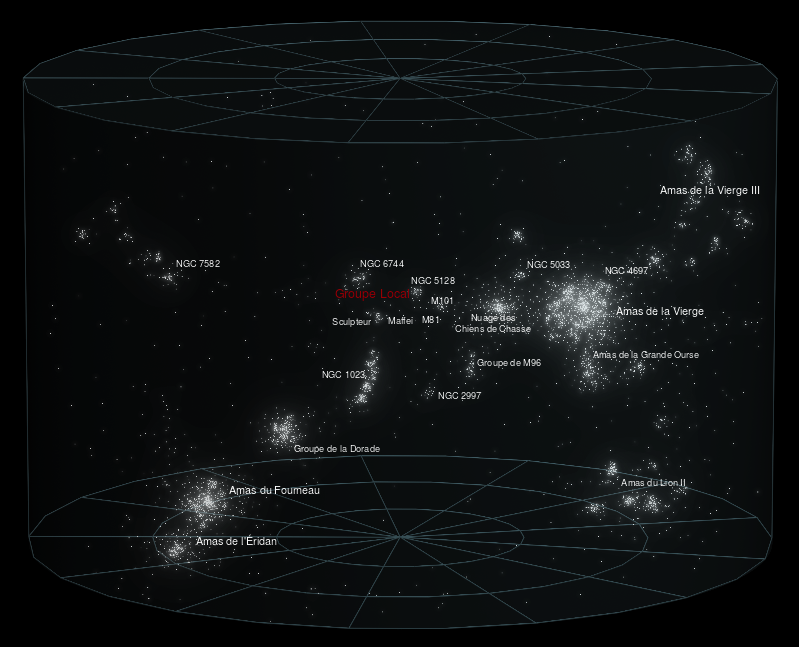}
\end{center}
\vspace*{-5mm} %\caption{Fig_cartographie}
\caption{Cartographie des galaxies du groupe local (en haut) et de l'Amas de la Vierge (en bas).}\label{Fig_GrLocal}

\end{figure}

Le groupe local est constitué d'environ 100 galaxies dans un volume d'environ 3 Mpc \citep{McConnachie12}.  Les deux membres principaux de ce groupe sont la galaxie d’Andromède (M31) et la Voie Lactée, chacune d’elles possédant son propre système de galaxies satellites (voir Fig.~\ref{Fig_GrLocal}a). Autour de la Voie Lactée, gravitent principalement les deux nuages de Magellan et les galaxies naines du Grand Chien, du Sagittaire, de la Petite Ourse, du Dragon, de la Carène, du Sextant, du Sculpteur, du Fourneau, du Lion I, du Lion II et du Toucan. Le système d’Andromède comprend M32, M33, M110, NGC 147, NGC 185, Andromeda I, Andromeda II, Andromeda III et Andromeda IV. La galaxie du Triangle (M33) est la troisième plus grande galaxie du groupe local. Les autres membres du groupe local sont gravitationnellement indépendants de ces larges sous-groupes.  Le filament du Sculpteur (voir Fig.~\ref{Fig_GrLocal}b), dont le membre le plus brillant et le plus massif est la galaxie du sculpteur située à 3.3 Mpc (à ne pas confondre avec la galaxie naine du Sculpteur qui est beaucoup plus proche), regroupe une vingtaine de galaxies situées au voisinage du pôle galactique sud, dans la constellation du Sculpteur \citep{kara03}. La Table~\ref{Tab.GL} est une compilation de plusieurs listes et de références qui indique, pour chaque galaxie du groupe local et du filament du sculpteur, une estimation de sa distance récente (jusqu'à 6.5 Mpc), la méthode utilisée pour déterminer cette distance, ainsi que sa métallicité {\it moyenne}, exprimée en [Fe/H]. A cette liste nous ajoutons M83 et NGC5253 liées gravitationnellement, ainsi que M82. Ces trois galaxies sont dirigées vers le centre de l'amas de la Vierge, un grand amas de galaxies situé entre 15 et 22 Mpc (voir Fig.~\ref{Fig_GrLocal}b) et dans lequel on trouve effectivement des galaxies hôtes de SNIa. Le site web\footnote{http://www.cbat.eps.harvard.edu/lists/Supernovae.html} du CBAT ({\it Central Bureau for Astronomical Telegrams}) liste l'ensemble des 6500 supernovae observées depuis 1885. Parmi celles-ci, on peut extraire environ 2800 SNIa dont la première référencée remonte à 1937. Parmi ces SNIa, quelles ont été les plus proches ? Historiquement d'abord (i.e. avant 1037), quatre ont été observées dans la Voie Lactée: (a) SN185, observée par les Chinois en 185 après JC et étudiée récemment par \citet{zhao06}, (b) SN1006, également observée par les Chinois et étudiée par \citet{winkler03}, (c) SN1572, observée par Tycho Brahé et (d) SN1604, observée par Kepler. Après 1937, les SNIa les plus proches observées sont SN1972E, SN2014J, SN1983N, et SN1937C, toutes situées à moins de 6.5 Mpc et repérées dans la Tab.~\ref{Tab.GL}.  Il est à noter que SN1972E est le prototype des SNIa pour avoir été observée pendant 1 an \citep{ardeberg73}. Cependant, \cite{riess16} indiquent qu'une SNIa pourra être utilisée comme étalon de distance si elle répond à différents critères: 1) sa mesure photométrique doit être moderne, c'est à dire basée sur des mesures CCD, 2) le début des observations doit commencer avant le maximum de brillance, 3) son rougissement doit être faible ($A_\mathrm{V} < 0.5$ mag), 4) elle doit être typique d'un point de vue spectroscopique, et enfin 5) la galaxie hôte doit nécessairement contenir des Céphéides observées par le HST. Ce dernier point implique de considérer des galaxies de type Sa à Sd, dont la distance est inférieure à 40 Mpc, dont l'inclinaison est inférieure à 75° et  dont  la taille apparente est supérieure à 1'.  \cite{riess16} ont trouvé seulement 19 SNIa qui répondent strictement à ces critères, indiquées dans la Tab.~\ref{Tab.GL}. J'ai rajouté dans la table des estimations de distance récentes de ces galaxies hôtes de SNIa, ainsi que leur métallicité (lorsque disponible). Ces SNIa sont d'une extrême importance vu qu'elle serviront à étalonner la relation des SNIa, qui elle contient plusieurs centaines de SNIa. 

\begin{figure}[htbp]
\begin{center}
\includegraphics[width=12cm]{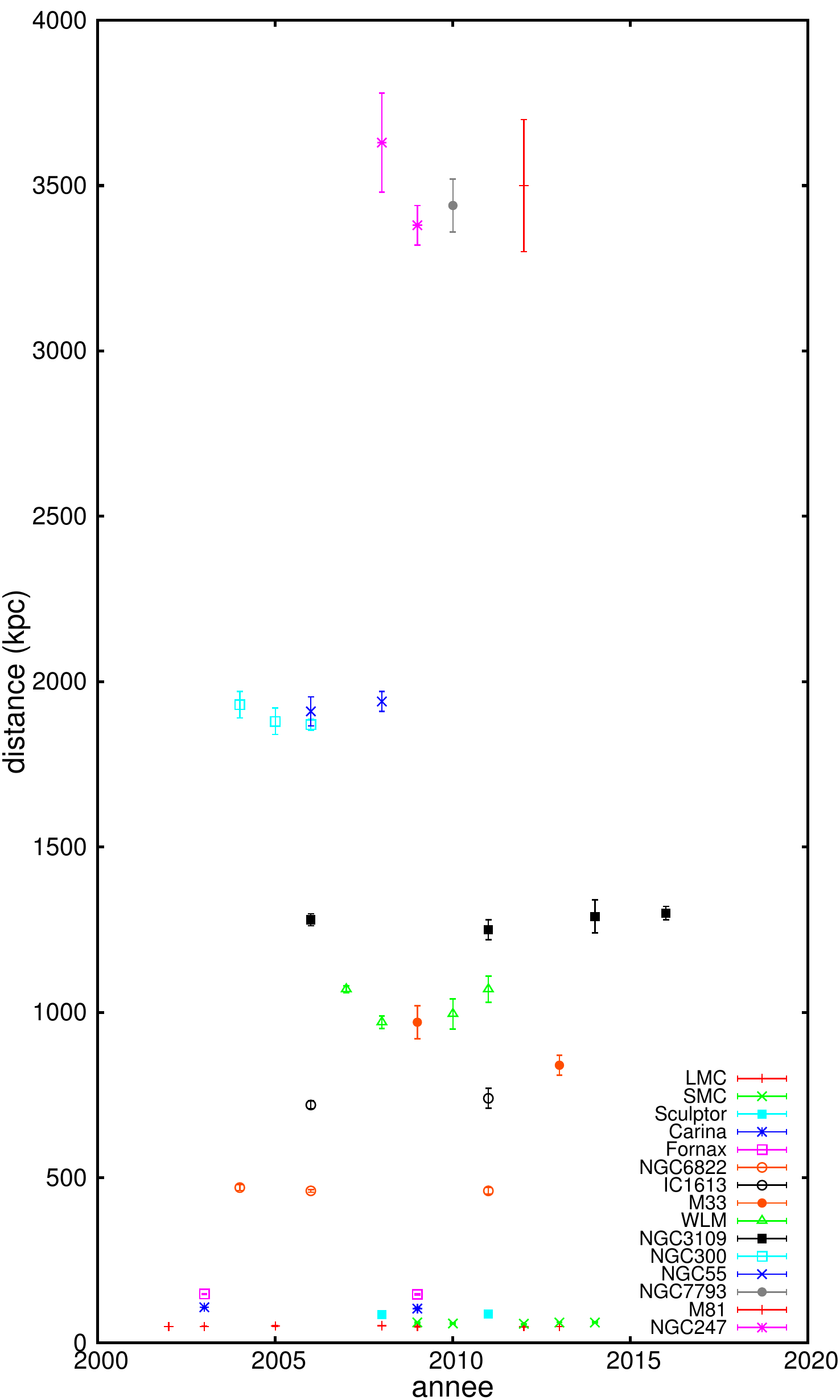}
\end{center}
\caption{Distances des différentes galaxies déterminées dans le cadre du projet Araucaria. Cette figure a été établie à partir de la Tab.~\ref{Tab.Araucaria}}\label{Fig_Araucaria}
%\vspace*{-5mm} \caption{Fig_cartographie}
\end{figure}

\begin{figure}[htbp]
\begin{center}
\includegraphics[width=12cm]{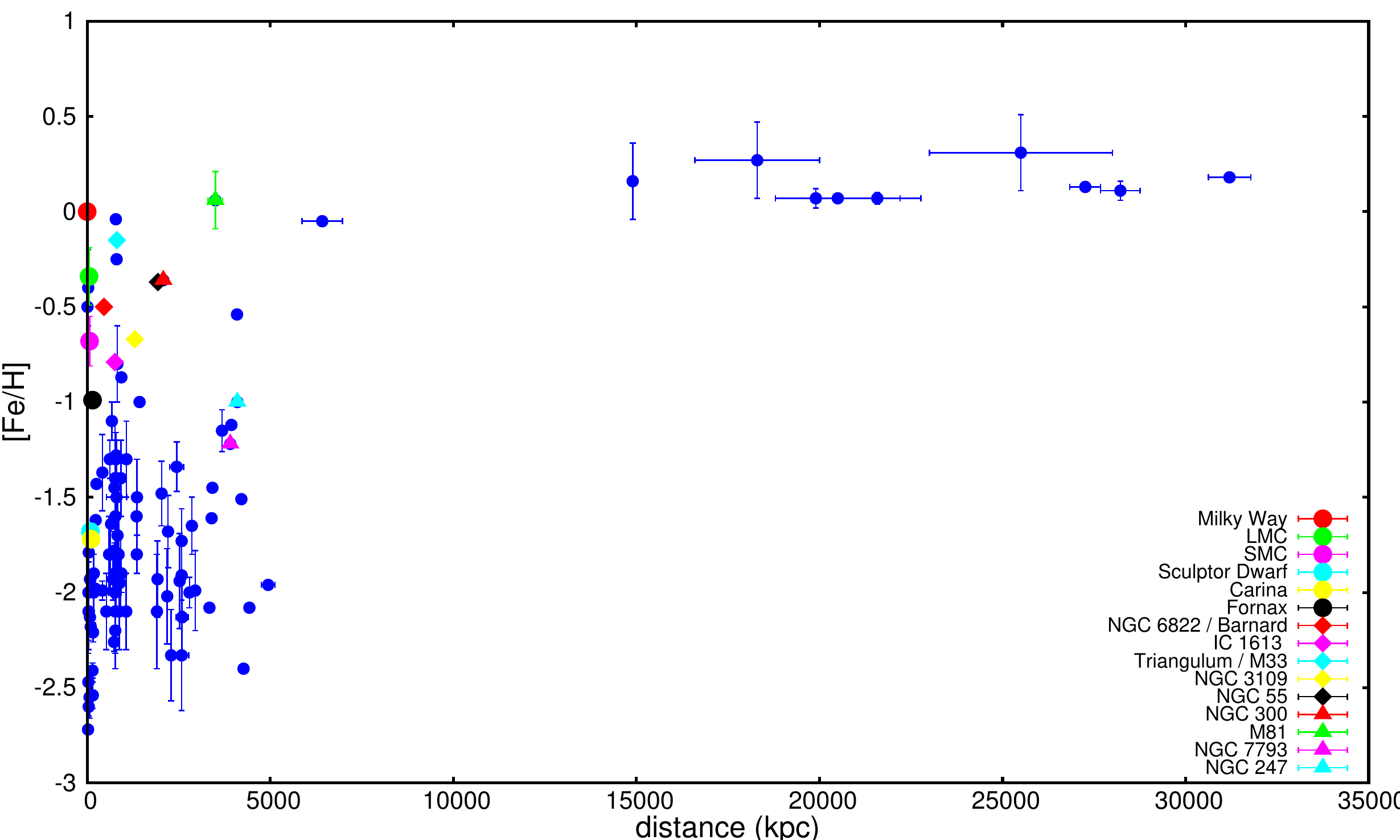}
\includegraphics[width=12cm]{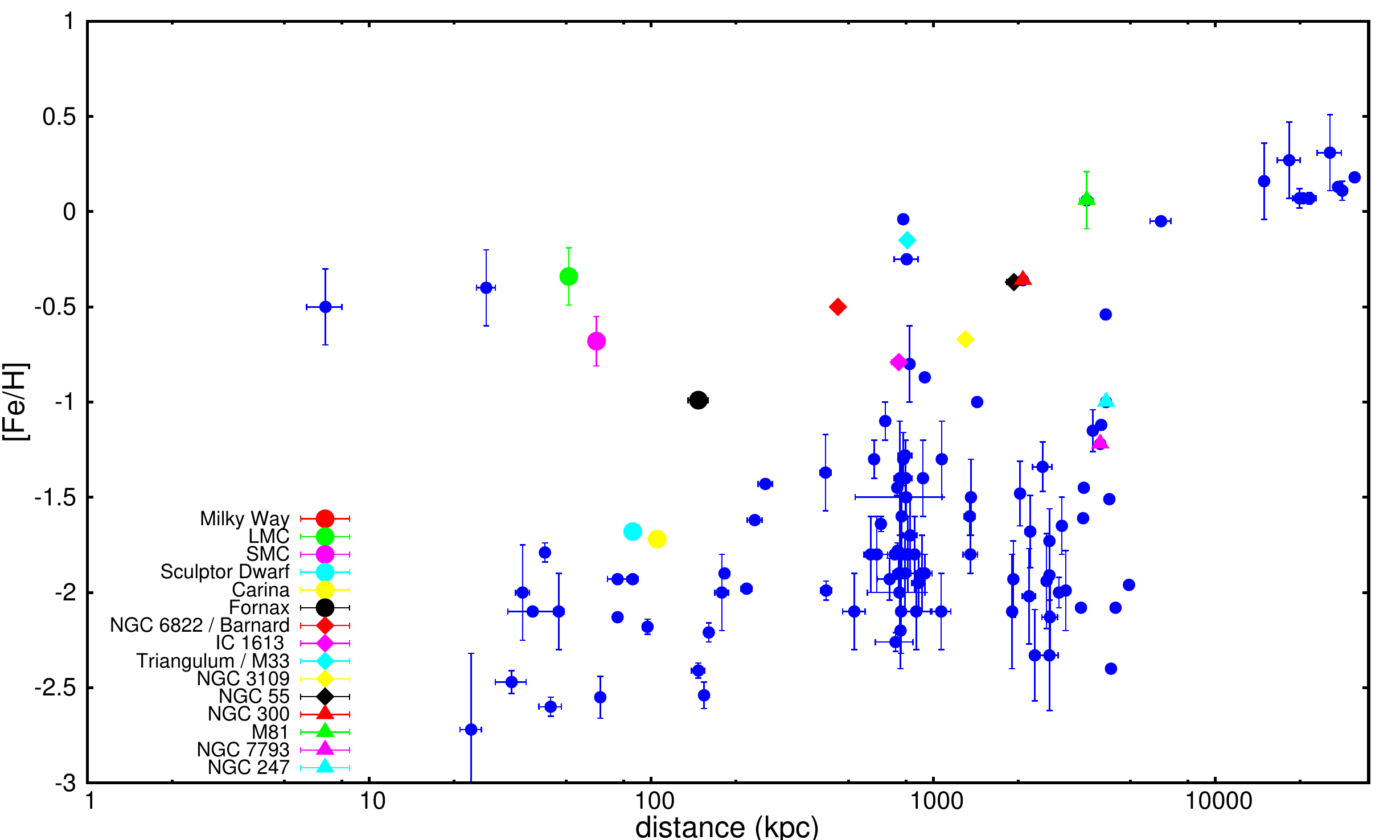}
\end{center}
\caption{Abondance en fer par rapport à l'hydrogène [Fe/H] en fonction de la distance des galaxies du groupe local (en $\log$ pour la figure du bas). Les galaxies qui ont fait l'objet d'étude dans le cadre du projet Araucaria sont indiquées par différentes couleurs.}\label{Fig_FeH}
%\vspace*{-5mm} \caption{Fig_cartographie}
\end{figure}

Le projet international et à long terme Araucaria\footnote{voir https://araucaria.camk.edu.pl/  pour les objectifs, les participants, et la liste de publications. Je suis membre du projet Araucaria depuis 2008, date de mon arrivée à Concepcion du Chili pour mon post-doc.} consiste à renforcer l'étalonnage des échelles de distance à l'aide de différents estimateurs : Céphéides, RR Lyrae, supergéantes rouges ou bleues et plus récemment, les binaires à éclipses. L'idée est de déterminer avec soin la distance d'un grand nombre de galaxies proches, dans le groupe local et dans  le Sculpteur, et de tester la dépendance de ces estimateurs avec les paramètres des galaxies hôtes (métallicité, âge des populations stellaires). Ainsi, dans la Tab.~\ref{Tab.GL}, j'ai identifié les galaxies pour lesquelles le projet Araucaria a apporté une contribution significative. Les valeurs de distance indiquées correspondent alors à la moyenne des différentes estimations de distance issues de différentes méthodes utilisées dans le cadre du projet Araucaria,  tandis que l'incertitude correspond à la dispersion des estimations. Le détail des distances obtenues dans le cadre du projet Araucaria sont indiquées dans la Table~\ref{Tab.Araucaria} et représentées sous forme graphique en fonction de l'année de publication dans la Fig.~\ref{Fig_Araucaria}. La Fig.~\ref{Fig_FeH} montre la distance des galaxies de la Table~\ref{Tab.GL}, en fonction de leur métallicité [Fe/H] (pour les galaxies du Groupe Local et du Sculpteur) et [O/H] pour les galaxies hôtes de SNIa. J'indique également sur cette figure les galaxies qui ont été étudiées jusqu'à présent dans le cadre du projet Araucaria. Cette figure montre clairement un point important: les galaxies hôtes de SNIa ont une métallicité très proche de celle de la Voie Lactée, alors que le LMC et le SMC sont clairement moins riches en métaux, ainsi que la majorité des galaxies situées à moins de 6.5 Mpc.

 Ainsi, il est tentant de conclure qu'avec {\it Gaia}, il n'y aura pas lieu de s'inquiéter de l'impact de la métallicité sur la relation \emph{PL} dans la mesure où la Voie Lactée et les galaxies hôtes de supernovae de type Ia ont environ la même abondance en métaux. Cependant, ce n'est pas si simple, car il est très difficile de mesurer la métallicité des Céphéides situées dans les galaxies hôtes de supernovae. Cette dernière est effectivement déduite {\it de manière indirecte} à partir de l'abondance de l'oxygène par rapport à l'hydrogène [O/H] dans les régions HII (zone de formation d'étoiles), et on fait ensuite l'hypothèse que ces régions HII et les Céphéides dans leur voisinage ont la même métallicité, ce qui n'est pas forcément le cas. En effet, l'abondance [O/H] est utilisée comme un diagnostic de l'évolution des galaxies du fait d'une relation entre l'abondance [O/H] des régions HII et la masse des galaxies \citep{lequeux79}. Cependant, cette relation n'est pas compatible pour les galaxies les plus proches, la Voie Lactée, le SMC et le LMC \citep{tremonti04} probablement du fait de raies en émission H et O  difficiles à étalonner dans les régions HII des galaxies lointaines \citep{kewley08} avec des écarts de 0.9 dex selon les méthodes utilisées. Ainsi, la métallicité des galaxies hôtes de supernova n'est pas encore clairement établie. Par ailleurs, il est intéressant de noter que \citet{rafelski14}, à une autre échelle, trouvent une relation (un peu dispersée certes) entre la métallicité et le décalage vers le rouge (une relation Zz), et ce jusqu'à environ $z=6$. C'est pourquoi, R. Kudritzki, F. Bresolin et M. Urbaneja de l'équipe Araucaria visent à étalonner la relation [O/H]-masse des galaxies en utilisant les supergéantes bleues et rouges. Ces étoiles sont effectivement suffisamment brillantes pour une détermination spectroscopique de leur abondance. La méthode a été appliquée aux supergéantes de NGC~300 à 1.9 Mpc \citep{kudritzki08} et NGC~3621 à 6.5 Mpc \citep{kudritzki14}. Pour aller plus loin, une méthode a été développée, s'appuyant non pas sur une supergéante individuelle, mais sur un nuage d'étoiles supergéantes: M83 à 4.3 Mpc \citep{gazak14} et les galaxies des Antennes (NGC 4038 / NGC 4039) à 20 Mpc \citep{lardo15}. Des comparaisons entre l'abondance des supergéantes bleues et rouges, ainsi que les régions HII sont également en cours \citep{bresolin16, davies17}, et un projet SSC (‘‘Super Stars Clusters'') se met actuellement en place pour appliquer ces méthodes aux galaxies hôtes de SNIa utilisées dans le cadre du projet SHOES. Il est également important d'avoir à l'esprit que les supergéantes  sont utilisées comme indicateurs de distance. Les rouges ont des phases d'évolution reconnaissables (où elles passent plus de temps), ‘‘red tip of giant branch'' ou ‘‘red clump stars''  (voir Table~\ref{Tab.GL} pour des références récentes) ce qui en fait de bons indicateurs de distances (magnitude absolue visible comprise entre -8 et -11). Concernant les supergéantes bleues, après leur sortie de la séquence principale, elles suivent une ligne de magnitude absolue relativement constante pour une masse donnée.  Il existe ainsi une relation entre la gravité de surface et la luminosité (''Flux Weighted Gravity - Luminosity Relationship (FGLR)'', \citep{kudritzki03}). Les supergéantes rouges permettent d'atteindre des distances de l'ordre de 30 Mpc \citep{jang15, jang17b} tandis que les bleues (moins brillantes) permettent d'atteindre environ 6.5 Mpc. Dans ce contexte en permanente évolution, il semble ainsi toujours pertinent de se poser la question de l'impact de la métallicité sur la relation \emph{PL}. 

%La méthode la plus utilisée pour déterminer l'abondance des galaxies hôtes des SNIa et des Céphéides est basée sur l'étude des régions HII et donne accès à l'abondance de l'oxygène par rapport à l'hydrogène. Nous allons ainsi maintenant nous intéresser à l'impact de la métallicité sur la relation \emph{PL} et son impact sur la constante de Hubble.%La distance est indiquée, ainsi que la méthode utilisée pour déterminer cette distance, ainsi que la métallicité moyenne de la galaxie. 
%Cette liste indique également les distances de ces galaxies (issues de références récentes), mais pas les méthodes utilisées pour déterminer ces distances. J'ai ainsi identifié et répertorié les méthodes de la manière suivante: rcs (pour ‘‘Red Clump Stars''), msf (pour ‘‘Main Sequence Fitting''), rtgb (pour ‘‘Red Tip of Giant Branch''), hbf (pour ‘‘Horizontal Branch Fitting''), rrl (pour ‘‘RR Lyrae stars''), tf (pour ''Tully-Fisher relation"), cep (pour ''Cepheids"), sbf (pour ‘‘Surface Brightness Fluctuations''), redshift pour les distances déduites du site web suivante: https://dso-browser.com/. Pour la métallicité, j'invite le lecteur à regarder les méthodes utilisées ainsi que les références associées dans \citet{McConnachie12}. Le Tab.~\ref{Tab.GL}  
%Les galaxies suivantes se trouvent dans l'amas de la Vierge : M49, M58, M59, M60, M61, M84, M85, M86, M87, M88, M89, M90, M91, M98, M99, et M100. 

\begin{table*}
\caption{\label{Tab.GL}  {\small Liste des galaxies issues de \cite{McConnachie12} pour le groupe local, de \cite{kara03} pour le filament du sculpteur, et de \cite{riess16} pour les galaxies hôtes de SNIa. La liste des SNIa a été également complétée à partir du site web du CBAT. $\mathrm{[Gr]}$  indique le groupe auquel la galaxie est gravitationnellement liée: $\mathrm{[G]} =$ Galaxy/Voie Lactée, $\mathrm{[A]} =$ M31,  $\mathrm{[L]} =$ Local Group, $\mathrm{[N]} =$ nearby neighbors. $\mathrm{[Type]}$ indique le type de la galaxie: $\mathrm{[S]} =$ Spiral, $\mathrm{[dSph]} =$ dwarf Spheroidal, $\mathrm{[Irr]} =$ Irregular,  $\mathrm{[dIrr]} =$ dwarf Irregular. La distance de la galaxie ($D$) est indiquée en kpc, ainsi que la méthode utilisée et la référence récente associée: rcs (Red Clump Stars), msf (Main Sequence Fitting), rtgb (Red Tip of Giant Branch), hbf (Horizontal Branch Fitting), rrl (RR Lyrae stars), tf (Tully-Fisher relation), cep (Cepheids), sbf (Surface Brightness Fluctuations), redshift pour les distances déduites à partir du décalage vers le rouge calculé sur le site web suivant https://dso-browser.com/. Les métallicités moyennes des galaxies du Groupe Local sont par défaut issues de \citet{McConnachie12} . On trouvera les méthodes utilisées, ainsi que les références associées dans ce papier. Pour les autres galaxies, des recherches au cas par cas ont été nécessaires. Toutes les galaxies dont la distance est inférieure à 6410 kpc ont des métallicités exprimées en [Fe/H] (sauf si une note indique le contraire). Pour les galaxies avec $d > 6410$ kpc, c'est à dire les galaxies hôtes de SNIa, il n'est pas possible de déduire la métallicité à partir de Céphéides individuelles. Les déterminations de métallicité sont basées sur l'étude des régions HII et donne accès à l'abondance de l'oxygène par rapport à l'hydrogène, et sont ainsi exprimées en 12+[O/H]. Pour la cohérence de la Table, j'ai fait une conversion pour ces galaxies en [O/H] en utilisant 12 + log(O/H) $= 8.69$ pour le Soleil. On fait alors l'hypothèse que [O/H] est équivalent ou relativement proche de [Fe/H], ce qui n'est pas forcément le cas.}}
\begin{center}
%\small
%\footnotesize
\tiny
\begin{tabular}{lccccccccc}
Name	&	Gr.	&	Type	&		D				&		&	Ref	&	[Fe/H]		&	Ref	&	SNIa	\\
 		&	 	&	 	&		kpc				&		&	 	&	[Sun]	&		&		\\
\hline	&		&		&						&		&		&		&		&		&		\\
La Voie Lactée	&	G  	&	S(B)bc   	&	$	0	_\mathrm{\pm	0}$	&		&	     	&	0	&	0	&		&		\\
Canis Major dSph                 	&	G  	&	dSph	&	$	7	_\mathrm{\pm	1}$	&	rcs	&\cite{bellazini06}&$	-0.50	_\mathrm{\pm	 0.20}$&\cite{McConnachie12}&		\\
Segue (I)                    	&	G  	&	dSph     	&	$	23	_\mathrm{\pm	2}$	&	msf	&\cite{belokurov07}&$	-2.72	_\mathrm{\pm	 0.40}$&\cite{McConnachie12}&		\\
Sagittarius dSph             	&	G  	&	dSph     	&	$	26	_\mathrm{\pm	2}$	&	rtgb	&\cite{monaco04}&$	-0.40	_\mathrm{\pm	 0.20}$&\cite{McConnachie12}&		\\
Ursa Major II                	&	G  	&	dSph     	&	$	32	_\mathrm{\pm	4}$	&	msf	&\cite{zucker06}&$	-2.47	_\mathrm{\pm	 0.06}$&\cite{McConnachie12}&		\\
Segue II                     	&	G  	&	dSph     	&	$	35	_\mathrm{\pm	2}$	&	hbf	&\cite{belokurov09}&$	-2.00	_\mathrm{\pm	 0.25}$&\cite{McConnachie12}&		\\
Willman 1                    	&	G  	&	dSph     	&	$	38	_\mathrm{\pm	7}$	&	msf	&\cite{willman06}&$	-2.10			$&\cite{McConnachie12}&		\\
Bootes II                    	&	G  	&	dSph     	&	$	42	_\mathrm{\pm	1}$	&	msf	&\cite{Walsh08}&$	-1.79	_\mathrm{\pm	 0.05}$&\cite{McConnachie12}&		\\
Coma Berenices               	&	G  	&	dSph     	&	$	44	_\mathrm{\pm	4}$	&	msf	&\cite{belokurov07}&$	-2.60	_\mathrm{\pm	 0.05}$&\cite{McConnachie12}&		\\
Bootes III                   	&	G  	&	dSph?    	&	$	47	_\mathrm{\pm	2}$	&	msf	&\cite{grillmair09}&$	-2.10	_\mathrm{\pm	 0.20}$&\cite{McConnachie12}&		\\
LMC                         	&	G  	&	Irr      	&	$	51	_\mathrm{\pm	2}$	&	{\it cf.} Tab.~\ref{Tab.Araucaria}	&	Araucaria Project	&$	-0.34	_\mathrm{\pm	0.15}$&\cite{luck98}&		\\
SMC                         	&	G  	&	dIrr     	&	$	64	_\mathrm{\pm	4}$	&	{\it cf.} Tab.~\ref{Tab.Araucaria}	&	Araucaria Project	&$	-0.68	_\mathrm{\pm	0.13}$&\cite{luck98}&		\\
Bootes (I)                   	&	G  	&	dSph     	&	$	66	_\mathrm{\pm	2}$	&	rrl	&\cite{dallora06}&$	-2.55	_\mathrm{\pm	 0.11}$&\cite{McConnachie12}&		\\
Draco                        	&	G  	&	dSph     	&	$	76	_\mathrm{\pm	6}$	&	rrl	&\cite{bonanos04}&$	-1.93	_\mathrm{\pm	 0.01}$&\cite{McConnachie12}&		\\
Ursa Minor                	&	G  	&	dSph     	&	$	76	_\mathrm{\pm	3}$	&	hbf	&\cite{carrera02}&$	-2.13	_\mathrm{\pm	 0.01}$&\cite{McConnachie12}&		\\
Sculptor                	&	G  	&	dSph     	&	$	86	_\mathrm{\pm	6}$	&	{\it cf.} Tab.~\ref{Tab.Araucaria} 	&	Araucaria Project	&$	-1.68	_\mathrm{\pm	 0.01}$&\cite{McConnachie12}&		\\
Sextans (I)                  	&	G  	&	dSph     	&	$	86	_\mathrm{\pm	4}$	&	sfh	&\cite{lee09}&$	-1.93	_\mathrm{\pm	 0.01}$&\cite{McConnachie12}&		\\
Ursa Major (I)               	&	G  	&	dSph     	&	$	97	_\mathrm{\pm	4}$	&	hbf	&\cite{okamoto08}&$	-2.18	_\mathrm{\pm	 0.04}$&\cite{McConnachie12}&		\\
Carina                     	&	G  	&	dSph     	&	$	105	_\mathrm{\pm	6}$	&	{\it cf.} Tab.~\ref{Tab.Araucaria}	&	Araucaria Project	&$	-1.72	_\mathrm{\pm	 0.01}$&\cite{McConnachie12}&		\\
Hercules                     	&	G  	&	dSph     	&	$	147	_\mathrm{\pm	8}$	&	hbf	&\cite{Aden09}&$	-2.41	_\mathrm{\pm	 0.04}$&\cite{McConnachie12}&		\\
Fornax                    	&	G  	&	dSph     	&	$	147	_\mathrm{\pm	12}$	&	{\it cf.} Tab.~\ref{Tab.Araucaria}	&	Araucaria Project	&$	-0.99	_\mathrm{\pm	 0.01}$&\cite{McConnachie12}&		\\
Leo IV                       	&	G  	&	dSph     	&	$	154	_\mathrm{\pm	6}$	&	rrl/dscu	&\cite{moretti09}&$	-2.54	_\mathrm{\pm	 0.07}$&\cite{McConnachie12}&		\\
Canes Venatici II            	&	G  	&	dSph     	&	$	160	_\mathrm{\pm	4}$	&	rrl	&\cite{creco08}&$	-2.21	_\mathrm{\pm	 0.05}$&\cite{McConnachie12}&		\\
Leo V                        	&	G  	&	dSph     	&	$	178	_\mathrm{\pm	10}$	&	hbf	&\cite{belokurov08}&$	-2.00	_\mathrm{\pm	 0.20}$&\cite{McConnachie12}&		\\
Pisces II                    	&	G  	&	dSph     	&	$	182			$	&	hbf	&\cite{belokurov10}&$	-1.90		     	$&\cite{McConnachie12}&		\\
Canes Venatici (I)           	&	G  	&	dSph     	&	$	218	_\mathrm{\pm	10}$	&	hbf	&\cite{martin08}&$	-1.98	_\mathrm{\pm	 0.01}$&\cite{McConnachie12}&		\\
Leo II                       	&	G  	&	dSph     	&	$	233	_\mathrm{\pm	14}$	&	rtgb	&\cite{bellazzini05}&$	-1.62	_\mathrm{\pm	 0.01}$&\cite{McConnachie12}&		\\
Leo I                        	&	G,L	&	dSph     	&	$	254	_\mathrm{\pm	15}$	&	rtgb	&\cite{bellazzini04}&$	-1.43	_\mathrm{\pm	 0.01}$&\cite{McConnachie12}&		\\
Phoenix                      	&	L,G	&	dIrr/dSph	&	$	415	_\mathrm{\pm	19}$	&	rtgb	&\cite{hidalgo09}&$	-1.37	_\mathrm{\pm	 0.20}$&\cite{McConnachie12}&		\\
Leo T                        	&	L,G	&	dIrr/dSph	&	$	417	_\mathrm{\pm	19}$	&	hbf	&\cite{irwin07}&$	-1.99	_\mathrm{\pm	 0.05}$&\cite{McConnachie12}&		\\
NGC 6822 / Barnard               	&	L,G	&	dIrr     	&	$	459	_\mathrm{\pm	17}$	&	{\it cf.} Tab.~\ref{Tab.Araucaria}&	Araucaria Project	&$	-0.5			$&\cite{hosek14} (a)&		\\
Andromeda XVI                	&	A,L	&	dSph     	&	$	525	_\mathrm{\pm	48}$	&	rtgb	&\cite{ibata07}&$	-2.10	_\mathrm{\pm	 0.20}$&\cite{McConnachie12}&		\\
Andromeda XXIV               	&	A  	&	dSph     	&	$	600	_\mathrm{\pm	33}$	&	hbf	&\cite{richardson11}&$	-1.80	_\mathrm{\pm	 0.20}$&\cite{McConnachie12}&		\\
NGC 185                      	&	A  	&	dE/dSph  	&	$	617	_\mathrm{\pm	26}$	&	rtgb	&\cite{McConnachie05}&$	-1.30	_\mathrm{\pm	 0.10}$&\cite{McConnachie12}&		\\
Andromeda XV                 	&	A  	&	dSph     	&	$	631	_\mathrm{\pm	58}$	&	rtgb	&\cite{ibata07}&$	-1.80	_\mathrm{\pm	 0.20}$&\cite{McConnachie12}&		\\
Andromeda II                 	&	A  	&	dSph     	&	$	652	_\mathrm{\pm	18}$	&	rtgb	&\cite{McConnachie05}&$	-1.64	_\mathrm{\pm	 0.04}$&\cite{McConnachie12}&		\\
Andromeda XXVIII             	&	A,L	&	dSph?    	&	$	661	_\mathrm{\pm	152}$	&	rtgb	&\cite{slater11}&$	     		     	$&\cite{McConnachie12}&		\\
NGC 147                      	&	A  	&	dE/dSph  	&	$	676	_\mathrm{\pm	28}$	&	rtgb	&\cite{McConnachie05}&$	-1.10	_\mathrm{\pm	 0.10}$&\cite{McConnachie12}&		\\
Andromeda X                  	&	A  	&	dSph     	&	$	701	_\mathrm{\pm	68}$	&	hbf	&\cite{zucker07}&$	-1.93	_\mathrm{\pm	 0.11}$&\cite{McConnachie12}&		\\
Andromeda XXIX               	&	A  	&	dSph     	&	$	731	_\mathrm{\pm	74}$	&	rtgb	&\cite{bell11}&$	-1.80		     	$&\cite{McConnachie12}&		\\
Andromeda XIV                	&	A,L	&	dSph     	&	$	735	_\mathrm{\pm	112}$	&	rtgb	&\cite{majewski07}&$	-2.26	_\mathrm{\pm	 0.05}$&\cite{McConnachie12}&		\\
Andromeda I                  	&	A  	&	dSph     	&	$	745	_\mathrm{\pm	24}$	&	rtgb	&\cite{McConnachie05}&$	-1.45	_\mathrm{\pm	 0.04}$&\cite{McConnachie12}&		\\
Andromeda III                	&	A  	&	dSph     	&	$	748	_\mathrm{\pm	24}$	&	rtgb	&\cite{McConnachie05}&$	-1.78	_\mathrm{\pm	 0.04}$&\cite{McConnachie12}&		\\
IC 1613                   	&	L  	&	dIrr     	&	$	755	_\mathrm{\pm	42}$	&	{\it cf.} Tab.~\ref{Tab.Araucaria} 	&	Araucaria Project	&$	-0.79			$&\cite{hosek14} (a)&		\\

\end{tabular}
\normalsize
\end{center}
\end{table*}

\begin{table*}
\caption{\label{Tab.Ho}  Suite du Tableau.}
\begin{center}
%\small
%\footnotesize
\tiny
\begin{tabular}{lccccccccc}
Name	&	Gr.	&	Type	&		D				&		&	Ref	&	[Fe/H]		&	Ref	&	SNIa	\\
 		&	 	&	 	&		kpc				&		&	 	&	[Sun]	&		&		\\
\hline	&		&		&						&		&		&		&		&		&		\\

Cetus                        	&	L  	&	dSph     	&	$	755	_\mathrm{\pm	24}$	&	rtgb	&\cite{McConnachie05}&$	-1.90	_\mathrm{\pm	 0.10}$&\cite{McConnachie12}&		\\
Andromeda XI                 	&	A  	&	dSph     	&	$	759	_\mathrm{\pm	175}$	&	rtgb	&\cite{collins10}&$	-2.00	_\mathrm{\pm	 0.20}$&\cite{McConnachie12}&		\\
Andromeda XXVI               	&	A  	&	dSph     	&	$	762	_\mathrm{\pm	42}$	&	rtgb	&\cite{richardson11}&$	-1.90	_\mathrm{\pm	 0.20}$&\cite{McConnachie12}&		\\
Andromeda VII                	&	A  	&	dSph     	&	$	762	_\mathrm{\pm	35}$	&	rtgb	&\cite{McConnachie05}&$	-1.40	_\mathrm{\pm	 0.30}$&\cite{McConnachie12}&		\\
Andromeda IX                 	&	A  	&	dSph     	&	$	766	_\mathrm{\pm	25}$	&	rtgb	&\cite{McConnachie05}&$	-2.20	_\mathrm{\pm	 0.20}$&\cite{McConnachie12}&		\\
Andromeda XXIII              	&	A  	&	dSph     	&	$	769	_\mathrm{\pm	46}$	&	rtgb	&\cite{richardson11}&$	-1.80	_\mathrm{\pm	 0.20}$&\cite{McConnachie12}&		\\
LGS 3 (b)	&	A  	&	dIrr/dSph	&	$	769	_\mathrm{\pm	25}$	&	rtgb	&\cite{McConnachie05}&$	-2.10	_\mathrm{\pm	 0.22}$&\cite{McConnachie12}&		\\
Andromeda V                  	&	A  	&	dSph     	&	$	773	_\mathrm{\pm	28}$	&	rtgb	&\cite{McConnachie05}&$	-1.60	_\mathrm{\pm	 0.30}$&\cite{McConnachie12}&		\\
Andromeda /M31                   	&	A  	&	Sb       	&	$	783	_\mathrm{\pm	25}$	&	rtgb	&\cite{McConnachie05}&$	-0.04		     	$&\cite{hosek14} (a)&		\\
Andromeda VI                 	&	A  	&	dSph     	&	$	783	_\mathrm{\pm	25}$	&	rtgb	&\cite{McConnachie05}&$	-1.30	_\mathrm{\pm	 0.14}$&\cite{McConnachie12}&		\\
Andromeda XVII               	&	A  	&	dSph     	&	$	794	_\mathrm{\pm	37}$	&	rtgb	&\cite{irwin08}&$	-1.90	_\mathrm{\pm	 0.20}$&\cite{McConnachie12}&		\\
Andromeda XXII               	&	A  	&	dSph     	&	$	794			$	&	rtgb	&\cite{martin09}&$	-1.80		     	$&\cite{McConnachie12}&		\\
IC 10                        	&	A  	&	dIrr     	&	$	794	_\mathrm{\pm	44}$	&	tf	&\cite{tully06}&$	-1.28		     	$&\cite{McConnachie12}&		\\
Leo A                        	&	L  	&	dIrr     	&	$	798	_\mathrm{\pm	44}$	&	rrl	&\cite{dolphin02}&$	-1.40	_\mathrm{\pm	 0.20}$&\cite{McConnachie12}&		\\
Andromeda XX                 	&	A  	&	dSph     	&	$	802	_\mathrm{\pm	273}$	&	rtgb	&\cite{mcconnachie08}&$	-1.50	_\mathrm{\pm	 0.10}$&\cite{McConnachie12}&		\\
M32                          	&	A  	&	cE       	&	$	805	_\mathrm{\pm	78}$	&	rrl	&\cite{fiorentino10}&$	-0.25		     	$&\cite{McConnachie12}&		\\
Triangulum / M33             	&	A  	&	Sc       	&	$	809	_\mathrm{\pm	22}$	&	{\it cf.} Tab.~\ref{Tab.Araucaria} 	&	Araucaria Project	&$	-0.15		     	$&\cite{hosek14} (a)&		\\
Andromeda XXV                	&	A  	&	dSph     	&	$	813	_\mathrm{\pm	45}$	&	rtgb	&\cite{richardson11}&$	-1.80	_\mathrm{\pm	 0.20}$&\cite{McConnachie12}&		\\
NGC 205                      	&	A  	&	dE/dSph  	&	$	824	_\mathrm{\pm	27}$	&	rtgb	&\cite{McConnachie05}&$	-0.80	_\mathrm{\pm	 0.20}$&\cite{McConnachie12}&		\\
Andromeda XXVII              	&	A  	&	dSph     	&	$	828	_\mathrm{\pm	46}$	&	rtgb	&\cite{richardson11}&$	-1.70	_\mathrm{\pm	 0.20}$&\cite{McConnachie12}&		\\
Andromeda XXI                	&	A  	&	dSph     	&	$	859	_\mathrm{\pm	51}$	&	rtgb	&\cite{martin09}&$	-1.80	_\mathrm{\pm	 0.20}$&\cite{McConnachie12}&		\\
Andromeda XII                	&	A,L	&	dSph     	&	$	871	_\mathrm{\pm	120}$	&	rtgb	&\cite{collins10}&$	-2.10	_\mathrm{\pm	 0.20}$&\cite{McConnachie12}&		\\
Tucana                       	&	L  	&	dSph     	&	$	887	_\mathrm{\pm	49}$	&	rrl	&\cite{bernard09}&$	-1.95	_\mathrm{\pm	 0.15}$&\cite{McConnachie12}&		\\
Andromeda XIII               	&	A  	&	dSph     	&	$	912	_\mathrm{\pm	42}$	&	rtgb	&\cite{collins10}&$	-1.90	_\mathrm{\pm	 0.20}$&\cite{McConnachie12}&		\\
Pegasus dIrr                 	&	L,A	&	dIrr/dSph	&	$	920	_\mathrm{\pm	30}$	&	rtgb	&\cite{McConnachie05}&$	-1.40	_\mathrm{\pm	 0.20}$&\cite{McConnachie12}&		\\
Andromeda XIX                	&	A  	&	dSph     	&	$	933	_\mathrm{\pm	56}$	&	rtgb	&\cite{mcconnachie08}&$	-1.90	_\mathrm{\pm	 0.10}$&\cite{McConnachie12}&		\\
WLM  (c)	&	L  	&	dIrr     	&	$	933	_\mathrm{\pm	34}$	&	{\it cf.} Tab.~\ref{Tab.Araucaria} 	&	Araucaria Project	&$	-0.87			$&\cite{hosek14} (a)&		\\
Sagittarius dIrr             	&	L  	&	dIrr     	&	$	1067	_\mathrm{\pm	88}$	&	rtgb	&\cite{momamy02}&$	-2.10	_\mathrm{\pm	 0.20}$&\cite{McConnachie12}&		\\
Aquarius                     	&	L  	&	dIrr/dSph	&	$	1072	_\mathrm{\pm	39}$	&	rtgb	&\cite{McConnachie05}&$	-1.30	_\mathrm{\pm	 0.20}$&\cite{McConnachie12}&		\\
NGC 3109              	&	N  	&	dIrr     	&	$	1300	_\mathrm{\pm	48}$	&	{\it cf.} Tab.~\ref{Tab.Araucaria} &	Araucaria Project	&$	-0.67	_\mathrm{\pm	0,13}$&\cite{hosek14} (a)&		\\
Antlia                       	&	N  	&	dIrr     	&	$	1349	_\mathrm{\pm	62}$	&	tf	&\cite{tully06}&$	-1.60	_\mathrm{\pm	 0.10}$&\cite{McConnachie12}&		\\
Andromeda XVIII              	&	L  	&	dSph     	&	$	1355	_\mathrm{\pm	81}$	&	rtgb	&\cite{mcconnachie08}&$	-1.80	_\mathrm{\pm	 0.10}$&\cite{McConnachie12}&		\\
UGC 4879                     	&	L  	&	dIrr/dSph	&	$	1361	_\mathrm{\pm	25}$	&	rtgb	&\cite{kopylov08}&$	-1.50	_\mathrm{\pm	 0.20}$&\cite{McConnachie12}&		\\
Sextans B                    	&	N  	&	dIrr     	&	$	1426	_\mathrm{\pm	20}$	&	tf	&\cite{tully06}&	     		     			&		\\
Sextans A                    	&	N  	&	dIrr     	&	$	1432	_\mathrm{\pm	53}$	&	tf	&\cite{tully06}&$	-1.00		     	$&\cite{hosek14} (a)&		\\

HIZSS 3(A)                   	&	N  	&	(d)Irr?  	&	$	1675	_\mathrm{\pm	108}$	&	rtgb	&\cite{silva05}&	     		     			&		\\
HIZSS 3B                     	&	N  	&	(d)Irr?  	&	$	1675	_\mathrm{\pm	108}$	&	rtgb	&\cite{silva05}&	     		     			&		\\
KKR 25                       	&	N  	&	dIrr/dSph	&	$	1905	_\mathrm{\pm	61}$	&	tf	&\cite{tully06}&$	-2.10	_\mathrm{\pm	 0.30}$&\cite{McConnachie12}&		\\
E410-005 / KK3   	&	S	&	?	&	$	1920			$	&	rtgb	&\cite{kara00}&						&		\\
E294-010       	&	S	&	?	&	$	1920			$	&	rtgb	&\cite{karachentsev02}&						&		\\
ESO 410 / G 005               	&	N  	&	dIrr/dSph	&	$	1923	_\mathrm{\pm	35}$	&	tf	&\cite{tully06}&$	-1.93	_\mathrm{\pm	 0.20}$&\cite{McConnachie12}&		\\
NGC 55                      	&	S	&	Irr      	&	$	1932	_\mathrm{\pm	107}$	&	{\it cf.} Tab.~\ref{Tab.Araucaria}	&	Araucaria Project	&$	-0.37	_\mathrm{\pm	0.03}$&\cite{kudritzki16}&		\\
IC 5152                      	&	S	&	dIrr     	&	$	1950	_\mathrm{\pm	45}$	&	tf	&\cite{tully06}&			     			&		\\
ESO 294 / G 010               	&	N  	&	dIrr/dSph	&	$	2032	_\mathrm{\pm	37}$	&	tf	&\cite{tully06}&$	-1.48	_\mathrm{\pm	 0.17}$&\cite{McConnachie12}&		\\
NGC 300               	&	N  	&	Sc       	&	$	2080	_\mathrm{\pm	57}$	&	{\it cf.} Tab.~\ref{Tab.Araucaria}	&	Araucaria Project	&$	-0.36			$&\cite{kudritzki08}&		\\
GR 8                         	&	N  	&	dIrr     	&	$	2178	_\mathrm{\pm	120}$	&	tf	&\cite{tully06}&	     		     			&		\\
KKR 3                        	&	N  	&	dIrr     	&	$	2188	_\mathrm{\pm	121}$	&	tf	&\cite{tully06}&$	-2.02	_\mathrm{\pm	 0.25}$&\cite{McConnachie12}&		\\
UKS 2323-326                 	&	N  	&	dIrr     	&	$	2208	_\mathrm{\pm	92}$	&	tf	&\cite{tully06}&$	-1.68	_\mathrm{\pm	 0.19}$&\cite{McConnachie12}&		\\
UA438          	&	S	&		&	$	2230			$	&	rtgb	&\cite{karachentsev02}&						&		\\
IC 3104                      	&	N  	&	dIrr     	&	$	2270	_\mathrm{\pm	188}$	&	rtgb	&\cite{karachentsev02}&	     		     			&		\\
UGC 9128                     	&	N  	&	dIrr     	&	$	2291	_\mathrm{\pm	42}$	&	tf	&\cite{tully06}&$	-2.33	_\mathrm{\pm	 0.24}$&\cite{McConnachie12}&		\\
IC 4662                      	&	N  	&	dIrr     	&	$	2443	_\mathrm{\pm	191}$	&	rtgb	&\cite{Karachentsev06}&$	-1.34	_\mathrm{\pm	 0.13}$&\cite{McConnachie12}&		\\
KKH 98                       	&	N  	&	dIrr     	&	$	2523	_\mathrm{\pm	105}$	&	tf	&\cite{tully06}&$	-1.94	_\mathrm{\pm	 0.25}$&\cite{McConnachie12}&		\\
UGC 8508                     	&	N  	&	dIrr     	&	$	2582	_\mathrm{\pm	36}$	&	rtgb	&\cite{Dalcanton09}&$	-1.91	_\mathrm{\pm	 0.19}$&\cite{McConnachie12}&		\\
DDO 125                      	&	N  	&	dIrr     	&	$	2582	_\mathrm{\pm	59}$	&	rtgb	&\cite{Dalcanton09}&$	-1.73	_\mathrm{\pm	 0.17}$&\cite{McConnachie12}&		\\
KKH 86                       	&	N  	&	dIrr     	&	$	2582	_\mathrm{\pm	190}$	&	rtgb	&\cite{Dalcanton09}&$	-2.33	_\mathrm{\pm	 0.29}$&\cite{McConnachie12}&		\\
DDO 99                       	&	N  	&	dIrr     	&	$	2594	_\mathrm{\pm	167}$	&	rtgb	&\cite{Dalcanton09}&$	-2.13	_\mathrm{\pm	 0.22}$&\cite{McConnachie12}&		\\
DDO 190                      	&	N  	&	dIrr     	&	$	2793	_\mathrm{\pm	39}$	&	rtgb	&\cite{Dalcanton09}&$	-2.00	_\mathrm{\pm	 0.08}$&\cite{McConnachie12}&		\\
NGC 4163                     	&	N  	&	dIrr     	&	$	2858	_\mathrm{\pm	39}$	&	rtgb	&\cite{Dalcanton09}&$	-1.65	_\mathrm{\pm	 0.15}$&\cite{McConnachie12}&		\\
DDO 113                      	&	N  	&	dIrr     	&	$	2951	_\mathrm{\pm	82}$	&	rtgb	&\cite{Dalcanton09}&$	-1.99	_\mathrm{\pm	 0.21}$&\cite{McConnachie12}&		\\
UGCA 86                      	&	N  	&	dIrr     	&	$	2965	_\mathrm{\pm	232}$	&	rtgb	&\cite{Karachentsev06}&	     		     			&		\\
DDO6           	&	S	&	dIrr	&	$	3340			$	&	rtgb	&\cite{kara03}&$	-2.08			$&\cite{kara03}&		\\
KDG2,E540-030  	&	S	&	dSph	&	$	3400			$	&	rtgb	&\cite{kara03}&$	-1.61			$&\cite{kara03}&		\\
M83	&	M83	&	S	&	$	3420	_\mathrm{\pm	250}$	&	cep	&\cite{ferrarese07}&$	0.30			$&\cite{calzetti99}&	SN1983N (j)	\\
E540-032 / FG24  	&	S	&	dSph/dIrr	&	$	3420			$	&	rtgb	&\cite{kara03}&$	-1.45			$&\cite{kara03}&		\\
M81	&	M81	&	S	&	$	3500	_\mathrm{\pm	200}$	&	{\it cf.} Tab.~\ref{Tab.Araucaria} 	&	Araucaria Project	&$	0.06	_\mathrm{\pm	0.15}$   (d)  & \cite{kennicutt98}& 	\\
M82 / NGC 3034	&	M81	&	Irr	&	$	3680,981595			$	&	cep	&\cite{shappee11}&$	-1.15	_\mathrm{\pm	0.11}$ (e) &\cite{durrell10}&	SN 2014J  (j)		\\
NGC 7793    	&	S	&	S	&	$	3910			$	&	{\it cf.} Tab.~\ref{Tab.Araucaria} 	&	Araucaria Project	&$	-1.22			$&\cite{kara03}&		\\
NGC 253	&	S	&	Sc	&	$	3940			$	&	rtgb	&\cite{kara03}&$	-1.12			$&\cite{kara03}&		\\
N625	&	S	&	Irr	&	$	4070			$	&	tf	&\cite{tully06}&						&		\\
E349-031 / SDIG  	&	S	&	dG	&	$	4100			$	&	tf	&\cite{tully06}&						&		\\
NGC 247 	&	S	&	S	&	$	4100			$	&	{\it cf.} Tab.~\ref{Tab.Araucaria} 	&	Araucaria Project	&$	-1.0			$&\cite{garcia08}&		\\
NGC 5253	&	M83	&	dIrr?	&	$	4092	_\mathrm{\pm	115}$	&	cep	&\cite{sandage94}&$	-0.54		(d)	$&\cite{kennicutt98}&	SN1972E  (j)	 	\\
Sc22           	&	S	&	dSph     	&	$	4210			$	&	rtgb	&\cite{kara03}&$	-1.51			$&\cite{kara03}&		\\
UA442          	&	S	&	Im	&	$	4270			$	&	rtgb	&\cite{kara03}&$	-2.40			$&\cite{kara03}&		\\
E245-05        	&	S	&	Irr	&	$	4430			$	&	rtgb	&\cite{kara03}&$	-2.08			$&\cite{kara03}&		\\
IC 4182	&	S	&	S	&	$	4943	_\mathrm{\pm	185}$	&	cep	&\cite{sandage92}&$	-1.96			$&\cite{kara03}&	SN1937C  (j)		\\
DDO226         	&	S	&	dIrr	&	$	4920			$	&	sbf	&\cite{jerjen98}&						&		\\
NGC 59	&	S	&	S	&	$	5300			$	&	sbf	&\cite{jerjen98}&						&		\\
E149-003       	&	S	&	?	&	$	6400			$	&	tf	&\cite{tully06}&						&		\\
\hline
\end{tabular}
\end{center}
\normalsize
\end{table*}

\begin{table*}
\caption{\label{Tab.Ho}  Suite du Tableau.}
\begin{center}
%\small
%\footnotesize
\tiny
\begin{tabular}{lccccccccc}
Name	&	Gr.	&	Type	&		D				&		&	Ref	&	[Fe/H]		&	Ref	&	SNIa	\\
 		&	 	&	 	&		kpc				&		&	 	&	[Sun]	&		&		\\
\hline	&		&		&						&		&		&		&		&		&		\\
M101	&		&		&	$	6420	_\mathrm{\pm	550}$	&	cep	&\cite{shappee11}&$	-0.05			$& \cite{shappee11}&	SN2011fe	\\
M66 (f) (g) 	&		&	S	&	$	10560	_\mathrm{\pm	600}$	&	rtgb	&\cite{lee13}&			&			&	SN 1989B  (j)	 	\\
M96	(g) &		&	S	&	$	10710	_\mathrm{\pm	610}$	&	rtgb	&\cite{lee13}&					&	&	SN 1998bu  (j)		\\
NGC 3972	&		&	S	&	$	12260			$	&	redshift	&	DSOB (i)	&			&			&	SN2011By	\\
NGC 4536	&		&	S	&	$	14900			$	&	redshift	&	DSOB (i)	&$	0.16	_\mathrm{\pm	0.20}$&\cite{Macri06}&	SN1981B	\\
NGC 3447	&		&	S	&	$	15000			$	&	redshift	&	DSOB (i)	&			&			&	SN2012ht	\\
NGC 4526	&		&	S	&	$	15600			$	&	redshift	&	DSOB (i)	&			&			&	SN1994D (k)	\\
NGC 1448	&		&	S	&	$	16260			$	&	redshift	&	DSOB (i)	&			&			&	SN2001el	\\
NGC 4424	&		&	S	&	$	16260			$	&	redshift	&	DSOB (i)	&			&			&	SN2012cg	\\
NGC 7250	&		&	Irr	&	$	16260			$	&	redshift	&	DSOB (i)	&			&			&	SN2013dy	\\
NGC 1365	&		&	S	&	$	18300	_\mathrm{\pm	1700}$	&	cep	&\cite{silbermann99}&$	0.27	_\mathrm{\pm	0.20}$&\cite{kennicutt98}&	SN2012fr	\\
Fornax A / NGC 1316	&		&	E	&	$	20800	_\mathrm{\pm	500}$	&	sbf	&\cite{cantiello13}&			&			&	SN1980N, SN2006dd, SN1981D  (k)		\\
NGC 4753	&		&	S	&	$	19900			$	&	redshift	&	DSOB (i)	&$	0.07	_\mathrm{\pm	0.05}$&\cite{Macri06}&	SN 1983G  (j)		\\
NGC 3982	&		&	S	&	$	20500	_\mathrm{\pm	1700}$	&	cep	&\cite{stetson01}&$	0.07			$&\cite{riess09a}&	SN1998aq	\\
NGC 2442	&		&	S?	&	$	20900			$	&	redshift	&	DSOB (i)	&			&			&	SN2015F	\\
NGC 4639	&		&	S	&	$	25500	_\mathrm{\pm	2500}$	&	cep	&\cite{saha97}&$	0.31	_\mathrm{\pm	0.20}$&\cite{Macri06}&	SN1990N	\\
NGC 4038 / NGC 4039 (h)	&		&	S	&	$	21580	_\mathrm{\pm	1190}$	&	rtgb	&\cite{jang15}&$	0.07	_\mathrm{\pm	0.03}$&\cite{lardo15}&	SN2007sr	\\
NGC 5584 (h) 	&		&	S	&	$	22490	_\mathrm{\pm	1240}$	&	rtgb	&\cite{jang15}&			&			&	SN2007af	\\
NGC 5917	&		&	?	&	$	26700			$	&	redshift	&	DSOB (i)	&			&			&	SN2005cf	\\
NGC 3021	&		&	S	&	$	27260	_\mathrm{\pm	420}$	&	rtgb	&\cite{Jang17a}&$	0.13			$&\cite{riess09a}&	SN1995al	\\
NGC 3370	&		&	S	&	$	28220	_\mathrm{\pm	540}$	&	rtgb	&\cite{Jang17a}&$	0.11	_\mathrm{\pm	0.05}$&\cite{Macri06}&	SN1994ae	\\
NGC 1309	&		&	S	&	$	31200	_\mathrm{\pm	580}$	&	rtgb	&\cite{Jang17a}&$	0.18			$&\cite{riess09a}&	SN2002fk	\\
UGC 9391	&		&	S	&	$	36100			$	&	redshift	&	DSOB (i)	&			&			&	SN2003du	\\
NGC 1015	&		&	S	&	$	37100			$	&	redshift	&	DSOB (i)	&			&			&	SN2009ig	\\
\hline
\end{tabular}
\end{center}
\tiny
Notes:

(a) {\it cf.} Table 10 de \cite{hosek14} pour une liste de galaxies avec leur métallicité et les références associées. 

(b)  LGS 3 = Local Group Suspect 3

(c) WLM =  Wolf-Lundmark-Melotte

(d) Pour M81 et NGC 5253, la métallicité est calculée à partir de l'abondance [O/H] donnée dans la Tab. 4 de  \cite{kennicutt98} 

(e) Pour M82, l'abondance correspond à [M/H] et non [Fe/H] \citep{durrell10}.

(f) M66 appartient  au triplet du Lion avec M65 et NGC 3628 

(g) Un résumé exhaustif et récent des estimations de distance de M66 et M96 se trouve dans \citet{lee13}, Tab. 3 et 4, respectivement. 

(h) Un résumé exhaustif et récent des estimations de distance de NGC4038/39 et NGC 5584 se trouve dans \citet{jang15}, Tab. 4 et 5, respectivement.

(i) La distance des galaxies lointaines est difficile à trouver dans la littérature et est donc calculée à partir du redshift en utilisant le site web suivant: https://dso-browser.com/. 

(j) SNIa rejetées dans le cadre de l'étude de \citet{riess16} pour différentes raisons (voir l'introduction du Chapitre~\ref{Chap_Araucaria}). 

(k) les 4 SNIa (SN1994D, SN1980N, SN2006dd, SN1981D) sont de qualités suffisantes pour être utilisées pour l'étalonnage des échelles de distance dans l'univers, mais ne disposent pas de détermination de distance indépendante suffisamment précise. 

\normalsize
\end{table*}

\newpage
\section{L'impact de la métallicité sur la relation \emph{PL}}\label{sect_plz}

Les propriétés physiques qui expliquent la relation entre la luminosité, la couleur d'une Céphéide et sa période sont bien comprises. En utilisant la loi de Stephan-Boltzmann,

\begin{equation}\label{Eq_Stephan}
L = 4 \pi R^2 \sigma T_\mathrm{eff}^4
\end{equation}

la luminosité L peut être calculée pour toutes les étoiles incluant les Céphéides. Exprimée en magnitude cette relation devient:

\begin{equation}\label{Eq_Sephan_mag}
M_\mathrm{bol} = -5 \log R - 10 \log T_\mathrm{eff} + C
\end{equation}

Or, la température effective $\log T_\mathrm{eff}$ peut être physiquement reliée à la couleur intrinsèque de l'étoile (i.e. (B-V)$_\mathrm{0}$), tandis le rayon dépend fondamentalement de la période de l'étoile, au travers d'une relation entre la période et la densité moyenne. Ainsi, dans sa forme linéaire pour les étoiles pulsantes, la relation de Stephan-Boltzmann peut prendre la forme d'une relation Période-Luminosité-Couleur $\emph{PLC}$ \citep{sandage58,sandage68}:

\begin{equation}\label{Eq_PLC}
M_\mathrm{V} = \alpha \log P + \beta (B-V)_\mathrm{0} + \gamma
\end{equation}

La pulsation de la Céphéide est possible du fait d'un changement d'opacité au niveau de la zone d'ionisation de l'hélium. Sur une portion du cycle de pulsation de l'étoile, cette couche d'ionisation est opaque aux radiations, ce qui génère une augmentation de chaleur et donc de pression, qui engendre à son tour une élévation des couches de gaz situées au dessus de la zone d'ionisation de l'hélium et finalement l'étoile se dilate. Puis, au fur et à mesure que le rayon de l'étoile augmente, la température de l'atmosphère diminue, et à un certain point, la couche d'hélium doublement ionisé se recombine et devient à nouveau transparente à la radiation, l'expansion s'arrête, l'étoile commence à se contracter et le cycle se répète. Ce cycle intervient seulement dans la bande d'instabilité ($\emph{IS}$ pour "Instability Strip") du diagramme HR. En effet, les étoiles les plus froides de la bande IS ont des cellules convectives qui s'enfoncent jusque dans la zone d'ionisation de l'hélium et qui perturbent le mécanisme d'opacité. A l'inverse, les Céphéides les plus chaudes de la bande IS ont une zone d'ionisation de l'hélium tellement haute dans l'atmosphère qu'elle ne permet pas de jouer le rôle de piston. De manière générale, les relations \emph{PLC} observées et théoriques s'accordent assez bien \citep{caputo08}. Cependant, l'atmosphère des Céphéides est affectée par un changement de l'abondance en métaux. Il en est de même de leur structure interne, ce qui affecte la relation entre la masse et le rayon. Ainsi on s'attend  à ce que la relation entre la couleur et la magnitude des Céphéides, ainsi que la relation période-luminosité (\emph{PL})  correspondante, dépende de la métallicité, que ce soit sa pente ou son point-zéro \cite{sandage08}. Cependant, prédire l'effet de métallicité sur la magnitude (et même son signe) s'est avéré extrêmement compliqué. Des études théoriques ont été menées: \citet{alibert99, sandage99, bono08, caputo08} et \cite{romaniello08, romaniello09}. Nous nous concentrerons ici sur les aspects empiriques. Deux tests ont été proposés pour tester la sensibilité de la relation \emph{PL} à la métallicité. 

Le premier test consiste à observer des Céphéides situées respectivement dans la partie interne et externe du disque d'une galaxie individuelle, de mesurer leur métallicité (ce qui donne un gradient de métallicité dans le disque de la galaxie) et de comparer les relations \emph{PL} obtenues (pente et point-zéro). Seulement, à un rayon donné de la galaxie, la métallicité est bien souvent déterminée à partir de l'abondance [O/H] et on fait alors l'hypothèse que les Céphéides ont la même métallicité, ce qui est loin d'être établi \citep{kudritski12, bresolin09}. Par ailleurs, on considère que d'autres facteurs n'entrent pas en jeu, comme une dépendance radiale du ‘‘crowding'', un changement de la loi d'extinction en fonction du rayon, etc... Cette approche a été adoptée pour M31 \citep{Freedman90}, M101 \citep{kennicutt98}, NGC 4258 \citep{Macri06} et M33 \citep{Scowcroft09}. 

Le second test consiste à comparer la distance d'une galaxie obtenue à l'aide de différents estimateurs (par exemple les Céphéides et les RTGB) pour ensuite chercher une corrélation avec la métallicité de la galaxie. Ainsi \citet{udalski01} ont comparé les distances des Céphéides et des RTGB dans la galaxie IC 1613 (de forte métallicité: [Fe/H]$=0.79$) et n'ont trouvé aucun effet de la métallicité: les distances obtenues étaient comparables. En revanche, en étudiant 10 galaxies contenant des Céphéides et des RTGB, \citet{sakai04}  ont trouvé que la différence de distance entre les Céphéides et les RTGB décroit effectivement vers zéro lorsque la métallicité de la galaxie augmente (on retrouve bien zéro dans le cas de IC 1613), mais augmente lorsque l'on se rapproche de la métallicité solaire et que l'effet est de l'ordre de $\delta (m-M) / \delta \mathrm{[O/H]} = -0.24 \pm 0.05$ mag.dex$^{-1}$.  Notons que \citet{gro14} ont trouvé un effet de $\delta (m-M) / \delta  \mathrm{[Fe/H]}= -0.8 \pm 0.3 $ mag.dex$^{-1}$ en utilisant seulement les Céphéides de la Voie Lactée et son gradient de métallicité, et $\delta (m-M) / \delta  \mathrm{[Fe/H]}= -0.27 \pm 0.8 $ mag.dex$^{-1}$ en considérant les Céphéides du LMC. Ceci dit, les choses ne sont pas si simples, car \cite{romaniello08}, en se basant sur des mesures de métallicités spectroscopiques (i.e. sur des étoiles individuelles et non des régions HII), trouvent un effet de la métallicité sur la relation \emph{PL} inverse à ce que nous venons de voir. Le tableau~\ref{Tab.metaux} tiré de \cite{romaniello08} et actualisé, fait une synthèse de l'impact de la métallicité sur la relation \emph{PL}. Même si ce sujet fait l'objet d'intenses débats et recherches, la valeur qui fait actuellement référence est celle de \citet{sakai04}.  Cela veut dire que si l'on utilise la relation \emph{PL} des Céphéides du LMC ([Fe/H]$=-0.34\pm0.15$, \cite{luck98} par exemple) pour déduire la distance d'une Céphéide se situant dans une galaxie hôte de SNIa de métallicité relativement élevée, similaire à celle de la Voie Lactée, alors son module de distance est diminué d'environ $ \Delta \mu = \mathrm{[Fe/H]_{LMC}} \Delta\mathrm{[O/H]} = 0.34*0.24 \simeq0.08$ magnitude, i.e. que la galaxie paraît  plus proche que ce qu'elle est en réalité. 

Ce que nous venons de voir concerne l'impact de la métallicité sur le point-zéro de la relation \emph{PL}, c'est-à-dire que la correction appliquée sur la magnitude est la même pour toutes les Céphéides, quelle que soit leur période. Mais on peut également se poser la question de l'impact de la métallicité sur la pente de la relation  \emph{PL}. Un exemple frappant se trouve dans la revue de \citet{jackson15}. Prenons le résultat de \cite{riess05} d'un côté avec $H_\mathrm{0}=73\pm4\pm5$\kms et de l'autre, le résultat de \cite{sandage06} avec $H_\mathrm{0}=63.3\pm1.9\pm5$\kms. Ces valeurs de $H_\mathrm{0}$ sont clairement incompatibles (voir également Fig.~\ref{Fig_Ho}), bien qu'elles reposent sur les mêmes galaxies de référence, c'est-à-dire: NGC 3370 (SN1994ae), NGC 3982 (SN1998aq), NGC 4639 (SN1990N) et  NGC 4536 (SN1981B). Dans le pire des cas, i.e. SN 1990N dans NGC 4639, les auteurs trouvent une différence de 20-25\% pour le module de distance: $\mu^0=31.74$ pour \cite{riess05} et $\mu^0=32.20$ pour \cite{sandage06}.  \citet{jackson15} indique que la différence ne vient pas de la relation \emph{PL} utilisée pour déterminer la distance de cette galaxie, mais de la façon dont la correction de la métallicité est effectuée. D'un côté, \cite{riess05} appliquent une correction identique à toutes les Céphéides (comme nous l'avons expliqué plus haut) avec une valeur de correction en terme de magnitude qui s'exprime de la manière suivante: $ \Delta \mu = -0.24 \Delta$[O/H]. De son côté, \cite{sandage06} considèrent une correction de la métallicité qui dépend de la période de la Céphéide: $ \Delta \mu = 1.67 (\log P - 0.933) \Delta$[O/H], où $\Delta$[O/H] est la métallicité des Céphéides observées dans NGC 4639 moins la métallicité du LMC. \citet{Macri06} ont montré que l'approche de  \cite{riess05} est probablement plus appropriée, mais la question reste épineuse et montre les difficultés liées à la métallicité \footnote{A titre de comparaison, il est maintenant établi que les étoiles pulsantes de type RR Lyrae présentent une relation période - métallicité - luminosité, qu'elle soit déterminée à partir d'amas globulaires \citep{sollima06,marengo17} ou du SMC \citep{muraveva15}. Les RR Lyrae permettent de déterminer des distances dans le groupe local jusqu'à quelques Mpc (voir Table~\ref{Tab.GL}) ce qui n'en fait pas une méthode compétitive. Cependant, elles sont bien plus utiles pour faire de l'archéologie galactique avec des études montrant les interactions entre le Voie Lactée et les galaxies naines du sagittaire \citep{alcock97,sesar17} ou encore la géométrie des nuages de Magellan \cite{jac16, jac17}.}. 

%Les premiers résultats obtenus dans le cadre du projet Araucaria ont montré que la pente de la relation \emph{PL} est indépendante de la métallicité sur l'intervalle allant de -1.0 à -0.2 dex dans les bandes $V$ et $K$ \cite{gieren05b}. L'extension de ce résultat à la métallicité de la Voie Lactée présenté dans \cite{gieren05} était alors basé sur une modification de la méthode de BW \citet{gieren97} (que nous présenterons dans la Sect.~\ref{Sect_BW}). 
Dans le cadre du projet Araucaria, nous avons utilisée récemment la méthode de Baade-Wesselink, ainsi qu'un nouvel étalonnage du facteur de projection (nous y reviendrons dans la Sect.~\ref{Chap_BW}), pour déterminer la distance de 70 Céphéides Galactiques et en déduire une relation \emph{PL} \citep{storm11a} (voir l'annexe \ref{storm11a}). De la même manière, nous avons établi une relation  \emph{PL}  basée sur l'observation de 36 Céphéides du LMC  \citep{storm11b} (voir l'annexe \ref{storm11b}). Il ressort de cette étude que 1) la pente de la relation \emph{PL} n'est pas sensible à la métallicité, 2) le point-zéro de la relation  \emph{PL} n'est pas sensible à la métallicité dans les bandes $V$, $J$ et $K$, mais 3) qu'un effet de  $0.23 \pm 0.10$ mag/dex est obtenu lorsque l'on considère l'index de Wesenheit\footnote{Si on adopte une loi d'extinction et qu'on l'applique à toutes les Céphéides, on peut utiliser les magnitudes et les couleurs observées pour en déduire l'extinction totale sur la ligne de visée. Ainsi, par exemple, si on a $V$ et $I$ (comme c'est souvent le cas pour les Céphéides observées par le HST) et qu'on fait l'hypothèse du rapport de l'absorption totale par rapport à une bande sélective $R_\mathrm{VI} = \frac{A_\mathrm{V}}{E(V-I)}$ \citep{cardelli89}, alors on obtient des magnitudes et couleurs corrigées de l'extinction, que l'on appelle magnitude Wesenheit \citep{madore82}. } (V, I). Ces résultats sont partiellement compatibles avec ceux obtenus par \cite{gro13}.

\begin{table*}
\caption{Impact de la métallicité sur le point zéro de la relation \emph{PL}. Tableau~\ref{Tab.metaux} tiré de \cite{romaniello08} et actualisé.}\label{Tab.metaux}  
\begin{center}
\tiny
\begin{tabular}{lccc}
$\frac{\delta \mu}{\delta \mathrm{[M/H]}}$	         &		&	Method  	&		Référence  \\
(mag/dex)	                                                                &		&	          	&		  		  \\
\hline
$-0.32 \pm 0.21$	& [Fe/H]   & Analyse de Céphéides dans 3 champs de M31 (bandes $BVRI$) &  \cite{Freedman90}	\\$-0.88 \pm 0.16$	& [Fe/H]   & Comparaison de Céphéides dans 3 champs de M31 et du LMC (bandes $BVRI$)  & \cite{gould94} 	\\
$-0.40 \pm 020$	& [O/H]   & Solution simultanée pour la distance pour 17 galaxies  (bandes $UBVRIJHK$) & \cite{kochanek97} 	\\										
$-0.44_\mathrm{-0.20}^{0.10}$ 	& [O/H]   & Comparaison des observations EROS des Céphéides du SMC et du LMC (bandes $VR$) & \cite{sasselov97} 	\\	
$-0.24 \pm 0.16$	& [O/H]   & Comparaison des observations du HST dans les champ interne et externe de M101 & \cite{kennicutt98} 	\\	
$-0.12 \pm 0.08$	& [O/H]   & Comparaison de 10 Céphéides Galactiques par la méthode RTGB & \cite{kennicutt98} 	\\$-0.20 \pm 0.20$	& [O/H]   & Valeur utilisée lors du HST ‘‘Key Projet''  & \cite{freedman01} 	\\				
0	& [Fe/H]   & Comparaison OGLE des Céphéides dans IC 1613 et MC  (bandes $VI$)  & \cite{udalski01} 	\\
0	& [O/H]   & Comparaison des méthodes PNLF et SBF  & \cite{ciardullo02} 	\\
$-0.24 \pm 0.05$	& [O/H]   & Comparaison de la distance de 17 RTGB  & \cite{sakai04} 	\\
$-0.21 \pm 0.19$	& [Fe/H]   & Méthode BW des Céphéides de la Galaxie et du SMC (bandes $VK$) & \cite{storm04a} 	\\
		$-0.23 \pm 0.19$	& [Fe/H]   & Méthode BW des Céphéides de la Galaxie et du SMC (bandes $I$) & \cite{storm04a} 	\\
$-0.29 \pm 0.19$	& [Fe/H]   & Méthode BW des Céphéides de la Galaxie et du SMC (indexe $W$) & \cite{storm04a} 	\\	
	$-0.27 \pm 0.08$	& [Fe/H]   & Compilation des distances et des métallicités de 53 Céphéides Galactiques et des nuages de Magellan & \cite{gro04} 	\\	
$-0.39 \pm 0.03$	& [Fe/H]   & Distance des galaxies hôtes de SNIa à l'aide des Céphéides & \cite{saha06} 	\\
$-0.29 \pm 0.09$	& [O/H]   & Céphéides dans NGC 4258 et gradient [O/H]  issu de \cite{zaritsky94} & \cite{Macri06} 	\\
$-0.10 \pm 0.03$	& [Fe/H]   & Moyenne  pondérée de Kennicutt, Macri \& Groenewegen & \cite{benedict07} 	\\
$-0.017 \pm 0.113$	& [O/H]   & Comparaison de la distance de 18 galaxies (Céphéides et RTGB) & \cite{Tammann08} 	\\
0	& [Fe/H]   & Comparaison des pentes entre les Céphéides Galactiques et du LMC & \cite{fouque07} 	\\
$0.05 \pm 0.03$	& [Fe/H]   & Relation théorique entre la période et l'index Wesenheit (V,I)  & \cite{bono08} 	\\	
$-0.17 \pm 0.31$      &  [O/H]   & Céphéides du LMC (0.45-8.0 $\mu$m)  &  \cite{Freedman11} \\
$-0.23 \pm 0.10$      &  [Fe/H]   & Distances BW des Céphéides de la Voie Lactée, du LMC et du SMC (Wesenheit (V,I))  &  \cite{storm11b} \\
0      &  [Fe/H]   & Distances BW des Céphéides de la Voie Lactée, du LMC et du SMC  (bandes $VJK$)  &  \cite{storm11b} \\
0      &  [Fe/H]   & Distances des Céphéides Galactiques HST et du LMC/SMC  (Wesenheit (V,I)) &  \cite{majaess11} \\
0      &  [Fe/H]   & Distances des Céphéides Galactiques HST et du LMC/SMC  (bande $K$ \& Wesenheit (V,K))  &  \cite{gro13} \\			
$-0.23 \pm 0.11$   &  [Fe/H]   & Distances des Céphéides Galactiques HST et du LMC/SMC  (bande $V$)  &  \cite{gro13} \\
$-0.33 \pm 0.12$ & [O/H] &  Céphéides dans M31  (bandes $VI$) & \cite{mager13} \\ 
\hline
\end{tabular}
\end{center}
%\tiny
%Notes: DEBS = Detached Eclipsing Binaries ; RCS = Red Clump Stars ; RRL = RR Lyrae stars ; CEP = Cepheids ; W VIR = W Vir stars (Type 2 Cepheids) ; RTGB = Red Tip of Giant Branch ; FGLR = Flux weighted Gravity--Luminosity Relationship ; 
\normalsize
\end{table*}

Pour résumer,  il semble admis actuellement par la plupart des chercheurs dans ce domaine que la pente de la relation \emph{PL} ne dépend pas de la métallicité dans l'infrarouge. En revanche, concernant l'impact de la métallicité sur le point-zéro de la relation \emph{PL}, la question reste ouverte et fait actuellement l'objet d'intenses recherches. Dans \cite{riess16}, l'erreur sur $H_\mathrm{0}$ due à la métallicité est estimée à 0.5\%\footnote{Il est intéressant de constater que dans leur dernière détermination de $H_\mathrm{0}$, \cite{riess16} ont directement ajusté la correction de la magnitude liée à la métallicité: il s'agit du paramètre  $Z_\mathrm{W}$ dans l'Eq.~2. Les valeurs obtenues de $\gamma$ sont ainsi des {\it outputs} et sont indiqués dans leur Table~8.} (leur Table 7).  Cette erreur tombe à zéro pour NGC4258, la galaxie hôte d'un mégamaser \cite{Humphreys13} dont la métallicité est proche de celle de la Voie Lactée. De même, il est également important de mentionner que {\it Gaia} fournira prochainement une relation \emph{PL} Galactique qui pourra être utilisée directement pour la détermination de distance des galaxies hôtes de SNIa, ce qui devrait réduire l'erreur liée à la métallicité sur $H_\mathrm{0}$, même si le problème de la détermination de la métallicité de ces céphéides lointaines reste difficile. Il est aussi utile de rappeler que des galaxies comme M31, M33, M81, M83 (étudiées dans le cadre du projet Araucaria) ont une métallicité très proche de celle de la Voie Lactée, ce qui en fait des cibles prioritaires, très intéressantes pour l'avenir. A plus long terme, la stratégie du projet Araucaria d'observer des galaxies aux métallicités très différentes devrait également permettre de mieux contraindre l'impact de la métallicité sur la relation \emph{PL} et ainsi fournir de nouvelles galaxies de références  pour la détermination de $H_\mathrm{0}$. En effet, la plus grosse contribution dans le budget d'erreur sur $H_\mathrm{0}$ vient de l'incertitude sur la distance des galaxies de références: le LMC et M31 pour environ 1.3\% \citep{riess16}. 

%\footnote{Les étoiles géantes rouge  de faible métallicité (pop II) ont une coupure bien définie dans leur luminosité, qui, en bande I, ne varie que faiblement avec d'autres paramètres comme l'âge, ce qui en fait d'excellent indicateurs de distance.} 

\section{Les binaires à éclipses et la distance du LMC}\label{s_EB_LMC}

\begin{figure}[htbp]
\begin{center}
\includegraphics[width=8cm]{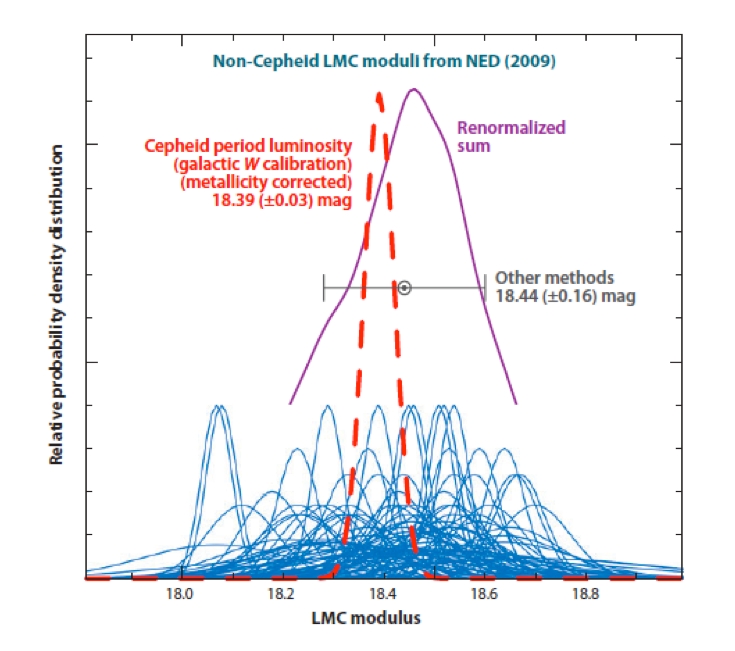}
\end{center}
\vspace*{-3mm} \caption{ \footnotesize Densités de probabilité
correspondant à l'estimation du module de distance du LMC obtenues à
partir de différentes méthodes. Le trait plein en violet correspond à la
résultante (figure tirée de \cite{freedman10}). Toutes méthodes
confondues, l'incertitude sur le module de distance du LMC est de
0.15 mag. La valeur retenue par les auteurs était de 18.5 $\pm$ 0.1
mag, ce qui entraînait alors une incertitude de 5\% sur $H_\mathrm{0}$. Actuellement, grâce à la distance du LMC déduite de la méthode des binaires à éclipses, la précision obtenue est de 2.2\% \citep{pietrzynski13} et l'impact sur $H_\mathrm{0}$ de 1.3\% \citep{riess16}. \vspace{0.2cm}}\label{Fig_LMC}
\end{figure}

Déterminer avec précision la distance de certaines galaxies de référence {\it par une méthode indépendante des Céphéides et si possible indépendante de la métallicité} est l'une des priorités du projet Araucaria. Cela permet à la fois de contraindre la relation \emph{PL} des Céphéides (utilisée ensuite pour la détermination de distance des SNIa) et d'estimer l'impact de la métallicité sur son  point-zéro (voir section précédente). Les galaxies de référence utilisées par \citet{riess16} se résument au LMC et à M31. 
Et, très clairement, la distance du LMC est actuellement la source d'erreur la plus importante sur l'étalonnage des échelles de distances dans l'univers et sur la détermination de $H_\mathrm{0}$ (1.3\% d'après la Table 7 de \cite{riess16}).   La Figure~\ref{Fig_LMC} montre les densités de probabilité correspondant à l'estimation du module de distance du LMC obtenues à partir de différentes méthodes jusqu'en 2010 \citep{freedman10}. Depuis 2010, la situation s'est bien améliorée grâce à la méthode des binaires à éclipses.

‘‘The Optical Gravitionnal Lensing Experiment'' OGLE\footnote{http ://ogle.astrouw.edu.pl/} \citep{udalski92} a observé environ 35 millions d'étoiles dans le champ du LMC pendant plus de 16 ans. En utilisant, ces données, notre groupe du projet Araucaria mené par G. Pietrzynski a détecté une douzaine de binaires à éclipses de masses intermédiaires avec des périodes orbitales s'échelonnant de 60 à 772 jours  \citep{pietrzynski13} (voir annexe~\ref{nature}). La distance du LMC est alors basée sur une formule assez simple: 

\begin{equation}\label{Eq_LMC}
d\mathrm{[pc]} = 1.337.10^{5} \frac{r\mathrm{[km]} }{\theta\mathrm{[mas]} }
\end{equation}

Le diamètre linéaire ($r$) peut être déterminé pour une binaire à éclipses détachée à partir de l'analyse de sa courbe de lumière et de son orbite spectroscopique avec une précision de l'ordre de 1\% \citep{andersen91}, tandis que le diamètre angulaire ($\theta$) peut être calculé à partir d'une relation brillance de surface - couleur, i.e. $ m_\mathrm{0}= S - 5 \log(\phi) $ où $S$ est la brillance de surface et  $m_\mathrm{0}$ est la magnitude intrinsèque dans une bande photométrique. La méthode fut décrite et utilisée pour la première fois par \cite{lacy77}. La principale source d'erreur de cette méthode provient quasi-totalement de notre méconnaissance de la relation brillance de surface - couleur. Celle-ci est relativement bien connue pour les étoiles tardives (à 2\%), mais très incertaine pour les étoiles de type spectral ABO (à 6-8\%). Nous reviendrons sur ce point dans le chapitre~\ref{Chap_EBs}. Ainsi la méthode est affectée par les sources d'incertitudes suivantes (voir \citet{pietrzynski13}): la précision sur l'amplitude des courbes de vitesses radiales ($K_\mathrm{1}$ et $K_\mathrm{2}$, environ 0.5\%), la dimension absolue (0.5\%), l'inclinaison (0.2\%), le rayon relatif (1\%), le point-zéro de la photométrie visible (0.6\%) et infrarouge (0.8\%), le rougissement (seulement 0.8\%, puisque que la relation brillance de surface - couleur est quasi-parallèle aux effets liés au rougissement \citep{barnes76}), et enfin l'erreur liée à l'étalonnage de la relation brillance de surface - couleur ~(2\%). Par ailleurs, la méthode est très peu sensible à la métallicité avec une correction de seulement 0.007 magnitude entre les binaires à éclipses de la Voie Lactée et du LMC  \citep{Thompson01, dibenedetto05}. En combinant ces incertitudes quadratiquement, nous avons déterminé une distance au LMC à 2.2\% de précision \citet{pietrzynski13}. Cette distance a été utilisée par \citet{riess16} et a fortement contribué à obtenir une précision de 2.4\% sur $H_\mathrm{0}$.

\begin{figure}[htbp]
\begin{center}
\includegraphics[width=6cm]{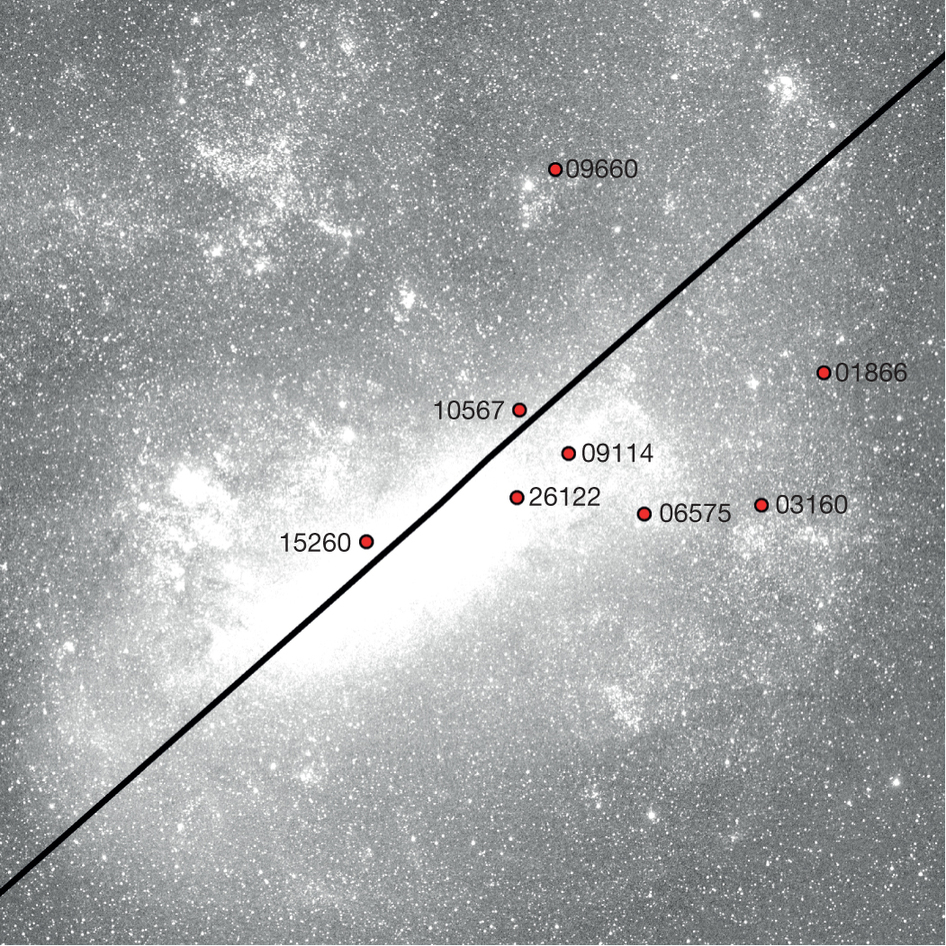}
\includegraphics[width=6cm]{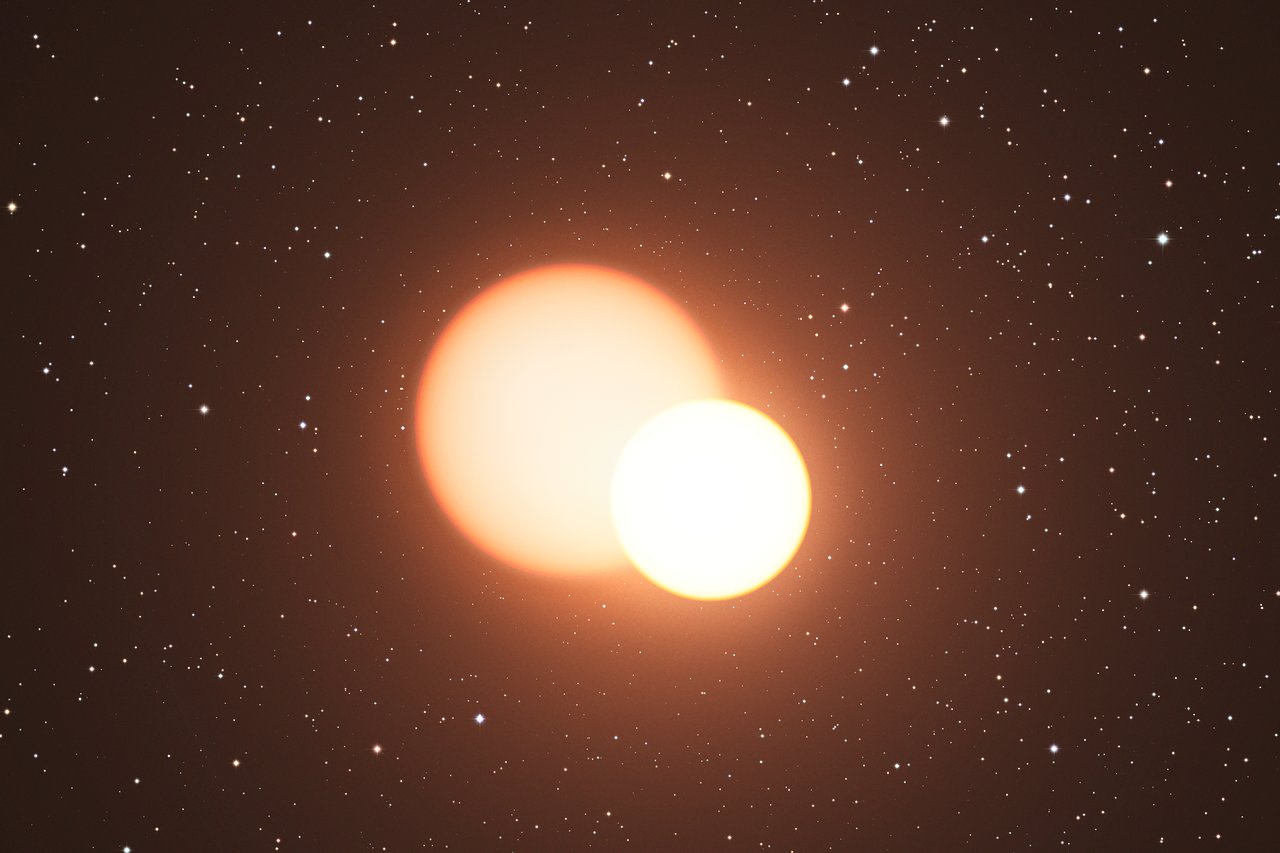}
\end{center}
\vspace*{-5mm} \caption{ \footnotesize   A gauche, la position des binaires à éclipses observées dans le LMC par \citet{pietrzynski13}. A droite, figure d'artiste d'un système double à éclipses. Plusieurs études ont montré que le LMC est incliné par rapport au plan du ciel d'environ 35 degrés (son inclinaison serait nulle s'il était vu de face) \citep{olsen02,vdm02,nikolaev04}. Les binaires à éclipses observées étaient proches du centre du LMC et de la ligne des noeuds, ce qui n'affecte pas l'estimation de la distance. Cependant, la situation est maintenant plus complexe depuis l'étude détaillée de \cite{jac16, jac17} de la géométrie et de la profondeur du LMC déduite de l'observation des Céphéides et des RR Lyrae avec OGLE \citep{soszynski15,soszynski16}.}
\end{figure}

Les binaires à éclipses ont jusqu'à présent été utilisées pour déterminer la distance de quelques galaxies jusqu'à environ 1 Mpc: LMC (2\%), SMC (3\%), M31 (4\%) et M33 (6\%). Afin d'obtenir de meilleures précisions et déterminer la distance de galaxies au-delà de 1 Mpc, il est nécessaire d'observer des binaires à éclipses de type précoces (plus brillantes) et donc d'étalonner la relation brillance de surface - couleur pour ces étoiles. Ce point est abordé dans le chapitre \ref{Chap_EBs}.  

\begin{table*}
\caption{Les galaxies dont la distance a été déterminée par la méthode des binaires à éclipses (voir aussi Tab.~\ref{Tab.Araucaria})}
\label{Tab_modeles}
\small
\begin{center}
\begin{tabular}{|l|c|c|c|c|c|c|}
\hline
   galaxie           & distance                     & reference             \\
                           &  $\mathrm{kpc \pm \sigma_{stat} \pm \sigma_{syst}} $    &                                 \\
                           \hline

Sagittarus Arm  &  $2.14 \pm 0.06 \pm 0.05 $ & \cite{Suchomska15} \\
  \hline
            LMC                       &  $50.30 \pm  0.53$  &   \cite{elgueta16} \\
            LMC                      &  $50.0 \pm 0.2 \pm 1.1$        &    \cite{pietrzynski13}         \\
            
            LMC &   $50.6 \pm 1.6 $   &   \cite{bonanos11}  \\
                        LMC                      &  $50.1 \pm 1.4$        &    \cite{pietr09b}         \\
                        LMC  &  $43.2 \pm 1.8$  & \cite{fitzpatrick03}  \\
%LMC    &  $43.2 \pm 1.8$   & \cite{fitzpatrick03} \\
LMC    &    $47.5\pm 1.8$  &  \cite{ribas02}  \\
                                        LMC   &               $50.7 \pm 1.2$  & \cite{fitzpatrick02} \\ 
                    LMC                    &   $47.9 \pm 1.6$       &    \cite{Nelson00}  \\
                    LMC   &               $45.7 \pm 1.6$  & \cite{guinan98} \\ 
                              LMC                    &   $41.7 \pm 6.2$       &    \cite{bell91}  \\
                            
 \hline                             
          
          SMC                     &   $62.1 \pm 1.9$       &    \cite{graczyk14}  \\
            SMC                     &   $58.3 \pm 0.5 \pm 1.3$       &    \cite{graczyk12}  \\
          SMC    &   	$60.6 \pm 1.0 \pm 2.8 $	           &   \cite{hilditch05} \\
            SMC      & $ 60.0 \pm 1.1 \pm 2.8 $   & \cite{harries03}  \\
            SMC                     &   $52.5 \pm 7.7$       &    \cite{bell93}  \\
\hline
            M31    &  $744 \pm 33 $  & \cite{vilardell10}  \\
            M31    & $772 \pm 44 $   & \cite{ribas05}\\
            
\hline
	M33  &  $964 \pm 54 $ & \cite{bonanos06b}  \\

\hline
\end{tabular}
\end{center}
\normalsize
\end{table*}

\newpage 
\section{Impact de la binarité et de l'environnement des Céphéides sur la relation \emph{PL}}\label{s_CSEPL}

\begin{figure}[htbp]
\begin{center}
\includegraphics[width=9cm]{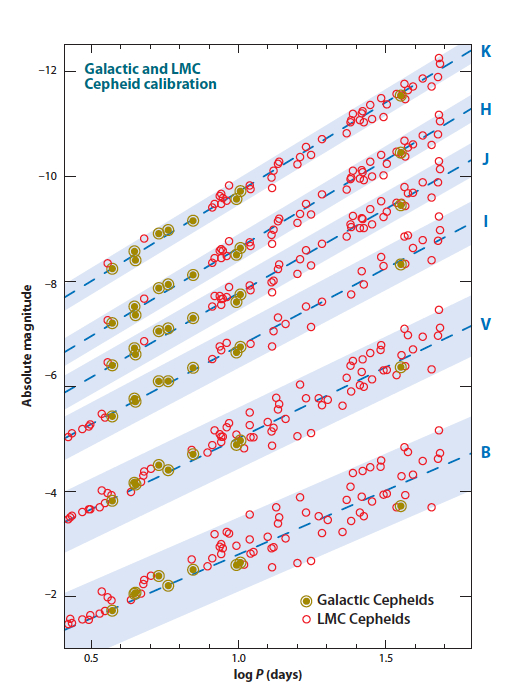}
\includegraphics[width=7cm]{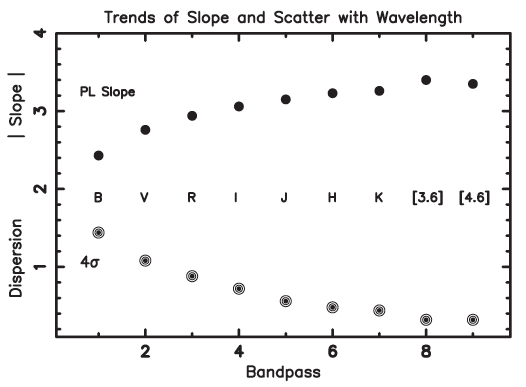}
\end{center}
\vspace*{-5mm} 
\caption{ \footnotesize   A gauche, Figure extraite de \cite{freedman10b} représentant la relation \emph{PL} dans différentes bandes photométriques: $B$, $V$, $I$, $J$, $H$ et $K$. On remarque que la pente augmente avec la longueur d'onde de référence de la bande photométrique, tandis que sa dispersion diminue, ce que l'on constate également sur la figure de droite tirée de \cite{madore12}. Actuellement, la relation \emph{PL} a une dispersion (mode fondamental) de 0.208, 0.146 et 0.077 magnitudes dans les bandes V, I et W$_\mathrm{I}$ de OGLE basé sur 4620 Céphéides (de différents modes de pulsation) des nuages de Magellan \citep{soszynski15}.  \citet{persson04} obtiennent une dispersion de 0.11 magnitude en bande K, tandis que les observations Spitzer  donnent 0.108 et 0.115 magnitude dans les bandes 3.6 et 4.5 $\mu$m, respectivement \citep{scowcroft11}.} \label{Fig_PL_bandes}
\end{figure}

D'après \cite{riess16}, la dispersion de la relation \emph{PL} aurait un impact de 0.7\% sur $H_\mathrm{0}$. La figure~\ref{Fig_PL_bandes} indique très clairement que la pente de la relation \emph{PL} augmente avec la longueur d'onde, tandis que sa dispersion diminue. On passe ainsi d'une dispersion de 0.27 mag dans la bande $V$ à une dispersion de 0.11 mag dans la bande $K$, tandis que la dispersion à 3.6$\mu$m, c'est-à-dire à la longueur d'onde de {\it Spitzer}, est également autour de 0.11 magnitude \citep{scowcroft11, freedman12}. Il est donc préférable en principe de se déplacer vers les hautes longueurs d'onde pour étalonner les échelles de distance dans l'univers à l'aide des Céphéides. Mais c'est sans prendre en compte l'impact de l'environnement des Céphéides et de la binarité. 

En effet, la plupart des Céphéides observées par interférométrie ou par d'autres techniques présentent une coquille, ou plus généralement un environnement qui contribue pour environ 0.06 mag dans le visible (voir Sect.~\ref{sect_CSE}) et jusqu'à 0.3 mag ou plus dans l'infrarouge thermique (3-8$\mu$m). La table~\ref{Tab.cse} fait la synthèse des mesures effectuées jusqu'à présent. L'environnement des Céphéides peut ainsi créer un biais positif sur la relation \emph{PL}: sur le point-zéro, mais aussi sur la pente dans la mesure où une relation entre la période et la contribution en flux de l'enveloppe n'est pas exclue. Ceci ne serait pas nécessairement un problème si la contribution en flux de l'environnement était la même pour toutes les Céphéides d'une période donnée. Or, ce dernier point est loin d'être évident. L'environnement des Céphéides dépend très probablement de la perte de masse, de l'état évolutif de la Céphéide (premier, deuxième ou troisième passage) et peut-être aussi de la température effective, c'est-à-dire la position de la Céphéide sur le bord bleu ou rouge de la bande d'instabilité. Ainsi un des objectifs majeurs de l'ANR {\it UnlockCepheids} pilotée par Pierre Kervella vise à établir des corrections photométriques dans différentes bandes pour l'ensemble des Céphéides de la bande d'instabilité. Il est à noter également que l'environnement des Céphéides peut avoir un impact sur la relation brillance de surface - couleur et donc potentiellement sur la méthode de Baade-Wesselink que nous aborderons dans la prochaine Section \ref{Sect_BW}.  Nous parlerons de la découverte récente d'un environnement visible autour de $\delta$~Cep dans la Sect.~\ref{sect_CSE}.

Par ailleurs, 60\% des Céphéides sont très probablement des binaires \citep{szabados03, neilson15b}. Au premier ordre, ces binaires n'ont pas d'impact sur la relation \emph{PL} dans la mesure où les compagnons sont généralement des étoiles de la séquence principale ayant une luminosité plusieurs centaines de fois plus faible que celle de la Céphéide. Néanmoins, plusieurs études récentes montrent que ces compagnons peuvent être détectés par interférométrie  et que leur contribution en flux n'est pas négligeable, que ce soit dans le visible (rapport de flux plus favorable) ou l'infrarouge  \citep{gallenne13a, gallenne14,gallenne16b}. Concernant V1334 Cyg, une contribution du compagnon en bande $H$ de 3\% a été obtenue \citep{gallenne14}. L'impact de la binarité sur la relation \emph{PL} constitue également un objectif de l'ANR {\it UnlockCepheids}, mais nous ne détaillerons pas ce point dans ce manuscrit. 

%Les processus dans le visible et dans l'infrarouge son différents. On s'attend à de l'émission thermique dans l'infrarouge mais à de la diffusion dans le visible.

\begin{table*}
\caption{ \label{Tab.cse} Les enveloppes de Céphéides détectées par interférométrie (VEGA, FLUOR, MIDI), imagerie (VISIR, NACO, {\it Spitzer}) ou via le calcul de SEDs sont indiquées dans cette table avec la contribution en flux de l'enveloppe (comparée à la photosphère) et sa taille lorsqu'elle a été calculée. Concernant les SEDs, les excès infrarouge calculés à partir des données VISIR dans les PAH1 (à 8.6$\mu$m)  et à différentes phases de pulsation sont assimilés à des rapports de flux entre enveloppe et photosphère \citep{gallenne12}. D'autres valeurs correspondant à d'autres domaines de longueur d'onde (PAH2, Si C) sont indiquées dans  \cite{gallenne12}, mais non reportées ici. Concernant les modèles géométriques ajustés dans les différentes études, on peut répertorier: des modèles de disque avec bord interne à partir de DUSTY (D), des modèle géométriques de gaussiennes (G), des modèles géométriques d'anneaux (A), ainsi que de simple disque (D). A ces informations, on peut rajouter que  \cite{marengo10b} ont observé 29 Céphéides avec {\it Spitzer} et n'ont pas trouvé d'indice d'excès infrarouge causé par un environnement de poussière chaude ($\simeq$ 500K), excepté pour RS~Pup, S Mus et $\delta$ Cep, tandis que sa présence est indiquée comme probable pour GH~Lup, $\ell$~Car, T~Mon, et X~Cyg \citep{barmby11}.  Il est intéressant de noter que l'excès infrarouge, pour certains auteurs, n'est pas dû à de la perte de masse (voir par exemple  \citealt{schmidt15}) et que les structures résolues autour de $\delta$~Cep, au lieu d'être des CSEs, seraient la contribution en flux d'une nébuleuse (dû au choc entre la perte de masse de l'étoile et le milieu interstellaire, telle que mise en évidence dans l'infrarouge par  \citet{marengo10} ou dans le domaine radio \citep{matthews12}. Enfin la complexité de l'environnement des Céphéides s'est encore accrue avec des observations XMM-Newton dans le domaine des rayons X \citep{engle14,ayres17}. Polaris est également une céphéide très particulière (voir l'annexe~\ref{Polaris}).}
\begin{center}\label{Tab_cse}
\setlength{\doublerulesep}{\arrayrulewidth}
\begin{tabular}{lllllll}
\hline
\hline
Céphéides                              &   Période                      &  instrument  & $\lambda$             &       $\frac{f_\mathrm{cse}}{f_\mathrm{\star}}$                            &    $\frac{\theta_\star}{\theta_\mathrm{cse}}$  & Référence                \\
			                      &    $\mathrm{jours}$    &                       &  $\mu$m             &                     $\mathrm{\%}$                                                                                  &                                                                                     &                         \\
\hline

FF Aql & 4.47  & VISIR/SED  & 8.6 & $1.6\pm2.8$  & - &   \cite{gallenne12}  \\
\hline
Polaris  & 3.97 & FLUOR & 2.2  & $1.5\pm0.4$ & $2.4\pm 0.1^{A}$ & \cite{merand06} \\
\hline
AX Cir  &  5.27   & VISIR/Image    &   8.6  &  $13.8 \pm 2.5$ & -    &   \cite{gallenne12}  \\ 
		&	  & VISIR/SED    &   8.6  &      $-0.8 \pm 2.9$ & -    &   \cite{gallenne12}  \\ 
\hline
$\delta$ Cep & 5.36 & VEGA & 0.7  & $7\pm1$  & $6.3^{DI}$  & \cite{nardetto16a} \\
  		&		& FLUOR & 2.2  & $1.5\pm0.4$  & $2.4^{A}$  & \cite{merand06} \\
\hline
X~Sgr   & 7.01     &  MIDI &   10.2                 &  7                 &   $12.6 \pm 3.8^{D}$  & \cite{gallenne13b} \\
    &     & VISIR/Image  & 8.6  &  $7.9 \pm 1.4$ &           -                     &   \cite{gallenne12}  \\ 
        &     & VISIR/SED/$\phi_1$  & 8.6  &  $14.1 \pm 5.4$ &           -                     &   \cite{gallenne12}  \\ 
                &     & VISIR/SED/$\phi_2$  & 8.6  &  $3.8 \pm 9.8$ &           -                     &   \cite{gallenne12}  \\ 
\hline
$\eta$~Aql & 7.18  & VISIR/SED/$\phi_1$  & 8.6 &  $0.38 \pm 3.6$ & - &  \cite{gallenne12}  \\
		&	      & VISIR/SED/$\phi_2$  & 8.6 &  $9.0 \pm 4.1$ & - &  \cite{gallenne12}  \\
\hline
W Sgr     & 7.59      & VISIR/Image  & 8.6  &  $3.8 \pm 0.6$ &            -                    &   \cite{gallenne12}  \\ 
		&	      & VISIR/SED/$\phi_1$  & 8.6  &  $10.4 \pm 4.6$ &            -                    &   \cite{gallenne12}  \\ 
		&	      & VISIR/SED/$\phi_2$  & 8.6  &  $18.2 \pm 3.3$ &            -                    &   \cite{gallenne12}  \\ 
\hline
$\kappa$~Pav & 9.06  &  VISIR/SED &  8.6 & $22.2 \pm 3.2$  &  & \cite{gallenne12}  \\
\hline
Y Oph   & 17.12   & FLUOR   &  2.2  & $5.0\pm 2.0$   &      	-		&    \cite{merand07}  \\
   &    & VISIR/Image   &  8.6  & $15.1\pm 1.4$   &      	-		&   \cite{gallenne12}  \\
      &    & VISIR/SED/$\phi_1$   &  8.6  & $6.8\pm 2.7$  &      	-		&   \cite{gallenne12}  \\
            &    & VISIR/SED/$\phi_2$   &  8.6  & $6.3\pm 2.7$   &      	-		&   \cite{gallenne12}  \\
\hline
T Mon & 27.02    & MIDI  & 10.2       &               19                  &     $16.9 \pm 3.8^{D}$                  &  \cite{gallenne13b}    \\
\hline
$\ell$~Car & 35.56  &  VINCI &   2.2   &     $4.2 \pm 0.2$            &          $1.9 \pm 1.5^{G}$       &    \cite{kervella06a}  \\
 &  &  MIDI   &  10.2   &        10        &         $3.0\pm1.1^{G}$                 & \cite{kervella06a}  \\
&   &  {\it Spitzer} &   70   &     $50$           &          -      &    \cite{kervella06a}  \\
\hline
U Car  & 38.81   & VISIR/Image  & 8.6  & $16.3\pm 1.4$   &      	-		&   \cite{gallenne12}  \\
	  &		  & VISIR/Image  & 8.6  & $32.1\pm 2.5$   &      	-		&   \cite{gallenne12}  \\
  	&		  & VISIR/Image  & 8.6  & $30.9\pm 6.2$   &      	-		&   \cite{gallenne12}  \\
\hline
RS Pup  & 41.44 & NACO  & 1.6  & $38\pm 17$  & - & \cite{gallenne11} \\ 
  &  & NACO  & 2.2 & $24\pm 11$  & - & \cite{gallenne11} \\ 
\hline
SV Vul & 45.00 & VISIR/SED  & 8.6   & $25.0\pm 2.5$ &  -  &  \cite{gallenne12} \\

\hline
\end{tabular}
\end{center}
\end{table*}

\begin{figure}[htbp]
\begin{center}
\includegraphics[width=15cm]{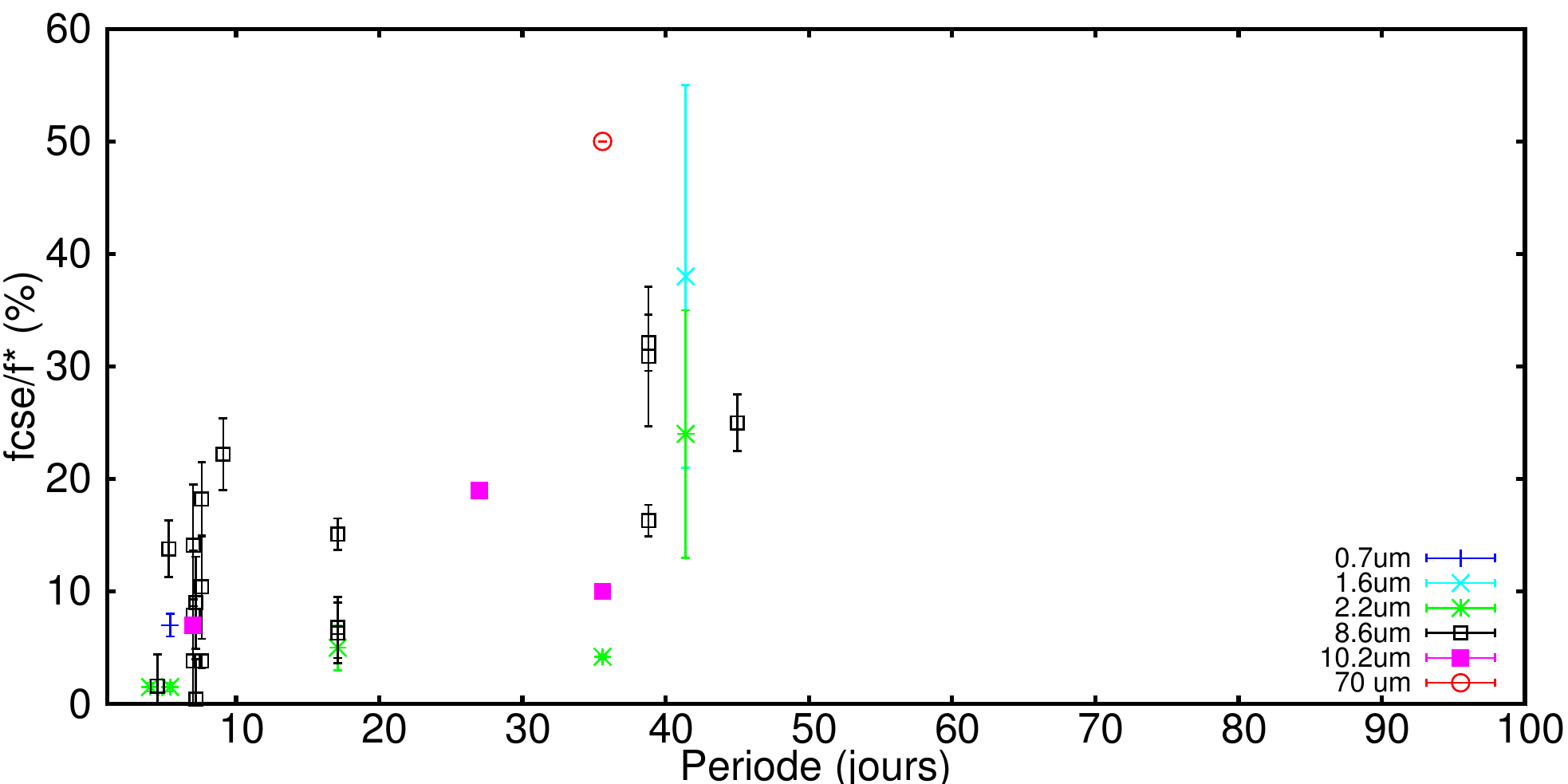}
\includegraphics[width=15cm]{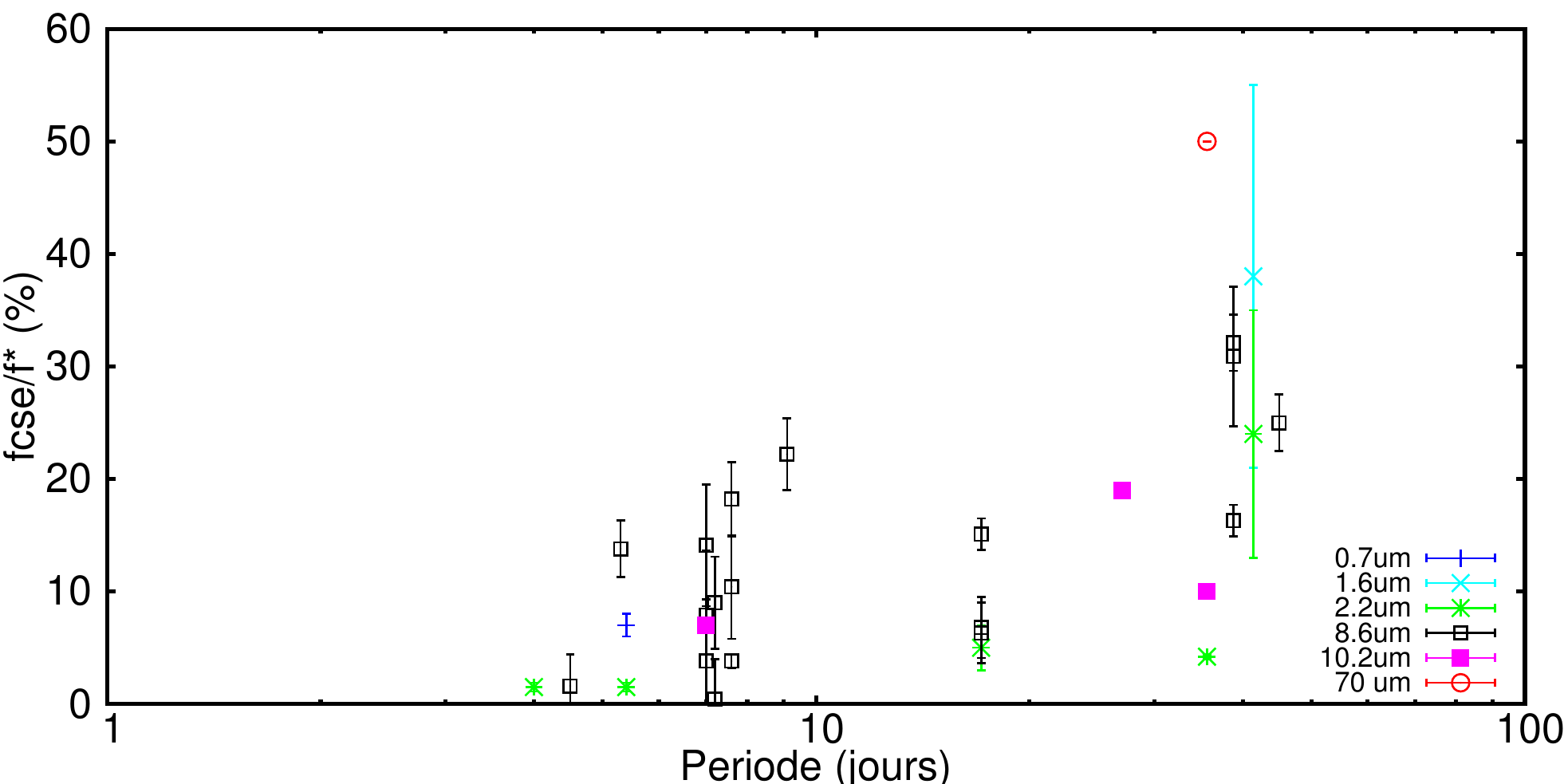}
\end{center}
\vspace*{-5mm} \caption{ \footnotesize  Le rapport du flux de l'enveloppe à celui de la photosphère (exprimé en pourcentage) est représenté en fonction de la période (en haut) et de son logarithme (en bas). Cette figure utilise les données listées dans la table~\ref{Tab_cse}.} \label{Fig_CSE}
\end{figure}

\newpage

%%%%%%%SECTION 
\section{La méthode de BW appliquée aux Céphéides}\label{Sect_BW}

%\subsection*{La méthode Baade-Wesselink}

Nous avons vu que \cite{riess16} se basent actuellement sur 15 Céphéides galactiques dont la distance a été déterminée par une mesure de parallaxe trigonométrique du {\it Hubble Space Telescope} pour contraindre $H_\mathrm{0}$ (en plus du LMC, M31 et NGC 4258). Les incertitudes sur ces distances vont de 4\% à
12\%  \citep{benedict07}. Dix-huit Céphéides longue-périodes supplémentaires situées à plus de $1$kpc sont actuellement observées par le HST dans le cadre du projet SHOES \citep{riess14}. Il est également intéressant de noter qu'en comparaison, la précision sur les distances {\it Hipparcos}  des Céphéides s'échelonne entre 5\% et 29\% \citep{vanleeuwen07}.  

\begin{figure}[h]
%\vspace*{-8mm}
\begin{flushleft}
\resizebox{0.45\hsize}{!}{\includegraphics[angle=0]{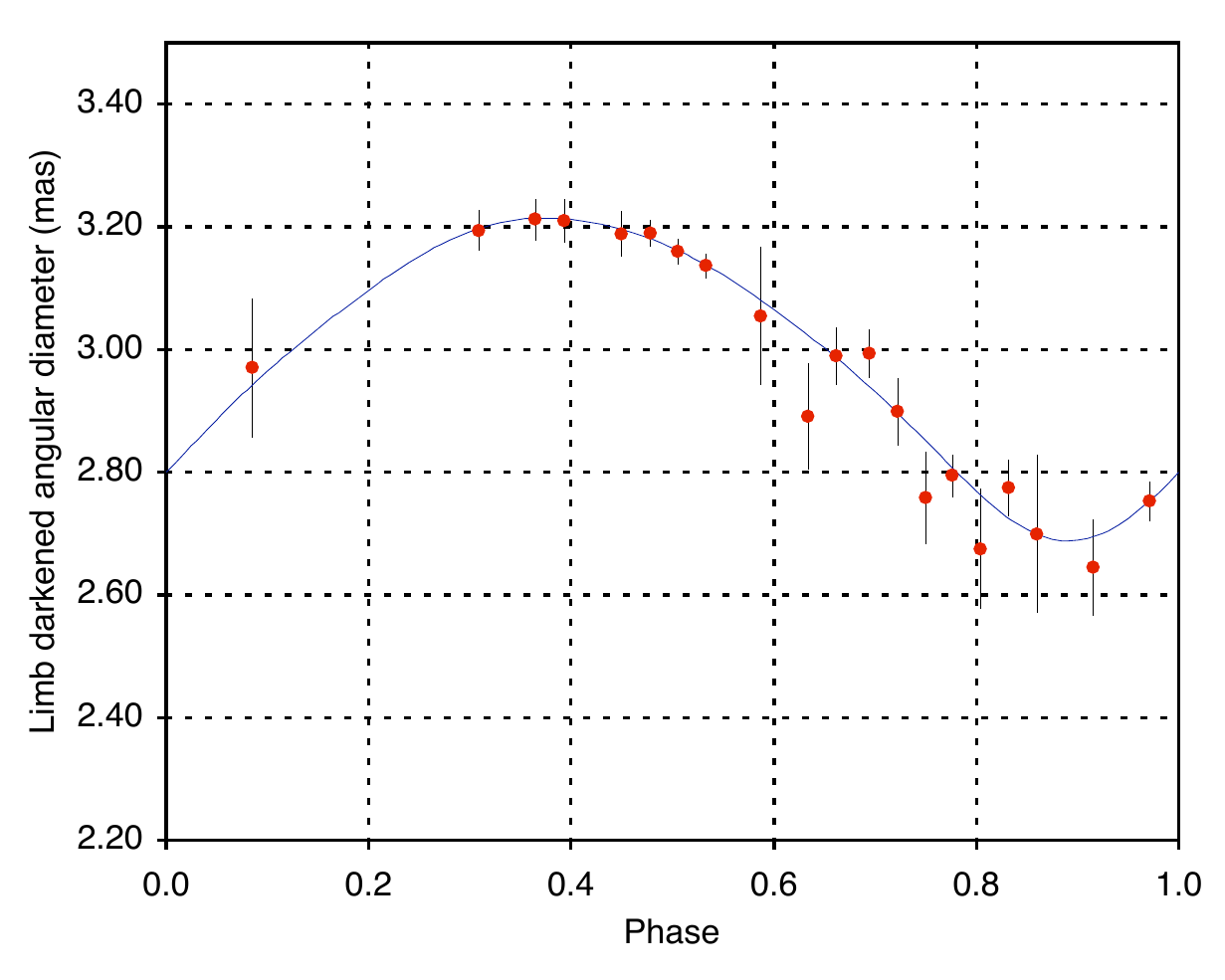}}
\end{flushleft}
%\vspace*{-8mm}
\begin{minipage}{0.52\hsize}
\caption{ \footnotesize  A droite : variation du profil spectral
d'un raie métallique en fonction de la phase de pulsation
(observation HARPS de $\beta$~Dor). En rouge est indiquée la
bi-gaussienne analytique permettant de mesurer l'asymétrie de la
raie spectrale. Une fois intégrée dans le temps, la vitesse pulsante
(= vitesse radiale x le facteur de projection) permet de déduire la
variation du rayon de l'étoile. Ci-dessus, courbe de diamètre
angulaire en fonction de la phase de pulsation de l'étoile
($\ell~$Car). Les points rouge correspondent aux mesures de
l'instrument VINCI/VLTI. La forme de la courbe bleue correspond à la
variation du rayon de l'étoile déduite de la spectroscopie, tandis
que son amplitude est liée à la distance de l'étoile.} \label{Fig_principeBW}
\end{minipage}
\vspace*{-110mm}
\begin{flushright}
\begin{minipage}{75mm}{
\resizebox{1.05\hsize}{!}{\includegraphics[angle=90]{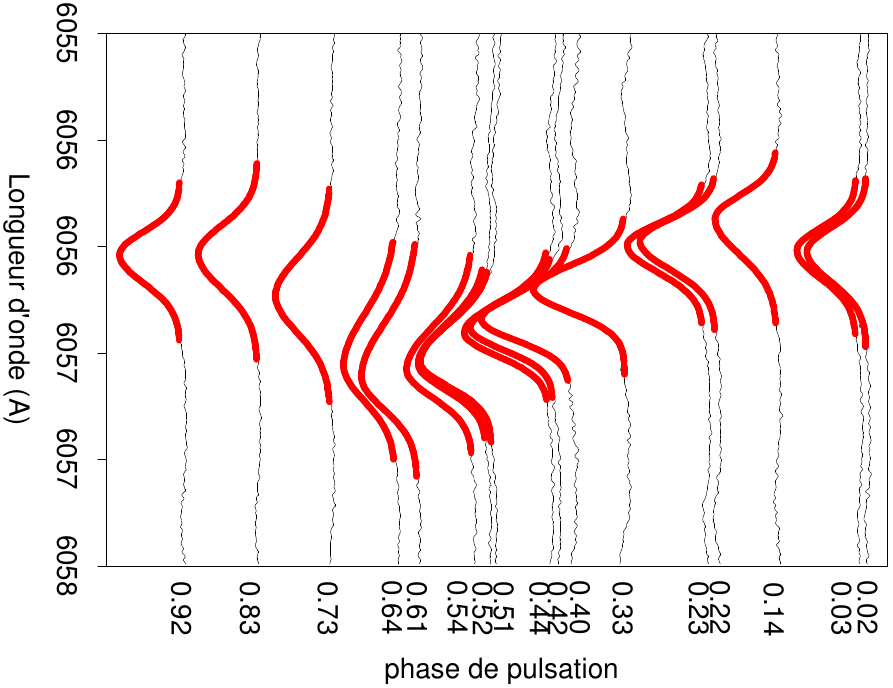} }}
\end{minipage}
\end{flushright}
%\vspace*{-5mm}
\end{figure}

A côté de cela, la méthode de Baade-Wesselink (BW) a permis de déterminer la distance de 70 Céphéides galactiques (à moins de 4 kpc) et de 36 Céphéides du LMC avec une précision de quelques pourcents \citep{fouque07, storm11a, storm11b}. La méthode a été décrite pour la première fois par \citet{lindermann18}\footnote{Publication écrite en Allemand}, puis plus tard par \citet{baade26, wesselink46}.  Le principe est simple : il s'agit de comparer les dimensions linéaire et angulaire de la Céphéide afin de déterminer sa distance au moyen d'une simple division (voir Fig. \ref{Fig_principeBW}). Les mesures photométriques (associées à une relation brillance de surface - couleur) fournissent la variation du diamètre angulaire {\it photosphérique} de l'étoile sur tout le cycle de pulsation, tandis que la variation du diamètre linéaire est déterminée par une intégration temporelle de la vitesse pulsante ($V_{\mathrm{puls}}$) de l'étoile. La détermination de cette dernière, à partir du décalage Doppler de la raie spectrale ($V_{\mathrm{rad}}$), est extrêmement délicate et fait intervenir ce que l'on appelle le facteur de projection, $p$, défini par $V_{\mathrm{puls}}=p V_{\mathrm{rad}}$.  Ce nombre résume à lui seul toute la physique de l'atmosphère de la Céphéide (assombrissement centre-bord, gradient de vitesse, ...).  En 1997, une deuxième version de la méthode de BW, non pas basée sur les relations brillance de surface - couleur, mais sur l'interférométrie a été tentée sur $\delta$ Cep dans le visible avec l'instrument  GI2T\footnote{Grand Interféromètre à 2 Télescopes situé sur la Plateau de Calern} \citep{mourard97}. Puis, quelques années plus tard, les premiers résultats concluants ont été obtenus dans l'infrarouge \citep{kervella99, lane00, kervella01}. Depuis lors, la méthode a été appliquée à 12 Céphéides au total, essentiellement dans l'infrarouge (voir Table \ref{Tab.hra}). Parmi ces 12 Céphéides, 4 ont été observées dans le visible, mais la variation du diamètre angulaire n'a été mesurée de manière significative que pour une seule Céphéide, $\ell$~Car \citep{davis09}. Une version plus récente de la méthode de BW (SPIPS) utilise l'ensemble des données disponibles, c'est-à-dire interférométriques, photométriques (dans de nombreuses bandes) et les courbes de vitesses radiales \citep{breitfelder15, merand15, breitfelder16}. 

La méthode de BW est un outil précieux. En effet, {\it Gaia} ne pourra pas déterminer la distance d'étoiles individuelles dans le LMC. En revanche, avec les futurs instruments E-ELT, JWST, HIRES,...  il sera possible de déduire la distance et la métallicité de Céphéides individuelles dans le groupe local, et ceci de manière totalement indépendante de la relation \emph{PL}. Il est intéressant de remarquer que, jusqu'à présent, les distances de BW n'ont pas été utilisées pour déterminer $H_\mathrm{0}$. La raison invoquée, à juste titre probablement par \cite{riess09a} est la suivante: {\it ‘‘We have not made use of additionnal distance measures to Galactic Cepheids based on the BW method (…) due to uncertainties in their projection factors''.} Ainsi, un des objectifs prioritaires, à la fois dans le cadre du projet Araucaria et de l'ANR  {\it UnlockCepheids}, est d'étalonner les facteurs de projection (à l'aide de {\it Gaia}) et rendre ainsi la méthode de BW robuste pour l'étalonnage des échelles de distance dans le groupe local. Mes contributions sont multiples au sein de ces projets, mais la plus évidente concerne probablement le facteur de projection et c'est ce que nous allons aborder maintenant.  

\begin{table*}
\caption{ \label{Tab.hra} Céphéides pour lesquelles la version interférométrique de la méthode de BW a été appliquée. Pour les premières tentatives (indiquées par un astérisque), la pulsation n'a pas été clairement résolue par l'interféromètre et une valeur moyenne du diamètre angulaire a été considérée. Dans le domaine visible, la méthode n'a été appliquée avec succès que pour une seule Céphéide, $\ell$~Car \citep{davis09} et une enveloppe a été découverte pour $\delta$~Cep (voir Sect.~\ref{sect_CSE}). Dans l'infrarouge, la pulsation a été détectée pour l'ensemble des Céphéides observées, même X~Sgr, une Céphéide atypique présentant des ondes de chocs \citep{mathias06}.}
\begin{center}
\setlength{\doublerulesep}{\arrayrulewidth}
\begin{tabular}{ll}
\hline
\hline
Cepheids                                                         & reference                                    \\
\hline
\multicolumn{2}{c}{visible} \\
$\delta$~Cep$^{\star}$      & \citet{mourard97}  \\
$\alpha$~UMi$^{\star}$, $\zeta$~Gem$^{\star}$, $\delta$~Cep$^{\star}$, $\eta$~Aql$^{\star}$   & \citet{nordgren00} \\
$\delta$~Cep$^{\star}$, $\eta$~Aql$^{\star}$  & \citet{armstrong01} \\
$\ell$~Car & \citet{davis09} \\
\hline
\multicolumn{2}{c}{H band} \\
$\zeta$~Gem & \citet{lane00} \\
$\zeta$~Gem, $\eta$~Aql & \citet{lane02} \\
$\kappa$~Pav & \citet{breitfelder15} \\
$\ell$~Car &   \citet{anderson16} \\
X~Sgr, W~Sgr, $\zeta$~Gem, $\beta$~Dor, $\ell$~Car &  \citet{breitfelder16} \\

\hline
\multicolumn{2}{c}{K band} \\
$\zeta$~Gem$^{\star}$ & \citet{kervella01} \\
X~Sgr$^{\star}$, $\eta$~Aql, W Sgr, $\zeta$~Gem$^{\star}$, $\beta$~Dor, Y Oph$^{\star}$, $\ell$~Car &\citet{kervella04a}  \\

 $\delta$~Cep & \citet{merand05} \\
 %$\alpha$~UMi, $\delta$~Cep & \citet{merand06} \\
Y Oph & \citet{merand07} \\
FF Aql, T Vul & \citet{gallenne12} \\
$\delta$~Cep, $\eta$~Aql & \citet{merand15} \\
\hline
%\multicolumn{2}{c}{N-band} \\
%$\ell$~Car, RS Pup &  \citet{kervella09} \\
%T Mon, X Sgr &  \citet{gallenne13b} \\
\end{tabular}
\end{center}
\end{table*}

\chapter{La méthode de Baade-Wesselink appliquée aux Céphéides: trois problématiques}\label{Chap_BW}

%%L'objectif principe de mes recherches est d'exploiter les expertises existantes en terme de modélisation
%hydrodynamique (et de développer au besoin les interfaces
%nécessaires et/ou certains modules manquants) afin de construire un
%modèle hydrodynamique de Céphéide pragmatique et robuste, dédié à la
%résolution de problèmes spécifiques, et incluant les mécanismes de
%la pulsation ainsi que les théories évolutives. Chaque stade de
%développement de ce modèle, et ceci constitue la ligne directrice de
%mon programme de recherche (à long terme), sera contraint par un ensemble
%d'observations multi-techniques, avec un va-et-vient permanent entre
%observation et modélisation afin d'accéder à une interprétation fine
%des Céphéides. Ce cadre théorique, allié à mon approche originale de
%l'étude de la dynamique atmosphérique, permettra de comparer les
%Céphéides à d'autres classes d'étoiles pulsantes telles que les RR
%Lyrae, $\delta$~Scuti et $\beta$~Céphéides. La comparaison des
%étoiles pulsantes entre elles permettra de renforcer les théories de
%la pulsation. 

Comme nous venons de le voir (Section~\ref{Sect_BW}), la méthode de BW de détermination de distance vise à combiner les dimensions angulaires et réelles, c'est-à-dire en km, de la variation de rayon d'une étoile pulsante. Le principe est le suivant. Il s'agit de faire un ajustement statistique classique dans lequel le ${\chi}^{2} $ à minimiser est défini de la manière suivante:

\begin{equation}
\label{chi2sum} {\chi}^{2} = \sum_{i}{\frac{(\theta_{\rm
obs}(\phi_{i}) - \theta_{\rm model}(\phi_{i}))^2}{\sigma_{\rm
obs}(\phi_{i})^2}},
\end{equation}

où

\begin{itemize}
\item $ \theta_{\rm obs}(\phi_{i})$ est le diamètre angulaire (prenant en compte l'assombrissement centre-bord de l'étoile) déduit soit des observations interférométriques, soit d'une relation brillance de surface - couleur. $\phi_{i}$ est la phase de pulsation correspondant au $i$-ème point de mesure 
\item $\sigma_{\rm obs}(\phi_{i})$ sont les incertitudes statistiques correspondant aux mesures de diamètre angulaire
\item  $\theta_{\rm model}(\phi_{i})$ est le modèle de diamètre angulaire défini de la manière suivante:
\begin{equation}\label{diam_mod}
\theta_{\rm model}(\phi_{i}) = \overline{\theta} +
9.3009\,\frac{p_\mathrm{cc-g}}{d} \left(\int RV_\mathrm{cc-g}(\phi_{i})d\phi_{i}\right) [{\rm mas}],
\end{equation}
\end{itemize}
où le facteur de conversion $9.3009$ prend en compte la valeur de rayon solaire donné dans \citet{prsa16}. $RV_\mathrm{cc-g}(\phi_{i})$ est la courbe de vitesse radiale (interpolée) correspondant, par exemple, à la méthode très utilisée de la la cross-corrélation (cc) combinée avec un ajustement gaussian (g) du profil moyen, notée (cc-g). Dans l'équation~\ref{diam_mod}, on force la moyenne de la courbe de vitesse radiale $RV_\mathrm{cc-g}$ à zéro de façon à s'affranchir de la vitesse du centre de masse de l'étoile $V_\mathrm{\star}$. Les paramètres $\overline{\theta}$ et $p_\mathrm{cc-g}$ sont le diamètre angulaire moyen (en mas) et le facteur de projection (associé à la méthode cc-g), respectivement. $d$ est la distance de l'étoile. Les quantités  $\overline{\theta}$ et $d$ sont ajustées de façon à réduire le ${\chi}^{2}$ dans le cas de la méthode de BW classique. On peut aussi ajuster $\overline{\theta}$ et $p_\mathrm{cc-g}$, et fixer $d$; il s'agit alors de la méthode de BW {\it inverse}.

La méthode de BW a déjà permis de déterminer la distance des Céphéides dans les nuages de Magellan \citep{storm11a, storm11b}. D'ici quelques années, il deviendra possible grâce à cette méthode et aux télescopes spatiaux et/ou de nouvelle génération (E-ELT, JWST, ...) de déterminer la distance des Céphéides dans le groupe local (sans passer par la relation \emph{PL}).  Mais avant cela, nous devons rendre la méthode plus robuste. Il est possible de lister tout un ensemble de limitations liées à cette technique, mais les trois principales sont le $p$-facteur, le $k$-facteur et l'impact de l'environnement de la Céphéide, que nous allons aborder maintenant. 

%\begin{enumerate}
%\item Le facteur de projection est la principale limitation de la méthode: s'il existe un biais de 5\% sur le facteur de projection, alors la distance de BW sera également biaisée à hauteur de 5\%.  Celui-ci dépend (par ordre d'importance): de l'assombrissement centre-bord de l'étoile (et donc de la phase de pulsation de l'étoile), de la dynamique atmosphérique, de la longueur d'onde de la méthode utilisée pour déduire la vitesse radiale, de la longueur d'onde de la raie considérée
%\begin{end}

%%%%%%%SECTION 
\section{Le $p$-facteur}\label{Sect_p}

Le facteur de projection ($p = \frac{V_\mathrm{puls}}{V_\mathrm{rad}}$; voir Section~\ref{Sect_BW}) est une quantité physique complexe qui a fait l'objet de nombreuses études depuis plus de 65 ans. L'approche initiale \citep{vanhoof52} consiste à considérer le facteur de projection comme une quantité purement géométrique prenant en compte un seul paramètre physique: l'assombrissement centre-bord de l'étoile. La vitesse radiale $V_\mathrm{rad}$ mesurée sur la ligne de visée est alors une intégration sur le disque de l'étoile $D$, de rayon $R$, du champ de vitesse pulsante $V_\mathrm{puls}$. Cette intégration est pondérée par la distribution d'intensité de l'étoile $I (x,y)$: 

\begin{equation}\label{Eq_pf_definition_xyEtacb}
V_\mathrm{rad}=  \int_{x,y \in D} \left[ V_\mathrm{puls}
\sqrt{1-\frac{(x^2+y^2)}{R^2}}\right] I(x,y) dx dy
\end{equation}

avec

\begin{equation}\label{Eq_pf_definition_xyEtacb2}
\int_{x,y \in D} I(x,y) dx dy = 1
\end{equation}

La façon la plus simple de modéliser la distribution d'intensité de l'étoile est de considérer un assombrissement centre-bord linéaire avec un paramètre unique $u$ compris entre 0 (disque uniforme) et 1 (disque assombri):

\begin{equation}\label{Eq_acb}
I(x,y)=I_0(1-u+u*\mu) = I_0(1-u+u \sqrt{1-(x^2+y^2)})
\end{equation}

Ainsi si l'on observe une raie spectrale dans le domaine visible, le facteur de projection géométrique (noté $p_\mathrm{0}$) s'écrit \citep{getting34} :

\begin{equation}\label{Eq_Getting}
p_\mathrm{0} = \frac{3}{2} - \frac{u_\mathrm{V}}{6} 
\end{equation}

où le coefficient $u_\mathrm{V}$, c'est-à-dire l'assombrissement centre-bord, dépend de la température effective, de la gravité de surface, de la métallicité et de la vitesse de microturbulence. On peut par exemple utiliser les tables de \citet{claret11} basées sur les modèles de \citet{kurucz79} pour déduire $u_\mathrm{V}$. On peut faire alors plusieurs remarques importantes: 

\begin{enumerate}
\item L'équation~\ref{Eq_Getting} {\it n'est valide que si l'on utilise le premier moment de la raie pour déduire la vitesse radiale}\footnote{Le premier moment de la raie $RV_{\mathrm c}$ (‘‘c'' pour {\it centroid}) est défini au sens mathématique par $RV_{\mathrm c} = \frac{\int_{\rm line} \lambda S(\lambda) d\lambda}{\int_{\rm line} S(\lambda) d\lambda}$, où S est le profil de la raie. L'intégration en longueur d'onde associée à la méthode du premier moment est formellement équivalente à l'intégration spatiale considérée dans l'Eq.~\ref{Eq_pf_definition_xyEtacb} pour déduire le facteur de projection géométrique. Si l'on utilise un ajustement gaussien pour déduire la position de la raie spectrale, ou encore la vitesse associée au minimum de la raie, la vitesse radiale obtenue est alors {\it de facto} différente (voir \cite{nardetto06a}), mais surtout le lien physique qui lie la vitesse radiale au facteur de projection géométrique est rompu et l'Eq.~\ref{Eq_Getting} ne peut plus être utilisée.}. Ceci a été montré dans ma thèse de Doctorat. 
\item Le facteur de projection géométrique vaut $p_\mathrm{0} =1.5$ dans le cas d'un disque uniforme. Une valeur supérieure à $1.5$ est {\it a priori} irréaliste dans le cas d'une Céphéide car équivalente à un éclaircissement centre-bord. 
\item Puisque la température et la gravité de surface varient avec la phase de pulsation, le facteur de projection géométrique est également supposé varier avec la phase de pulsation. Ce dernier est cependant considéré comme {\it constant} sur le cycle de l'étoile. Nous reviendrons sur cette hypothèse par la suite. 
\item Les valeurs du facteur de projection géométrique de $\delta$ Cep que l'on trouve dans la littérature sont directement liées à la valeur de l'assombrissement centre-bord {\it moyen} considéré: $p_\mathrm{0} = \frac{24}{17}= 1.415$ ($u_\mathrm{V}=0.60$,  \citet{getting34}), $p_\mathrm{0} = 1.375$ ($u_\mathrm{V}=0.75$, \citet{vanhoof52}), $p_\mathrm{0}=1.360$ ($u_\mathrm{V}=0.80$, \citet{burki82}). La valeur $p=1.36$ \citep{burki82} a été largement utilisée  dans la littérature pendant près de 20 ans. Ces valeurs sont résumées dans la table~\ref{tab_history}. Aux références indiquées dans cette table, nous pouvons rajouter d'autres études qui visaient à l'époque à relier le facteur de projection à l'asymétrie ou à la largeur de la raie, de façon à en déduire sa variation avec la phase \citep{parsons72, karp75c, hindsley86, albrow94}. 
\end{enumerate}

Récemment, \citet{neilson12} a utilisé un modèle statique avec un transfert de rayonnement en géométrie sphérique pour déduire le facteur de projection géométrique en fonction de la période et dans différentes bandes spectrales. Il obtient ainsi dans le visible et dans le cas de $\delta$~Cep une valeur autour de $p_\mathrm{0}=1.33$. Il faut également noter qu'il est formellement possible de se passer du facteur de projection en ajustant directement un modèle (statique ou même éventuellement hydrodynamique) sur les profils de raies spectrales. Dans ce cas, la vitesse pulsante est alors un {\it output}. Ainsi dans les études de \citet{gray07} et \citet{hadrava09a, hadrava09b}, les modèles ajustés sont statiques et il n'est donc pas étonnant que le facteur de projection déduit {\it a posteriori} (en comparant la vitesse pulsante issue du code et la vitesse radiale observée) ait une valeur très proche des valeurs des facteurs de projection géométriques dont nous avons discuté plus haut. Le point faible de ces études est que l'ajustement n'est effectué que sur une ou quelques raies spectrales. Or, c'est justement en comparant différentes raies spectrales que toute la finesse liée à la dynamique atmosphérique des Céphéides apparaît. 

En effet, les Céphéides ne pulsent pas de manière quasi-hydrostatique. L'interaction entre l'enveloppe pulsante et l'atmosphère est extrêmement complexe \citep{sanford56,bell64, karp75a, karp75b, karp75c, sasselov90,  wallerstein15}. Ainsi pour affiner la description du facteur de projection, il devient indispensable de décrire la pulsation de l'atmosphère avec un modèle hydrodynamique (voir Tab.~\ref{Tab_modeles})). A ce jour, le facteur de projection a été étudié avec seulement trois modèles de ce type. Le premier considère un {\it piston} à la base de l'atmosphère (c'est-à-dire que la vitesse radiale est un {\it input}) \citep{sabbey95} tandis que le second est {\it self-consistent} \citep{nardetto04}. \citet{sabbey95} ont ainsi obtenu une valeur moyenne du facteur de projection de $p=1.34$ (voir également \citet{marengo02, marengo03}). Cependant cette valeur a été obtenue avec la méthode du bisecteur pour déduire la vitesse radiale (à partir des profils théoriques), ce qui rend la comparaison avec les autres études difficile. La méthode la plus courante utilisée dans la littérature pour déduire la courbe de vitesse radiale est l'ajustement d'une gaussienne sur un profil moyen cross-correlé ($RV_\mathrm{cc-g}$). %, ‘‘cc'' pour Cross-Correlation et ‘‘g'' pour Gaussienne). 
En considérant cette méthode, le facteur de projection, que l'on peut qualifier alors de {\it dynamique}, est environ 10\% plus faible que dans le cas purement géométrique, avec une valeur autour de $p=1.25 \pm 0.05$ \citep{nardetto09}. Très récemment, un grand pas en avant a été réalisé par \citet{vasilyev17}. Ce groupe a modélisé une Céphéide de période 2.8 jours ($T_\mathrm{eff}= 5600$ K, $\log g = 2.0$), soit très proche de $\delta$~Cep, en utilisant un code {\it bi-dimensionnel incluant la convection}. Avant cela, d'autres modèles multi-dimensionnels 2D ou 3D incluant une description de la convection ont vu le jour \citep{mundprecht13, Mundprecht15, geroux11, geroux13, geroux14,  geroux15}, mais ces derniers n'incluaient pas de transfert de rayonnement et ne permettaient donc pas de calculer un profil de raie, ni le facteur de projection. Ainsi, après plus d'une décennie de développement, il est maintenant possible grâce au modèle de \citet{vasilyev17} d'avoir une idée de l'impact de la convection non seulement sur le $p$-facteur,  mais aussi sur le $k$-facteur, que nous aborderons dans la Sect.~\ref{Sect_k}. Ainsi, selon la force de la raie considérée, ils obtiennent un facteur de projection de $p \simeq 1.23 - 1.27$, ce qui est parfaitement cohérent avec le modèle hydrodynamique 1D et sans convection décrit dans \citet{nardetto04}. \footnote{ Cependant, la granulation pose probablement un autre problème. Des variations de cycle à cycle ont été observées dans quatre Céphéides, deux Céphéides de courte période et de faible amplitude (pulsant dans le premier harmonique), QZ Normae et V335 Puppis,  ainsi que dans deux céphéides de longue période $\ell$~Carinae et RS Puppis \citep{anderson14}. Ces variations dans les vitesses radiales (de l'ordre de quelques pourcents) ne sont pas synchronisées avec les variations de diamètre angulaire que l'on peut mesurer par interférométrie, ce qui peut avoir un impact sur la variation du facteur de projection de l'ordre de 5\% \citep{anderson16}. Ces longues séries de données vélocimétriques sur $\ell$~Car ont également permis d'étudier les variations de cycle à cycle du gradient de vitesse dans l'atmosphère de l'étoile \citep{anderson16a, anderson16b}.}.

%mais aussi par la modélisation. Il faut encourager le travail fait par certains groupes qui visent à développer des modèles de pulsations 2D ou 3D prenant en compte la convection. Le point clef qui est souvent mis de côté par les modélisateurs est de développer un code de transfert de rayonnement dans l'atmosphère du modèle de façon à prédire des raies spectrales. Sans ces raies spectrales théoriques, il n'est pas possible d'étudier le $p$-facteur ou le $k$-facteur. 

En utilisant une méthode basée uniquement sur des observations, \citet{merand05} ont appliqué la méthode de BW {\it inverse} à partir d'observations interférométriques infrarouge. Le facteur de projection est ajusté tandis que la distance est connue avec une précision d'environ 4\% à partir d'une mesure de parallaxe du HST ($d=274 \pm 11$pc; \citet{benedict02}). Ils ont trouvé: $p=1.273 \pm 0.021 \pm 0.05$ en considérant $RV_\mathrm{cc-g}$ pour la vitesse radiale. La première incertitude est liée à l'ajustement, tandis que la seconde est liée à l'incertitude sur la distance. \citet{gro07} a trouvé $p=1.245\pm0.030\pm0.050$ en utilisant presque la même distance ($273$ au lieu de $274$ pc), un jeu de données différent pour la vitesse radiale et une méthode d'ajustement légèrement différente. Récemment,   \citet{merand15} ont développé une méthode intégrée (SPIPS), l'ont appliquée à $\delta$~Cep (en combinant interférométrie et photométrie) et ont obtenu $p=1.29\pm0.02\pm0.05$. Toutes ces valeurs s'accordent remarquablement bien avec les résultats du modèle hydrodynamique \citep{nardetto09}. Une méthode légèrement différente consiste à appliquer la méthode de BW {\it inverse} aux Céphéides lointaines en utilisant la relation brillance de surface - couleur dans l'infrarouge (voir \citealp{fouque97, kervella04b} pour le principe\footnote{Il s'agit en fait de la méthode de BW version photométrique dans laquelle les bandes V et K sont utilisées. La méthode SPIPS quant à elle utilise l'ensemble des données disponibles: photométriques (dans toutes les bandes ou presque) et interférométriques.}) afin d'en déduire une relation \emph{Pp}. Dans cette méthode, on considère que la distance BW de chaque Céphéide du LMC est la même (la géométrie du LMC n'est pas prise en compte). Cela permet de contraindre la pente de la relation \emph{Pp}. En revanche, le point zéro de la relation \emph{Pp} est fixé soit en utilisant les Céphéides dans les amas galactiques \citep{gieren05}, soit en utilisant les parallaxes HST des Céphéides proches obtenues par  \citet{vl07} \citep{laney09, storm11b}, ou une combinaison des deux \citep{gro13}. \citet{laney09} ont utilisé également les  {\it High-Amplitude $\delta$ Scuti Stars} (HADS) pour déduire une relation $Pp$ (nous y reviendrons dans la Sect.~\ref{Sect_HADS}). Les facteurs de projection obtenus par \citet{laney09} et \citet{gro13} sont cohérents avec les valeurs interférométriques tandis que les facteurs de projection de \citet{gieren05} et \citet{storm11b} sont significativement plus élevés (voir Table~\ref{tab_history}). Une autre approche novatrice et purement observationnelle utilise les Céphéides appartenant à un système binaire à éclipses. Nous reviendrons sur le principe dans le chapitre sur les binaires à éclipses. Le facteur de projection obtenu pour une Céphéide de période $P=3.80$ jours (c'est-à-dire comparable à $\delta$ Cep) est de $p=1.21\pm0.04$ \citep{pilecki13}. Enfin, pour être complet, signalons que des approches semi-théoriques \footnote{\cite{molinaro11} ont développé une nouvelle version de la méthode de Baade-Wesselink, s'appuyant sur l'approche CORS \citep{caccin81}, dont l'objectif est de déterminer des rayons moyens à partir de relations brillance de surface - couleur théoriques (c'est à dire basées sur des modèles d'atmosphères). Ces rayons sont alors utilisés pour déterminer des relations période - rayon et période - luminosité à partir de deux estimations du facteur de projection $p=1.36$ et $p=1.27$. Les auteurs déduisent de cette étude que la valeur $p=1.27$ est la plus appropriée, la valeur $p=1.36$ donnant des incohérences de l'ordre de $2\sigma$ sur les relations. Dans le même ordre d'idée, \citet{marconi17} ont utilisé des modèles d'atmosphère pour reproduire les courbes photométriques des Céphéides du SMC, ainsi que les courbes de vitesses (déduites à partir de la dérivée du rayon théorique) pour en déduire des valeurs du facteur de projection, dont les valeurs vont de $1.16$ à $1.40$.} font également intervenir le facteur de projection \citep{molinaro11,marconi17}.

\begin{table*}
\begin{center}
\caption{\label{tab_history} Liste non exhaustive des facteurs de projection de la littérature dans le cas de $\delta$~Cep. La méthode utilisée pour déduire la vitesse radiale est indiquée, et {\it cc-g} correspond à un ajustement gaussien du profil déduit de la cross-corrélation. Certaines valeurs sont déduites à partir d'une relation entre la période et le facteur de projection en considérant une période de $P=5.366208$ jours \citep{engle14}.}
\setlength{\tabcolsep}{4pt}
\begin{tabular}{|l|l||l|}
\hline
method &  $p$ & reference \\
\hline
\multicolumn{3}{|c|}{Modèles géométriques}  \\
\hline
  centroid & 1.415 & \citet{getting34} \\
    centroid & 1.375   &\citet{vanhoof52}\\
      centroid & 1.360  &\citet{burki82} \\
      centroid & 1.328  &\citet{neilson12} \\
      \hline
\multicolumn{3}{|c|}{Modèles hydrodynamiques}  \\
\hline
  bisector & 1.34 & \cite{sabbey95} \\
    Gaussian &1.27 $\pm$ 0.01  & \citet{nardetto04} \\
%    centroid & 1.33  $\pm$ 0.02  & \citet{nardetto07}  \\
       cc-g ($Pp$) & 1.25 $ \pm$ 0.05   &\citet{nardetto09}  \\
       \hline
\multicolumn{3}{|c|}{Observations}  \\
\hline
     cc-g & 1.273 $\pm$ 0.021 $\pm$ 0.050 &\citet{merand05} \\
     cc-g & 1.245 $\pm$ 0.030 $\pm$ 0.050 &\citet{gro07}  \\
     cc-g &1.290 $\pm$ 0.020 $\pm$ 0.050 &\citet{merand15} \\
     cc-g & 1.239 $\pm$ 0.034 $\pm$ 0.023 & \citet{nardetto17} \\
\hline    
     cc-g ($Pp$)  & 1.47 $\pm$ 0.05  &\citet{gieren05}  \\
     cc-g  ($Pp$)  &1.29 $\pm$ 0.06  &\citet{laney09} \\
     cc-g  ($Pp$) &1.41 $\pm$ 0.05  &\citet{storm11b} \\
     cc-g   ($Pp$)  &1.325 $\pm$ 0.03   &\citet{gro13}  \\   
\hline
\end{tabular}
\end{center}
\end{table*}

Cette petite revue sur le facteur de projection de $\delta$ Cep illustre les différentes approches actuellement utilisées. De mon côté, avec de nombreux collaborateurs, j'ai progressé par étapes:
\begin{enumerate}
\item Comme déjà mentionné, j'ai d'abord étudié le facteur de projection de façon globale et purement théorique avec le modèle hydrodynamique développé par \citet{fokin94}. J'ai ainsi trouvé pour $\delta$ Cep un p-facteur de $p=1.27 \pm 0.01$ \citep{nardetto04}. J'ai également montré durant ma thèse de Doctorat que la variation du facteur de projection (et donc de l'assombrissement centre-bord) avec la phase n'a pas d'impact sur la distance. Cela produit simplement un décalage en phase de la courbe de rayon de 0.02, ce qui n'affecte pas l'amplitude de la courbe et donc la distance \citep{nardetto06b}. 
\item En 2007, à l'aide d'une approche semi-théorique, et en rupture avec les méthodes académiques, j'ai {\it décomposé} le concept du facteur de projection avec une approche cohérente et généralisable à l'ensemble des Céphéides du diagramme Hertzsprung-Russell (HR). J'ai ainsi découvert une relation entre la période des Céphéides et le facteur de projection (relation \emph{Pp}).  La décomposition du facteur du projection dans le cas de $\delta$ Cep est illustrée par la figure \ref{Fig_p} et le papier correspondant se trouve dans l'annexe~\ref{nardetto07}. 

%Du fait de la variation en température et en gravité de surface des Céphéides dans la bande d'instabilité, il existe une relation entre la période et la facteur de projection géométrique \citep{neilson12}. 

\begin{figure}[htbp]
\begin{center}
\includegraphics[width=18cm]{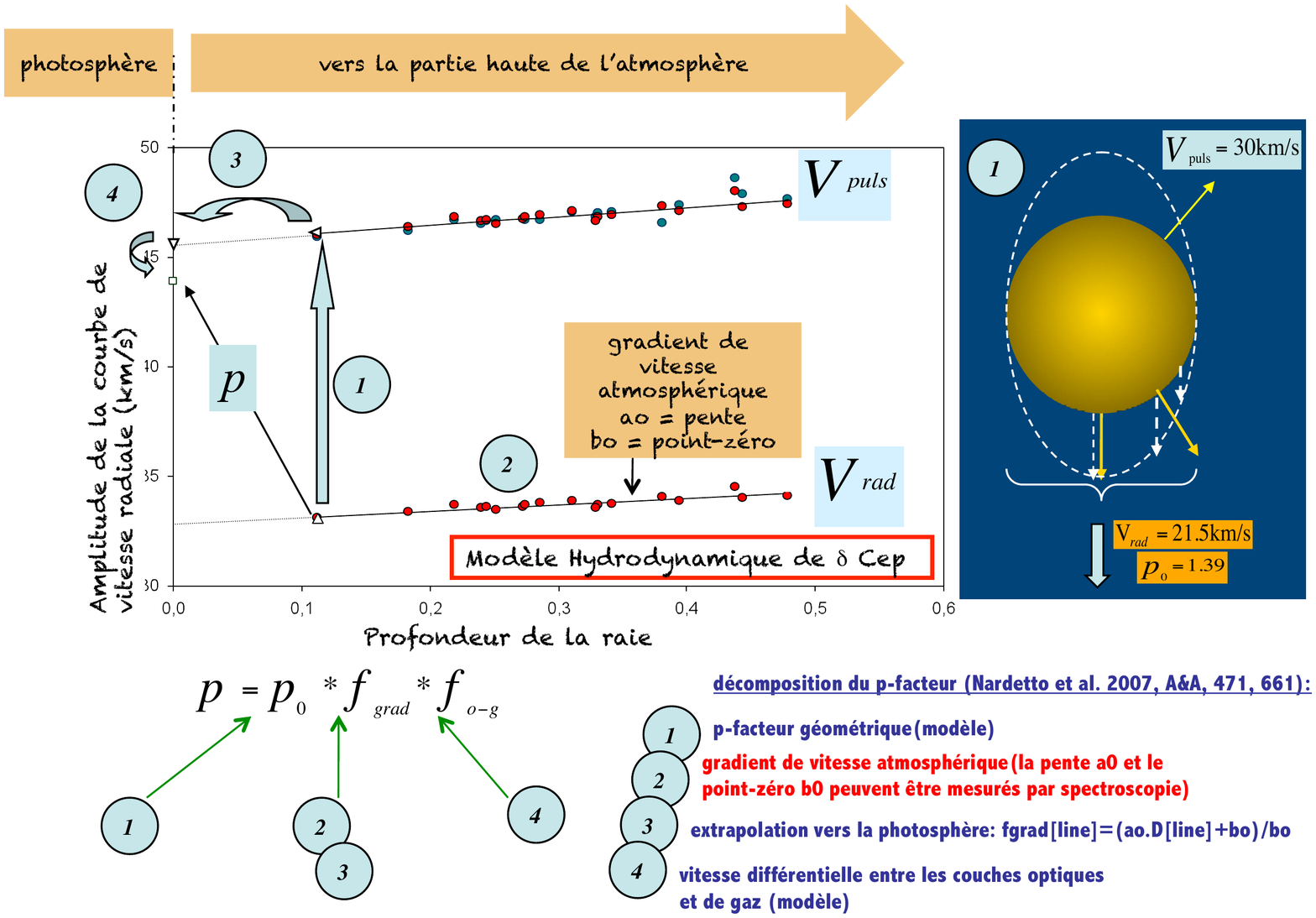}
\end{center}
\vspace*{-15mm} \caption{ \footnotesize   Décomposition du facteur de projection illustrée dans le cas de $\delta$ Cep et à partir du modèle hydrodynamique {\it uniquement}. On procède de la manière suivante. Il faut d'abord déterminer le facteur de projection géométrique $p_\mathrm{0}$. Une Céphéide qui pulserait avec un disque ayant une distribution d'intensité uniforme aurait invariablement un facteur de projection de 1.5.  Du fait de l'assombrissement centre bord, la vitesse radiale résultante sur la ligne de visée est inférieure à la vitesse pulsante (par exemple 21.5 \kms au lieu de 30 \kms; voir encart) ce qui donne un facteur de projection géométrique {\it moyen} (on néglige la variation de l'assombrissement centre-bord et du p-facteur avec la phase) pour $\delta$ Cep entre 1.36 et 1.39, selon la longueur d'onde considérée dans le domaine visible (car l'assombrissement centre-bord dépend de la longueur d'onde).  Ce facteur de projection géométrique permet de passer (étape 1 sur la figure) de l'amplitude de la courbe de vitesse radiale pour une raie donnée (quantité par définition indépendante de la phase et tirée des profils théoriques du modèle; points rouge en bas sur la figure) à l'amplitude de la courbe de vitesse pulsante correspondante (points rouges en haut), dont la valeur s'accorde effectivement bien avec les valeurs de référence issues directement des couches dans l'atmosphère du modèle hydrodynamique (points verts en haut sur la figure). Cet accord confirme la cohérence de la décomposition du facteur de projection. Il faut ensuite prendre en compte le gradient de vitesse dans les couches de l'atmosphère de l'étoile. Celui-ci est illustré par la relation entre l'amplitude des courbes de vitesses radiales ($\Delta RV_\mathrm{c}$, obtenues par la méthode du premier moment; ceci est indispensable sinon cela ne fonctionne pas du fait des biais associés aux autres méthodes) et la profondeur de la raie ($D$, prise au rayon minimum, sinon la décomposition n'est plus cohérente). La relation s'écrit ainsi: $\Delta RV_\mathrm{c} = a_\mathrm{0} D + b_\mathrm{0}$ et peut être directement déduite des observations spectroscopiques (étape 2 sur la figure). La correction sur le facteur de projection géométrique dépendra ainsi de la raie spectrale considérée et s'écrit: $ f_\mathrm{grad} = \frac{a_\mathrm{0} D [line] + b_\mathrm{0}}{b_\mathrm{0}} $.  Ce qu'il faut comprendre ici, c'est que l'on compare l'amplitude de la courbe de vitesse radiale associée à la raie  ($a_\mathrm{0} D[line] + b_\mathrm{0}$) à l'amplitude de la courbe de vitesse radiale associée à la photosphère $b_\mathrm{0}$. Car justement, il ne faut pas perdre de vue l'objectif final de déduire l'amplitude de la courbe de vitesse pulsante associée à la photosphère de façon à obtenir (après intégration temporelle) la variation de rayon photosphérique, celle-là même qui peut être comparée de manière {\it cohérente} à la variation de diamètre angulaire de l'étoile déduite de l'interférométrie ou de la photométrie (à partir du {\it continuum}, c'est-à-dire l'émission du corps noir). Ainsi,  une extrapolation vers la photosphère (étape 3 sur la figure) est nécessaire, et plus la raie se forme haut dans l'atmosphère ($D$ élevé), plus la correction $ f_\mathrm{grad}$ est importante et plus le facteur de projection est réduit (jusqu'à 3\%). Il reste une dernière subtilité: le rayonnement issu du {\it continuum} de l'étoile est absorbé par les éléments chimiques présents dans le gaz qui constitue l'atmosphère. La vitesse radiale peut donc être associée à la vitesse du gaz atmosphérique. Ce n'est pas le cas des observations interférométriques ou photométriques, qui elles sont sensibles à la couche optique que constitue la photosphère de l'étoile. En effet, le gaz peut très bien traverser cette couche optique, et la vitesse différentielle entre les deux doit être prise en compte et l'impact sur le facteur de projection corrigé. Cette correction est notée $f_\mathrm{o-g}$  et peut valoir quelques pourcents selon la Céphéide considérée (étape 4 sur la figure). Finalement, le facteur du projection qui relie très exactement l'amplitude de la courbe de vitesse radiale pour une raie donnée, à la vitesse pulsante de la couche optique constituant la photosphère de l'étoile s'écrit: $ p = p_\mathrm{0} f_\mathrm{grad} f_\mathrm{o-g} $ . Cette décomposition du facteur de projection, dont la cohérence interne a été validée par le code hydrodynamique \citep{nardetto07} a été confirmée dix ans plus tard par une mesure directe du gradient de vitesse dans l'atmosphère de $\delta$ Cep \citep{nardetto17}.} \label{Fig_p}
\end{figure}

\item Dans \citet{nardetto07}, la relation \emph{Pp} obtenue ne s'applique qu'à une raie spécifique, ou peut éventuellement être déduite pour différentes raies à l'aide d'une ‘‘recette''. C'est la raison pour laquelle en 2009, j'ai calculé la correction à appliquer lorsqu'on utilise la méthode de la cross-corrélation (‘‘cc-g'') qui est largement utilisée par la communauté \citep{nardetto09} (voir l'annexe~\ref{paperV}) . L'avantage de la méthode est sa sensibilité (elle permet notamment d'accéder aux Céphéides des nuages de Magellan). Cependant l'inconvénient majeur est qu'elle mélange l'information liée à de nombreuses raies, ce qui induit des hypothèses particulières sur le facteur de projection. De plus, le fait d'utiliser un ajustement gaussien génère également un biais sur le facteur de projection.  Ainsi, le facteur de projection est réduit de l'ordre de 5\% quelle que soit la période de la Céphéide par rapport à un facteur de projection qui serait déduit à partir d'une raie individuelle et la méthode du premier moment.
\item En 2011, j'ai montré de manière purement théorique que le facteur de projection ne dépend pas de la métallicité de l'étoile \citep{nardetto11b} (voir l'annexe~\ref{nardetto11b}), ce qui est un point important à prendre en compte lorsqu'on étudie les Céphéides des nuages de Magellan \citep{storm11a, storm11b}. 
\item Enfin, en 2017, j'ai comparé pour la première fois le modèle hydrodynamique de $\delta$ Cep avec des observations HARPS-N d'une qualité exceptionnelle \citep{nardetto17} (voir l'annexe~\ref{nardetto17}). Ainsi le modèle re-normalisé\footnote{Dans \citet{nardetto17}, il a été nécessaire de mettre à l'échelle le modèle hydrodynamique présenté dans \citet{nardetto04}. En effet, le modèle prédisait des amplitudes de courbes de vitesses radiales inférieures de l'ordre de 7\% à celles obtenues à partir des observations HARPS-N. Ainsi, l'idée est de simplement multiplier la courbe de la vitesse pulsante par un facteur de 7\%, ce qui augmente alors l'amplitude de la courbe de vitesse radiale de 7\% ainsi que la variation de rayon de 7\%, tandis que la facteur de projection lui-même reste inchangé par définition. Le fait que le modèle hydrodynamique ne reproduise pas l'amplitude ne manière convenable est attribué au fait qu'il ne prend pas en compte la convection.}  reproduit de façon remarquable: (1) le gradient de vitesse atmosphérique déduit des données HARPS-N, (2) la variation de diamètre angulaire établie avec l'instrument FLUOR sur l'interféromètre CHARA, et (3) les facteurs de projection purement observationnels déduits de la méthode de BW inverse (prenant la distance du HST comme référence). Ces résultats semblent indiquer que la décomposition du facteur de projection telle que présentée dans l'article de 2007 est adéquate (voir Fig.~\ref{Fig_harps-n}). 
 \end{enumerate}

\begin{figure}
\centering
%\sidecaption
\includegraphics[width=8cm,clip]{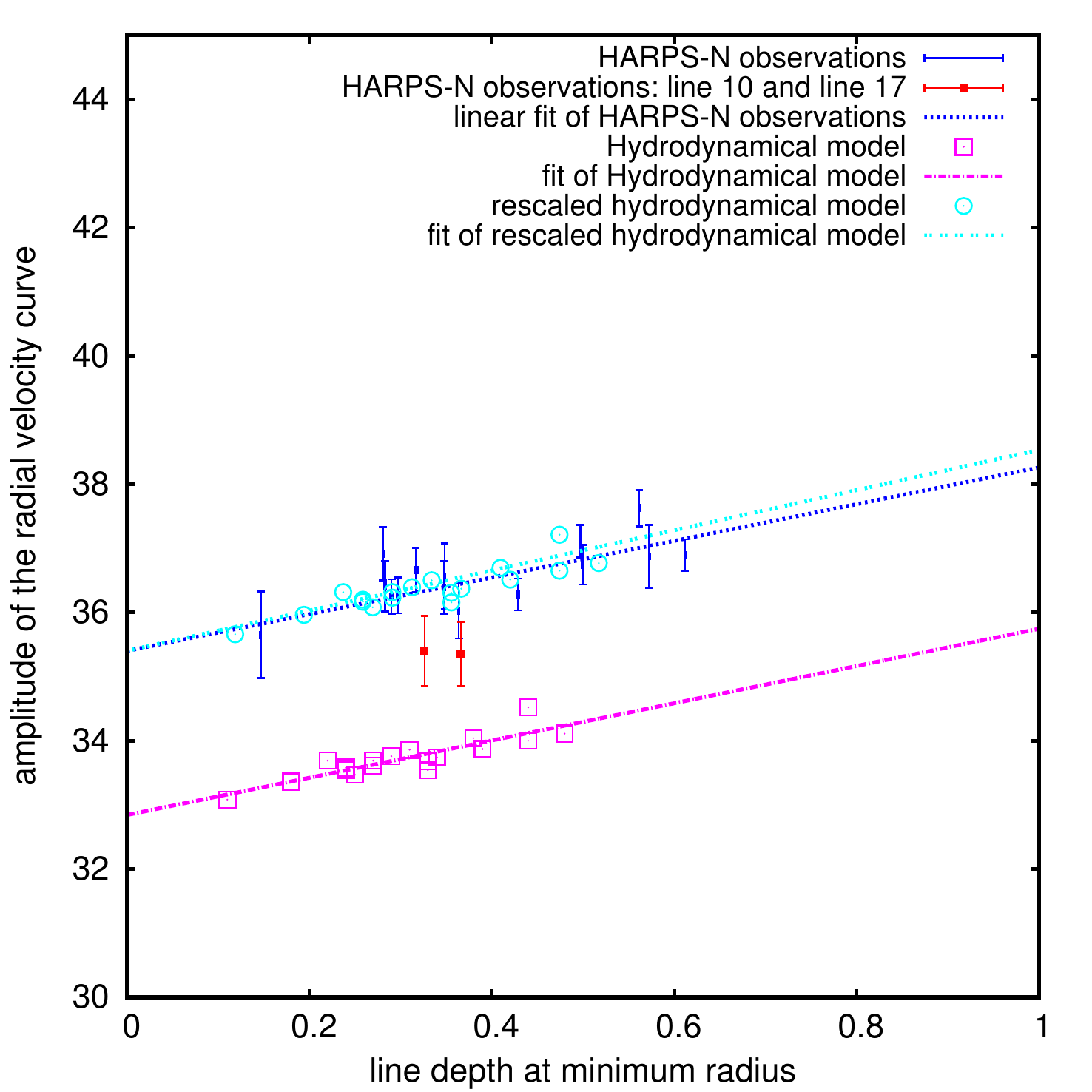}
\includegraphics[width=8cm,clip]{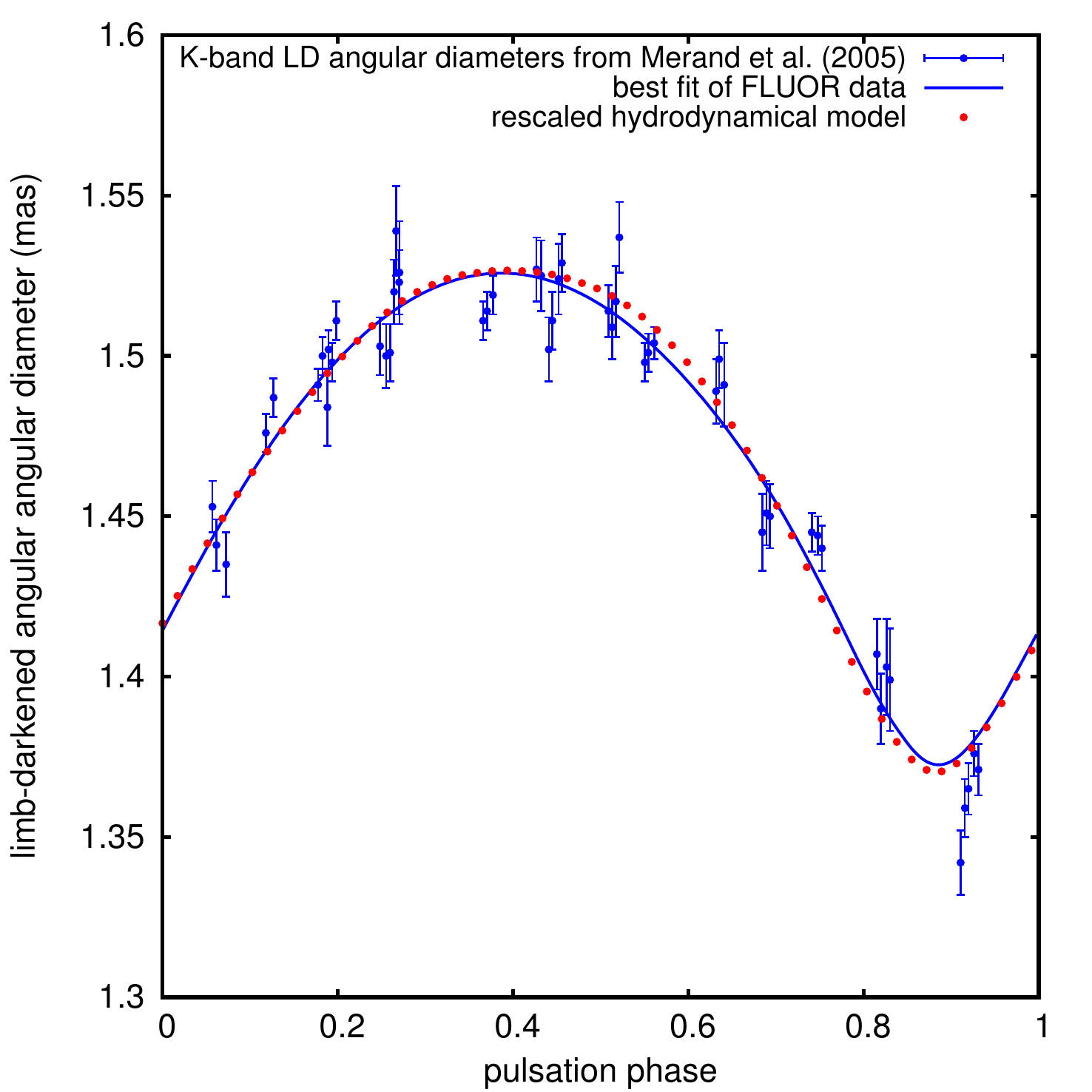}
%\centering
%\sidecaption
%\vspace{-1cm}
\includegraphics[width=8cm,clip]{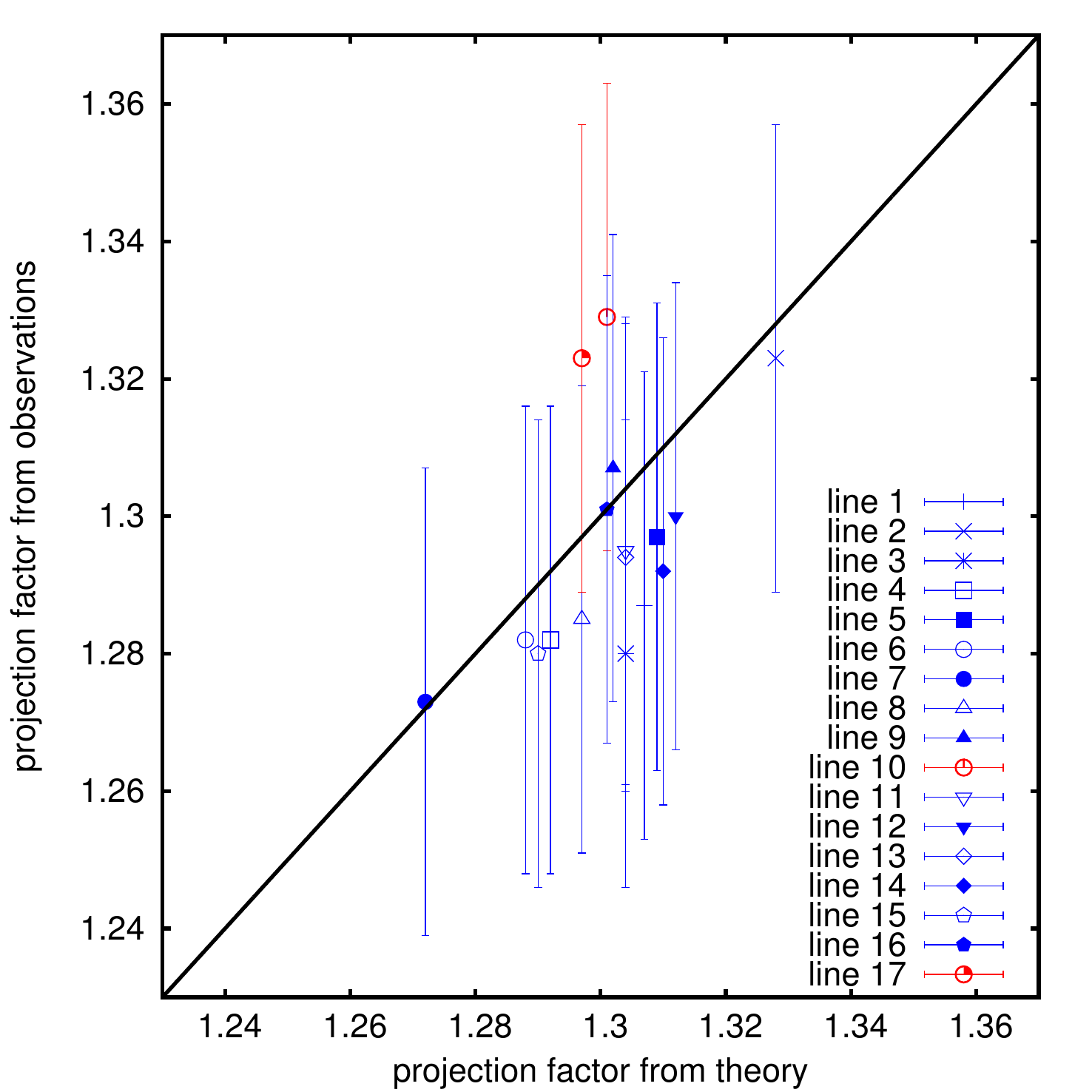}
\caption{ \footnotesize   Figure en haut à gauche: Comparaison du gradient de vitesse dans l'atmosphère de $\delta$ Cep déduit des observations HARPS-N (bleu foncé), du modèle hydrodynamique (magenta) et du modèle hydrodynamique redimensionné (bleu clair). Figure en haut à droite: comparaison de la variation du diamètre angulaire déduite des données FLUOR/CHARA (en bleu) et du modèle hydrodynamique redimensionné (en rouge). Figure du bas: comparaison des facteurs de projection déduits de la méthode de BW inverse (observation) et du modèle hydrodynamique (théorie).  Sur ces trois figures, l'accord entre les observations et le modèle est remarquable, ce qui confirme la décomposition du facteur de projection décrite dans \cite{nardetto17}.}
\label{Fig_harps-n}       % Give a unique label
\end{figure}

\begin{figure}
\centering
%\sidecaption
\includegraphics[width=8cm,clip]{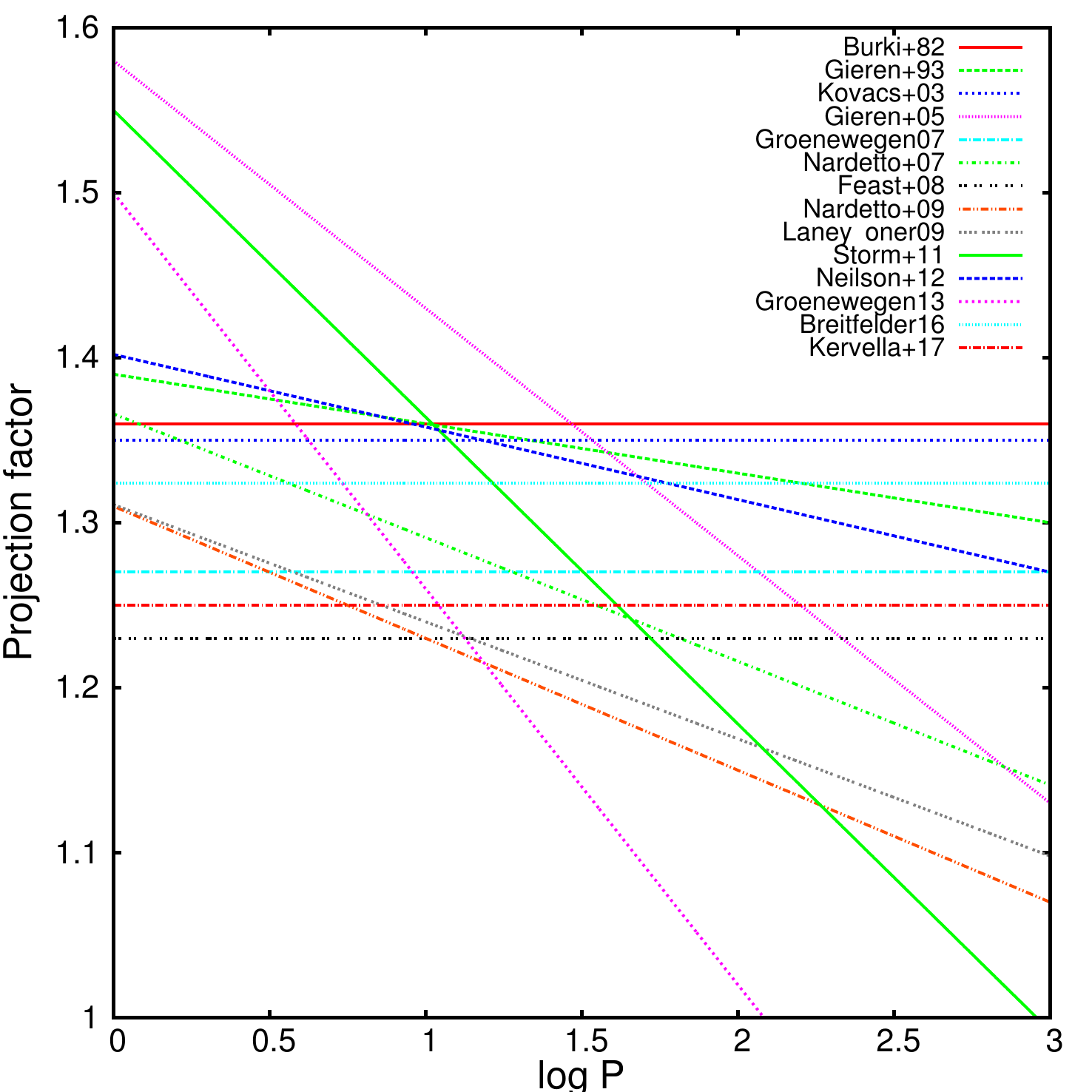}
%\centering
%\sidecaption
%\vspace{-1cm}
\includegraphics[width=8cm,clip]{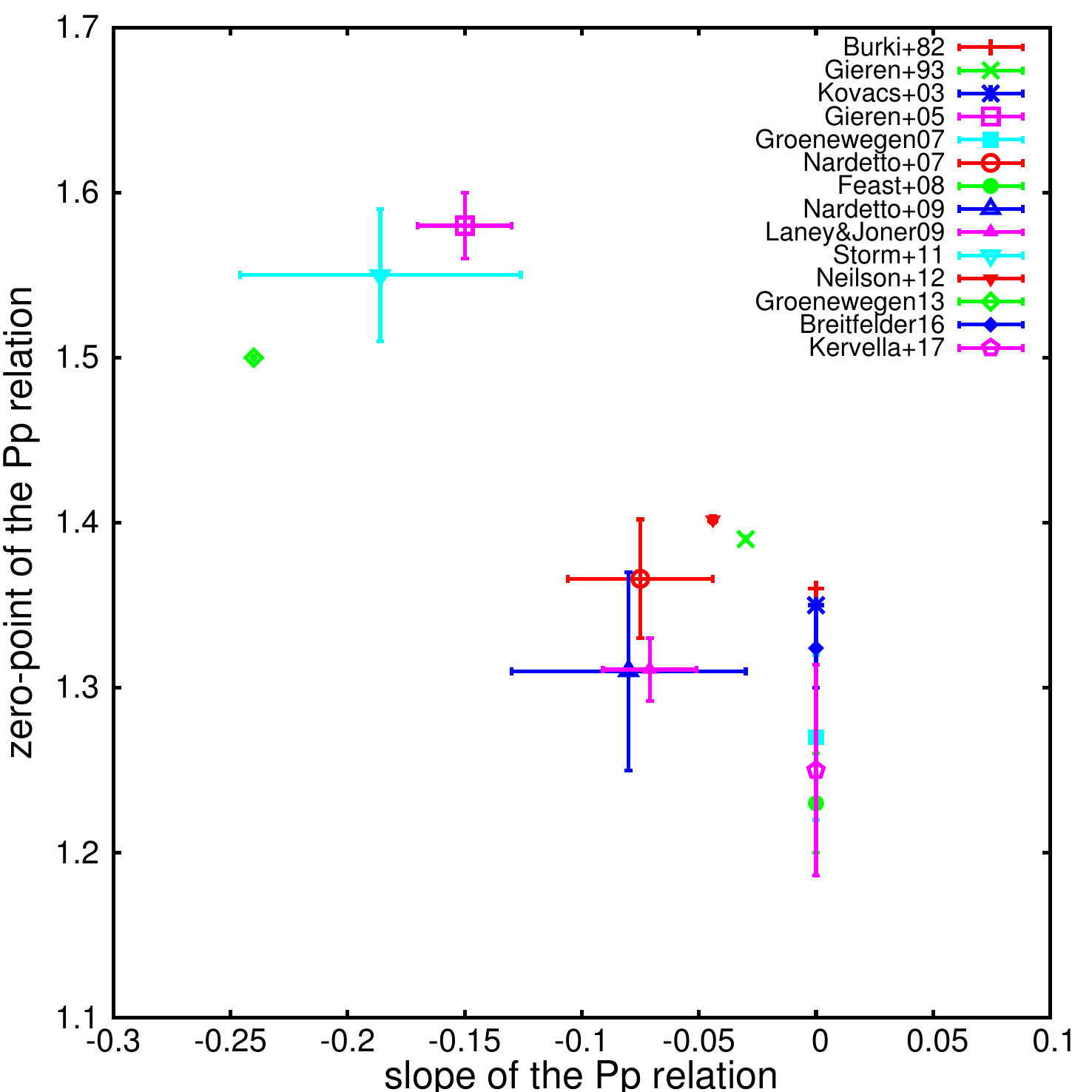}
\caption{ \footnotesize   A gauche, les relations $Pp$ disponibles à ce jour. A droite, le point-zéro de ces relations est représenté en fonction de leur pente.}
\label{Fig_Pp1}       % Give a unique label
\end{figure}

Ainsi le facteur de projection {\it dynamique} de $\delta$~Cep semble bien établi et compris physiquement avec une valeur (pour la méthode {\it cc-g}) autour de 1.25. Cependant, plusieurs points restent à résoudre:

\begin{enumerate}
\item Pourquoi les valeurs basées sur la relation brillance de surface - couleur {\it inverse} dans l'infrarouge sont significativement supérieures \citep{gieren05, storm11b, gro13} à 1.25. Il est à noter tout de même que la valeur de \citet{merand15} qui est hybride (basée sur l'interférométrie et la photométrie) donne une valeur intermédiaire. Est-ce que la relation brillance de surface - couleur est sensible à l'environnement des Céphéides ? Par exemple, il est intéressant de noter que les deux versions de la méthode de Baade-Wesselink basées respectivement sur l'interférométrie et les relations brillance de surface - couleur donnent des résultats cohérents pour la Céphéide longue période $\ell$~Car \citep{kervella04d}, mais pas pour $\delta$ Cep \citep{ngeow12}. Par ailleurs, toutes ces valeurs relativement plus élevées du facteur de projection sont déduites de relations \emph{Pp} basées sur les Céphéides du LMC.  Est-ce donc un problème lié à la métallicité ou à la géométrie du LMC ? 
\item Mais le problème est ensuite plus complexe, comme on le voit sur les figures \ref{Fig_Pp1}ab qui illustrent les différentes relations $Pp$ disponibles à ce jour. Des désaccords importants sont toujours présents et débattus. Or ces relations ont un impact significatif sur les distances. En effet, si l'on utilise un facteur de projection constant (traditionnellement $p=1.36$ pour toutes les étoiles), au lieu de la relation donnée par \citep{nardetto09} par exemple, alors on peut introduire une différence de l'ordre de 0.10 et 0.03 mag respectivement sur la pente et le point-zéro de la relation \emph{PL}, ce qui correspond à une différence de l'ordre de 10\% en terme de distance. 
\end{enumerate}

\begin{figure}
\centering
%\sidecaption
\includegraphics[width=15cm,clip]{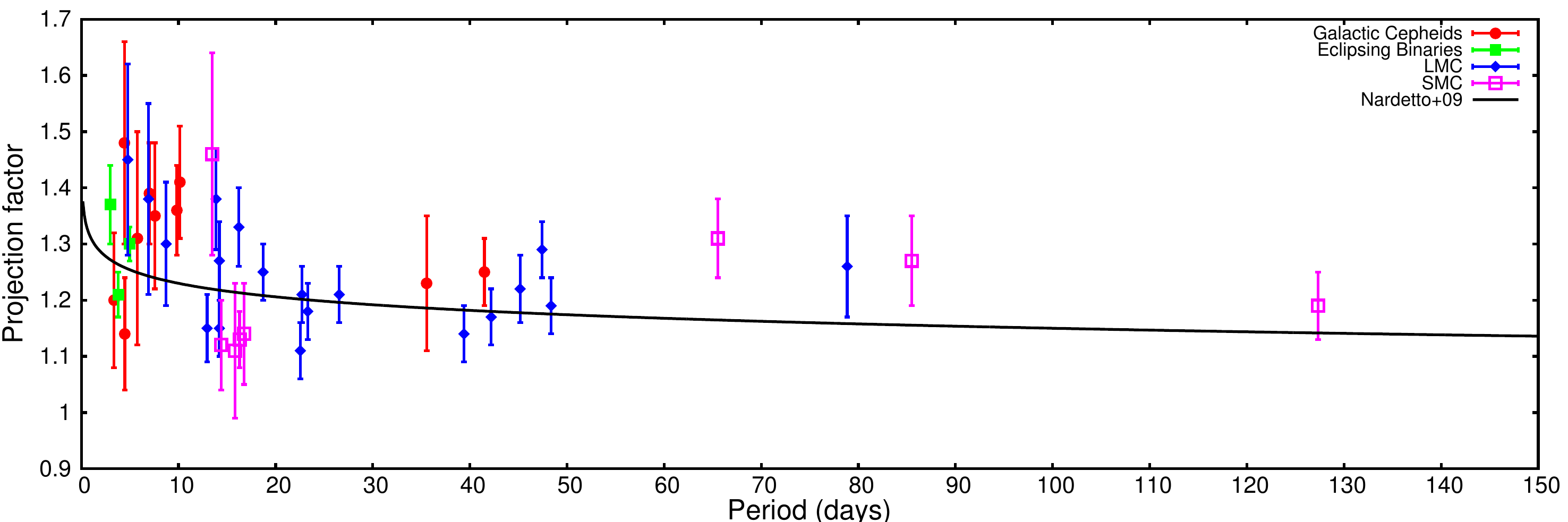}
%\centering
%\sidecaption
%\vspace{-1cm}
\includegraphics[width=15cm,clip]{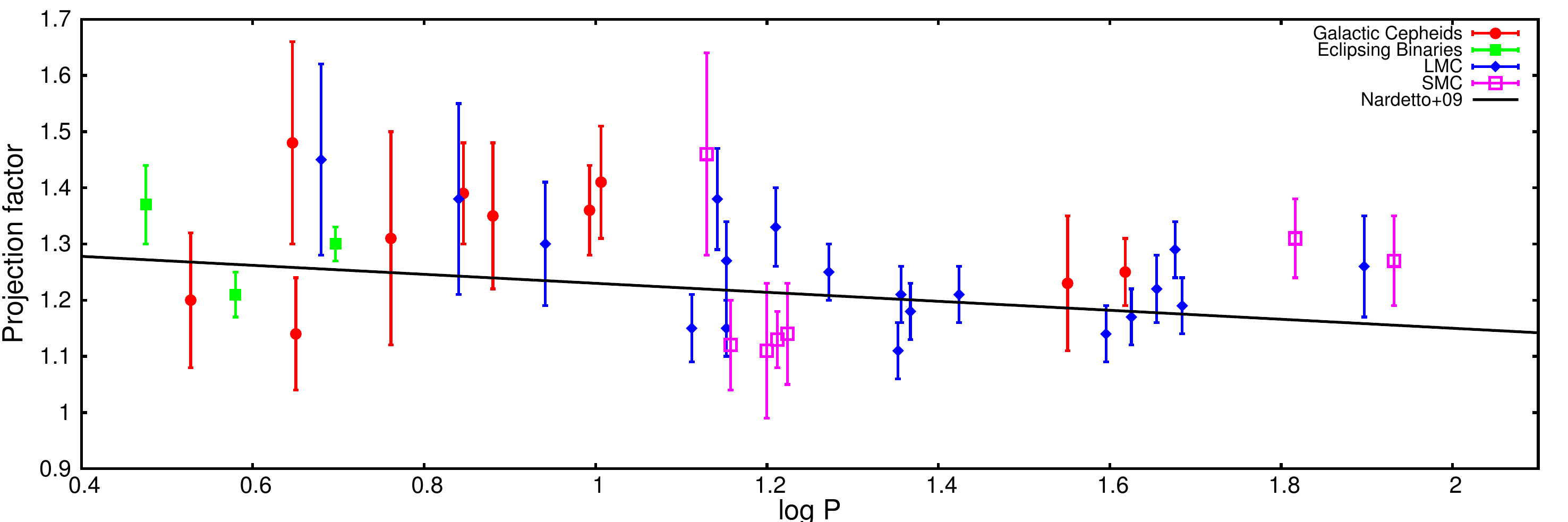}
\caption{ \footnotesize   Le facteur de projection est représenté en fonction du logarithme de la période. Ces valeurs de $p$ sont purement observationnelles, i.e. soit déduites des méthodes de BW inverses appliquées aux Céphéides galactiques \citep{breitfelder16}, du SMC, et du LMC \citep{gallenne17}, soit des binaires à éclipses dont l'une des composantes est une Céphéide (voir Sect.~\ref{Sect_EBp}). Avec Gaia, nous serons en mesure de déduire environ 300 $p$-facteurs de ce type avec une précision de l'ordre de 3\%. Notre connaissance du facteur de projection, en particulier sa décomposition, sera alors très utile pour interpréter ces résultats. }
\label{Fig_Pp2}       % Give a unique label
\end{figure}

L'apport de {\it Gaia} sera considérable pour répondre à ces questions. En effet, en appliquant la méthode de BW inverse aux parallaxes {\it Gaia} (DR2, Avril 2018), nous aurons très prochainement accès au facteur de projection de 300 Céphéides avec une précision de 3\%. Par ailleurs, la distance du LMC et du SMC deviennent maintenant suffisamment précises pour appliquer les méthodes de BW inverse aux Céphéides de ces galaxies. Le travail a commencé, représenté par la figure \ref{Fig_Pp}ab. Dans ce contexte,  et à terme, la décomposition du facteur de projection sera un outil précieux pour comprendre et analyser l'ensemble de ces facteurs de projection.

% récupérer les facteurs de projection de Storm, EBs, BW inverse + bilan des relation Pp. 

\newpage

%%%%%%%SECTION 
\section{Le $k$-facteur}\label{Sect_k}

\begin{figure}[htbp]
\begin{center}
\includegraphics[width=14cm]{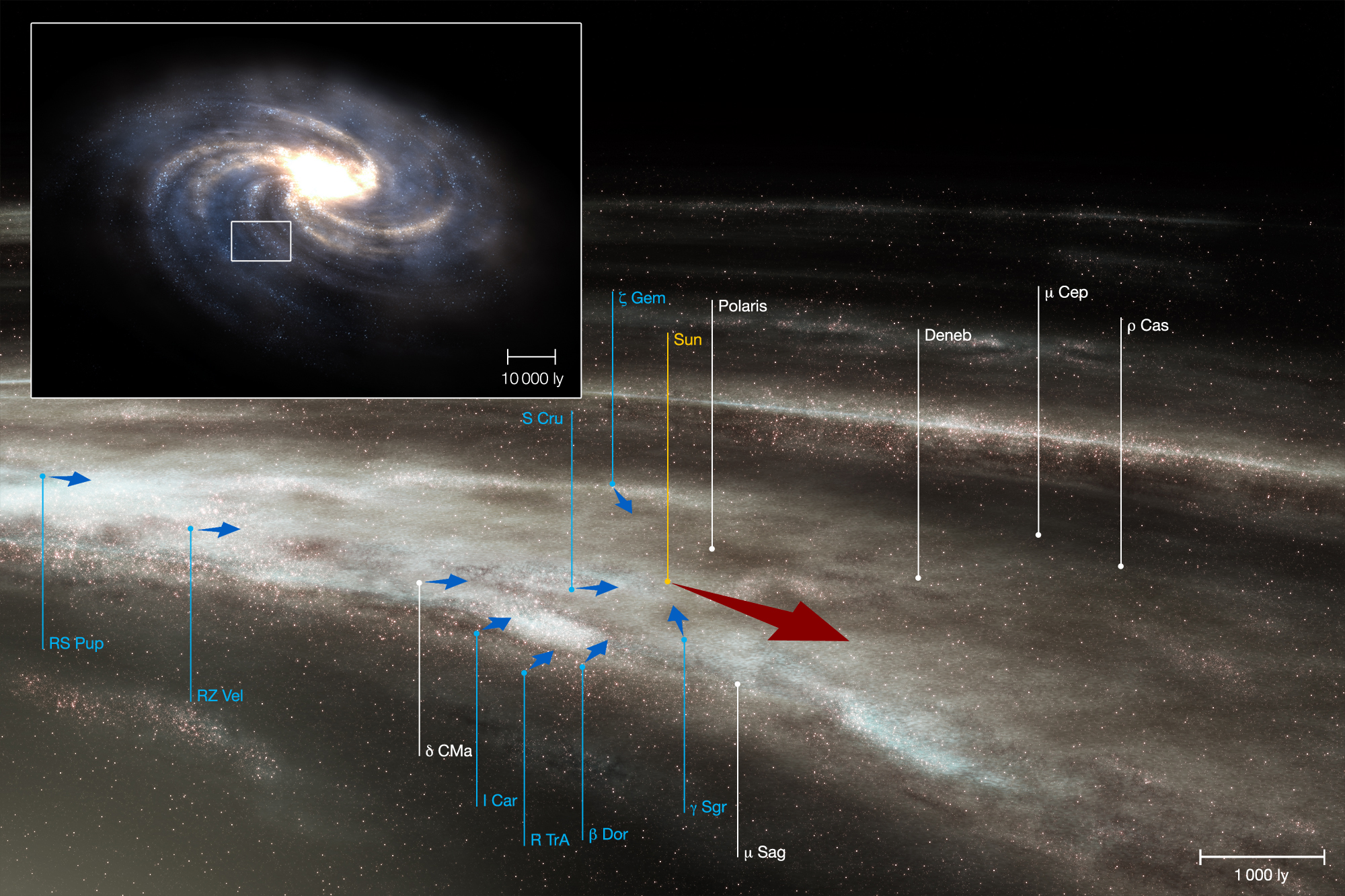}
\end{center}
\vspace*{-5mm} \caption{ \footnotesize   Depuis 1939 et jusqu'à récemment, on s'interrogeait sur le fait que les Céphéides de notre voisinage semblaient tomber vers le Soleil avec une vitesse résiduelle de l'ordre de 1 ou 2 \kms. Etait-ce le résultat d'un mouvement réel des Céphéides au sein de notre Voie Lactée ou une propriété intrinsèque liée à leur dynamique atmosphérique ? Grâce à l'étude de l'asymétrie des raies spectrales, nous avons montré qu'il s'agissait un phénomène intrinsèque aux Céphéides \citep{nardetto08a} [Crédit - ESO Press Release: Pinning down the Milky Way's spin (19 September 2008)]. Très récemment,  \citet{vasilyev17} ont montré en utilisant un modèle bidimensionnel d'une Céphéide courte période qu'il s'agit d'un phénomène lié à la granulation.} \label{Fig_keso}
\end{figure}

Si l'on considère un mouvement axi-symétrique de rotation de la Voie Lactée, les Céphéides présentent un mouvement résiduel d'approche vers le Soleil d'environ 2 \kms (Fig.~\ref{Fig_keso}). Ce mouvement résiduel, appelé le $k$-facteur, a été pour la première fois observé par \citet{joy39}. Depuis 1939, la question est de savoir si le $k$-facteur est effectivement lié à un mouvement réel des Céphéides au sein de la Voie Lactée \citep{camm38, camm44, parenago45, stibbs56, caldwell87, moffett87b, wilson91, pont94}  ou s'il s'agit d'un phénomène intrinsèque lié à la dynamique atmosphérique de ces étoiles \citep{wielen74,butler96}. La réponse est naturellement venue en étudiant les détails de la méthode de BW.

\begin{figure}[htbp]
\begin{center}
\includegraphics[width=16cm]{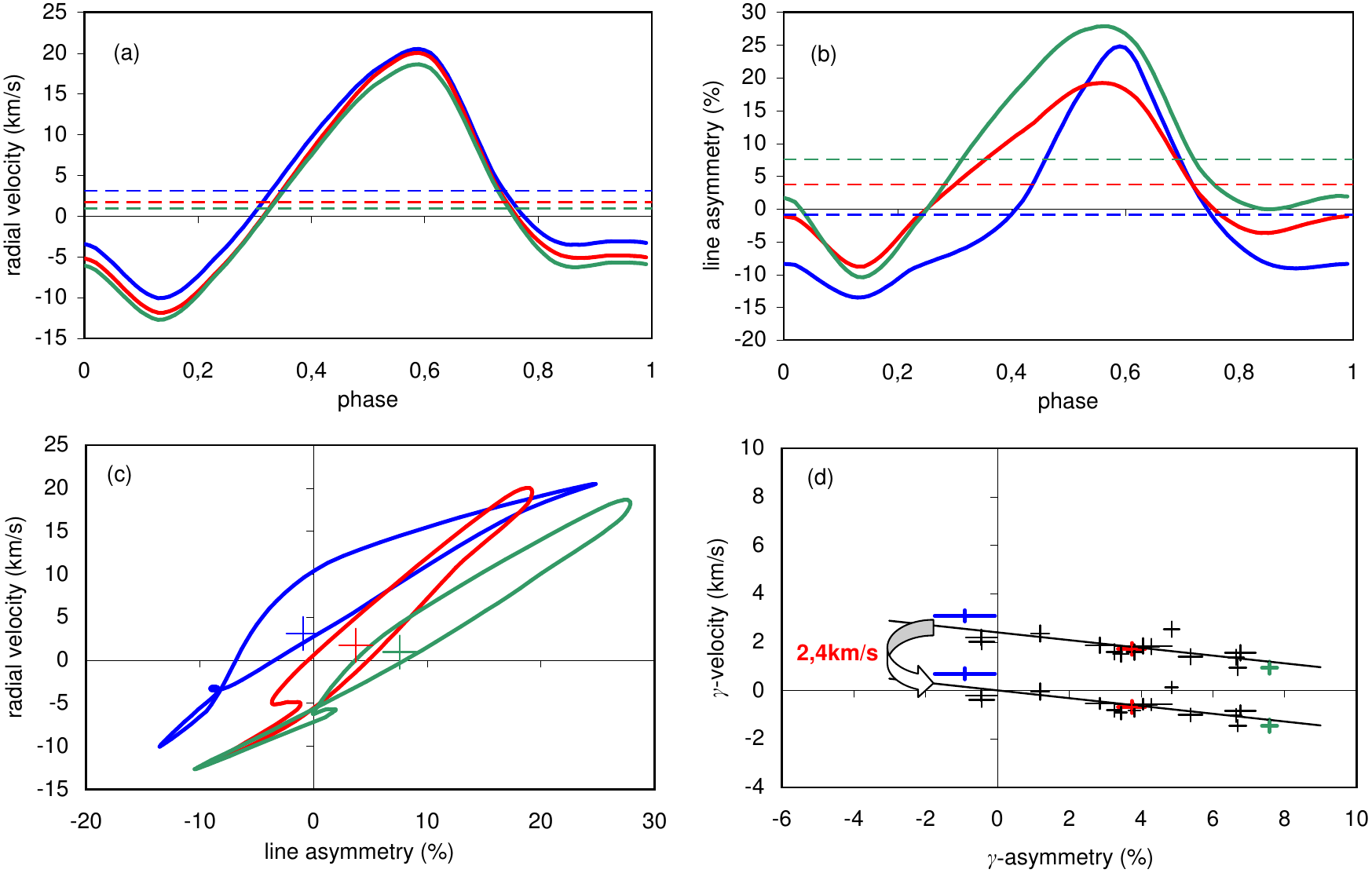}
\end{center}
\vspace*{-5mm} \caption{ \footnotesize  Figure explicative du $k$-facteur des Céphéides (voir le texte).} \label{Fig_k}
\end{figure}

En effet, lorsqu'on applique la méthode de BW, il faut corriger la courbe de vitesse radiale  $RV_\mathrm{cc-g}$  de la vitesse du centre de masse de l'étoile $V_\mathrm{\star}$ (Eq.~\ref{diam_mod}) et donc forcer sa moyenne à zéro de façon à ce que l'intégration temporelle donne bien la variation de rayon. Or, si l'on compare les vitesses radiales associées aux premiers moments de trois raies spectrales comme illustré sur la Fig~\ref{Fig_k}a dans le cas de $\beta$~Dor, on s'aperçoit que ces moyennes (notées $\gamma$-vitesses) sont différentes d'une raie à l'autre, de l'ordre de quelques pourcents. Ceci est surprenant. En effet, les amplitudes des trois courbes peuvent éventuellement être différentes du fait du gradient de vitesse atmosphérique comme nous l'avons déjà vu dans la section précédente. En revanche, un élément chimique donné à l'origine d'une raie spectrale en absorption dans l'atmosphère de la Céphéide devrait en principe retrouver sa position initiale après un cycle de pulsation, et les trois courbes devraient avoir une moyenne identique (correspondant à la vitesse du centre de masse de l'étoile), et par extension, cela devrait être aussi le cas pour $RV_\mathrm{cc}$ qui repose sur des centaines de raies spectrales. Mais ce n'est manifestement pas le cas. Un problème se pose alors: comment véritablement séparer la pulsation de l'étoile de la vitesse de son centre de masse ? Et à quoi correspondent physiquement  ces différences de vitesses moyennes ? Ceci a bien sûr son importance dans le cadre de la méthode de BW.

Nous avons alors proposé une approche qui permet de calculer la vitesse du centre de masse de l'étoile, basée sur l'asymétrie des raies \citep{nardetto08a} (voir le papier correspondant dans l'annexe \ref{nardetto08a}) . L'asymétrie\footnote{Il existe plusieurs façons de déterminer l'asymétrie d'une raie spectrale. La méthode que nous avons développée dans \citet{nardetto06a} (voir Eq. 1) consiste à ajuster deux demi gaussiennes sur la raie spectrale et de comparer leur largeur à mi-hauteur.} des trois raies est représentée sur la Fig~\ref{Fig_k}b. Il est intéressant de remarquer que les valeurs moyennes de ces trois courbes d'asymétrie (notées $\gamma$-asymétries) sont anti-corrélées aux valeurs moyennes des courbes de vitesses radiales. La Fig~\ref{Fig_k}c représente la vitesse radiale en fonction de l'asymétrie de la raie spectrale sur le cycle de l'étoile. On observe des boucles (dont la forme est complexe et reliée à la dynamique de l'atmosphère de la Céphéide) et dont les barycentres (les trois croix de couleurs) sont alignés. On peut effectivement généraliser ce résultat en prenant une vingtaine de raies parfaitement ‘‘propres'', c'est-à-dire non contaminées par d'autres raies stellaires ou telluriques (Fig~\ref{Fig_k}d). Dans la mesure où un modèle quasi-statique de Céphéide ne présente pas de résidus de vitesses radiales ni d'asymétrie (point à l'origine du graphique), on fait alors l'hypothèse suivante: {\it une raie spectrale dont la $\gamma$-asymétrie est nulle doit nécessairement avoir une $\gamma$-vitesse nulle}. En conséquence, le point zéro de la relation entre la $\gamma$-vitesse et la $\gamma$-asymétrie correspond à la vitesse du centre de masse de l'étoile. Cette valeur vaut $V_\mathrm{\star} = 9.8 \pm 0.1$ \kms dans le cas de $\beta$~Dor \citep{nardetto08a}. Or, si l'on considère la valeur correspondante issue de la base de données de \citet{fernie95}, on obtient <$RV_\mathrm{cc-g}$> = 7.4 \kms. La différence entre les deux vaut donc (toujours dans le cas de $\beta$~Dor) $2.4\pm 0.1$ \kms. On trouve le même comportement pour les huit Céphéides étudiées \citep{nardetto08a} avec une moyenne de $1.8 \pm 0.2$ \kms (en prenant comme point référence les courbes cross-corrélées de la littérature). Cette valeur a ensuite été révisée à $1.0 \pm 0.2$ \kms en considérant les courbes de vitesses radiales cross-corrélées déduites des données HARPS, i.e. de manière cohérente en utilisant le même jeu de données \citep{nardetto09}. 

Ainsi, si l'on prend en compte la dynamique atmosphérique des Céphéides, c'est-à-dire si l'on détermine la vitesse du centre de masse des Céphéides avec la méthode proposée ci-dessus, alors les Céphéides ne semblent plus s'approcher du soleil avec une vitesse radiale résiduelle de l'ordre de 1-2\kms. Il faut alors chercher l'explication du $k$-facteur non pas dans le mouvement des Céphéides au sein de la Voie Lactée avec un mouvement non axi-symétrique par exemple \citep{pont94} mais en étudiant la physique des Céphéides. Plusieurs pistes ont alors été évoquées pour expliquer le $k$-facteur: 1) la position des zones de formation des raies dans l'atmosphère des Céphéides n'est pas strictement cyclique ou alors, 2) le décalage vers le bleu observé est dû à des mouvements de convection, de la granulation. Récemment, une réponse a une nouvelle fois été apportée par \citet{vasilyev17} à partir de leur modèle bi-dimensionnel incluant la convection. Il semblerait que le $k$-facteur soit principalement dû à la granulation \footnote{Il est intéressant d'avoir en tête que le $k$-facteur, qui est  donc mesuré à partir des raies spectrales, est lui aussi affecté par le facteur de projection.}. Ils trouvent effectivement une valeur du $k$-facteur moyen (i.e. calculé à partir de nombreuses raies spectrales de forces différentes) entre -0.5 et -1.0 \kms, ce qui semble compatible avec la valeur obtenue par \citet{nardetto09} à partir de la méthode de la cross-corrélation. La question importante à laquelle il faudra répondre dans les années à venir est de quantifier l'impact de ce $k$-facteur sur la distance de BW des Céphéides.

\newpage

%%%%%%%SECTION 
\section{L'environnement des Céphéides}\label{sect_CSE}

Il semblerait que la plupart des Céphéides disposent d'une enveloppe circumstellaire dont la contribution en flux est corrélée à leur période de pulsation (voir Fig.~\ref{Fig_CSE}). Ces enveloppes ont un double impact: (1) elles ont un effet sur la relation \emph{PL} des Céphéides (voir Sect.~\ref{s_CSEPL}) et (2) elles affectent également, dans le cadre de la méthode de BW,  l'estimation du diamètre angulaire (interférométrique ou photométrique). Or, la taille, la forme et la nature chimique (gaz et/ou poussière) de ces enveloppes sont très mal connues. Un des {\it work-packages} de l'ANR {\it UnlockCepheids} dont je suis responsable vise précisément à caractériser les enveloppes des Céphéides de façon à les intégrer de façon cohérente dans l'approche SPIPS. L'idée {\it in fine} est de relier les caractéristiques des enveloppes à la position des étoiles  dans la bande d'instabilité de façon à fournir des facteurs correctifs dans différentes bandes photométriques, avec l'objectif, par exemple, de débiaiser la future relation \emph{PL} qui sera établie à partir des observations du JWST. 

\begin{figure}[htbp]
\begin{center}
\includegraphics[width=16cm]{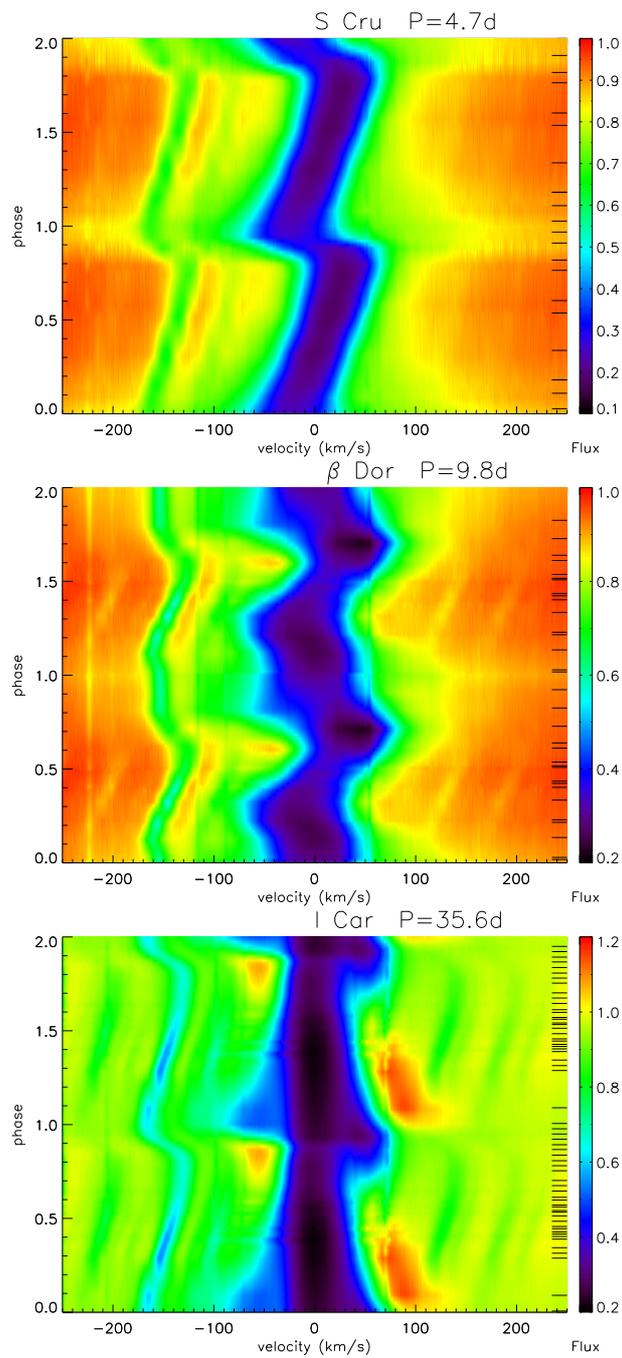}
\end{center}
\vspace*{-5mm} \caption{ \footnotesize   Profils H$\alpha$ des Céphéides en fonction de la période de pulsation. Plus la période de la Céphéide est importante, plus le profil moyen est asymétrique vers le bleu, signature probable d'une perte de masse \citep{nardetto08b}}\label{Fig_Ha}
\end{figure}

\begin{figure}[htbp]
\begin{center}
\includegraphics[width=16cm]{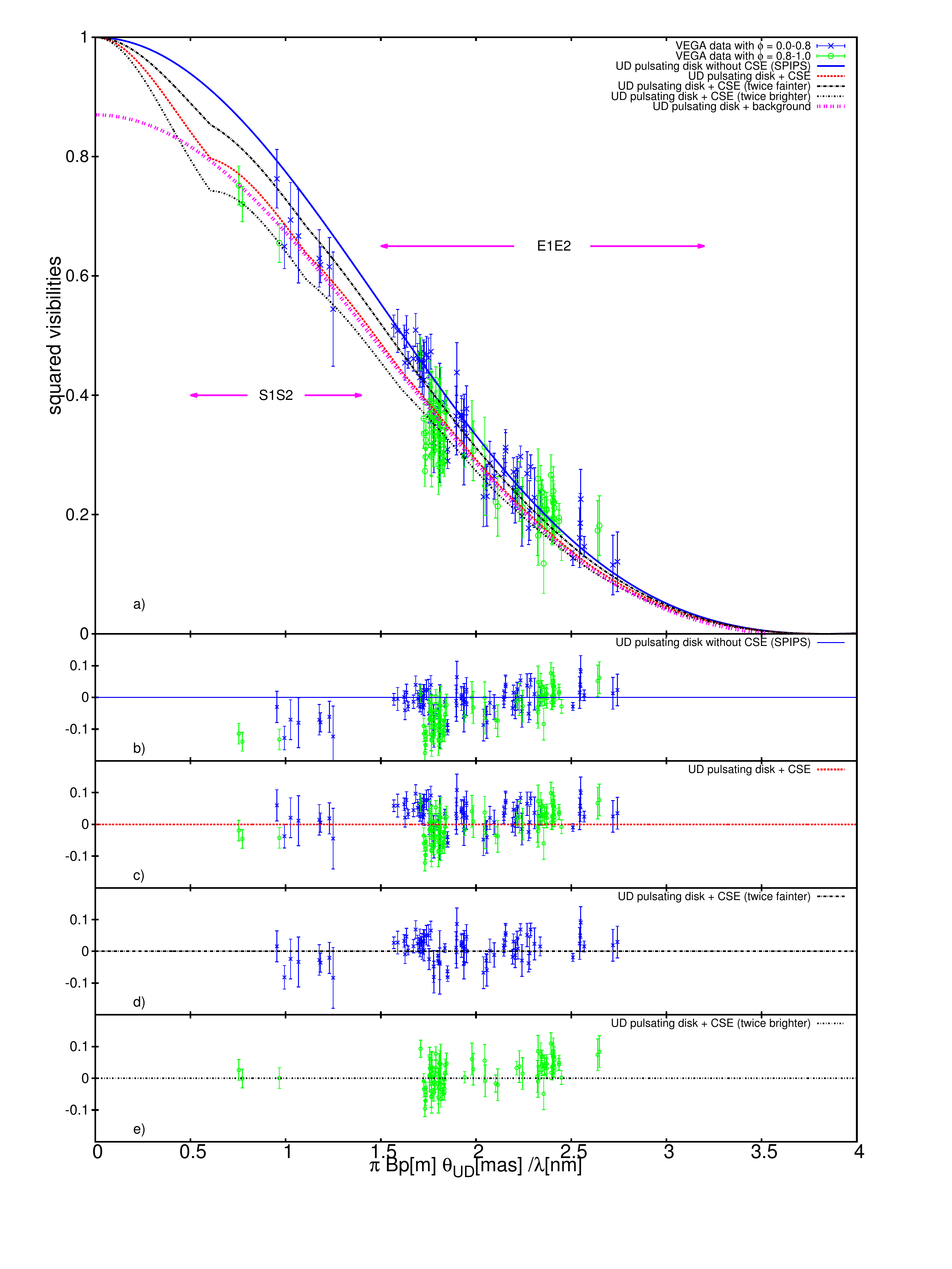}
\end{center}
\vspace*{-15mm} \caption{ \footnotesize  a) Les visibilités carrées obtenues avec l'instrument VEGA/CHARA sur $\delta$ Cep sont représentées en fonction de  $x = \frac{\pi B_p \mathrm{[m]} \theta_\mathrm{UD} \mathrm{[mas]} }{\lambda \mathrm{[nm]}}$ de façon à permettre la comparaison entre des données provenant de différentes phases de pulsation (i.e. avec des  $\theta_\mathrm{UD}$ et des longueurs d'onde effectives différentes). Cinq modèles sont représentés: (1) un disque pulsant uniforme (UD)  de référence issu de l'analyse SPIPS combinant plusieurs bandes photométriques (ligne bleue), (2) le disque pulsant UD SPIPS + une CSE (ligne en pointillés rouge), (3) le disque pulsant UD SPIPS + une CSE 2 fois plus brillante (ligne en pointillés), (4) le disque pulsant UD SPIPS + une CSE 2 fois moins brillante (ligne en tirés), (5) le disque pulsant UD SPIPS + une contribution en flux d'un fond couvrant l'ensemble du champ de vue de l'interféromètre (ligne en magenta). Les résidus entre les observations et les modèles 1 à 4 sont indiqués dans les panels b), c), d) et e), respectivement. Dans les panels d) et e), seules les données correspondant aux intervalles de phase $\phi=0.0-0.8$ et $\phi=0.8-1.0$ sont représentées, respectivement. On peut conclure de l'analyse de ces données que $\delta$ Cep est effectivement entourée d'une enveloppe ou d'une contribution en flux d'arrière-plan, et qu'il existe deux régimes $\phi=0.0-0.8$ et $\phi=0.8-1.0$, respectivement, qui pourraient être l'indice d'une réverbération au moment du rayon minimum de l'étoile.}\label{Fig_DC_VEGA}
\end{figure}

Ces enveloppes ont été découvertes dans l'infrarouge, par interférométrie, imagerie ou par un calcul de SED (voir le Tableau récapitulatif~\ref{Tab.cse}). Dans le visible, des indices d'enveloppe ont été rapportées par deux études. La première porte sur des séries de profils H$\alpha$ de huit Céphéides \citep{nardetto08b} (voir le papier correspondant dans l'annexe~\ref{nardetto08b}). Plus la période d'une Céphéide donnée est élevée, plus ses profils présentent une asymétrie résiduelle vers le bleu importante, avec des vitesses correspondantes (moyennées sur l'ensemble du cycle) de quelques~\kms pour les plus courtes période à -20 ~\kms pour les longues périodes. Ces vitesses pourraient être le résultat d'une perte de masse de l'étoile. Pour $\ell$~Car, non seulement le profil H$\alpha$ est asymétrique, mais il présente également une composante en absorption statique correspondant une vitesse nulle, que l'on attribue à une enveloppe visible statique (voir Fig.~\ref{Fig_Ha}). Cependant, $\ell$~Car a également été observée par interférométrie visible avec SUSI  \citep{davis09} (voir Tableau~\ref{Tab.hra}) et bien que la pulsation ait été mesurée pour cette étoile, aucune enveloppe n'a été détectée (seulement une variation d'assombrissement centre-bord avec la phase).  Entre 2008 et 2014, mes collaborateurs et moi-même avons décidé d'observer les Céphéides avec VEGA/CHARA. Or, 15860 cycles après sa découverte par J. Goodrick en 1783, $\delta$~Cep, l'étoile prototype qui donne son nom aux étoiles de type Céphéides, avait encore quelques secrets à nous livrer.  Nous avons ainsi mis en évidence la présence d'un environnement visible autour de cette étoile \citep{nardetto16a} (voir l'annexe~\ref{nardetto16a}). Du fait de l'inadéquation entre le diamètre angulaire de $\delta$~Cep (environ 1.4 mas) et la configuration de bases de l'interféromètre CHARA (les plus courtes bases sont de 30 mètres environ),  il n'a pas été possible de caractériser véritablement l'environnement. Il peut s'agir d'une enveloppe avec un diamètre de l'ordre de $8.9\pm3.0$ mas avec une contribution en flux de l'ordre de $7\pm1$\%. Ou alors, il s'agit un rayonnement d'arrière-plan remplissant le champ de vue de l'interféromètre, mais contribuant toujours à hauteur de 7\% en terme de flux (voir Fig.~\ref{Fig_DC_VEGA}). Notons que le compagnon de $\delta$ Cep récemment découvert \citep{anderson15a} n'est pas détectable par l'instrument VEGA. Un deuxième aspect ressort de cette étude:  il semblerait que $\delta$~Cep s'éloigne d'un modèle de pulsation quasi-statique autour du rayon minimum. La variation de l'assombrissement centre-bord de l'étoile ne peut pas expliquer les données, car l'effet serait d'un ordre de grandeur plus faible d'après les modèles. Notre meilleure hypothèse serait l'existence d'un phénomène de réverbération: au rayon minimum, l'étoile est chaude, brillante, et son flux pourrait se refléter sur l'environnement, ce qui modifierait sa contribution en flux. Il est difficile de conclure pour le moment. 

Cette découverte de l'environnement visible des Céphéides n'aura pas d'impact {\it direct} sur l'étalonnage des distances dans l'univers dans la mesure où la relation \emph{PL},  dans le {\it visible}, est très dispersée, au contraire de sa contrepartie infrarouge qui, elle, est utilisée pour les distances. En revanche, ces données visibles apporteront très probablement des contraintes très fortes sur les modèles d'enveloppes, et c'est justement à partir de ces modèles que les contributions infrarouges pourront être estimées de manière exacte. 

%%%%%%%SECTION 
\section{Une piste de recherche: la comparaison avec les autres types d'étoiles pulsantes}\label{Sect_HADS}

Le $p$-facteur, le $k$-facteur et la caractérisation de l'environnement des Céphéides sont des problèmes complexes. Procéder par comparaison avec d'autres types d'étoiles pulsantes, dont le mode radial est dominant, est une approche intéressante. Il s'agit de mettre en évidence certaines tendances dans le diagramme HR : existe-il une relation entre la position de l'étoile pulsante et les gradients de vitesse? Les effets d'asymétrie des raies sont-ils communs à toutes les étoiles pulsantes ?  La relation \emph{Pp} existe-t-elle pour d'autres types d'étoiles pulsantes que les Céphéides ?  Une telle étude comparative permet de placer les Céphéides dans un contexte plus vaste et de mieux comprendre les mécanismes de la pulsation, et de la perte de masse, en lien également avec l'environnement de ces étoiles. 

\begin{figure}[h]
%\vspace*{-8mm}
\begin{flushleft}
\resizebox{0.45\hsize}{!}{\includegraphics[angle=90]{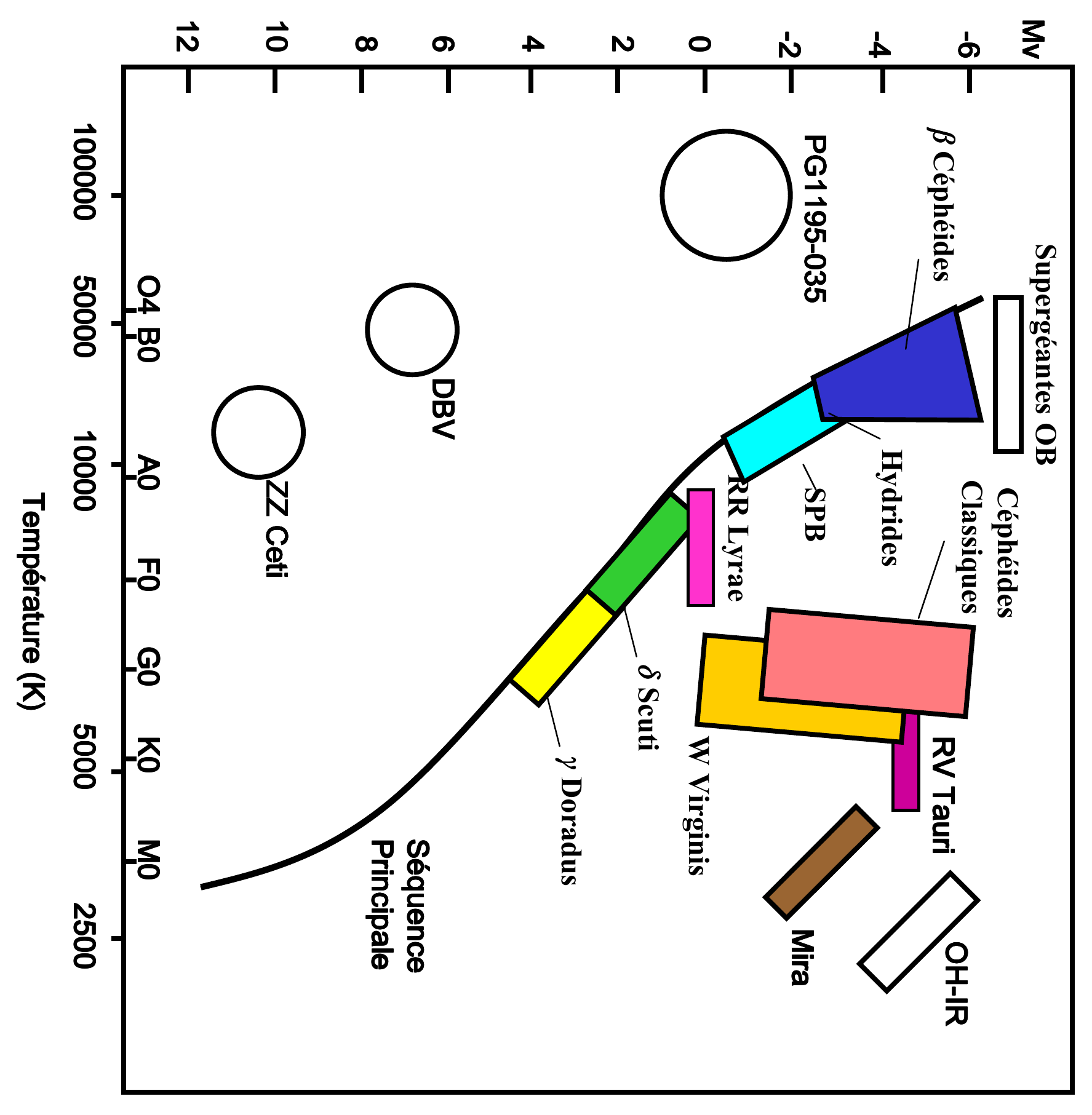}}
\end{flushleft}
%\vspace*{-8mm}
\begin{minipage}{0.52\hsize}
\end{minipage}
\vspace*{-87mm}
\begin{flushright}
\begin{minipage}{90mm}{
\resizebox{0.80\hsize}{!}{\includegraphics[angle=0]{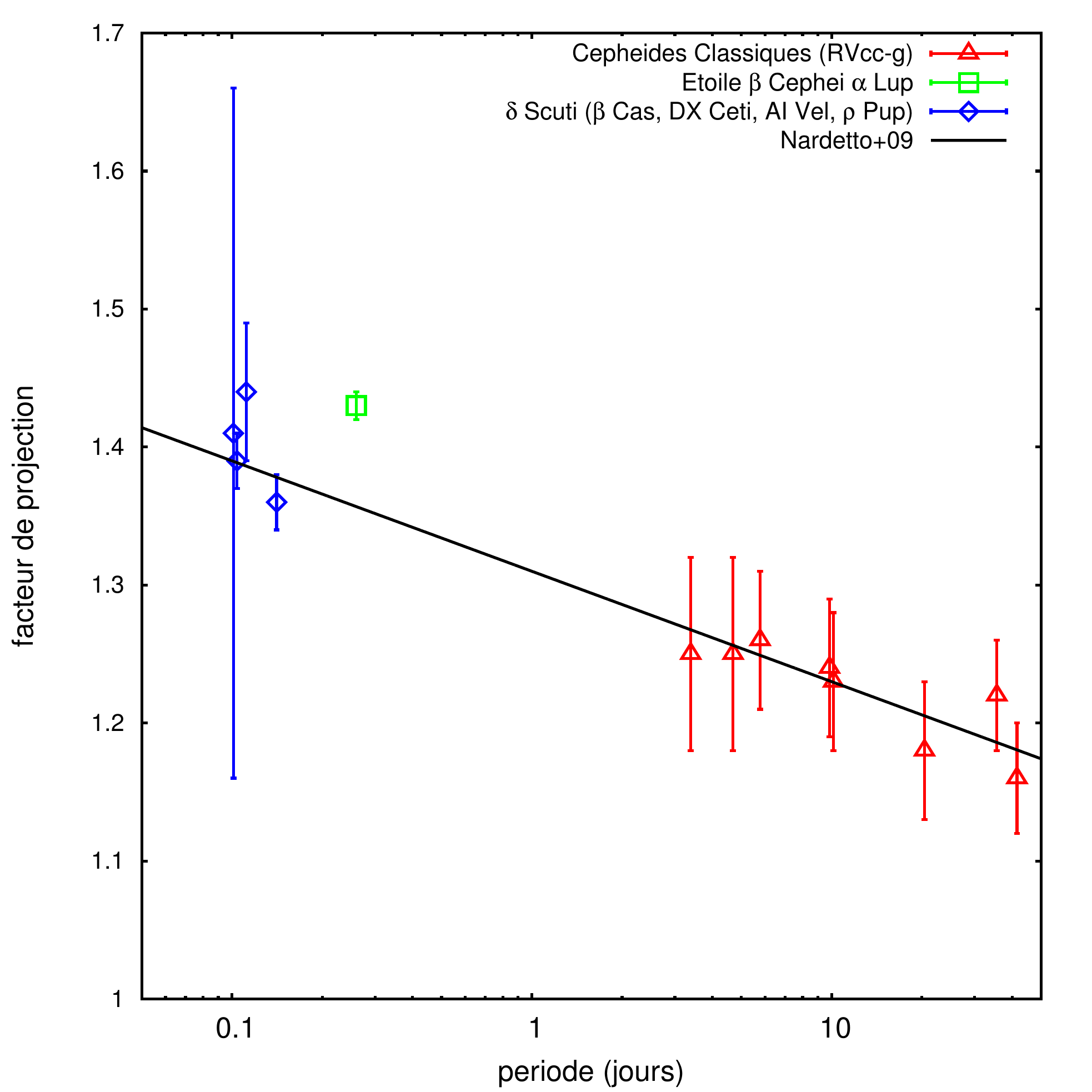} }}
\end{minipage}
\caption{ \footnotesize  La relation \emph{Pp} obtenue pour différents types d'étoiles pulsantes. La dispersion obtenue pour les étoiles de type $\delta$~Scuti est liée à l'impact de la rotation, qui du fait de l'assombrissement gravitationnel qu'elle génère, affecte le facteur de projection.}\label{Fig_Pp}
\end{flushright}
\vspace*{-5mm}
\end{figure}

D'abord quelques remarques sur les $\delta$ Scuti et les RR Lyrae (Fig.~\ref{Fig_Pp}). Les $\delta$ Scuti ont une période comprise entre 0.5 et 5 heures et peuvent pulser dans différents modes: fondamental, multipériodique radial, ou non radial.  \citet{dzi77} a d'abord montré que la méthode de BW permettait de distinguer les modes radiaux et non-radiaux. Puis la méthode de BW a été modifiée de façon à prendre en compte les modes non radiaux en incluant un facteur de projection dépendant de $l$ \citep{balona79, balona79a, balona79c, stamford81, hatzes96}. Les spécialistes du domaine se sont ensuite intéressés aux $\delta$ Scuti à forte amplitude qui pulsent dans un mode radial fondamental, les ‘‘High Amplitude $\delta$ Scuti Stars'' (ou HADS). L'application de la méthode de BW pour ces étoiles permet de déterminer leur rayon et donc, en principe, leur relation période-luminosité \citep{wilson93, milone94, wilson98}. Mais la méthode la plus utilisée pour étalonner la relation \emph{PL} reste les mesures de parallaxes d’Hipparcos \citep{petersen98}. \citet{fernie92} montre ainsi que les relations \emph{PL} des $\delta$ Scuti et des Céphéides sont compatibles. \citet{mcnamara07} ont déterminé la pente de la relation \emph{PL} des $\delta$ Scuti des nuages de Magellan grâce aux observations photométriques OGLE (voir également  \citet{mcnamara97, petersen99} et \citet{poleski10} pour les résultats d'OGLE III). Les $\delta$ Scuti, trop faibles, ne peuvent pas véritablement être utilisées pour déterminer des distances dans le groupe local, en revanche, elles s'avèrent très intéressantes car elles sont beaucoup plus nombreuses et proches du Soleil que les Céphéides et il y a donc moins de problème lié au rougissement. Les RR Lyrae, quant à elles, présentent une relation période - métallicité - luminosité \citep{sollima06,marengo17,muraveva15, benedict11}.  La méthode de BW a été appliquée à ce type d'étoiles à plusieurs reprises \citep{cacciari92, carney92, liu90, carrillo95}. Récemment, \citet{jurcsik17} ont déterminé la distance de l'amas globulaire M3 en appliquant la méthode de BW à 26 RR Lyrae ne présentant pas l'effet Blazhko\footnote{L'effet Blazhko a été pour la première fois mis en évidence par Sergei Blazhko en 1907 sur l'étoile RW Dra. Il s'agit d'un phénomène propre aux RR Lyrae et caractérisé par une modulation de la courbe de luminosité en amplitude et en période.} et en utilisant une valeur du facteur de projection de $p=1.35$\footnote{Cette valeur tirée de \cite{nardetto04} n'est pas appropriée car elle a été calculée pour la détermination de la vitesse radiale qui correspond au pixel minimum de la raie. Mais par un heureux hasard, la période des RR Lyrae étant autour de 0.5 jours, la relation \emph{Pp} de \citet{nardetto09} fournit une valeur de $p=1.33$ (compatible avec la méthode de cross-correlation), ce qui explique probablement les résultats cohérents en terme de distance de M3. Mais des études sur le facteur de projection des RR Lyrae sont nécessaires.}. La distance obtenue, autour de 10.5kpc, est tout à fait compatible avec d'autres méthodes. Appliquer la méthode BW aux RR Lyrae avec (ou même sans) effet Blazhko devrait apporter dans les années à venir des contraintes sur la dynamique atmosphérique de ces étoiles.

\begin{figure}[h]
%\vspace*{-8mm}
\begin{flushleft}
\resizebox{0.55\hsize}{!}{\includegraphics[angle=0]{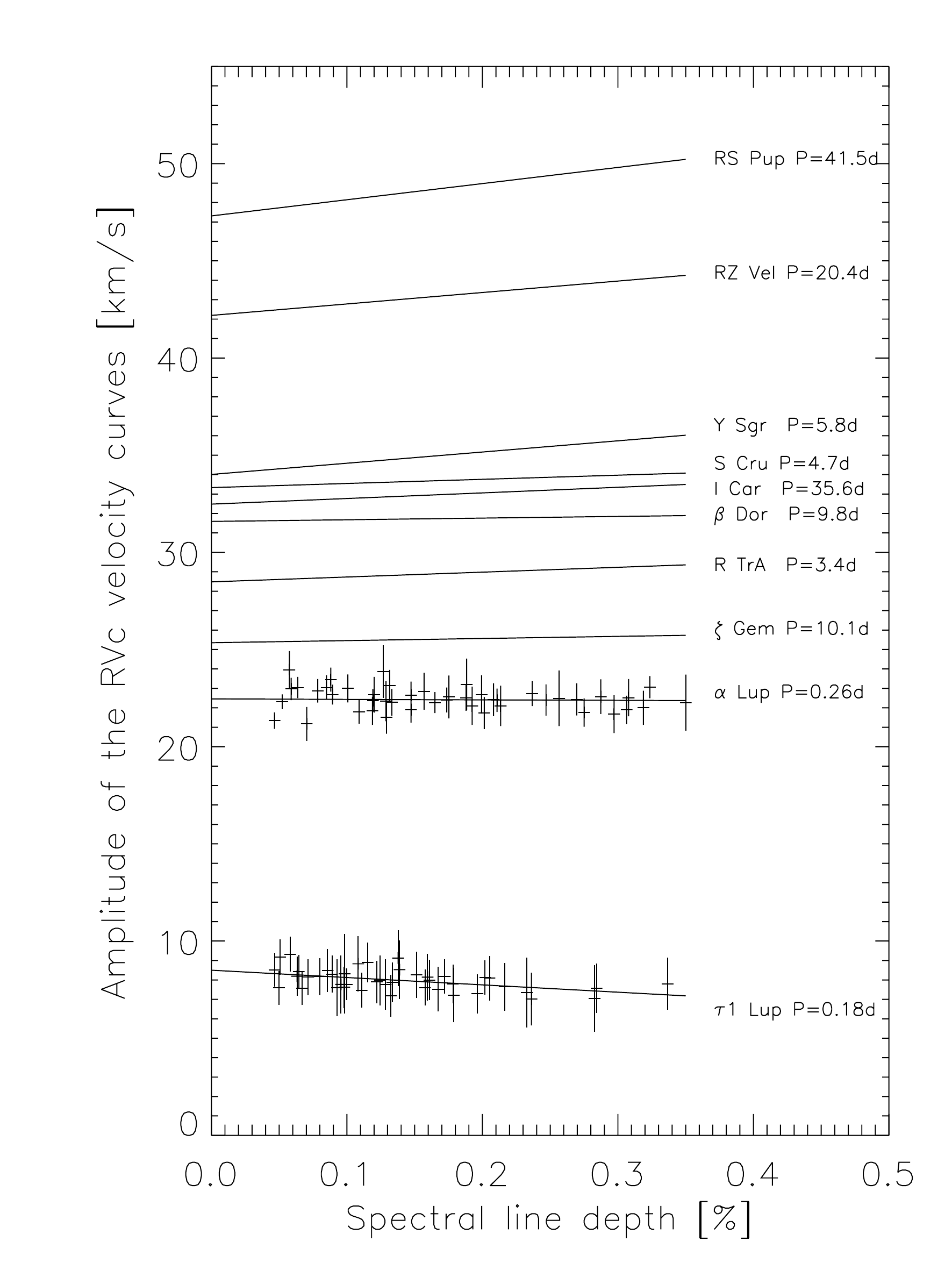}}
\end{flushleft}
%\vspace*{-8mm}
\begin{minipage}{0.52\hsize}
\end{minipage}
\vspace*{-120mm}
\begin{flushright}
\begin{minipage}{75mm}{
\resizebox{1.1\hsize}{!}{\includegraphics[angle=0]{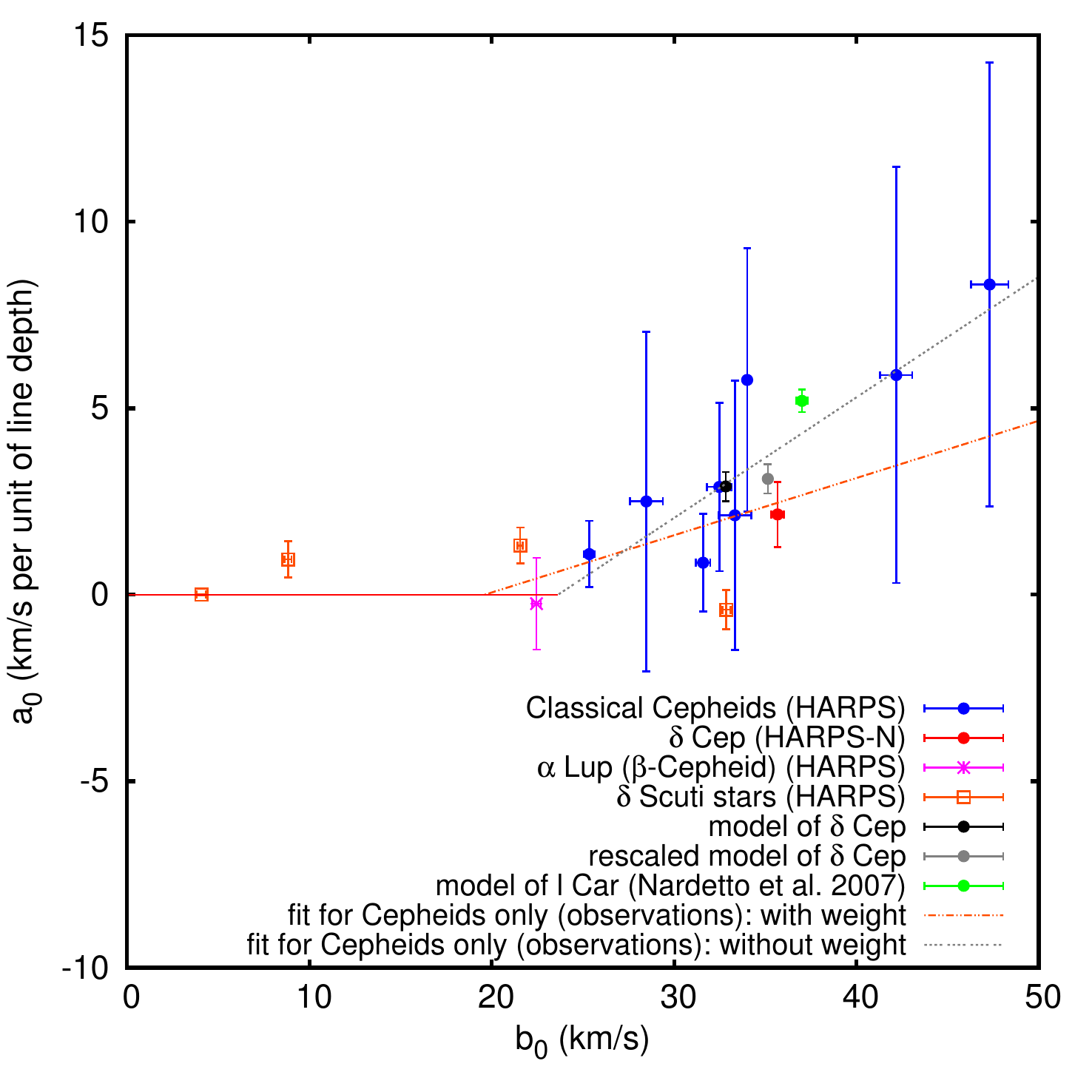} }}
\end{minipage}
\end{flushright}
\vspace*{25mm}
\caption{ \footnotesize  A gauche : l'amplitude des courbes de vitesse radiale associées à différentes raies se formant à différents niveaux dans l'atmosphère donne une bonne indication du gradient de vitesse dans l'atmosphère (ou une partie de l'atmosphère). La profondeur de raie {\it zéro} correspond à la photosphère de l'étoile. Les droites ($a_0 \Delta RV_c + b_0$) sont tirées de \citet{nardetto07} pour les Céphéides,  et de \citet{nardetto13} pour les deux $\beta$ Céphéides $\alpha$~Lup et $\tau^{1}$~Lup. Les données HARPS sont indiquées par des croix pour les $\beta$~Céphéides. A droite : la pente du gradient de vitesse ($a_0$) en fonction de l'amplitude de la courbe de vitesse au niveau de la photosphère ($b_0$). Ces données correspondent aux références suivantes: Céphéides en bleu \citep{nardetto07}, $\delta$~Cep en rouge  \citep{nardetto17}, $\alpha$~Lup \citep{nardetto13}, étoiles de type $\delta$ Scuti \citep{guiglion13, nardetto14} tandis que les valeurs théoriques sont tirées des modèles de $\delta$~Cep  \citep{nardetto04, nardetto17}  et de $\ell$~Car \citep{nardetto07}. Cette relation est intéressante pour estimer le facteur de projection des étoiles pulsantes dont le mode radial est dominant. }\label{Fig_Grad}
\end{figure}

\begin{figure}[htbp]
\begin{center}
\includegraphics[width=12cm]{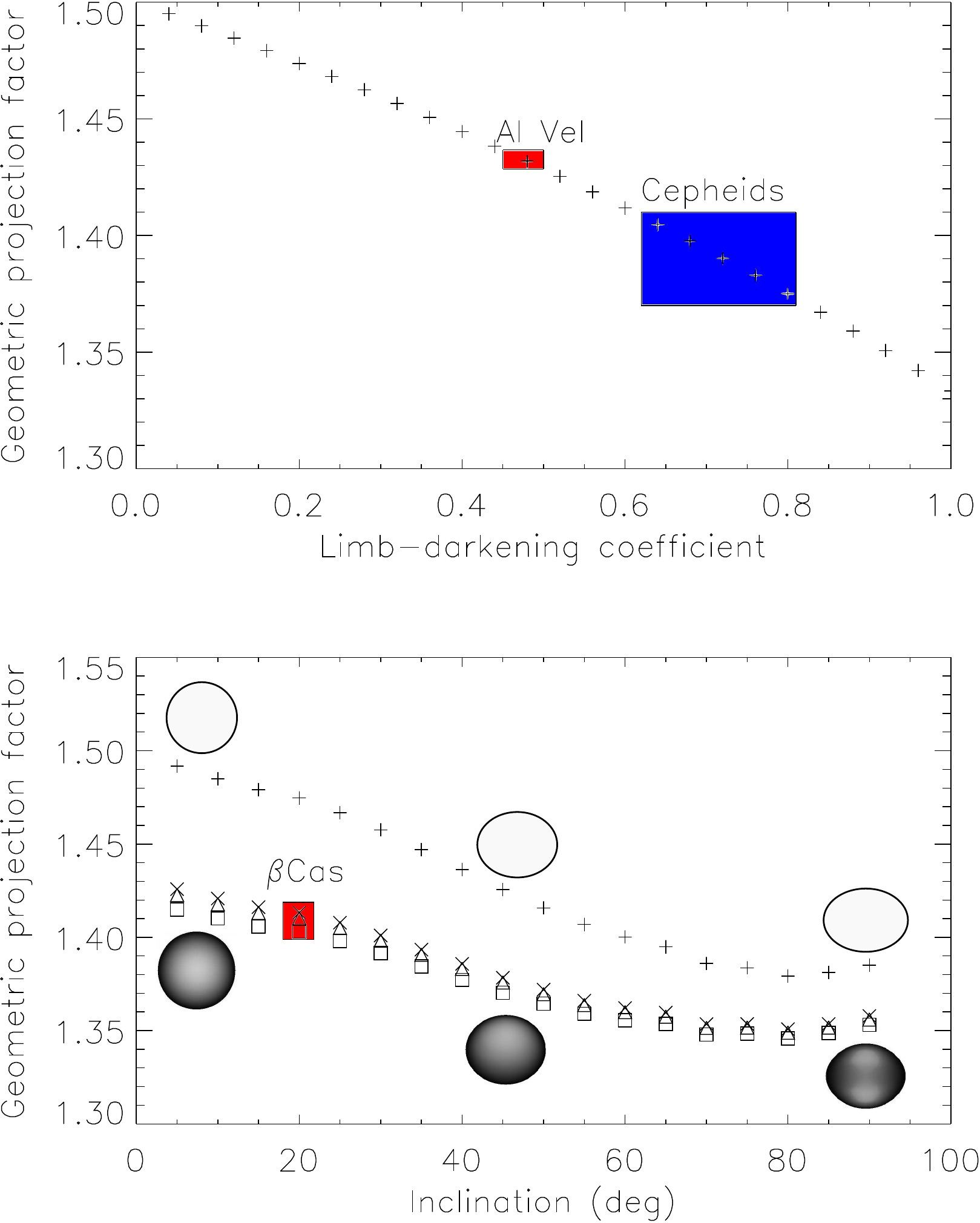}
\end{center}
\vspace*{-5mm} \caption{ \footnotesize  Le facteur de projection géométrique est sensible à l'assombrissement centre bord des étoiles. Pour les $\delta$ Scuti dont la rotation est négligeable (comme AI Vel), on obtient un facteur de projection géométrique $p_\mathrm{0}$ autour de 1.43, un peu plus élevé  que celui des Céphéides autour de 1.37-1.41 du fait d'un assombrissement centre-bord moins prononcé (en haut). Si la rotation est importante comme dans le cas de $\beta$ Cas (autour de 220 \kms), il faut alors prendre en compte l'assombrissement gravitationnel \citet{domiciano02, domiciano12} et le facteur de projection géométrique diminue alors avec l'inclinaison de l'étoile, autour de $i=20$ deg pour $\beta$ Cas et ne change pas véritablement d'une longueur d'onde à l'autre: $\lambda= 6000$ \AA\, (carrés), $\lambda= 6500$ \AA\,  (triangles), $\lambda= 700$ \AA\,  (croix).}\label{Fig_BetaCas}
\end{figure}

\newpage
Ainsi, ce projet de comparaison de la dynamique atmosphérique des étoiles pulsantes entre elles a débouché sur une première étude concernant les $\beta$-Céphéides\footnote{Les $\beta$-Céphéides sont des étoiles pulsantes de type spectral B qui ne sont pas utilisées pour les distances mais dont la dynamique atmosphérique est tout à fait intéressante à titre de comparaison (voir par exemple le travail effectué pendant mon stage de L3 \citet{fokin04}).} $\alpha$~Lup et $\tau^{1}$~Lup \citep{nardetto13} (voir le papier correspondant dans l'annexe~\ref{nardetto13}).  L'objectif de ce travail combinant observations spectroscopiques HARPS et modélisation hydrodynamique est de comparer la dynamique atmosphérique des $\beta$ Céphéides avec celle des Céphéides, et ce à trois niveaux : valeur du facteur de projection, valeur du gradient de vitesse dans l'atmosphère et, enfin, le $k$-facteur (et les effets d'asymétrie de raies). Pour $\alpha$ Lup, les modes non-radiaux étant négligeables, j'ai ainsi calculé un facteur de projection théorique de l'ordre de $p=1.43\pm0.01$. Par ailleurs, $\alpha$ Lup présente des raies spectrales dont l'asymétrie moyenne sur l'ensemble du cycle de pulsation est négative et de l'ordre de -7\%, ce qui se traduit par une vitesse radiale résiduelle d'éloignement (décalage vers le rouge) de $k=+2.2 \pm 0.8$ \kms (alors qu'elle est comprise entre $0$ et $-2$ \kms pour les Céphéides; voir Sect.~\ref{Sect_k}).   Pour $\tau^{1}$ Lup, les modes non-radiaux étant non négligeables, la modélisation hydrodynamique de l'étoile à l'aide du code de \citet{fokin94} n'est pas possible.  En revanche, $\tau^{1}$~Lup semble présenter un gradient de vitesse atmosphérique négatif: la vitesse associée aux couches les plus profondes est supérieure à la vitesse des couches de la partie haute de l'atmosphère. Quelle en est la signification ? Est-ce le résultat des modes non-radiaux ? Je me suis également intéressé aux étoiles pulsantes de type $\delta$~Scuti. Ainsi dans \citet{guiglion13}\footnote{Guillaume Guiglion a fait sont stage de M2 en 2012 avec moi sur ce sujet. Le papier correspondant se trouve dans l'annexe~\ref{guiglion13}.}, nous avons tout d'abord estimé l'impact de la rotation (non négligeable pour certaines de ces étoiles) et de l'assombrissement gravitationnel sur  le facteur de projection de $\beta$ Cas en utilisant le code de \citet{domiciano02, domiciano12} (voir Fig.~\ref{Fig_BetaCas}). Nous avons également rédigé une section sur le problème complexe de l'impact des modes non-radiaux sur le facteur de projection. Nous avons ensuite continué ce travail pour deux autres  $\delta$ Scuti, $\rho$ Pup et DX Ceti \citep{nardetto14} (voir l'annexe~\ref{nardetto14}). Nous avons ainsi mesuré à partir d'observations HARPS des gradients de vitesse atmosphérique autour de zéro pour ces deux étoiles à partir desquels il a été possible de déduire des valeurs du facteur de projection, en faisant l'hypothèse que la décomposition du facteur de projection développé dans \citet{nardetto07} est correcte. Les $k$-facteurs obtenus sont quant à eux de $-0.5 \pm 0.1$ \kms pour $\rho$ Pup et $0.0 \pm 0.1$ \kms pour AI Vel. Ainsi ces différentes études sur les étoiles pulsantes ont mené à plusieurs conclusions:
\begin{enumerate}
\item Il semble exister une relation \emph{Pp} unique s'appliquant aux Céphéides et aux étoiles pulsantes de type $\delta$~Scuti (Fig.~\ref{Fig_Pp}), mais ceci devra être confirmé par {\it Gaia}. Pour les  $\delta$~Scuti, il faut prendre en compte l'assombrissement gravitationnel si l'étoile tourne plus vite qu'une centaine de \kms. Le facteur de projection de $\alpha$ Lup (la $\beta$-céphéide) est quant à lui plus proche des étoiles de type $\delta$ Scuti que des Céphéides. 
\item L'intérêt autour du $k$-facteur est relancé depuis que \citet{vasilyev17} ont montré qu'il trouvait son origine dans la granulation des Céphéides. Plus une raie se forme haut dans l'atmosphère, plus le décalage vers le bleu est réduit ($k$ proche de zéro). A l'inverse, une raie se formant proche de la photosphère est très décalée vers le bleu ($k$ très négatif). Or, il est très intéressant de noter que l'on obtient une valeur du $k$-facteur autour de -1 \kms pour les Céphéides, +2 \kms pour la $\beta$ Céphéide, et entre -0.5 et 0.0 \kms pour les deux $\delta$ Scuti étudiées. Cela semble indiquer que les $\delta$ Scuti ont un comportement proche de celui des Céphéides en terme de convection (même si elle semble un peu moins marquée), ce qui n'est pas vraiment étonnant, alors que la $\beta$ Céphéide aurait un régime très différent, probablement sans convection. 
\item Il existe une relation entre la période des étoiles et les gradients de vitesse dans l'atmosphère (Fig.~\ref{Fig_Grad}). Plus précisément, le gradient de vitesse atmosphérique, ou du moins, le gradient de vitesse dans la zone de l'atmosphère sondée par les raies spectrales, est nul pour des amplitudes de vitesses inférieures à environ 20 \kms puis augmente de façon proportionnelle avec l'amplitude de vitesse (voir Fig.~\ref{Fig_Grad}). Etant donné que le gradient de vitesse atmosphérique est un facteur correctif contribuant pour quelques pourcents au facteur de projection, cette tendance peut s'avérer très intéressante pour prédire le facteur de projection des étoiles pulsantes en général. 
\item Enfin, en étudiant l'environnement de l'étoile prototype des $\beta$ Céphéides, nous avons découvert à l'aide de l'instrument visible VEGA/CHARA un environnement visible contribuant pour $23 \pm 2$\% du flux. Mais la situation semble {\it a priori} différente de celle de $\delta$ Cep dans la mesure où un champ magnétique pourrait être responsable du confinement d'un disque en co-rotation autour de l'étoile \citep{nardetto11a} (voir l'annexe~\ref{nardetto11a}).
\end{enumerate}

Ainsi, comparer les étoiles pulsantes entre elles est certainement une bonne façon de comprendre la physique de la pulsation, de la perte de masse et de l'environnement. Rappelons que ''The Optical Gravitational Lensing Experiment'' (OGLE\footnote{http://ogledb.astrouw.edu.pl/~ogle/OCVS/}) a découvert depuis 1992 plus d'un million d'étoiles variables (céphéides, céphéides de type 2, céphéides anormales, RR Lyrae, $\delta$ Scuti, Miras, binaires à éclipses, étoiles de type R CrB) dans notre Galaxie et dans les nuages.  Parmi ces étoiles, on trouve des Céphéides double modes (F/1O, 1O/2O, 1O/3O, 2O/3O), triple modes (F/1O/2O, 1O/2O/3O) et même des Céphéides pulsant dans des modes non-radiaux ou encore une Céphéide en fin de pulsation sortant de la bande d'instabilité. Cette diversité d'étoiles est un fabuleux laboratoire pour comprendre la dynamique atmosphérique des étoiles pulsantes.

\newpage
Pour conclure ce chapitre, nous pouvons ainsi résumer les limitations de la méthode de Baade-Wesselink:

\begin{enumerate}
\item Le facteur de projection est la principale limitation de la méthode: s'il existe un biais de 5\% sur le facteur de projection, alors la distance de BW sera également biaisée à hauteur de 5\%.  Celui-ci dépend (par ordre d'importance): de l'assombrissement centre-bord de l'étoile (et donc de la phase de pulsation de l'étoile), de la dynamique atmosphérique (i.e. de la profondeur de la raie spectrale considérée et du gradient de vitesse associé), de la méthode utilisée pour déduire la vitesse radiale, et des longueurs d'ondes considérées pour la raie spectrale (spectroscopie) et le continuum (interférométrie ou photométrie) et enfin de la granulation de l'étoile (variations possible de cycle à cycle pour certaines Céphéides proches du bord rouge de la bande d'instabilité).
\item La courbe de vitesse radiale (peu importe la méthode considérée) est corrigée de sa valeur moyenne (le $k$-facteur) en faisant l'hypothèse que la zone de formation de la raie se retrouve à la même position dans l'atmosphère de l'étoile d'un cycle à l'autre. Ce n'est pourtant pas le cas. L'impact de cette hypothèse sur la distance n'est pas clair, probablement faible, mais nécessite des investigations. 
\item L'environnement de l'étoile a un impact sur les mesures interférométriques et photométriques. Corriger la méthode de BW de l'enveloppe est un objectif de l'ANR. 
\end{enumerate}

L'objectif à terme est de rendre la méthode de BW robuste de façon à en faire un outil précieux (si ce n'est pas déjà le cas) pour les déterminations de distance dans le groupe local. 

%%%%%%%SECTION 

%%%%%%%%%%%%%%%%%%%
%%%%%%%CHAPTER III
%%%%%%%%%%%%%%%%%%%
\chapter{Les binaires à éclipses et l'interférométrie}\label{Chap_EBs}

% Sagittarus Arm => Suchomska15 => K1 and K4 and luminosity class II % SB of di Benedetto 05
% LMC => Pietrzynski13 => late-type => SB of di Benedetto 05
% SMC => Graczyk => late-type => SB of di Benedetto 05
% M31 et M33 % early-type et adjustment de modèle pour récupérer la magnitude absolue

Les binaires à éclipses permettent de déterminer la distance des galaxies proches (LMC, SMC, M31 et M33) avec une bonne précision (voir Tableau~\ref{Tab_modeles}). Le principe est relativement simple (voir Sect.~\ref{s_EB_LMC}). Les paramètres de la binaire, et en particulier les rayons des deux composantes, sont déduits des données spectroscopiques et photométriques. Parallèlement, le diamètre angulaire des deux composantes est une fonction de la magnitude et de la couleur des étoiles (relation brillance de surface - couleur). La combinaison des rayons linéaire et angulaire donne la distance du système (voir Eq.~\ref{Eq_LMC}). La principale source d'erreur de cette méthode provient quasi-totalement de notre méconnaissance de la relation brillance de surface - couleur. De nombreuses études ont d'abord utilisé les binaires à éclipses de type précoce, plus brillantes et faciles à détecter, pour déterminer la distance du LMC et du SMC. Mais dans ce cas, la précision sur la relation brillance de surface - couleur, 6 à 8\% actuellement (voir Sect.~\ref{s_SBs}), limite la précision que l'on peut espérer obtenir sur la distance des nuages. Les binaires à éclipses de type spectral tardif sont quant à elles extrêmement difficiles à détecter. Dans l'étude de \cite{pietrzynski13} par exemple, 16 ans d'observations de OGLE ont été nécessaires pour détecter une douzaine de binaires à éclipses. Ainsi, la précision de 2.2\% obtenue sur la distance du LMC est directement liée ou presque à la précision que l'on a actuellement sur la relation brillance de surface - couleur des étoiles de type K (classe III), c'est-à-dire 0.04 magnitude \citep{dibenedetto05}. A retenir: un rms de 0.022 magnitude sur la relation brillance de surface - couleur ($S_\mathrm{V}$) correspond à une précision sur le diamètre angulaire de 1\%. Pour les mêmes raisons, la précision sur la distance du SMC est aujourd'hui autour de 3\% \citep{graczyk14}. Outre le LMC et le SMC, quelques binaires à éclipses de type précoce ont été détectées dans M31 et M33. Du fait de l'imprécision sur la relation brillance de surface - couleur (autour de 10\% à l'époque), les auteurs ont préféré s'en tenir à des modèles stellaires pour déduire la magnitude absolue des étoiles et ainsi leur distance. Les précisions sur les distances obtenues sont ainsi de 4.4\% pour M31 \citep{vilardell10} et de 5.5\% pour M33 \citep{ribas05}. Ces valeurs ont été utilisées par \citet{riess16} pour déterminer la constante de Hubble (voir Sect.~\ref{Sect_d}).

La relation brillance de surface - couleur est donc un outil extrêmement utile pour la détermination de distance des binaires à éclipses, mais pas seulement. Nous avons déjà vu qu'elle est utilisée pour l'application de la méthode de BW afin de déterminer la variation de diamètre angulaire de la Céphéide (voir Sect.~\ref{Sect_BW}).  Mais de manière plus générale encore, les diamètres angulaires issus de la relation brillance de surface - couleur, combinés aux mesures de parallaxes Gaia permettront dans les années à venir de déduire le rayon de milliers, voire centaines de milliers d'étoiles. Parmi celles-ci, se trouvent des étoiles hôtes d'exoplanètes dont les mesures de transit photométriques par la mission spatiale PLATO permettront de déduire le rapport $\frac{R_\mathrm{p}}{R_\star}$ avec une précision de l'ordre de 1\% et donc par suite, le rayon de la planète $R_\mathrm{p}$ ainsi que sa densité. De la même manière, une mesure de rayon stellaire indépendante pose des contraintes importantes dans le cadre de l'étude des étoiles astérosismiques.

Ainsi, contraindre les relations brillance de surface - couleur est aujourd'hui une priorité. L'objectif est d'atteindre 1\% de précision pour les étoiles tardives (afin d'accéder à une précision également de 1\% sur la distance du LMC et du SMC) et de quelques pourcents (3-5\%) pour les étoiles précoces de façon à accéder aux galaxies du groupe local. Pour atteindre ces deux objectifs, il faut déterminer le diamètre angulaire des étoiles proches et dans ce domaine, l'interférométrie est la méthode principalement utilisée. 

%%%%%%%SECTION 
\section{Les mesures de diamètre angulaire par interférométrie et l'instrument VEGA/CHARA}

Les fentes de Young (ou interférences de Young) désignent en physique une expérience qui consiste à faire interférer deux faisceaux de lumière issus d'une même source, en les faisant passer par deux petits trous percés dans un écran opaque. Cette expérience fut réalisée pour la première fois par Thomas Young en 1801 et permit de comprendre la nature ondulatoire de la lumière. En termes simples, une source ponctuelle monochromatique produit un réseau de franges très contrasté. Imaginons un second point source situé à côté du précédent: deux réseaux de franges légèrement décalés vont se superposer et diminuer ainsi le contraste des franges. On comprend alors intuitivement le lien entre la distribution spatiale d'un objet et le contraste des franges d'interférence (théorème de Van-Cittert-Zernike). Un interféromètre mesure le contraste et la position (ou phase) des franges d'interférence et permet ainsi de remonter au diamètre angulaire de l'étoile observée. 

L'histoire de l'interférométrie commence avec l'expérience de \cite{michelson1881}, puis de Michelson \& Morley en 1887 qui démontre que la vitesse de la lumière est constante quelque soit le référentiel considéré et que l'éther n'existe pas (Prix Nobel de Physique de 1907). En 1921, au Mont Wilson, situé en Californie sur les hauteurs de Los Angeles, Michelson applique le principe optique de l'interférométrie à l'astrophysique en plaçant deux miroirs de 10 cm à l'extrémité d'une poutre de 6 mètres. Ce dispositif (le premier interféromètre) lui permet de mesurer, et ceci pour la première fois, le diamètre angulaire d'une étoile, Bételgeuse. Il obtient un diamètre de l'ordre de 50 millisecondes d'arc ou {\it mas} \citep{michelson21}. Il y a alors curieusement une longue période de calme (environ 50 ans) puis \citet{brown67, brown67b} développent un concept d'interféromètre à intensité et mesurent le diamètre angulaire de 15 étoiles (de 0.7 mas à 6.5 mas). A la même époque,  \citet{gezari72} mesurent le diamètre angulaire de 9 étoiles (autour de 10-40 mas) avec le télescope de 5 mètres du Mont Palomar (A Labeyrie est alors deuxième auteur de cette étude). Toujours avec le télescope du Mont Palomar, \citet{Labeyrie74} découvrent des compagnons autour de 12 étoiles. Puis, seulement 1 an plus tard, mais de l'autre côté de l'Atlantique, au Plateau de Calern, \citet{Labeyrie75} mesure le diamètre angulaire de l'étoile Vega avec l'Interféromètre à 2 Télescopes (I2T)\footnote{Le diamètre angulaire est indiqué comme étant inférieur à 5 mas dans ce papier.}. Quelques années plus tard, \citet{blazit77} mesurent avec l'I2T les diamètres de Capella A et B (respectivement $5.2\pm1.0$ mas et $4.0\pm2.0$ mas). La même année, les premières mesures de diamètre angulaires par occultation lunaire voient le jour \citep{africano77}. Dix-sept ans plus tard, le ‘‘Grand Interférométre à 2 Télescopes'' (GI2T) est mis en service \citep{mourard94} avec les premières publications \citep{Mourard89}. Cette histoire est forcément incomplète et insiste sur l'aventure niçoise.  En 2016, soit 95 ans après la première mesure de diamètre angulaire par \citet{michelson21}, mais seulement 50 ans après les mesures de \citet{brown67b}, \cite{duvert16} ont répertorié 1150 mesures de diamètre angulaires. Ces mesures concernent 627 étoiles, ce qui correspond en moyenne à la mesure du diamètre angulaire de 12 étoiles par an depuis 1967. Les diamètres (uniformes) s'échelonnent entre 0.215 mas (HD209458; \citet{boyajian15}) et 41 mas (Bételgeuse; \citet{montarges14}) et ont été mesurés par interférométrie optique, interférométrie d'intensité et par occultation lunaire. Il est intéressant de noter que l'étoile la plus faible observée dans le visible et pour laquelle un diamètre angulaire a été déterminée a une magnitude de V=7.7, alors que l'étoile la plus faible observée en infrarouge a une magnitude K=6.3. De même, dans cet échantillon de 627 étoiles, l'étoile la plus éloignée est à 4750 parsecs environ (HD 184283) et la plus proche est proxima du Centaure. 

\begin{figure}[htbp]
\begin{center}
\includegraphics[width=10cm]{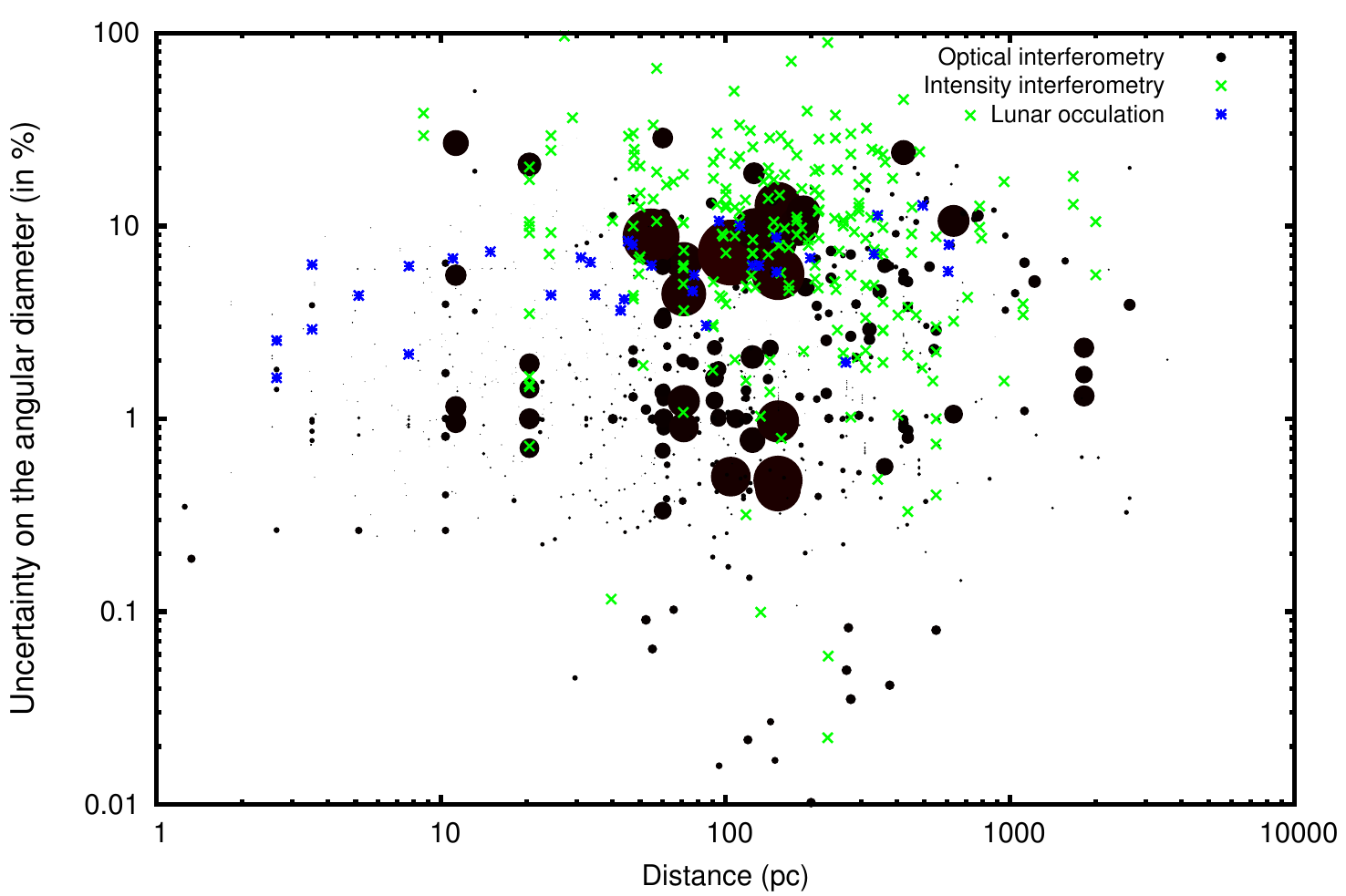}
\includegraphics[width=10cm]{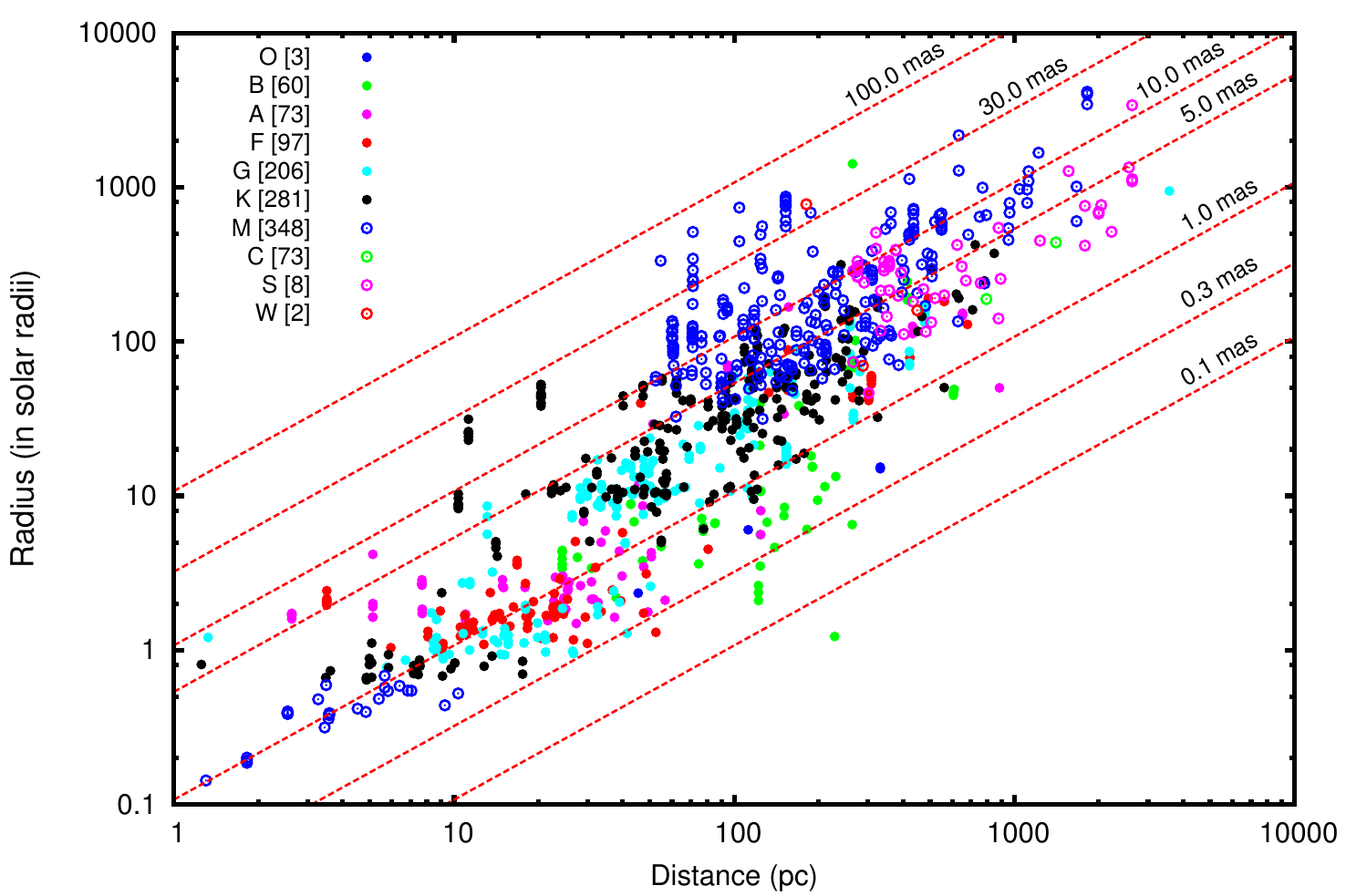}
\includegraphics[width=10cm]{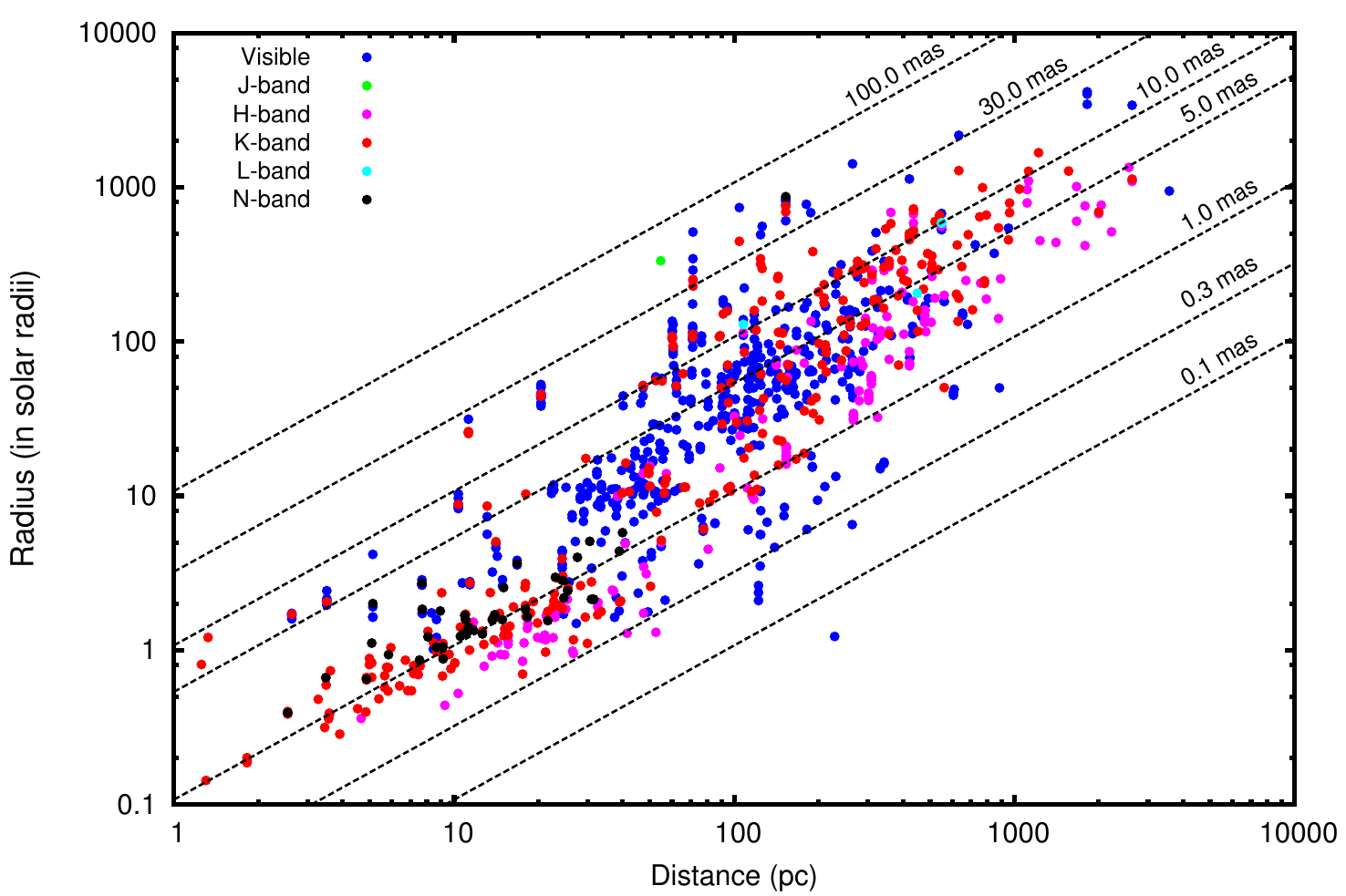}
\end{center}
\vspace*{-5mm} \caption{ \footnotesize  Etoiles observées par haute résolution angulaire d'après la base de données du JMDC \cite{duvert16}.}\label{Fig_JMDC}
\end{figure}

La figure \ref{Fig_JMDC}-haut représente la précision obtenue sur les 1150 diamètres mesurés (en pourcentage) en fonction de leur distance Hipparcos avec un code de couleur différent pour l'interférométrie optique, l'interférométrie d'intensité et la méthode des occultations lunaires. Pour l'interférométrie optique, la taille des disques noirs est proportionnelle au diamètre angulaire de l'étoile observée (à titre indicatif). La précision moyenne obtenue pour l'interférométrie optique (sur 491 objets) est de 2.9\% (c'est à dire en prenant en compte toutes les mesures depuis 1981), mais on constate que de nombreuses mesures ont une précision autour de 0.2\% de précision (les quelques points en dessous de 0.1\% de précision posent tout de même des questionnements). Les moyennes obtenues pour  l'interférométrie d'intensité et la méthode des occultations lunaires sont de 10.8\% et 6.0\%, respectivement, soit significativement au dessus. Aujourd'hui, une précision  en dessous de 1\% sur le diamètre angulaire est standard en interféromètrie optique quelles que soient la distance ou la magnitude de l'étoile, dans les limites actuelles mentionnées plus haut. 

La figure \ref{Fig_JMDC}-milieu montre le rayon associé aux 1150 mesures de diamètre angulaire (en combinant ces diamètres avec la distance Hipparcos) en fonction de la distance de l'étoile pour les différents types spectraux. Des lignes correspondant à différentes résolutions spatiales en mas sont également indiquées. Pour les types spectraux O à M, on distingue deux groupes dans le nuage de points, les séquences principales ($R<10\Rsolar$ et $d<300pc$) et les étoiles évoluées ($R>10\Rsolar$ et $d>300pc$). On peut tirer deux conclusions intéressantes de ce diagramme. D'abord, les étoiles de type O, B, A, F ont été beaucoup moins observées (environ 20\% des observations) et ce pour la raison qu'il faut une résolution spatiale inférieure à 0.3 mas (ou une magnitude limite plus importante) pour étudier ces étoiles, plus particulièrement pour les étoiles O et B de la séquence principale. On voit ici l'importance d'aller vers de plus basses résolutions spatiales (plus grandes bases combinées à des longueurs d'ondes plus faibles) pour espérer contraindre la relation brillance de surface - couleur des étoiles de type précoce efficacement. L'interférométrie visible est effectivement la méthode la plus appropriée pour mesurer le diamètre angulaire des étoiles avec un rayon inférieur à quelques rayons stellaires et à plus de 100 pc comme l'illustre la Figure \ref{Fig_JMDC}-bas. Il ne faut pas oublier que ceci n'est possible qu'en augmentant la magnitude limite ainsi que la sensibilité des instruments actuels. 

Ainsi l'instrument VEGA situé au foyer de l'interféromètre CHARA joue un rôle pionnier dans ce domaine. En effet, le 'Center for High Angular Resolution Astronomy' (CHARA, USA) est un réseau interférométrique situé à l'observatoire du Mount Wilson, en Californie, qui comprend 6 télescopes de 1~mètre sur des bases allant de 30 à 330 mètres. L'instrument VEGA\footnote{\url{http://www-n.oca.eu/vega/en/news/index.htm}}, placé au foyer de CHARA, a été développé au sein du Laboratoire Lagrange. De juin 2012 à septembre 2016, en remplacement de Denis Mourard mais avec son aide, je me suis occupé de l'organisation autour de l'instrument, de son exploitation astrophysique et de sa maintenance. Du fait de ses résolutions spatiales et spectrales, VEGA/CHARA offre une large gamme de possibilités en terme d'objets d'étude:  étoiles jeunes (AB~Aur), étoiles astérosismologiques (HD 49933, $\gamma$~Equ), étoiles à exoplanètes (13~Cyg, ...), étoiles Be ($\gamma$ Cas, 48 Per, $\psi$ Per,...), étoiles à vent (Deneb, Rigel)...  et comme nous l'avons vu Céphéides ($\delta$~Cep). Trente-cinq publications sont repertoriées\footnote{\url{https://www-n.oca.eu/vega/en/publications/index.htm}} depuis le lancement de VEGA\footnote{Depuis 2010, VEGA observe à distance (en mode télécommande depuis Nice) pour environ 70\% des nuits affectées et nous disposons d'environ 50 nuits par an. De nombreuses données sont ainsi recueillies chaque année pour satisfaire environ une trentaine de programmes scientifiques.  L'instrument est labellisé depuis le début de 2013 comme service d'observation SNO2 de l'INSU et est étroitement lié au SNO5-JMMC, pôle national pour l'exploitation des données interférométriques optiques. }. Ainsi, avec l'instrument VEGA/CHARA, nous avons fourni le diamètre angulaire de 8 étoiles de type précoces afin d'étalonner la relation brillance de surface - couleur. 

\newpage
%%%%%%%SECTION 
\section{La relation brillance de surface - couleur des étoiles }\label{s_SBs}

\begin{figure}[htbp]
\begin{center}
\includegraphics[width=17cm]{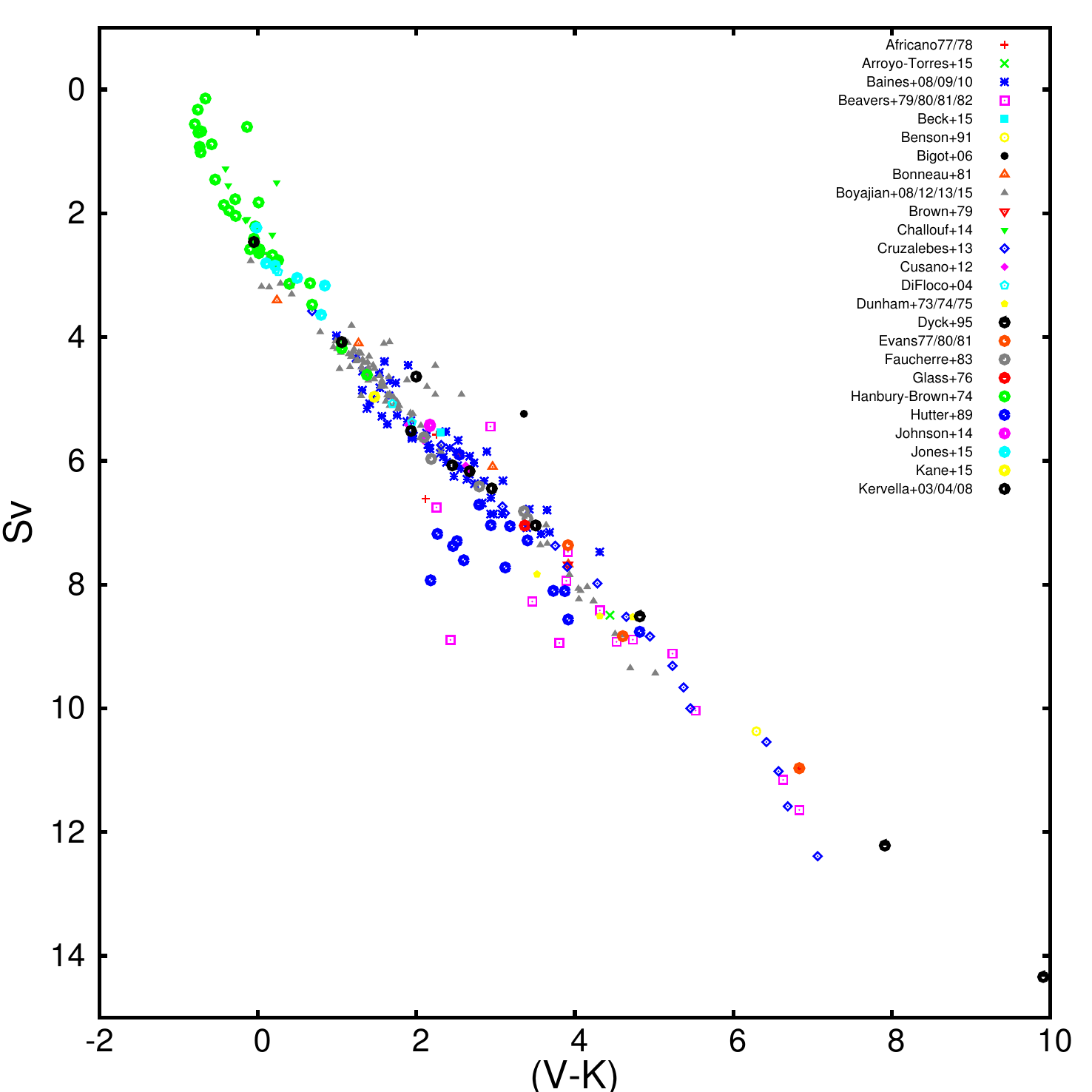}
\end{center}
\vspace*{-5mm} \caption{ \footnotesize  Relation brillance de surface - couleur réalisée à partir des 558 mesures de diamètres angulaires  (hors binarité et variabilité) listées dans la base de données JMDC de \citet{duvert16}. Les magnitudes V et K sont tirées respectivement de SIMBAD et du catalogue 2MASS.}\label{Fig_SB_JMDC}
\end{figure}

\begin{figure}[htbp]
\begin{center}
\includegraphics[width=17cm]{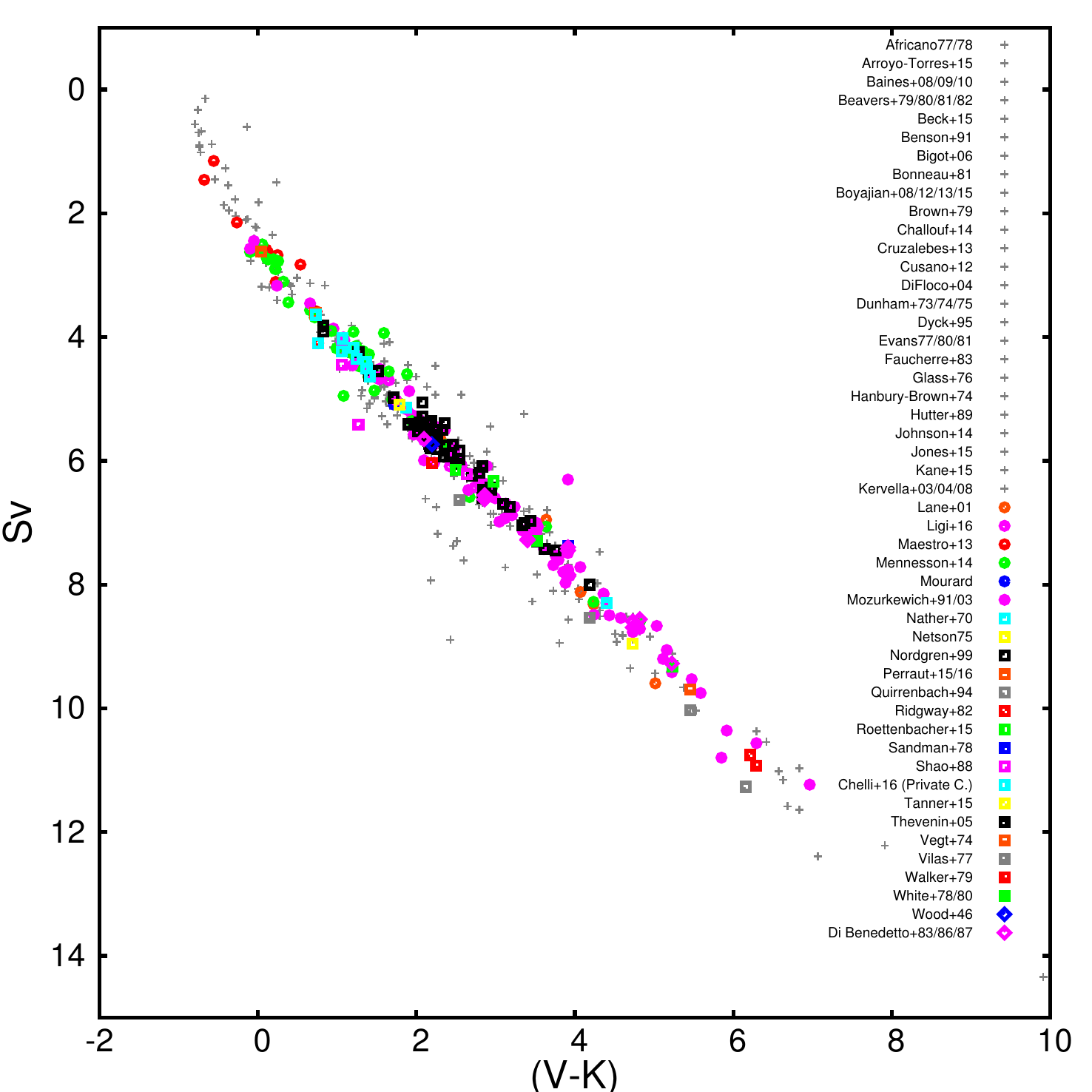}
\end{center}
\vspace*{-5mm} \caption{ \footnotesize  Suite de la Figure~\ref{Fig_SB_JMDC}.}\label{Fig_SB_JMDC2}
\end{figure}

Etalonner la relation brillance de surface - couleur n'est pas une chose aisée. Si l'on prend par exemple les 1150 mesures de diamètres angulaires répertoriées dans la base de données du JMDC \citep{duvert16} et que l'on retire les étoiles potentiellement contaminées par un compagnon et les étoiles variables, on obtient les figures~\ref{Fig_SB_JMDC} et \ref{Fig_SB_JMDC2}. Le graphique a été divisé en deux pour améliorer la lisibilité. Il contient 558 mesures de diamètre. Les magnitudes K sont tirées du catalogue 2MASS et les magnitudes V de SIMBAD. Pour rappel, la brillance de surface - couleur est donnée par l'équation $S_\mathrm{V}=V-5\log \theta_\mathrm{LD}$\footnote{Les diamètres assombris sont donnés dans la base JMDC et leur détermination n'est pas homogène du point de vue de la méthode.}. Dans ce graphique, l'extinction n'est pas prise en compte, mais il s'avère que la prendre en compte, en utilisant une formule du type $A_\mathrm{V}=\frac{0.8}{\pi}$ où $\pi$ est la parallaxe en secondes d'arc \citep{dibenedetto98, dibenedetto05} ne réduit pas la dispersion globale des points. Ceci reste valable si l'on utilise des approches plus élaborées à l'aide des cartes de poussières de \cite{schlegel98} par exemple. Ainsi faire un tri selon la méthode employée (interférométrie optique, d'intensité ou occultation lunaire), E(B-V), [Fe/H], le type spectral ou la classe de l'étoile, sur la distance, la précision sur le diamètre, ou le diamètre lui-même, ou encore la précision sur la magnitude K n'améliore pas significativement la dispersion des points. 

\begin{figure}[htbp]
\begin{center}
\includegraphics[width=11.5cm]{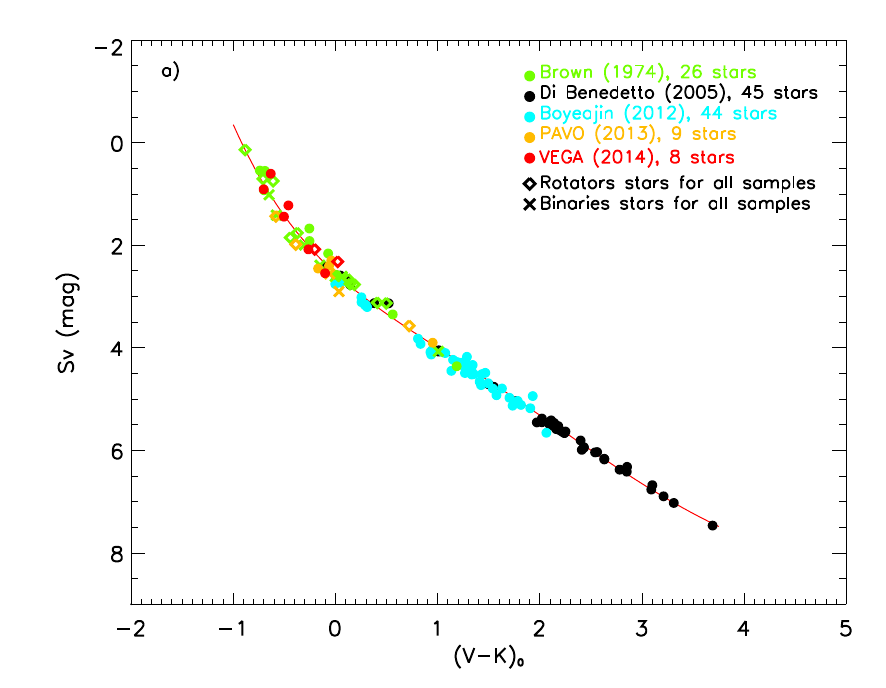}
\end{center}
\vspace*{-5mm} \caption{ \footnotesize  Relation brillance de surface - couleur issue du travail de thèse de Mounir Challouf \citep{challouf14}.}\label{Fig_challouf14}
\end{figure}

Un moyen de s'en sortir, est d'utiliser des sous-échantillons (comme le font la plupart des auteurs) et d'appliquer une méthode minutieuse au niveau du calcul de l'extinction. Un exemple est illustré par le papier de \cite{challouf14} (voir l'annexe \ref{challouf14}). Ce travail effectué dans le cadre de la Thèse de Mounir Challouf visait à observer des étoiles de type précoce avec l'instrument VEGA/CHARA afin précisément de contraindre la relation brillance de surface - couleur pour ce type d'étoiles. Chemin faisant, nous avons décidé de reprendre également l'étalonnage pour les étoiles de type tardif.  Ainsi 8 étoiles O, B, A ont été observées avec VEGA avec une précision sur le diamètre angulaire de 1.5\%. Ensuite, un très gros travail de Mounir a consisté à 1) calculer l'extinction des étoiles (à partir de 7 méthodes différentes; voir Section 4.1 du papier), 2) effectuer une sélection rigoureuse étoile par étoile (pas de binaires, pas de variabilité, pas d'environnement, pas de vent). Seules les étoiles dont la rotation est rapide et dont on sait qu'elles peuvent avoir un effet sur la relation brillance de surface - couleur ont été conservées pour augmenter la statistique des étoiles précoces. L'effet de cette rotation a été estimé ultérieurement dans un second papier (voir l'annexe~\ref{challouf15} et la Section~\ref{s_rot}).  Ce travail a permis d'étalonner la relation brillance de surface couleur pour différentes classes d'étoiles avec une précision de l'ordre de 
0.16 mag (ou 7\% en terme de diamètre) pour les étoiles précoces, et 0.04 magnitude comme précédemment pour les étoiles tardives. Ce résultat est illustré par la Figure~\ref{Fig_challouf14}. 

Comment se place ce résultat par rapport à la littérature ? Dans l'annexe~\ref{Annexe_SB}, j'actualise le Tableau I.1 de la Thèse de Mounir Challouf et l'agrémente de notes. Le premier constat est qu'il est extrêmement difficile de comparer les résultats dans la mesure ou les auteurs utilisent des définitions différentes de la relation brillance de surface - couleur. Ainsi j'ai relevé, selon les publications, les définitions suivantes:  

\begin{enumerate}

\item $S_\mathrm{V}=V-5\log \theta_\mathrm{LD}=\sum a_\mathrm{k} (V-K)^k $

\item $F_\mathrm{V}=4.2207 - 0.1 S_\mathrm{V} = \alpha + \beta (V-K)$

\item $\log \theta_\mathrm{LD} = d_1 + c_11 (V-K) - 0.2 V$ 

\item $ \theta_\mathrm{LD} (V=0) = 10^{A + B (V-K)}$ 

\item $\Phi_\mathrm{V} = \frac{\theta}{9.305*10^{\frac{-V}{5}}} = \sum z_\mathrm{k} (V-K)^k$

\end{enumerate}

\begin{figure}[htbp]
\begin{center}
\includegraphics[width=17cm]{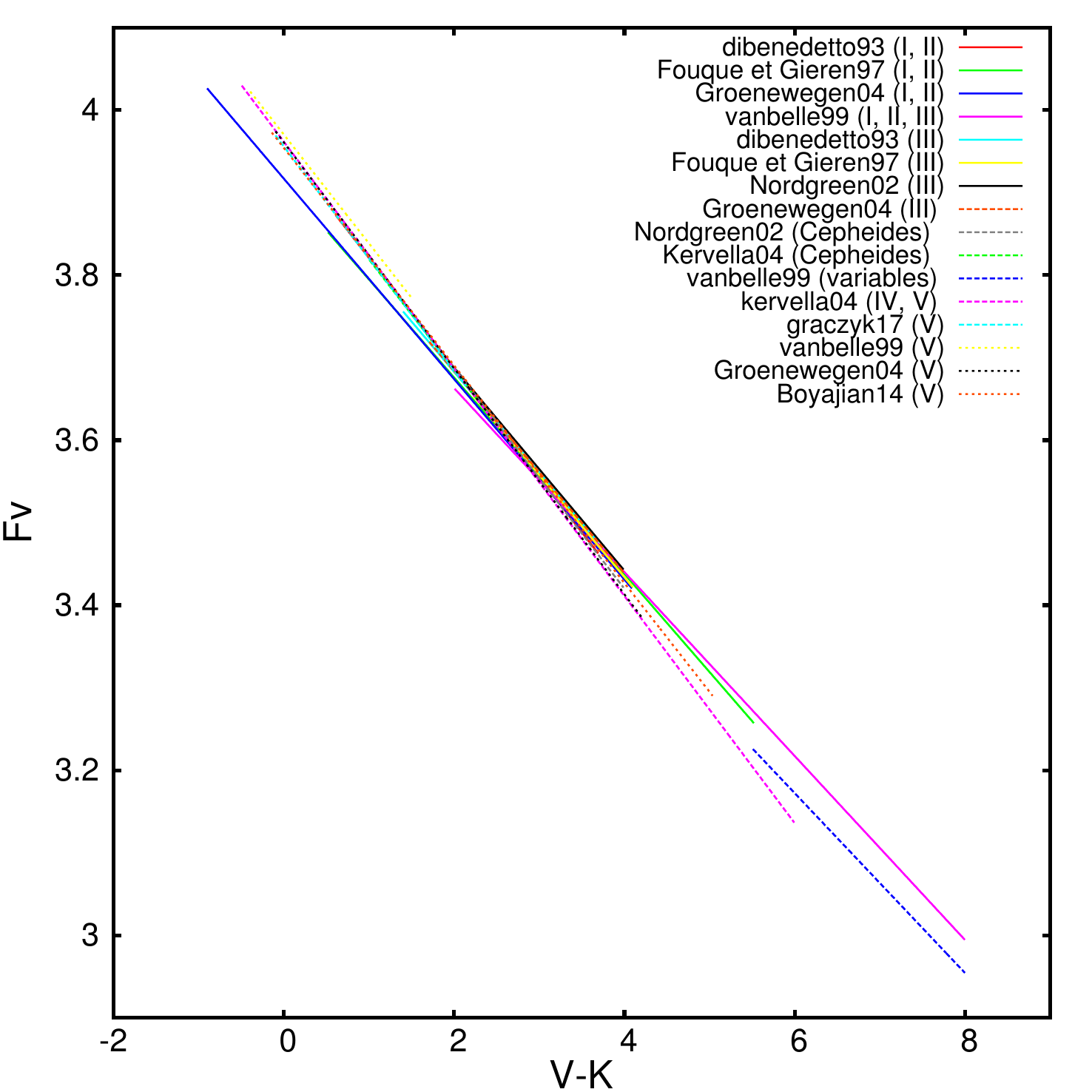}
\end{center}
\vspace*{-5mm} \caption{ \footnotesize  Relations brillance de surface - couleur linéaires réalisées à partir des observations à haute résolution angulaire. Les segments de droite correspondent aux domaines de validité des relations. }\label{Fig_CompSB0}
\end{figure}

\begin{figure}[htbp]
\begin{center}
\includegraphics[width=8cm]{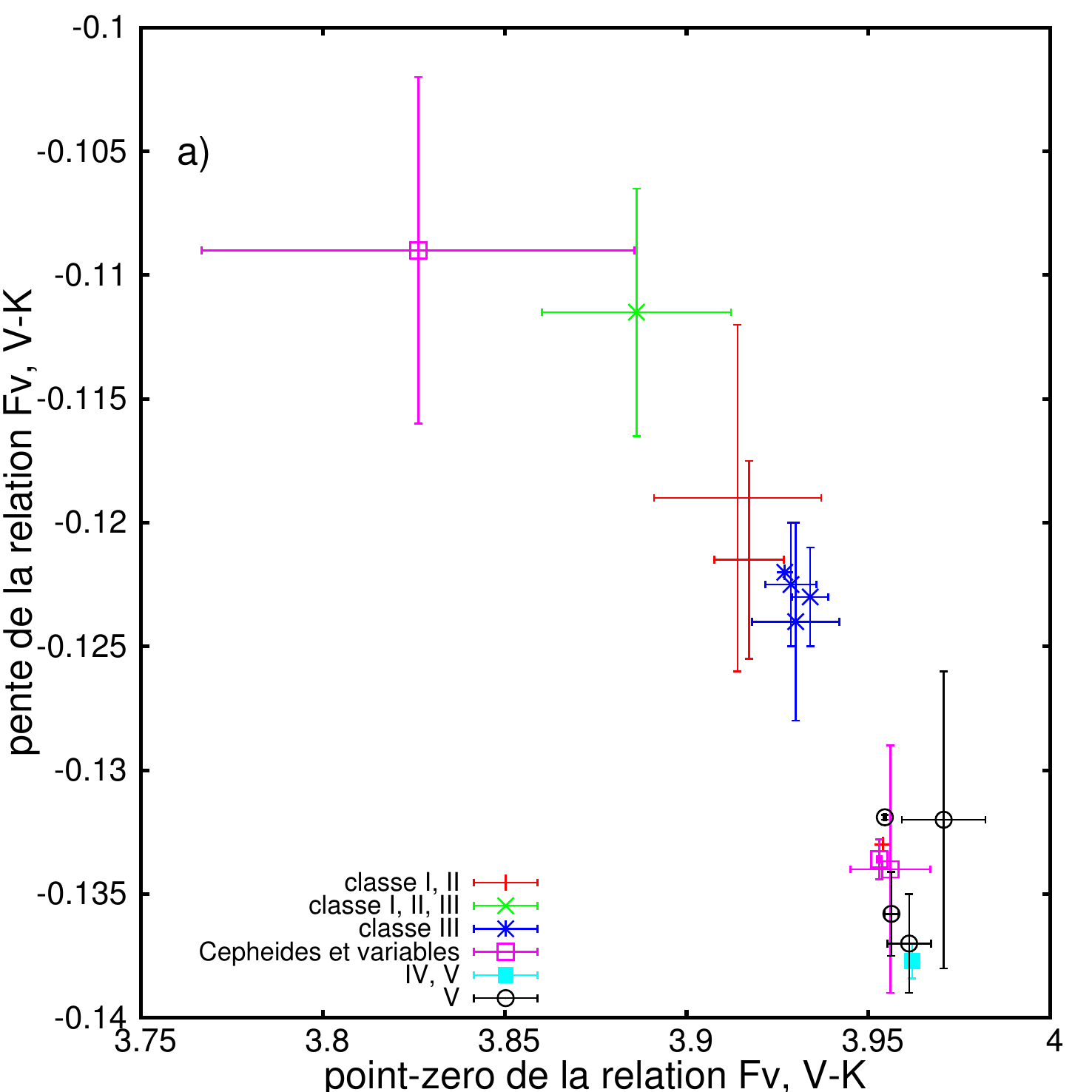}
\includegraphics[width=8cm]{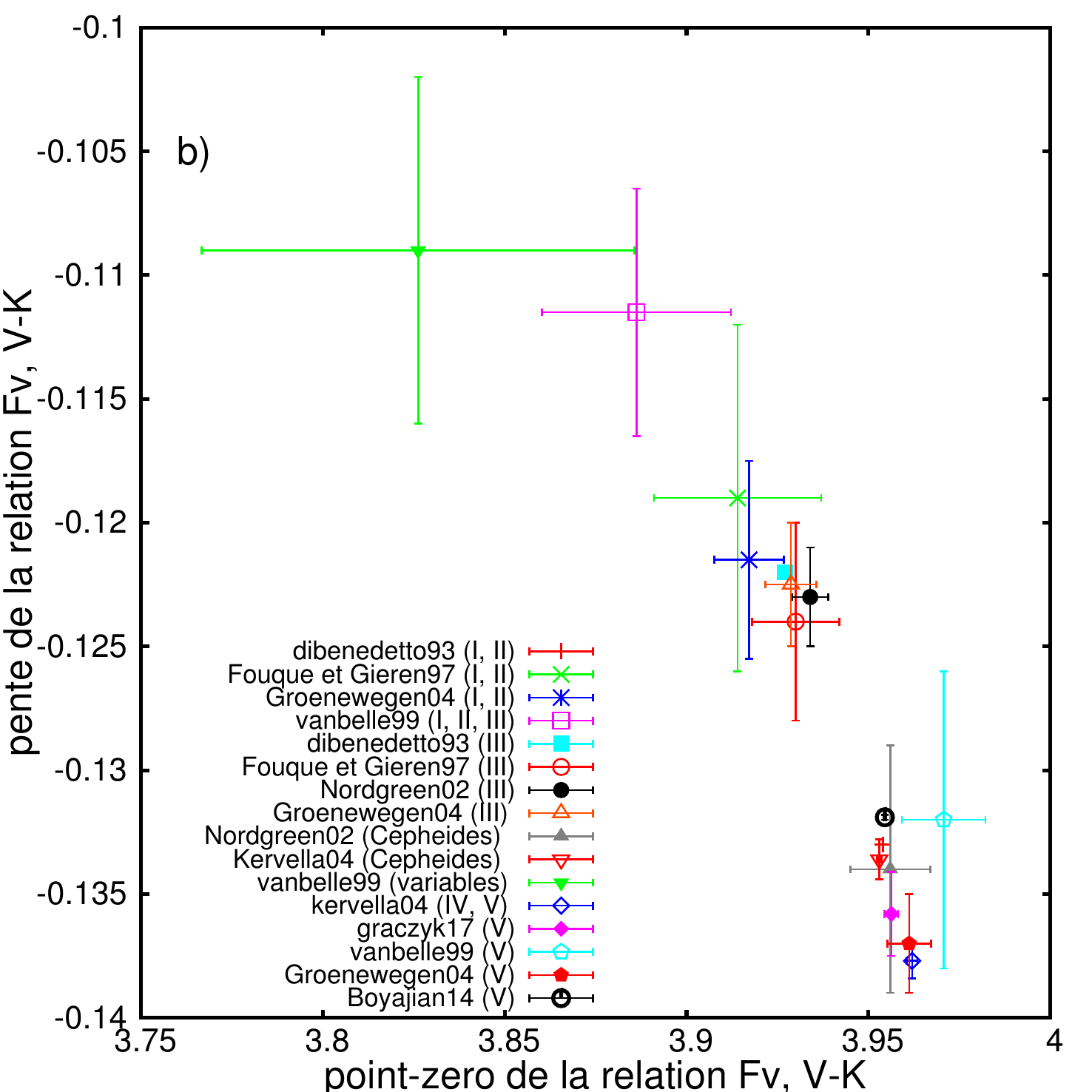}
\includegraphics[width=8cm]{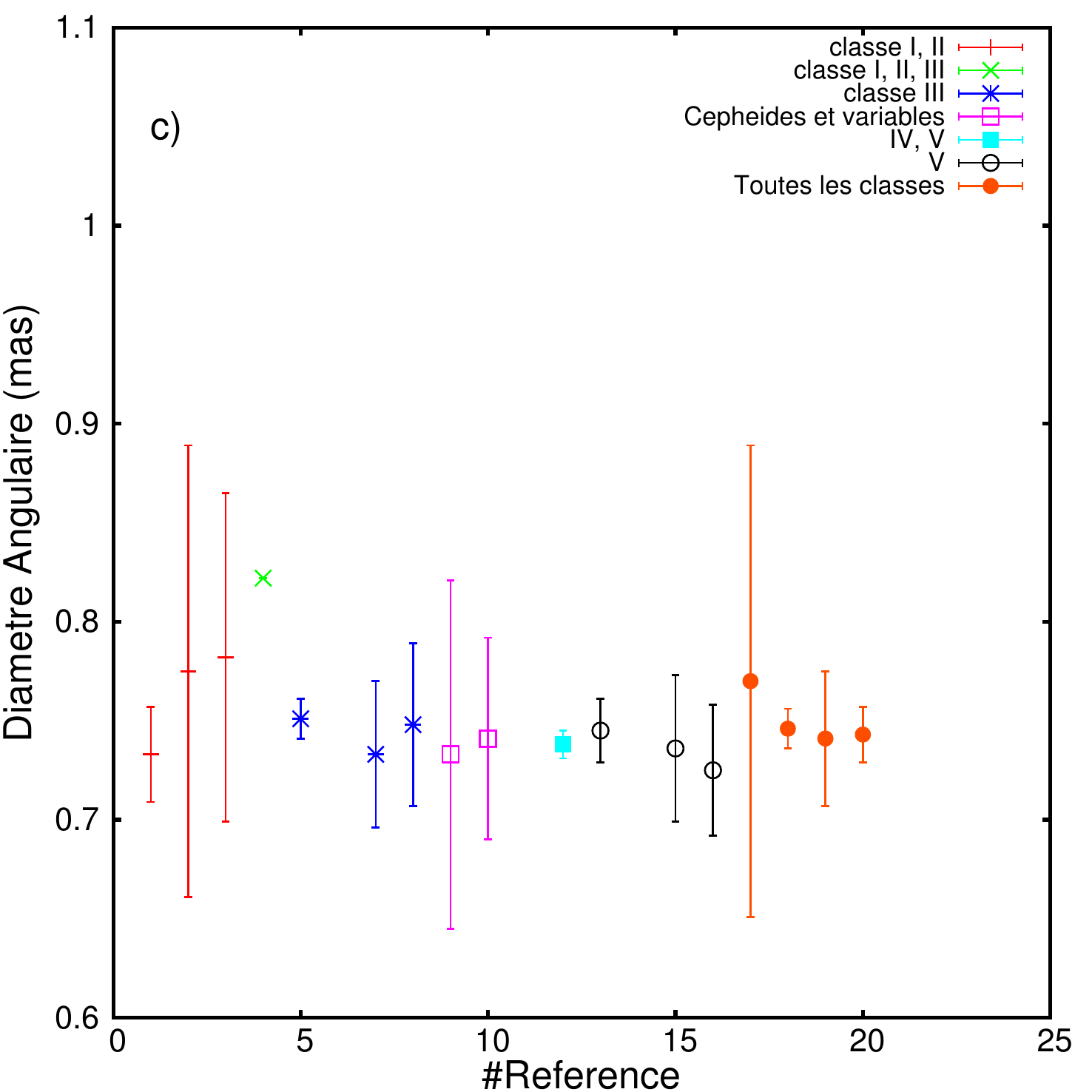}
\includegraphics[width=8cm]{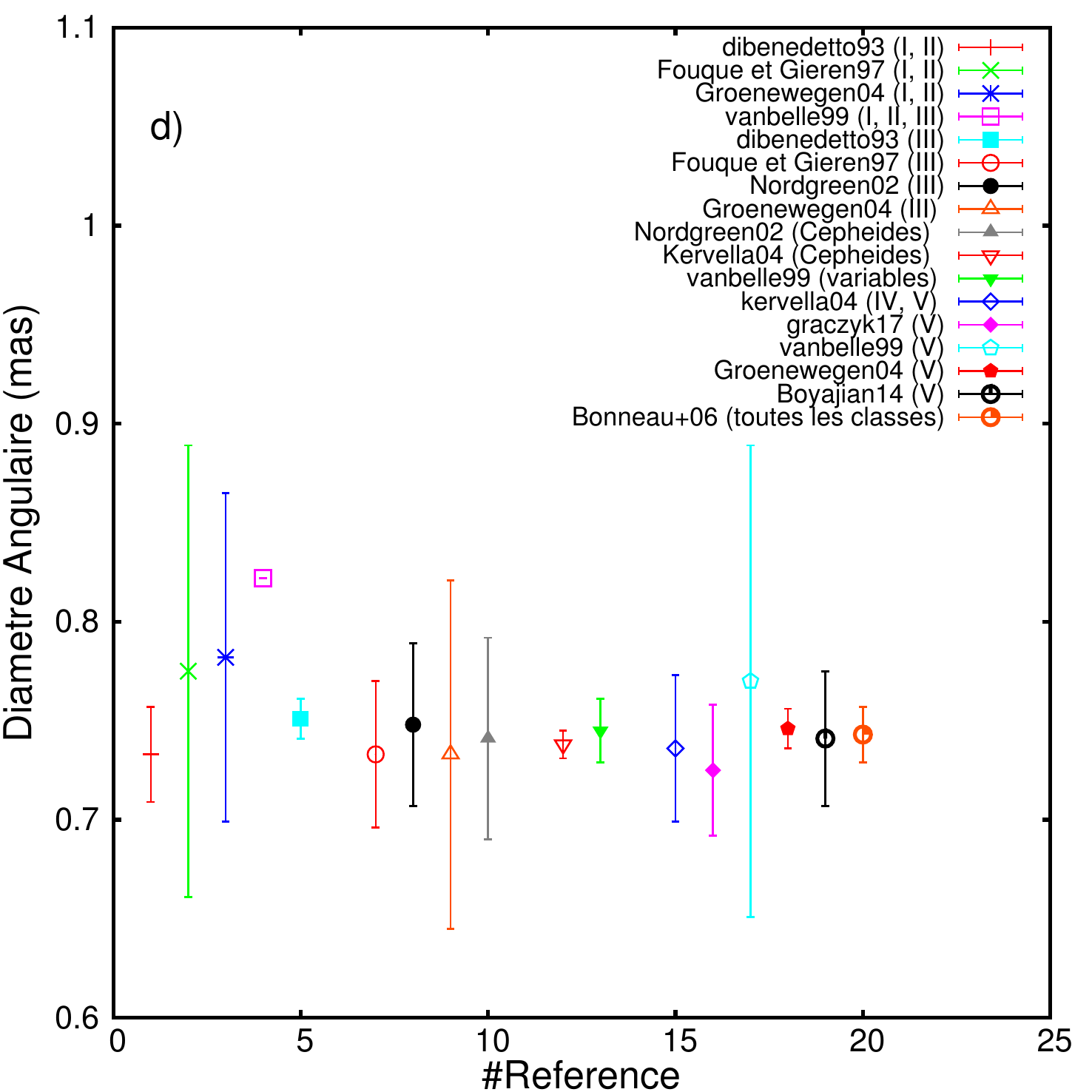}
\end{center}
\vspace*{-5mm} \caption{ \footnotesize  En haut, comparaison des pentes et point-zéros des relations brillance de surface - couleur linéaire: code de couleur par classe a), code de couleur par référence b). En bas: les relations brillance de surface couleur  linéaires et non-linéaires sont utilisées pour estimer le diamètre angulaire d'une étoile théorique qui aurait une magnitude V de 6 et une couleur $V-K$ de 2 et sont présentées par classe c) et par référence d). Dans ces figures, les références suivantes sont exclues car non valides sur le domaine de couleur V-K=2: \cite{fouque97} (classe III, ref 6), \cite{vanbelle99} (variables, ref 11) et \cite{vanbelle99}  (classe V, ref 14). La ref 4, bien qu'incompatible avec les autres estimations de diamètre angulaire est valide sur le domaine de couleur considéré dans cette figure. }\label{Fig_CompSB}
\end{figure}

Dans ces formules, l'indice '0' signifiant habituellement que les magnitudes V ou (V-K) sont corrigées de l'extinction n'est pas indiqué par commodité. Par ailleurs, selon les publications l'indice k vaut 0, 1, 2, ou même 5. En utilisant des formules de passage, on peut comparer les relations brillance de surface - couleur  linéaires (les relations non-linéaires sont exclues à ce niveau, car les transformations sont complexes, mais on y reviendra plus loin) en prenant en compte leur intervalle de validité. Ceci est illustré par la Fig.~\ref{Fig_CompSB0}. On remarque immédiatement que l'ensemble des relations sont assez proches sur l'intervalle $2 < V-K < 4$. En revanche, en dehors de cette intervalle $V-K<2$ et $V-K>4$, on note des divergences. Mais ces différences peuvent être en partie\footnote{Et non "en parti expliquées" comme me l'a très justement fait remarquer mon collègue Frédéric Morand lors de sa Hautement Dense Relecture (HDR) du manuscrit à l'occasion d'observations VEGA nuageuses: "les étoiles ne font pas de politique". Je le remercie ici en catimini pour son aide plus que précieuse ! Cela évitera au lecteur de subir par exemple des "occupations lunaires", car bien sûr, comme le précise Frédéric: "la Lune n'est pas occupée ... ou alors seulement à tourner autour de la Terre !".} expliquées par le fait qu'il faut faire également une distinction selon la classe des étoiles (indiquées entre parenthèses dans la légende de la figure). Ainsi, on peut représenter ces données différemment. Dans les figures~\ref{Fig_CompSB}-ab je représente la pente $\beta$ de la relation  $F_\mathrm{V}= \alpha + \beta (V-K)$ en fonction de son point zéro $\alpha$ pour les différents auteurs (Fig.~\ref{Fig_CompSB}-a) et selon la classe des étoiles pour laquelle la relation est censée être utilisée (Fig.~\ref{Fig_CompSB}-b). On peut faire plusieurs remarques: 

\begin{enumerate}
\item Il semble exister des relations brillance de surface couleur associées à chaque classe d'objet, en particulier (I, II), (III) et (IV, V). Si ce résultat est confirmé, cela pourrait indiquer que les étoiles de classes (IV, V) sont plutôt compatibles avec les propriétés d'un corps noir, tandis que  classes III et (I, II) s'en éloigneraient du fait très probablement d'une activité. 
\item Les relations pour les classes I, II sont comparativement peu précises \footnote{Curieusement, la relation de \cite{dibenedetto93} valable pour les classes I, II et donnée sans précision (petite croix rouge sur la figure \ref{Fig_CompSB}-b) est plutôt compatible avec les autres relations associées à la classe V.}. 
\item Les relations pour les classes III sont à l'inverse précises, compatibles entre elles, ce qui est encourageant pour l'établissement d'une relation à 1\% de précision pour la détermination de distance des binaires à éclipses de type tardif, qui sont effectivement, dans la plupart des cas, de classe III.
\item La relation brillance de surface - couleur des Céphéides (typiquement de classe III) est plutôt compatible avec les relations établies pour les classes V. Le point de \cite{vanbelle99} (variables) concerne quant à lui les étoiles de type S, Mira, ... 
\item Les relations pour les classes V présentent des incohérences. Parmi ces relations, la plus précise est celle de \cite{boyajian14} comprenant 124 étoiles (l'incertitude est comprise dans le point). La relation établie par \cite{kervella04}, qui inclut peu d'étoiles, mais proches ($d<15 pc$), vient juste après en terme de précision sur les coefficients de la relation. Les relations de \cite{kervella04}  et \cite{boyajian14} sont incompatibles. Cependant, il faut voir deux choses. Il ne faut pas confondre la précision sur $\alpha$ et $\beta$ et la dispersion de la relation (ou rms). Ainsi, la relation de \cite{boyajian14} contenant de nombreuses étoiles est précise mais présente une dispersion non négligeable qui induit une erreur sur le diamètre de 4.6\%. A l'inverse, la relation de \cite{kervella04} est établie sur beaucoup moins d'étoiles, présente une dispersion faible, ce qui correspond finalement à une précision sur le diamètre angulaire de 1\%. Par ailleurs, comparer les pentes et les points zéros des relations est intéressant, mais pas forcément la meilleure façon de procéder, surtout si l'on veut inclure les relations non-linéaires dans la comparaison.
\end{enumerate}

\begin{figure}[htbp]
\begin{center}
\includegraphics[width=8cm]{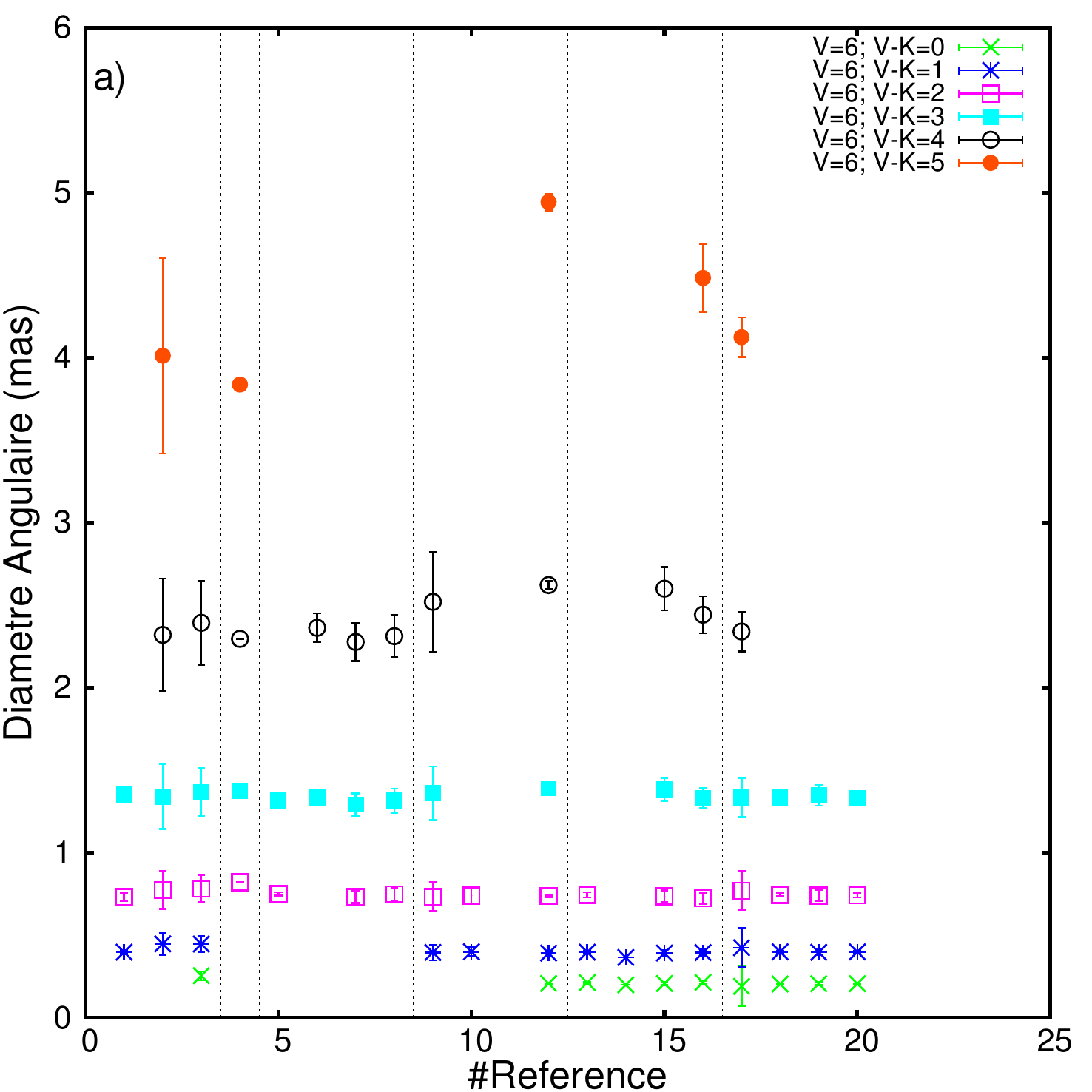}
\includegraphics[width=8cm]{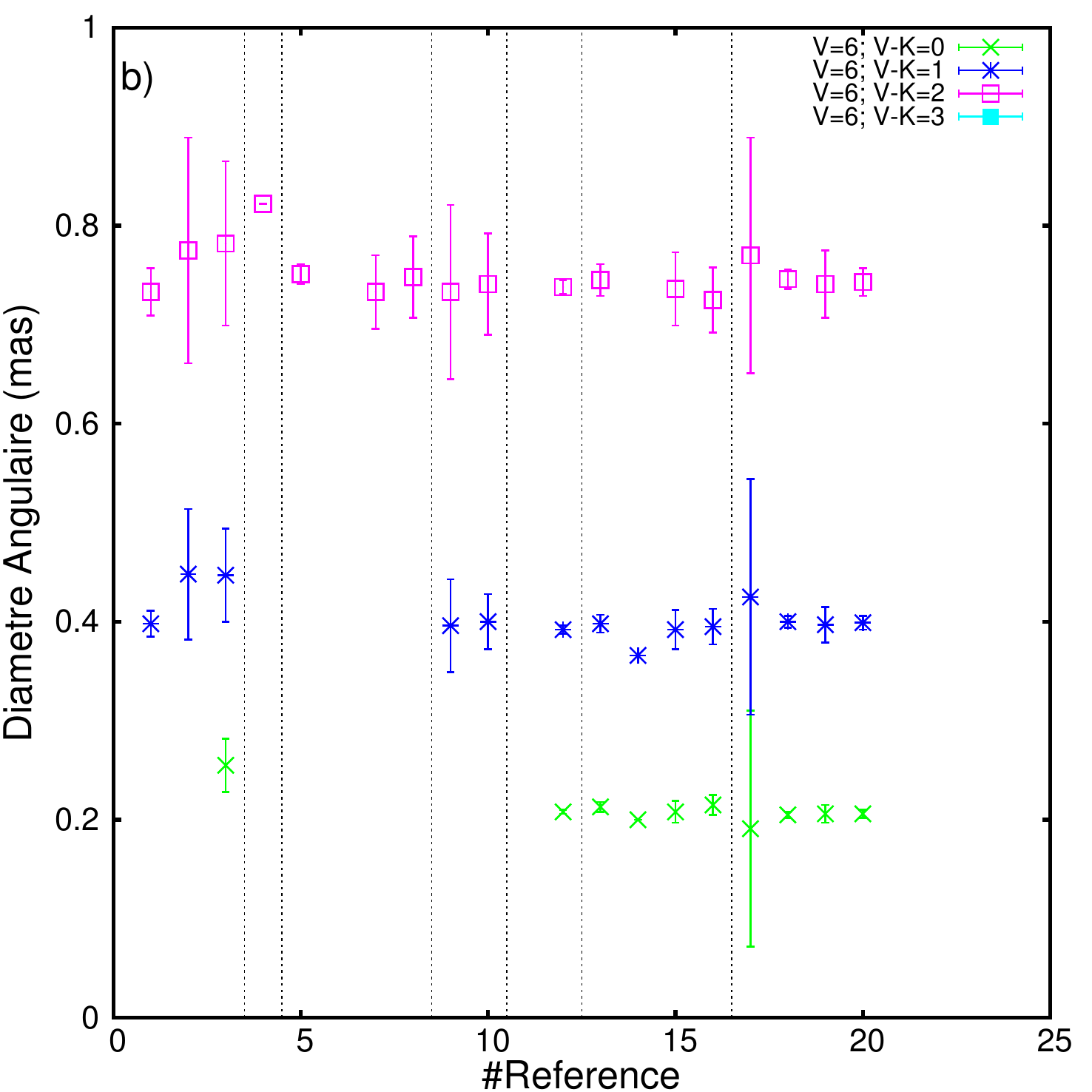}
\end{center}
\vspace*{-5mm} \caption{ \footnotesize  Même chose que la figure~\ref{Fig_CompSB}cd mais pour différentes valeurs de V-K: de 0 à 5 à gauche et seulement de 0 à 3 à droite (zoom). Pour chaque $V-K$ ne figurent que les estimations de diamètres compatibles avec le domaine de validité de la relation brillance surface couleur utilisée. Les lignes verticales en pointillés séparent les valeurs associées aux différentes classes selon la légende de la Fig.~\ref{Fig_CompSB}c.}\label{Fig_CompSB2}
\end{figure}

\begin{figure}[htbp]
\begin{center}
\includegraphics[width=5cm]{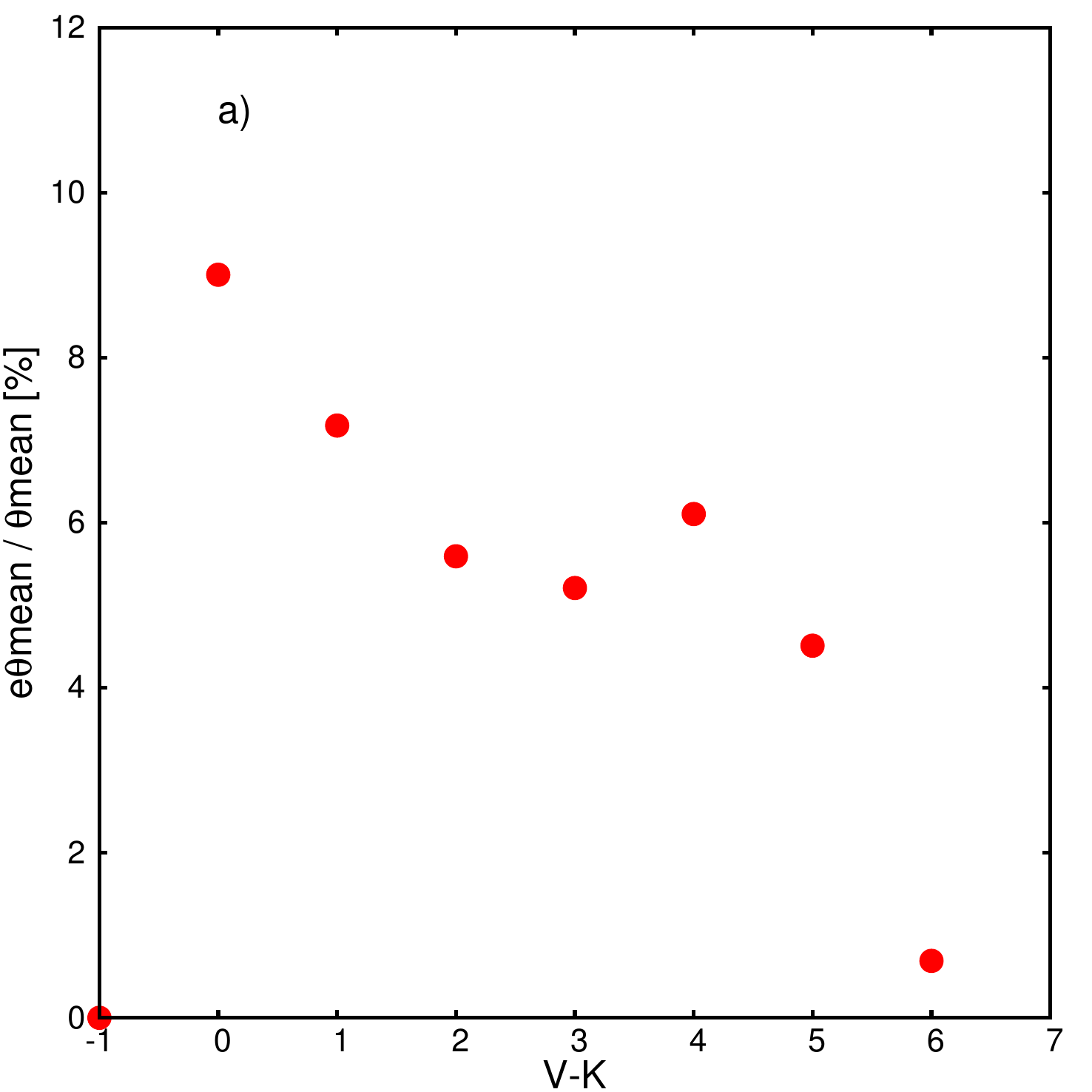}
\includegraphics[width=5cm]{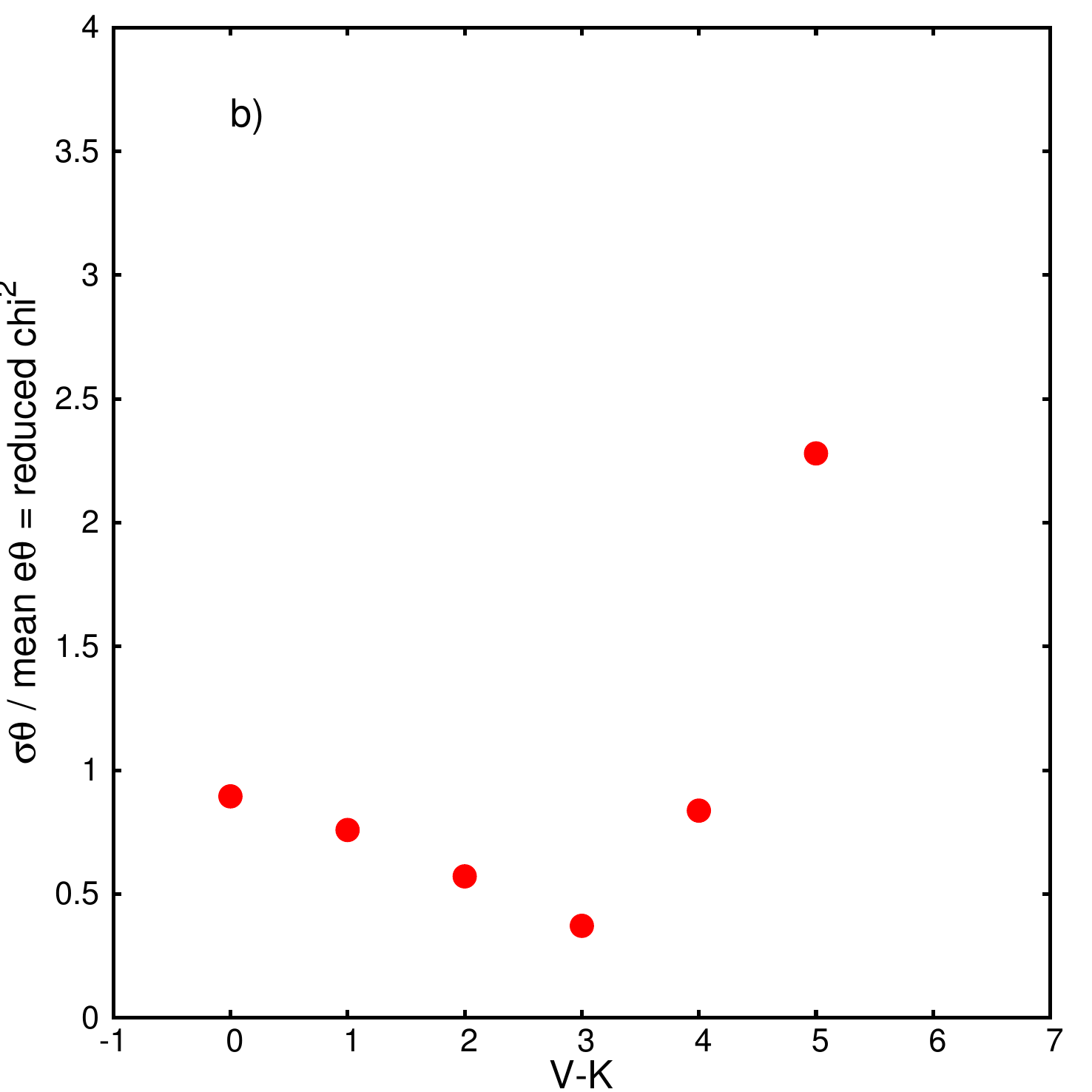}
\includegraphics[width=5cm]{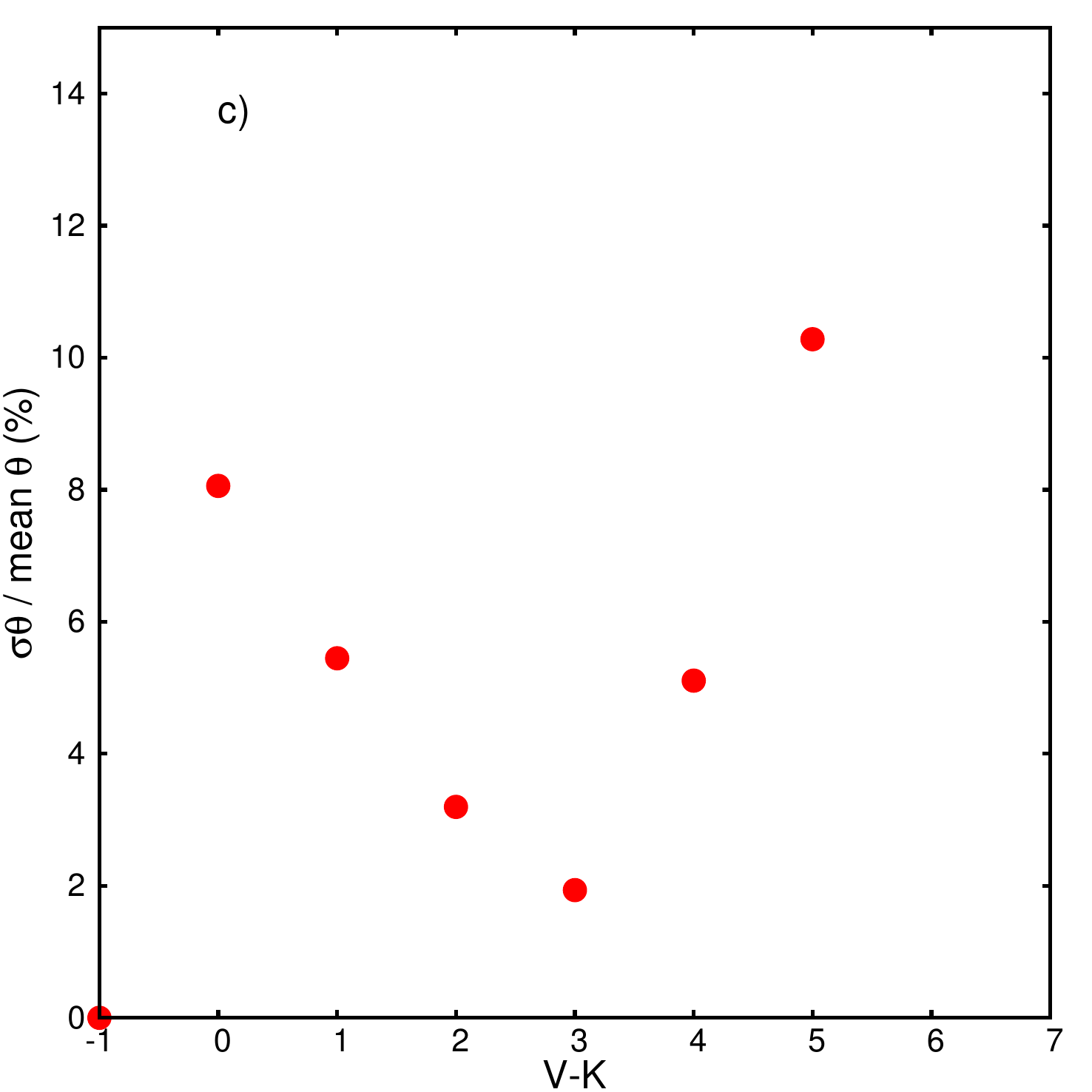}
\end{center}
\vspace*{-5mm} \caption{ \footnotesize  Analyse statistique à partir de la figure~\ref{Fig_CompSB2}. A gauche, il s'agit de l'erreur moyenne sur le diamètre, comparée au diamètre moyen pour différents $V-K$. Cela correspond donc à l'erreur relative en pourcentage telle qu'elle est estimée dans les publications. On prend ici en compte les domaines de validité en $V-K$ des relations. En revanche, toutes les classes sont considérées. Le diagramme du milieu teste la validité des estimations d'erreurs dans la littérature en comparant les estimations entre elles, et notamment leur dispersion, par rapport à l'erreur correspondante qui est annoncée. On obtient en quelque sorte une estimation du $\chi^2$ réduit basé sur l'ensemble de la littérature. Des incohérences importantes sont ainsi observées à V-K=5. Le dernier diagramme à droite donne l'estimation la plus réaliste des erreurs sur les diamètres que l'on peut espérer avoir actuellement, puisqu'on compare la dispersion sur les diamètres issus de la littérature avec le diamètre moyen correspondant. }\label{Fig_CompStats}
\end{figure}

Ainsi, une autre façon de comparer les relations (linéaires et non linéaires) des différents auteurs est de comparer directement les diamètres prédits en considérant une étoile typique donnée, par exemple $V-K=2$, $V=6$ et $K=4$ et en prenant bien sûr en compte l'intervalle de validité de la relation en V-K donné par l'auteur. J'obtiens alors les figures Fig.~\ref{Fig_CompSB}-cd selon la même distinction (classe d'objet et auteurs) que précédemment. Le point important est qu'ici les incertitudes sur les diamètres sont déduits non pas des incertitudes sur les relations brillance de surface - couleur (quelque soit la définition) mais sur la dispersion de la relation (rms). 
%A part trois  déterminations qui sont incohérentes \cite{vanbelle99} (variables), \cite{vanbelle99} (I, II, III) et \cite{dibenedetto93} (III), déjà soulignées, les autres valeurs de diamètre angulaire sont compatibles ce qui semble encourageant. 
Les valeurs de diamètre angulaire sont compatibles  et ceci est également vrai pour les diamètres estimés avec les relations de \cite{kervella04} et \cite{boyajian14}. Cependant, avec $V-K$ autour de 2, nous nous sommes placés dans un régime favorable (voir Fig.~\ref{Fig_CompSB0}). Ainsi, la figure~\ref{Fig_CompSB2} représente la même figure mais pour différentes valeurs de V-K allant de 0 à 5 sur la figure de gauche, et pour plus de clarté, seulement de 0 à 3 sur la figure de droite. Pour mieux apprécier ces résultats, on peut considérer trois estimateurs statistiques, représentés en fonction de V-K sur les figures Fig.~\ref{Fig_CompStats}abc:
\begin{enumerate}
\item Le rapport de l'erreur moyenne sur le diamètre sur le diamètre moyen (telle que déduite des différentes sources bibliographiques) pour un V-K donné. Cette quantité correspond finalement à l'erreur sur le diamètre annoncée par la littérature. On peut ainsi espérer obtenir une précision sur le diamètre d'une étoile précoce ($V-K=0$) de 9\% alors que cette valeur descend progressivement jusqu'à moins de 1\% pour une étoile tardive ($V-K=6$) (voir Fig.~\ref{Fig_CompStats}a).
\item Le rapport de la dispersion des mesures de diamètre sur l'erreur moyenne correspondante, toujours pour un V-K donné. Cela correspond en quelque sorte à un $\chi^2$ réduit. Une valeur de 1 (comme pour les étoiles avec $V-K=0$) veut dire qu'il y a une cohérence entre la précision et la dispersion des mesures. En revanche, les étoiles avec $V-K=3$ ont une précision sur le diamètre probablement sous-estimée, tandis que c'est inverse pour les étoiles tardives avec $V-K=5$, où les précisions annoncées dans la littérature sont probablement deux fois trop grandes (voir Fig.~\ref{Fig_CompStats}b). 
\item Enfin, le rapport de la dispersion des mesures de diamètres sur la valeur moyenne donne la précision réelle que l'on peut espérer obtenir actuellement à partir des relations brillance de surface couleur (sans considération de classe). Elle est autour de 8\% pour les étoiles précoces ($V-K=0$), 2\% pour les étoiles avec $V-K=3$ et 10\% pour les étoiles tardives ($V-K=5$) (voir Fig.~\ref{Fig_CompStats}c). Il est important de noter que cette valeur de 10\% reflète le fait qu'il n'existe pas de relation brillance de surface couleur universelle et qu'il faut prendre en compte la classe et donc très probablement l'activité de l'étoile pour gagner en précision. \end{enumerate}

%Néanmoins,

On peut alors tirer quelques conclusions d'ordre général:

\begin{enumerate}
\item Comme déjà indiqué, il semble clair qu'il existe des relations brillance de surface - couleur spécifiques à trois grands groupes de classes: (I, II), III et (IV, V). Ces relations sont actuellement très peu précises pour les classes (I, II), relativement précises pour les classes III, et incohérentes pour les classes (IV, V). Du point de vue du type spectral, il semble important d'augmenter la précision sur les relations brillance de surface - couleur des étoiles de type précoce (O, B, A, F), actuellement autour de 8\%, afin d'accéder à une précision de l'ordre de 3-5\%. Ceci est nécessaire pour déduire la distance des galaxies M31 et M33 par la méthode des binaires à éclipses. Pour avancer sur cette thématique, l'interférométrie visible semble la meilleure piste. Cependant, pour cela, il faut prendre en compte l'impact de la rotation et des vents. D'un autre côté, il semble que l'activité des étoiles tardives, par exemple la granulation en lien avec un effet de classe, génère des incohérences (à 2.5$\sigma$) dans les estimations de diamètre des étoiles tardives. Il faut également noter que les étoiles évoluées tardives (classe I, II, III) sont généralement trop résolues par interférométrie, ce qui peut aussi poser des difficultés dans l'estimation du diamètre, car les mesures sont alors très sensibles à de nombreux effets physiques. L'imagerie devrait apporter des clefs importantes dans ce domaine. 
\item  On note que les échantillons de diamètres d'étoiles considérés sont plutôt faibles: 42 pour \cite{kervella04}, 124 pour \cite{boyajian14} et au maximum 239 pour  \cite{vanbelle99}  par exemple (voir Tab.~\ref{Tab.SB}), par rapport à un échantillon de diamètres potentiel de 558 (voir Fig.~\ref{Fig_SB_JMDC}) qui, comme déjà expliqué, ne prend pas en compte les binaires et les variables. Cela veut dire que beaucoup d'étoiles sont exclues sur la base de critères différents selon les auteurs. Ainsi, les relations obtenues pour un domaine de validité en V-K et pour une classe donnée sont-elles comparables pour autant ? N'y a-t-il pas des effets de sélection. Comment alors obtenir des relations robustes utilisables pour les milliers et centaines de milliers d'étoiles observées avec {\it Gaia} et PLATO ? Aussi, pour augmenter la précision et l'exactitude de ces relations, il semble important de doubler au minimum l'échantillon d'étoiles pour lesquelles on a une mesure de diamètre angulaire, et l'idéal, serait de le faire en utilisant un seul et même instrument de façon à réduire autant que possible les erreurs systématiques. Cette approche devrait aussi permettre de comprendre les incohérences à 2.5$\sigma$ observées pour les étoiles tardives. 
%\item Comme déjà indiqué, les relations sont peu précises pour les classes I et II. Ceci n'a pas vraiment d'impact sur l'intervalle central $0 < V-K < 2$ car finalement toutes les relations sont assez équivalentes dans ce régime.  En revanche, l'ensemble de la littérature semble indiquer que les classes I, II ont des relations brillance de surface spécifiques lorsque l'on considère des étoiles précoces ($V-K < 1$) ou tardives ($V-K > 4$). Or, les étoiles précoces (toutes classes confondues) ont été peu observées par interférométrie (20\% de l'échantillon) du fait du fort pouvoir de résolution nécessaire (environ 0.3 mas) pour une estimation de diamètre. Par ailleurs, ces étoiles sont très actives avec bien souvent de fortes rotations et du vent. De l'autre côté du diagramme, les étoiles tardives de classe I, II sont souvent trop résolues et sujettes à une forte activité (granulation) ce qui complique les mesures de diamètre angulaire. %Pour résoudre ces difficultés, il faut observer des étoiles plus petites angulairement, et donc plus faibles en terme de magnitude. Il s'agit donc d'aller dans le visible et augmenter la sensibilité des instruments. 
\item Il faut avoir aussi à l'esprit que l'activité des étoiles (granulation, taches, rotation, vent et environnement) jouent un rôle majeur et contribuent à rompre l'hypothèse de base sur laquelle repose l'établissement des relations brillance - couleur de surface, à savoir qu'une étoile est un corps noir. C'est probablement une des raisons qui explique la forte dispersion obtenue lorsque l'on considère tout l'échantillon. Si l'on veut établir une relation brillance de surface - couleur, il faut donc choisir des étoiles qui sont de bons corps noirs. A l'inverse, si l'on veut utiliser une relation brillance de surface - couleur (dans le contexte {\it Gaia} ou PLATO) pour estimer le diamètre angulaire d'une étoile quelconque, il faut avoir à l'esprit que le diamètre estimé correspondra à celui d'un corps noir sans activité. Pour améliorer la relation brillance de surface - couleur dans le cadre de la détermination de distance des binaires à éclipses de type tardif, une solution intéressante est de considérer uniquement des étoiles de type K et de classe III de façon à réduire au maximum les divergences entre l'objet utilisé pour étalonner la relation brillance de surface - couleur et l'objet auquel on applique la relation. Ce travail est en cours avec l'instrument Pionier/VLTI. 
\item Afin d'étalonner efficacement une relation brillance de surface couleur, disposer d'une photométrie V et K (ou autre) homogène et précise semble enfin un point clef. 
\end{enumerate}

Ainsi, pour augmenter la précision et l'exactitude sur les relations brillance de surface - couleur, il faut augmenter/doubler l'échantillon d'étoiles observées par interférométrie, avec un seul et même instrument, en prenant bien garde de considérer un échantillon d'étoiles homogène en terme de classe et de type spectral. Il faudra en particulier augmenter l'échantillon des étoiles de type spectral O, B, A, F et de classe I, II. L'impact de l'activité stellaire est également un point clef car elle est probablement à la source des différences observées entre les relations. 

Dans les années à venir, l'instrument SPICA\footnote{Une ERC déposée par D. Mourard.} devrait permettre d'avancer sur de nombreux points. Il s'agit d'un instrument à 6 télescopes, fibré, muni d'une caméra de nouvelle génération dans le visible et tirant parti des nouvelles optiques adaptatives du réseau interférométrique CHARA. L'objectif scientifique est de déterminer les paramètres fondamentaux (dont le rayon) de 200 étoiles à exoplanètes, 200 étoiles astérosismiques, ainsi que 600 étoiles ‘‘standards'' afin d'améliorer la précision et l'exactitude des relations brillance de surface - couleur.  Parmi cet échantillon de 1000 étoiles, 200 étoiles seront imagées afin d'étudier l'activité stellaire plus en détail. Les relations brillance de surface - couleur qui découleront de ce travail seront un outil précieux dans le contexte {\it Gaia} Gaia et PLATO, car elles permettront de déduire le rayon d'étoiles non résolues et trop faibles pour l'interférométrie. Ce projet aura également des répercutions importantes pour l'étalonnage des échelles de distances dans l'univers. 

Avant de clore cette section, il est intéressant de souligner que nous avons adopté une méthode un peu différente récemment dans le cadre du projet Araucaria pour étalonner la relation brillance de surface - couleur. La méthode consiste à inverser la méthode des binaires à éclipses à partir des parallaxes {\it Gaia}. Effectivement, si la distance des binaires à éclipses galactiques est connue, et que d'autre part on dispose des rayons individuels des composantes par photométrie et spectroscopie, alors il est possible d'en déduire le diamètre angulaire des étoiles et ainsi contraindre la relation brillance de surface - couleur. Ce travail original, qui a fait l'objet d'une publication \citep{graczyk17} (voir l'annexe \ref{graczyk17}), donne des résultats compatibles avec les relations déduites par interférométrie. D'autres études suivront avec les futures données {\it Gaia} avec l'objectif d'atteindre 1\% de précision sur la relation. \\[0.2cm]

\newpage
%%%%%%%SECTION 

\section{L'impact de la rotation sur la relation brillance de surface - couleur}\label{s_rot}

\begin{figure}[htbp]
\begin{center}
\includegraphics[width=11cm]{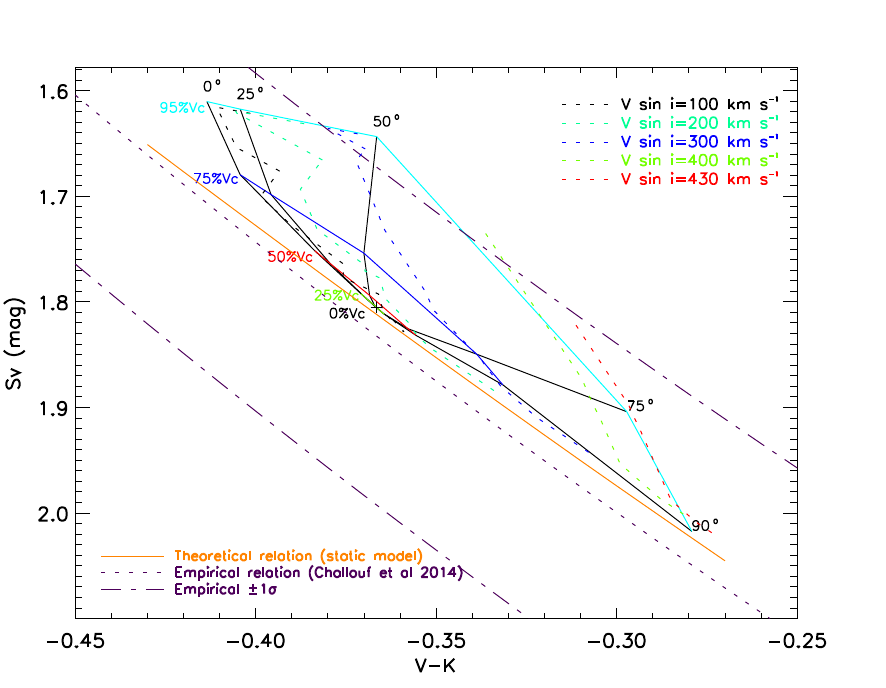}
\includegraphics[width=11cm]{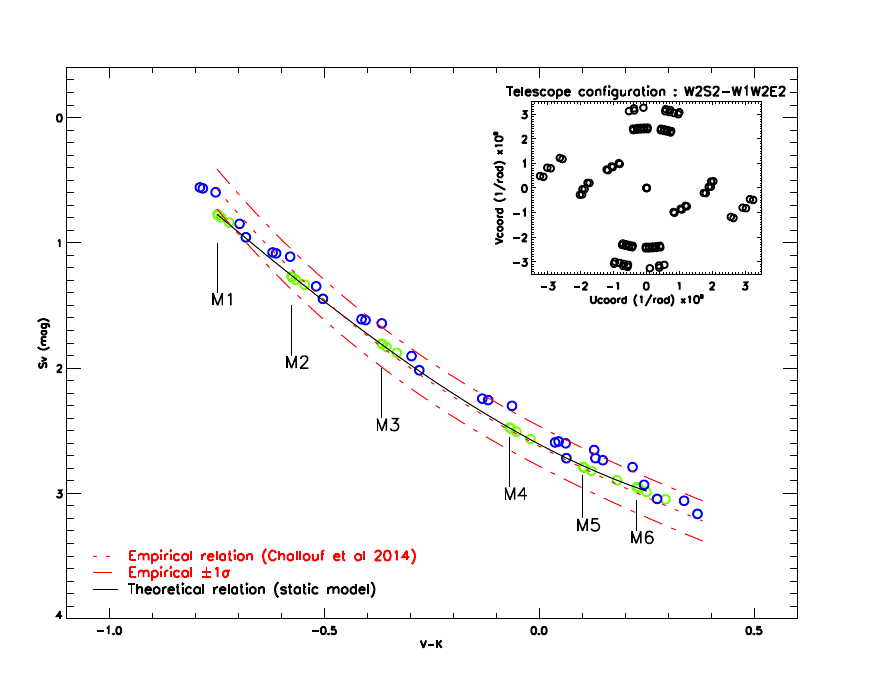}
\end{center}
\vspace*{-5mm} \caption{ \footnotesize   Illustration de l'impact de la rotation rapide sur l'étalonnage de la relation brillance de surface - couleur des étoiles de type précoce (voir le texte pour les explications).}\label{Fig_rot}
\end{figure}

Nous avons vu que l'activité d'une étoile peut avoir un impact sur sa brillance de surface - couleur. Dans l'étude menée par M. Challouf en 2015, nous avons utilisé le code CHARRON \citep{domiciano02} pour quantifier l'impact de la rotation rapide sur la relation brillance de surface - couleur des étoiles précoces. Ainsi nous avons calculé 6 modèles statiques de départ (M1 à M6) ayant des couleurs V-K s'échelonnant entre -1 et 0.5. A partir des 6 modèles statiques, nous avons considéré plusieurs valeurs de vitesse de rotation (0, 25, 50, 75 et 95\% de la vitesse de rotation critique) et plusieurs valeurs de l'inclinaison de l'axe de rotation par rapport à la ligne de visée (0, 25, 50, 75 et 90 degrés). Tous ces modèles (6x5x5=150 au total) ont leur axe équatorial selon l'axe Est-Ouest sur le plan du ciel. Nous avons alors simulé des observations CHARA (avec trois configurations différentes) et l'analyse a été menée {\it comme si} les étoiles étaient des étoiles statiques, c'est à dire sphériques, sans aplatissement dû à la rotation et sans assombrissement gravitationnel, afin d'en déduire finalement la relation brillance de surface - couleur. Ainsi, les valeurs de couleurs (V-K) et la brillance de surface - couleur (telle que définie précédemment) sont biaisées par la rotation et l'inclinaison. La figure~\ref{Fig_rot}-haut illustre ces biais pour le modèle statique de départ M3 ayant une valeur de $V-K=-0.3$ et pour une configuration de CHARA donnée (W2S2-W1W2S2). On a ainsi sur ce graphique 25 modèles correspondant à différentes valeurs de la rotation (exprimées en pourcentage de la vitesse critique $V_\mathrm{c}$) et différentes valeurs d'inclinaison. Les lignes en pointillés correspondent à des valeurs de $V_\mathrm{rot} \sin i $ identiques. La ligne en pointillé qui traverse la figure correspond à la relation empirique obtenue par \cite{challouf14} (Sect. précédente) avec sa dispersion ("dot-dashed lines").  La ligne orange illustre la relation brillance de surface - couleur ajustée sur l'ensemble des 6 modèles statiques qui sont représentés sur la figure~\ref{Fig_rot}-bas. Sur cette figure, les points bleus correspondent à des inclinaisons différentes (0, 25, 50, 75 et 90 degrés) pour une vitesse de rotation à 95\% de la vitesse critique, alors que les points verts correspondent à différentes vitesses de rotations pour une inclinaison donnée et à 90 degrés (‘‘edge-on''). 

Les conclusions de ce travail sont les suivantes:
\begin{enumerate}
\item La vitesse de rotation des étoiles a un impact sur le point-zéro ($\Delta a_\mathrm{0}$) de la relation brillance de surface - couleur et sur sa dispersion ($\sigma$), mais pas sur sa pente (par rapport à la relation brillance de surface - couleur calculée sur les modèles statiques). Ainsi, si on considère des étoiles qui tournent à moins de 50\% de la vitesse critique, l'effet est relativement faible avec un impact de 0.01 magnitude sur  $\Delta a_\mathrm{0}$ et $\sigma$. Si les étoiles tournent très vite, à plus de 75\% de la vitesse critique, alors l'effet peut monter à 0.08 magnitude sur $\Delta a_\mathrm{0}$ et 0.04 magnitude sur $\sigma$. 
\item Par ailleurs,  l'inclinaison a un impact essentiellement sur la couleur V-K de l'étoile. Ainsi $i< 50°$ (resp.  $i> 50°$) fait que l'étoile apparait plus rouge (resp. plus bleue).
\item Si on considère les 150 modèles, on obtient  $\Delta a_\mathrm{0}=0.03$ mag et $\sigma=0.04$ mag. Ces valeurs sont peu sensibles aux configurationx de CHARA considérées. 
\item La dispersion de ces 150 modèles  est compatible avec la relation empirique \cite{challouf14} qui présente une dispersion de 0.16 magnitude. 
\item Enfin, et c'est la conclusion la plus importante, si l'on désire étalonner la relation brillance de surface - couleur pour les étoiles précoces avec une précision de 0.02 magnitude (ou 1\% en terme diamètre), il faut considérer des étoiles dont la vitesse de rotation est inférieure à 50\% de la vitesse critique, ou à défaut (si la rotation n'est pas connue), des étoiles dont le $V_\mathrm{rot} \sin i$ est inférieur à 100\kms. Si l'on cherche une précision et exactitude meilleure que 0.05 magnitude (ou 4-5 \%), ce qui serait déjà un énorme progrès, il faut rejeter les étoiles qui tournent à plus de 75\% de la vitesse critique.
\end{enumerate}

%%%%%%%SECTION 
\section{Contraindre le p-facteur et le k-facteur des Céphéides à l'aide des binaires à éclipses}\label{Sect_EBp}

Les binaires à éclipses sont des objets astrophysiques fascinants car non seulement ce sont de bons indicateurs de distance comme nous venons de le voir (à condition de disposer d'une bonne relation brillance de surface - couleur), mais ils permettent également de déterminer la masse des deux composantes du système avec une excellente précision. Ainsi, si par chance on détecte une binaire à éclipses dont l'une des composantes est une étoile pulsante, on peut en déduire sa masse (et de nombreux autres paramètres), ce qui est particulièrement intéressant pour ce type d'objet, en particulier pour tenter de réconcilier les modèles de pulsation et d'évolution dont on sait qu'ils présentent des désaccords à hauteur de 10\%. Cependant, très peu d'étoiles pulsantes sont connues pour appartenir à des systèmes avec des éclipses. L'une des plus étudiées est AB Cas, un système binaire de type Algol contenant une $\delta$~Scuti \citep{rodriguez98}. \cite{rodriguez01} listent également 9 $\delta$ Scuti appartenant à des binaires à éclipses. Pendant un temps BM Cas était supposée être une Céphéide galactique dans une binaire à éclipses \citep{thiessen56}, mais il est apparu que ce n'était pas le cas \citep{fernie97}. Il faut attendre 2002 pour que trois Céphéides du LMC soient confirmées comme appartenant à des systèmes binaires à éclipses dans le cadre du projet MACHO \citep{alcock02}. Cependant, ces détections sont purement photométriques et les masses des Céphéides n'ont pas pu être déterminées. Les auteurs donnent néanmoins des valeurs d'assombrissement centre-bord (intéressantes pour la détermination du facteur de projection) mais assez imprécises. 

Ainsi, dans le cadre du projet Araucaria, une binaire à éclipses détachée à deux raies (SB2) comprenant une Céphéide OGLE-LMC-CEP0227 ($P_\mathrm{puls} = 3.80$ d, $P_\mathrm{orb} = 309$ d) a été détectée dans le LMC par OGLE \citep{soszynski08b}. La masse de la Céphéide a ainsi été déterminée pour la première fois avec une précision de 1\% \citep{Pietr10} et la valeur obtenue est en accord avec la masse issue des modèles de pulsation, tandis que la masse tirée des modèles d'évolution est significativement plus élevée. Le même exercice fut ensuite réalisé sur OGLE-LMC-CEP-1812, un objet similaire avec une précision sur la masse de la Céphéide de 1.5\% \citep{pietrzynski11}. Un an plus tard, de la même manière, une binaire à éclipses comprenant une RR Lyrae a été découverte dans le LMC avec une période orbitale de $15.24$ jours. Nous avons déterminé sa masse (0.26 masses solaires) avec une précision de 6\% \citep{Pietr12} (voir l'annexe~\ref{pietrzynski12}). A partir de modèles, nous avons montré que le système devait à l'origine être un système binaire proche composé de deux étoiles de 1.4 et 0.8 masses solaires, respectivement, et une période orbitale de 2.9 jours. Ainsi, un transfert de masse pendant environ 5.5 millions d'année a probablement généré le système actuel, plaçant  l'une des deux étoiles dans la bande d'instabilité des RR Lyrae. Seulement 0.2\% des RR Lyrae devraient être contaminées par un système comme celui-ci, ce qui implique que les distances déduites des RR Lyrae ne devraient pas être affectées de manière significative par ces systèmes atypiques. 

\begin{figure}[htbp]
\begin{center}
\includegraphics[width=18cm]{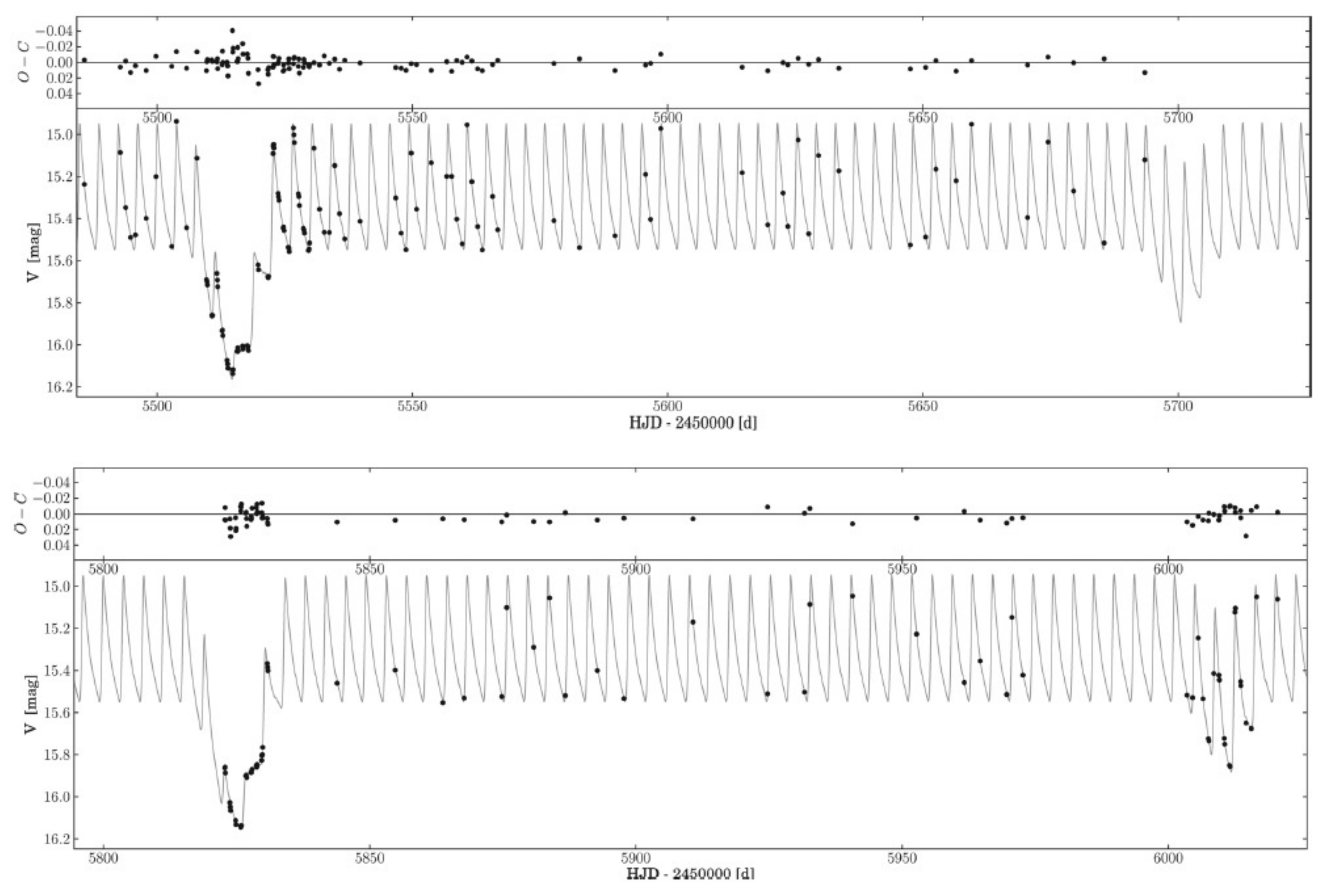}
\end{center}
\vspace*{-5mm} \caption{ \footnotesize   Binaire à éclipses composée d'une étoile standard et d'une Céphéide. Les points sont les observations en bande V (en haut) et en bande V (en bas), tandis que le trait noir représente le modèle. }\label{Fig_ebcep1}
\end{figure}

\begin{figure}[htbp]
\begin{center}
\includegraphics[width=11cm]{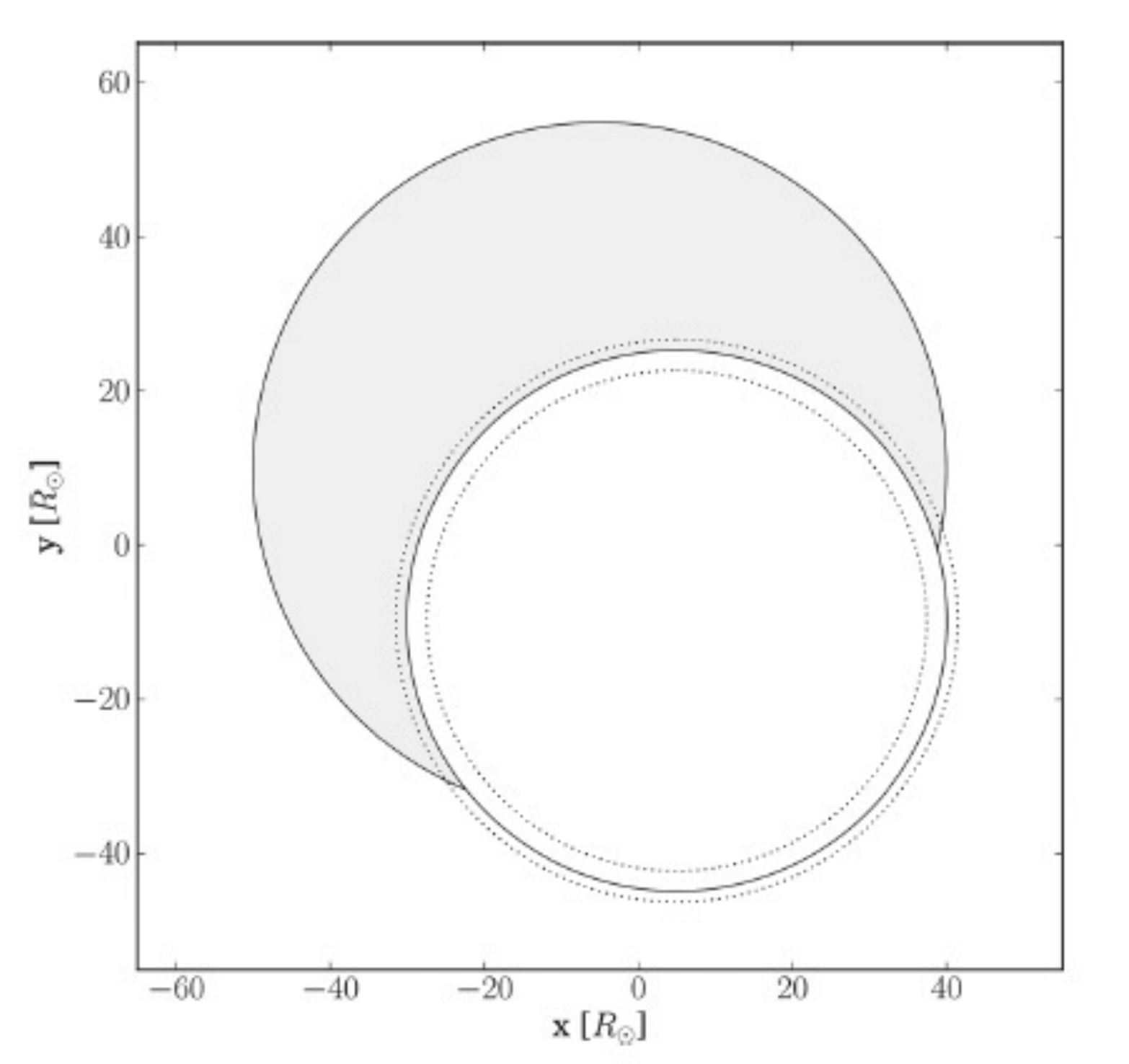}
\end{center}
\vspace*{-5mm} \caption{ \footnotesize   Représentation de la binaire à éclipses. La pulsation de la Céphéide, au premier plan, est représentée par les pointillés. }\label{Fig_ebcep2}

\end{figure}

Mais l'histoire ne s'arrête pas là, dans la mesure où, de façon assez inattendue, les binaires à éclipses incluant une étoile pulsante permettent également de déduire deux quantités physiques dont nous avons déjà parlé, à savoir le $p$-facteur et le $k$-facteur. Ce résultat remarquable a été obtenu par notre équipe du projet Araucaria menée par B. Pilecki. L'objet OGLE-LMC-CEP-0227 a été observé de manière intense au sol afin de récolter 1045 mesures en bande I, 317 en bande V (Télescope géré par le groupe OGLE à Varsovie) ainsi qu'en bande K (instrument SOFI à la Silla) et à 3.6 et 4.5$\mu$m avec le télescope {\it Spitzer}. Des observations effectuées avec les spectrographes MIKE, HARPS et UVES (123 spectres en tout) ont permis d'apporter des contraintes importantes supplémentaires dans la mesure où cet objet est de type SB2. Un modèle spécifique pour ce type d'objet a été développé. Ainsi, sur la figure~\ref{Fig_ebcep1}, on voit clairement dans les observations photométriques en bandes I et V (points noirs) la pulsation de la Céphéide ainsi que les éclipses primaires et secondaires. Le trait noir représente le modèle. Ce qui est intéressant, c'est que le passage de la Céphéide devant son compagnon permet, grâce aux observations photométriques, de déterminer directement la variation de son rayon (Fig.~\ref{Fig_ebcep2}). La dérivée temporelle de cette variation de rayon permet d'en déduire la courbe de vitesse pulsante. Comme l'étoile est SB2, on peut en déduire la courbe de vitesse radiale associée à la Céphéide et donc le facteur de projection qui, rappelons-le, est le rapport de la vitesse pulsante à la vitesse radiale. La valeur ainsi obtenue pour le facteur de projection est de $p=1.21 \pm 0.03(stat.) \pm 0.04(syst.)$ pour une période de pulsation de 3.8 jours \citep{pilecki13} (voir l'annexe~\ref{pilecki13}). Ce résultat est absolument compatible avec la relation \emph{Pp} discutée dans le chapitre précédent \citep{nardetto09} qui donne un facteur de projection, pour cette période, de $p=1.21$. A ce résultat remarquable s'ajoute la question du $k$-facteur des Céphéides. Grâce à la nature SB2 du système, il est possible de comparer la valeur moyenne de la courbe de vitesse radiale de la Céphéide à la vitesse du centre-de-masse du système binaire. La différence entre ces deux vitesses donne le k-facteur et l'on obtient $k=0.59$ \kms, une valeur cohérente avec les valeurs décrites dans \citet{nardetto08a}. 

Le même type d'approche a été effectuée sur un autre système OGLE LMC562.05.9009 ($P_\mathrm{puls} = 2.988$ d, $P_\mathrm{orb} = 1550$ d)  et le facteur de projection est cette fois beaucoup plus élevé, $p=1.37 \pm 0.07$, ce qui apporte un peu de confusion \citep{gieren15}. Suite à la détermination de masse de l'objet OGLE-LMC-CEP-1812, l'objet a été réétudié pour le calcul du facteur de projection, et nous avons obtenu $p=1.26\pm0.09$ [B. Pilecki, private communication]. Enfin, récemment, le même travail a été effectué sur un système comprenant une céphéide de type I et SB1 (OGLE-LMC-T2CEP-098; $P_\mathrm{puls} = 4.974$ d, $P_\mathrm{orb} = 397.2$ d). Dans ce cas, il y a une corrélation entre la masse de la Céphéide et le p-facteur. Le résultat est donc moins robuste et nécessite des hypothèses, mais nous avons conclu sur un $p$-facteur de $1.30 \pm 0.03$ \citep{pilecki17}. Ainsi, avec cette méthode des binaires à éclipses Céphéides, nous avons quatre valeurs du facteur de projection pour des Céphéides courtes périodes s'échelonnant entre 1.21 et 1.37. Pour finir, mentionnons un dernier résultat du groupe Araucaria avec la découverte de la toute première binaire à éclipses composée de deux Céphéides \citep{gieren14}. Mais pour ce système, l'analyse est trop délicate pour le calcul du facteur de projection.

%%%%%%%%%%%%%%%%%%%
%%%%%%%CHAPTER II
%%%%%%%%%%%%%%%%%%%

%%%%%%%%%%%%%%%%%%%
%%%%%%%CHAPTER IV
%%%%%%%%%%%%%%%%%%%
\chapter{Synthèse et perspectives}

\begin{figure}[h]
%\vspace*{-8mm}
\begin{flushleft}
\resizebox{0.45\hsize}{!}{\includegraphics[angle=0]{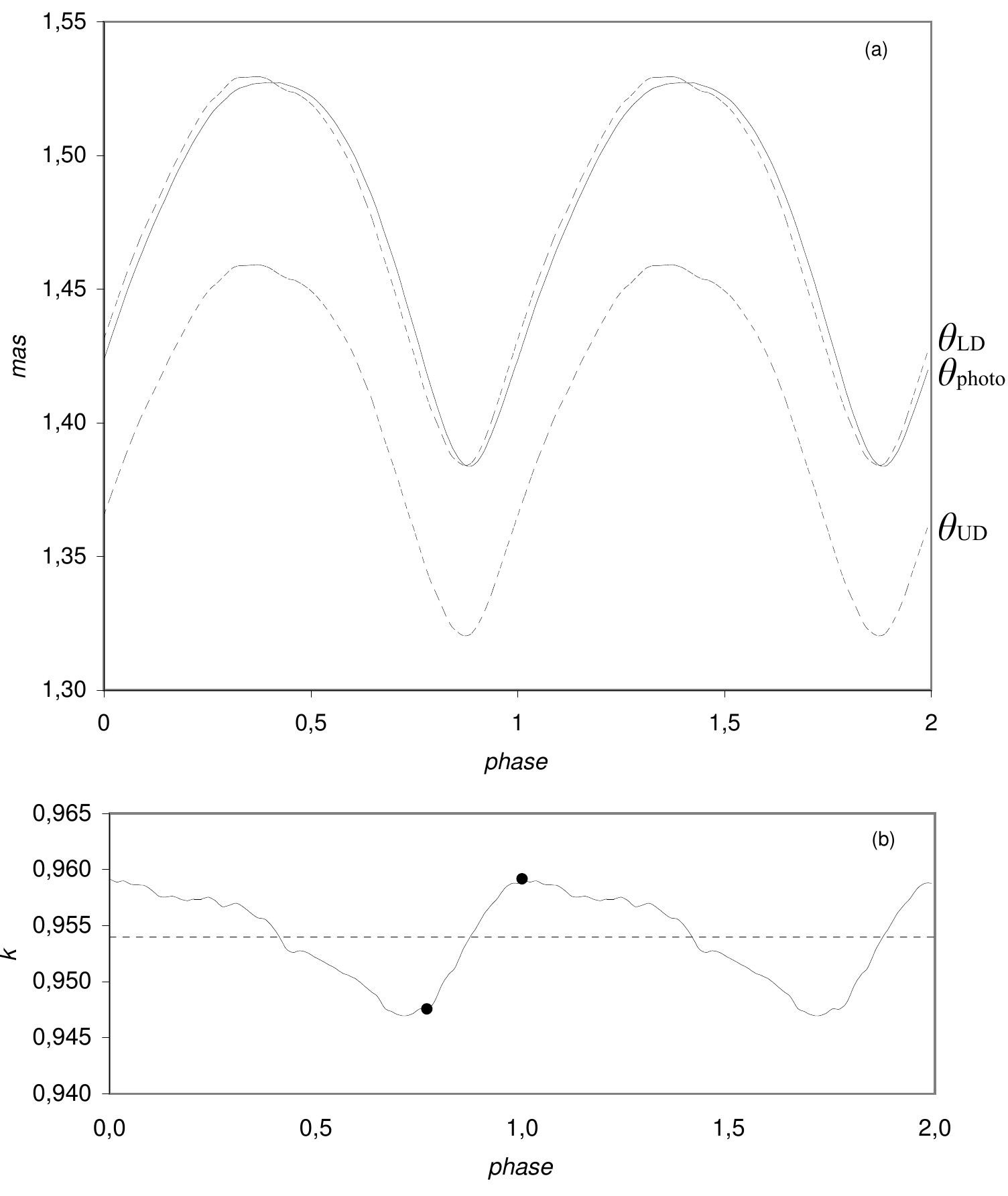}}
\end{flushleft}
%\vspace*{-8mm}
\begin{minipage}{0.52\hsize}
\end{minipage}
\vspace*{-100mm}
\begin{flushright}
\begin{minipage}{90mm}{
\resizebox{0.81\hsize}{!}{\includegraphics[angle=0]{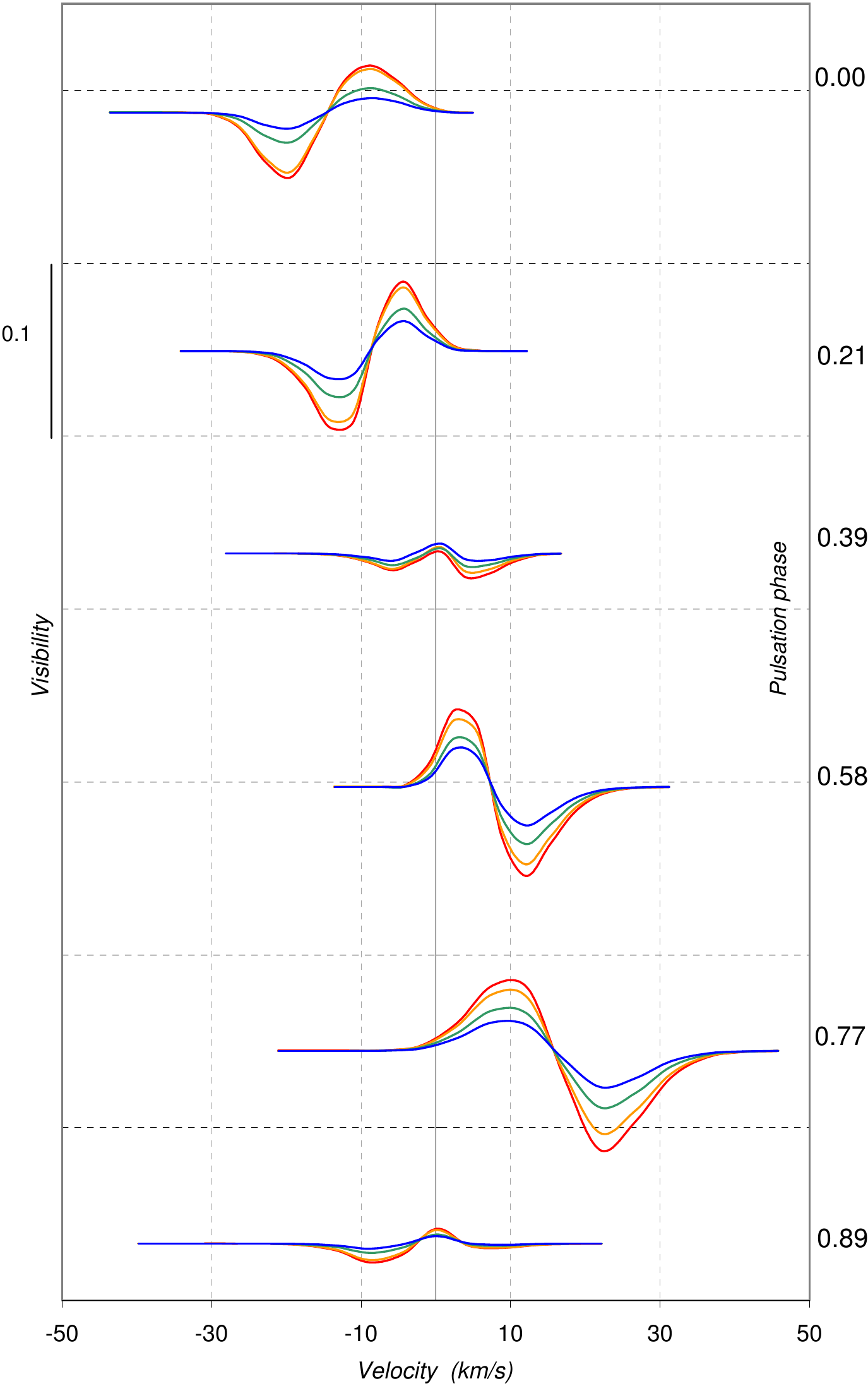} }}
\end{minipage}
\caption{ \footnotesize  Les points clefs des observations interférométriques futures: la mesure directe de l'assombrissement centre-bord (à gauche) et la spectro-interférométrie (à droite). Sur le diagramme de gauche, $\theta_\mathrm{LD}$ ne prend pas en compte la variation d'assombrissement de $\delta$ Cep, tandis que $\theta_\mathrm{photo}$, déduit du modèle hydrodynamique, prend en compte cette variation. On s'aperçoit que l'amplitude de la courbe de diamètre angulaire reste inchangée, et donc on peut conclure que la variation de l'assombrissement centre-bord n'a pas d'impact sur la distance de BW. mais cela reste à prouver avec des mesures directes de variation d'assombrissement centre bord par interférométrie visible, d'autant qu'une telle mesure permet de déduire directement le facteur de projection géométrique de la Céphéide. A droite, on représente les courbes de visibilité dans différentes raies de profondeurs différentes (différentes couleurs) au cours de la pulsation. Ces résultats théoriques ont été publié à l'époque de ma thèse \citet{nardetto06b} et attendent toujours des contraintes observationnelles.}\label{Fig_HRAS}
\end{flushright}
\vspace*{-5mm}
\end{figure}

Résumons. Pour comprendre la nature de l'énergie noire, il faut mesurer le taux d'expansion de l'univers, c'est-à-dire la constante de Hubble ($H_\mathrm{0}$), avec une bonne précision, soit 2\%, voire 1\%. La mesure de $H_\mathrm{0}$ qui utilise le rayonnement de fond cosmologique, c'est à dire à $z=1100$, est déterminée conjointement avec une vingtaine de paramètres ce qui pose le problème délicat des dégénérescences. D'un autre côté, la méthode qui utilise l'étalonnage des échelles des distances mesure le $H_\mathrm{0}$ localement ($z<0.15$) et est probablement très sensible à la physique des objets considérés (Céphéides, SNIa, ...). La tension actuellement de $3.4\sigma$ entre ces deux estimations est soit liée à une erreur systématique non prise en compte dans l'une ou l'autre méthode, soit elle correspond à un résultat physique à interpréter. Plusieurs pistes d'analyses sont possibles mais l'objectif qui vise à renforcer l'échafaudage des distances dans l'univers est de nos jours extrêmement pertinent. Pour cela, le chemin le plus court actuellement utilisé par la communauté est d'abord d'étalonner la relation période-luminosité des Céphéides, puis de l'utiliser afin de déterminer la distance d'une vingtaine de SNIa situées à environ 5-40 Mpc dans des galaxies qui ont une métallicité proche de celle de la Voie Lactée. Une fois la relation des SNIa étalonnée, on peut finalement mesurer $H_\mathrm{0}$. Les points clefs qui ont été identifiés et qui permettraient d'améliorer la précision et/ou d'extraire une erreur systématique sont les suivants et au nombre de trois. 

D'abord, la relation période-luminosité est sensible à la métallicité des Céphéides considérées, que ce soit sa pente ou son point-zéro, et ceci peut avoir un effet de 1\%, voire plus sur $H_\mathrm{0}$. Il est communément admis que la pente est toutefois peu sensible à cette métallicité lorsqu'on observe dans l'infrarouge. En revanche, l'impact sur le point-zéro n'est pas clair du tout. Un façon de s'affranchir de cela est d'étalonner la relation période-luminosité en ne considérant que les Céphéides de la Voie Lactée, dont la métallicité est effectivement proche de la métallicité des galaxies hôtes de SNIa. Mais ce n'est pas totalement satisfaisant dans la mesure où la métallicité des galaxies lointaines est déduite des nuages moléculaires et non à partir des Céphéides elles-mêmes. Ainsi, pour quantifier l'impact de la métallicité sur le point-zéro de la relation période-luminosité, il faut déterminer la distance de plusieurs galaxies du groupe local dont la métallicité est effectivement significativement différente.  Ce travail permettrait également d'étendre le nombre de galaxies de référence pour l'étalonnage de la relation. C'est bien l'objectif du projet Araucaria: déterminer la distance de galaxies à métallicité élevée (M31, M33, M82, et M83) de façon à en faire de nouvelles références, mais aussi de galaxies à métallicité faible (LMC, SMC, ...) de façon à évaluer l'impact de la métallicité sur la relation. A ce niveau, le problème revient à déterminer la distance des galaxies du groupe local de façon précise, en utilisant une autre méthode que la relation période-luminosité de façon à pouvoir l'étalonner justement. Une méthode récente et prometteuse est de découvrir \emph{•}t d'utiliser des binaires à éclipses dans ces galaxies. La distance du LMC a pu ainsi être déterminée avec une précision de 2.2\% et celle du SMC avec une précision de 3\%. Pour augmenter, la précision sur ces deux distances, il va falloir augmenter la précision sur la relation brillance de surface - couleur des étoiles tardives. Pour aller plus loin en terme de distance, il va falloir très probablement découvrir de nouvelles binaires à éclipses brillantes (donc précoces) dans des galaxies du groupe local. Il faudra ensuite augmenter la précision et l'exactitude de la relation brillance de surface - couleur pour les étoiles précoces. Le meilleur moyen de faire cela est d'utiliser l'interférométrie. Le projet SPICA sur CHARA vise précisément à déterminer la diamètre angulaire d'environ 1000 étoiles pour étalonner cette relation, non seulement pour les distances des binaires à éclipses, mais aussi pour estimer le rayon des étoiles et des planètes en lien avec PLATO et {\it Gaia}. Prendre en compte l'activité des étoiles (taches, bulles convectives, vent, environnement, rotation) semble maintenant incontournable pour estimer les relations brillance de surface - couleur (qui font l'hypothèse que les étoiles sont des corps noirs), mais aussi pour détecter et déterminer les paramètres des étoiles et des planètes.

Ensuite, outre l'impact de la métallicité sur la relation période-luminosité, sa dispersion est suffisamment importante pour avoir un impact d'environ 1-2\% (et même probablement plus) sur $H_\mathrm{0}$. Cette dispersion est en partie intrinsèque, liée à la largeur de la bande d'instabilité, mais elle est aussi due au rougissement, à la binarité ou à l'environnement. Evaluer l'impact de la binarité et de l'environnement, respectivement, sont des objectifs de l'ANR UnlockCepheids. Ainsi, avec toute une gamme d'instruments photométriques, spectroscopiques (VISIR) et interférométriques (MATISSE, Gravity, VEGA, SPICA, ...), nous allons établir des SEDs pour toutes les Céphéides galactiques proches (environ 50) en dissociant bien la contribution de l'étoile et celle de l'environnement. L'objectif est de fabriquer un outil qui permette d'estimer l'impact de l'environnement d'une Céphéide sur la relation période-luminosité, et ce quelle que soit sa position dans la bande d'instabilité. On cherchera donc prioritairement une relation entre la période et l'environnement, mais pas seulement. Par ailleurs, il faut noter que la nature des enveloppes des Céphéides est peu comprise (gaz, type de poussières) et que pour cela, l'interférométrie à haute résolution spectrale et dans les bandes L, M, N sera particulièrement utile pour faire de la minéralogie (MATISSE, code DUSTY). Pour ce qui est de la spectro-interférométrie, les modèles hydrodynamiques sont prêts depuis 2005 (voir Fig.~\ref{Fig_HRAS})... même s'il faudra prévoir quelques développements pour prendre en compte l'environnement. De même, la découverte récente que les enveloppes des Céphéides rayonnent également significativement dans le visible ouvre une voie de recherche intéressante.

Enfin, de manière générale, toutes les méthodes précises et exactes qui permettent de déterminer la distance des galaxies du groupe local sont à regarder avec intérêt. A ce titre, un objectif est de rendre la méthode de Baade-Wesselink suffisamment robuste pour déduire la distance des Céphéides {\it individuelles} dans les galaxies du groupe local. Pour cela, le point clef est de caractériser et de comprendre la dynamique atmosphérique des Céphéides. Le lien entre le facteur de projection, la distance des Céphéides et $H_\mathrm{0}$ est étroit et c'est bien la raison pour laquelle les distances de BW ne sont généralement par utilisées par la communauté pour contraindre $H_\mathrm{0}$. Cependant, des progrès conceptuels significatifs ont été effectués sur le facteur de projection. La décomposition du facteur de projection semble correcte. Cependant, un ingrédient doit manquer. Bientôt nous pourrons inverser la méthode de BW pour l'ensemble des Céphéides galactiques pour lesquelles on aura une parallaxe {\it Gaia}. Il y a fort à parier que la relation entre la période et le facteur de projection sera hautement dispersée. Il faudra alors se baser sur la décomposition du facteur de projection pour d'abord comprendre la physique cachée derrière ces valeurs, puis pour construire un outil ou des concepts qui permettent d'estimer le $p$-facteur d'une façon précise et exacte quel que soit la période de la Céphéide ou sa position dans la bande d'instabilité (ce travail est en cours). Comprendre la dynamique atmosphérique est un ensemble: cela passe aussi par l'étude du $k$-facteur, grâce à la spectroscopie optique et infrarouge. Une autre piste intéressante pour comprendre la physique de l'atmosphère des Céphéides est de procéder par comparaison en étudiant d'autres types d'étoiles pulsantes dont le mode radial est dominant (HADS, RR Lyrae, $\beta$-Céphéides, ...). 

Ces travaux permettront une exploitation des instruments futurs, tels que le JWST et l'ELT. Aujourd'hui, on peut atteindre des magnitudes limites d'environ 28 pour le HST (2.4 mètres dans l'espace) ou pour des télescopes de 8m au sol. Cela correspond pour des magnitude absolues d'environ -5 pour les Céphéides à des distances entre 30 et 40 Mpc. Avec le JWST et l'ELT, nous pourrons pousser ces limites à 100 Mpc et fort probablement augmenter le nombre de SN1a utilisées en tant que calibrateurs. 

Ainsi les perspectives de ces recherches sont claires; elles s'insèrent à l'interface de trois grands projets: le projet Araucaria (les distances dans le groupe local), l'ANR UnlockCepheids (dispersion de la relation période-luminosité des Céphéides) et le projet SPICA (contraindre la relation brillance de surface - couleur pour les binaires à éclipses). A côté de cela, je continuerai à sonder l'atmosphère des Céphéides de façon à améliorer notre connaissance du facteur de projection. Une bien belle aventure en perspective ...

\addcontentsline{toc}{chapter}{bibliographie}

\bibliography{bibtex_nn} % your references in file: Yourfile.bib

\begin{appendix}
\chapter{Tableau récapitulatif des valeurs de la constante de Hubble depuis 2000.}

\begin{table*}
\caption{\label{Tab.Ho}  Liste des valeurs de la constante de Hubble obtenues par différentes méthodes ou hypothèses depuis 2000. Ces valeurs sont représentées sous forme graphique dans la Figure~\ref{Fig_Ho}.}
\begin{center}
\begin{tabular}{lcc}
\hline
\hline
$H_\mathrm{0}$        & année &       référence   \\
    km/s/Mpc          &              &                            \\
    \hline  
\multicolumn{3}{c}{{\it Céphéide et SNIa: projet CHP}}  \\
$68^{5}_\mathrm{-5}$&$2000.9$&\cite{gibson00}\\
$71^{6}_\mathrm{-6}$&$2001.4$&\cite{freedman01}\\
$73^{4}_\mathrm{-4}$&$2010.3$&\cite{freedman10}\\
$74^{2}_\mathrm{-2}$&$2012.8$&\cite{freedman12}\\
\multicolumn{3}{c}{{\it Céphéide et SNIa: projet SHOES}} \\
$65^{2}_\mathrm{-2}$&$2000.6$&\cite{filippenko00}\\
$73^{5}_\mathrm{-5}$&$2005.6$&\cite{riess05}\\
$74^{4}_\mathrm{-4}$&$2009.4$&\cite{riess09a}\\
$74^{2}_\mathrm{-2}$&$2011.3$&\cite{riess11}\\
$73^{2}_\mathrm{-2}$&$2016.6$&\cite{riess16}\\
\multicolumn{3}{c}{{\it Céphéide et SNIa: Sandage}} \\
$62^{5}_\mathrm{-5}$&$2006.2$&\cite{sandage06}\\
\multicolumn{3}{c}{{\it Masers}} \\
$69^{7}_\mathrm{-7}$&$2013.3$&\cite{reid13}\\
$72^{3}_\mathrm{-3}$&$2013.7$&\cite{Humphreys13}\\
\multicolumn{3}{c}{{\it Galaxies: TF et FJ}} \\
$76^{1}_\mathrm{-1}$&$2000.9$&\cite{jensen01}\\
$69^{4}_\mathrm{-4}$&$2001.4$&\cite{tonry01}\\
$71^{6}_\mathrm{-6}$&$2002.9$&\cite{Mei03}\\
$73^{4}_\mathrm{-4}$&$2001.1$&\cite{watanabe01}\\
$65^{6}_\mathrm{-6}$&$2001.5$&\cite{Hendry01}\\
$74^{6}_\mathrm{-6}$&$2006.7$&\cite{masters06}\\
$63^{5}_\mathrm{-5}$&$2006.8$&\cite{Dunn06}\\
\multicolumn{3}{c}{{\it RTGB}} \\
$71^{2}_\mathrm{-2}$&$2017.2$&\cite{Jang17a}\\
\multicolumn{3}{c}{{\it Rayonnement de fond cosmologique}} \\
$72^{5}_\mathrm{-5}$&$2003.7$&\cite{spergel03}\\
$70^{5}_\mathrm{-5}$&$2003.8$&\cite{spergel03}\\
$71^{4}_\mathrm{-3}$&$2003.8$&\cite{spergel03}\\
$73^{3}_\mathrm{-3}$&$2007.5$&\cite{spergel07}\\
$72^{3}_\mathrm{-3}$&$2009.2$&\cite{hinshaw09}\\
$71^{1}_\mathrm{-1}$&$2009.2$&\cite{hinshaw09}\\
$71^{2}_\mathrm{-2}$&$2011.2$&\cite{Jarosik11}\\
$70^{1}_\mathrm{-1}$&$2011.2$&\cite{Jarosik11}\\
$70^{2}_\mathrm{-2}$&$2013.0$&\cite{Bennett14}\\
$69^{1}_\mathrm{-1}$&$2013.1$&\cite{Bennett14}\\
$67.4^{1.4}_\mathrm{-1.4}$&$2014.2$&\cite{planck16_XIII}\\
$67.9^{1.5}_\mathrm{-1.5}$&$2014.2$&\cite{planck16_XIII}\\
$67.3^{1.2}_\mathrm{-1.2}$&$2014.2$&\cite{planck16_XIII}\\
$67.3^{1.2}_\mathrm{-1.2}$&$2014.2$&\cite{planck16_XIII}\\
$67.9^{1.0}_\mathrm{-1.0}$&$2014.2$&\cite{planck16_XIII}\\
$67.8^{0.8}_\mathrm{-0.8}$&$2014.2$&\cite{planck16_XIII}\\
\hline
\end{tabular}
\end{center}
\end{table*}

\begin{table*}
\caption{\label{Tab.Ho}  Suite du Tableau.}
\begin{center}
\begin{tabular}{lcc}
\hline
\hline
$H_\mathrm{0}$        & année &       référence   \\
    km/s/Mpc          &              &                            \\
    \hline  
\multicolumn{3}{c}{{\it Lentilles gravitationnelles}} \\
$71^{6}_\mathrm{-6}$&$2000.8$&\cite{Giovi01}\\
$64^{4}_\mathrm{-4}$&$2000.8$&\cite{Giovi01}\\
$48^{4}_\mathrm{-4}$&$2001.8$&\cite{GilMerino02}\\
$62^{4}_\mathrm{-4}$&$2001.8$&\cite{GilMerino02}\\
$71^{5}_\mathrm{-5}$&$2001.9$&\cite{Wucknitz02}\\
$73^{8}_\mathrm{-8}$&$2002.3$&\cite{kochanek02}\\
$48^{3}_\mathrm{-3}$&$2003.4$&\cite{kochanek04}\\
$71^{3}_\mathrm{-3}$&$2003.4$&\cite{kochanek04}\\
$75^{7}_\mathrm{-6}$&$2003.4$&\cite{Koopmans03}\\
$78^{6}_\mathrm{-6}$&$2003.9$&\cite{wucknitz04}\\
$68^{8}_\mathrm{-8}$&$2006.7$&\cite{Oguri07}\\
$62^{8}_\mathrm{-4}$&$2009.2$&\cite{paraficz09}\\
$70^{5}_\mathrm{-5}$&$2010.2$&\cite{suyu10}\\
$66^{6}_\mathrm{-4}$&$2010.3$&\cite{paraficz10}\\
$62^{6}_\mathrm{-4}$&$2010.7$&\cite{courbin10}\\
$69^{6}_\mathrm{-6}$&$2014.1$&\cite{sereno14}\\
 \multicolumn{3}{c}{{\it Sunyaev-Zeldovich}} \\
$58^{4}_\mathrm{-3}$&$2003.7$&\cite{sereno03}\\
$77^{10}_\mathrm{-8}$&$2005.9$&\cite{Bonamente06}\\
$76^{10}_\mathrm{-8}$&$2006.6$&\cite{Bonamente06}\\
$74^{4}_\mathrm{-4}$&$2006.9$&\cite{cunha07b}\\
 \multicolumn{3}{c}{{\it BAO}} \\
$67^{3}_\mathrm{-3}$&$2011.8$&\cite{beutler11}\\
 \multicolumn{3}{c}{{\it Lymann $\alpha$}} \\
$74^{3}_\mathrm{-3}$&$2012.7$&\cite{chavez12}\\
\hline
\end{tabular}
\end{center}
\end{table*}

\chapter{L'étalonnage des échelles de distance dans l'univers.}

Figure and legend (in English) extracted from \citet{degrijs13}: Updated, present-day distance ladder, based on an original idea by Ciardullo (2006). Light orange: Methods of distance determination associated with active star formation (`Population I', intermediate- and high-mass stars). Light green: Distance tracers associated with `Population II' objects/low-mass stars. Blue: Geometric methods. Red: Supernovae (SNe) Ia, the planetary nebulae (PNe) luminosity function (PNLF) and surface-brightness fluctuations (SBF) are applicable for use with both Populations I and II. Light brown: Methods of distance or H0 determination which are not immediately linked to a specific stellar population. Dashed boxes: Proposed methods. Solid, dashed arrows: Reasonably robust, poorly established calibrations. B--W: Baade--Wesselink. RRL: RR Lyrae. RSGs/FGLR: Red supergiants/flux-weighted gravity--luminosity relationship. TRGB: Tip of the red-giant branch. GCLF: Globular cluster (GC) luminosity function. SZ: Sunyaev--Zel'dovich. CMB/BAO: Cosmic microwave background/baryon acoustic oscillations. Colour--magnitude relation: Refers to galactic colours and magnitudes.

\begin{figure}[htbp]
\begin{center}
\includegraphics[width=15cm]{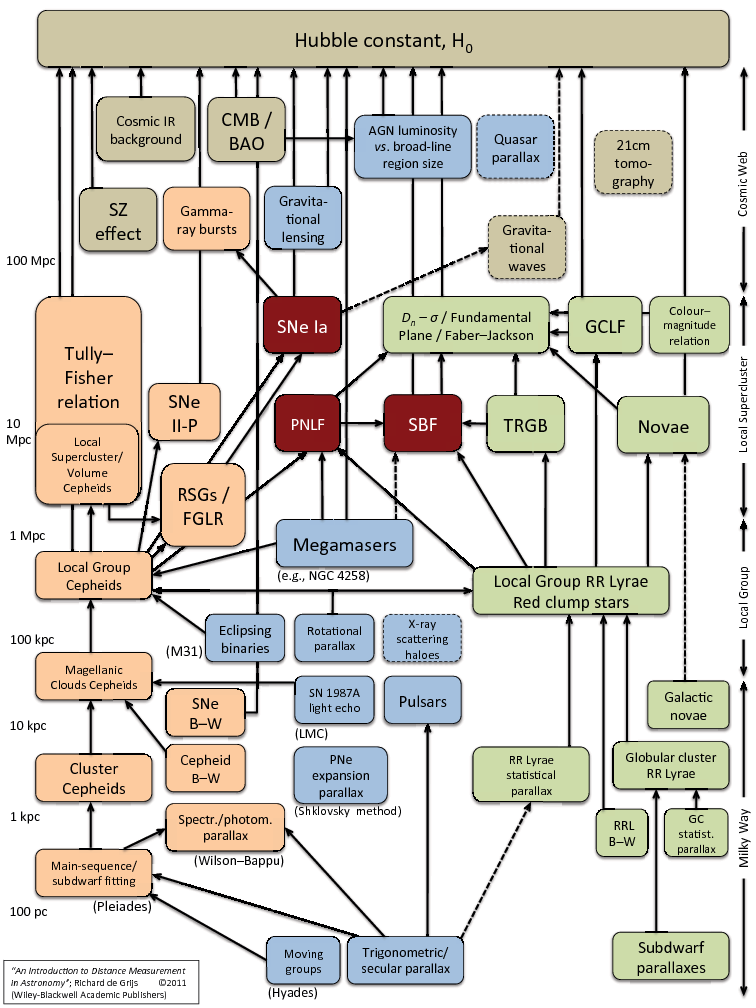}
\end{center}
\vspace*{-5mm} \caption{L'échafaudage des distances dans l'univers}\label{Fig_distances}
\end{figure}

\chapter{Distance des galaxies déterminées dans le cadre du projet Araucaria}

\begin{table*}
\caption{\label{Tab.Araucaria}  Distances et modules de distance des galaxies obtenus dans le cadre du projet Araucaria à partir de différentes méthodes.}
\begin{center}
\tiny
\begin{tabular}{lccccccc}
d	         &	$ed_\mathrm{stat}$	&	$ed_\mathrm{syst}$ 	&		MD				&	$eMD_\mathrm{stat}$	&	$eMD_\mathrm{syst}$	&	Méthode 		&	Ref		\\
$\mathrm{[kpc]}$		&	 $\mathrm{[kpc]}$	           	&	   $\mathrm{[kpc]}$                	&		$\mathrm{[mag]}$	          	&	$\mathrm{[mag]}$   				&	$\mathrm{[mag]}$			&				&	  		\\
\hline													
\multicolumn{8}{c}{{\bf LMC}}	\\														       
50.0	&	0.2	&	1.1	&	18.49	&	0.01	&	0.05	&	DEBS	&	\cite{pietrzynski13}	\\
49.4	&	0.5	&		&	18.47	&	0.02	&		&	RCS	&	\cite{laney12}	\\
50.1	&	1.4	&		&	18.50	&	0.06	&		&	EBS	&	\cite{pietr09b}	\\
52.0	&	0.7	&	2.6	&	18.58	&	0.03	&	0.11	&	RRL	&	\cite{Szewczyk08}	\\
51.5	&	0.9	&		&	18.56	&	0.04	&		&	CEP	&	\cite{gieren05}	\\
50.1	&	0.2	&		&	18.50	&	0.01	&		&	RCS	&	\cite{pietr03}	\\
49.8	&	0.2	&	1.0	&	18.49	&	0.01	&	0.05	&	RCS	&	\cite{pietr02}	\\
\multicolumn{8}{c}{{\bf SMC}}	\\	
62.1	&	1.9	&		&	18.97	&	0.03	&	0.04	&	DEBS	&	\cite{graczyk14}	\\
62.1	&	0.5	&		&	18.97	&	0.02	&		&	RCS	&	\cite{pietr03}	\\
58.3	&	0.5	&	1.3	&	18.83	&	0.02	&	0.05	&	DEBS	&	\cite{graczyk12}	\\
58.9	&	1.9	&	1.9	&	18.85	&	0.07	&	0.07	&	WVIR	&	\cite{Cie10}	\\
62.2	&	0.9	&	3.4	&	18.97	&	0.03	&	0.12	&	RRL	&	\cite{Szewczyk09}	\\
\multicolumn{8}{c}{{\bf Sculpteur}}	\\
87.9	&	3.2	&	4.0	&	19.72	&	0.08	&	0.1	&	RTGB	&	\cite{gorski11}	\\
85.9	&	0.8	&	4.7	&	19.67	&	0.02	&	0.12	&	RRL	&	\cite{pietr08}	\\
\multicolumn{8}{c}{{\bf Carlina}}	\\
104.2	&	1.4	&	5.8	&	20.09	&	0.03	&	0.12	&	RTGB	&	\cite{pietr09}	\\
107.9	&	0.7	&		&	20.17	&	0.02	&		&	RCS	&	\cite{pietr03}	\\
\multicolumn{8}{c}{{\bf Fornax}}	\\
147.2	&	2.0	&	8.1	&	20.84	&	0.03	&	0.12	&	RTGB	&	\cite{pietr09}	\\
148.5	&	0.9	&		&	20.86	&	0.01	&		&	RCS	&	\cite{pietr03}	\\
\multicolumn{8}{c}{{\bf NGC 68}}	\\
460	&	10	&	20	&	23.31	&	0.05	&	0.1	&	RTGB	&	\cite{gorski11}	\\
460	&	4	&	0	&	23.31	&	0.02	&				&	CEP	&	\cite{gieren06}	\\
470	&	9	&	11	&	23.34	&	0.04	&	0.05	&	CEP	&	\cite{pietr04}	\\
\multicolumn{8}{c}{{\bf IC 1613}}	\\
740	&	30	&	30	&	24.36	&	0.08	&	0.09	&	RTGB	&	\cite{gorski11}	\\
720	&	10	&	2000	&	24.291	&	0.035	&		&	CEP	&	\cite{pietr06c}	\\
\multicolumn{8}{c}{{\bf M33}}	\\
840	&	30	&		&	24.62	&	0.07	&		&	CEP	&	\cite{gieren13}	\\
970	&	50	&		&	24.93	&	0.11	&		&	FGLR	&	\cite{Urbaneja09}	\\
\multicolumn{8}{c}{{\bf WLM}}	\\
1070	&	40	&	60	&	25.14	&	0.09	&	0.12	&	RTGB	&	\cite{gorski11}	\\
995	&	46	&		&	24.99	&	0.1	&		&	FGLR	&	\cite{urbaneja08}	\\
970	&	19	&		&	24.924	&	0.042	&		&	CEP	&	\cite{gieren08b}	\\
1070	&	10	&	30	&	25.144	&	0.03	&	0.07	&	CEP	&	\cite{pietr07}	\\
\multicolumn{8}{c}{{\bf NGC 3109}}	\\
1300	&	20	&		&	25.571	&	0.024	&		&	CEP	&	\cite{soszynski06}	\\
1290	&	50	&		&	25.55	&	0.09	&		&	FGLR	&	\cite{hosek14}	\\
1250	&	30	&	50	&	25.49	&	0.05	&	0.09	&	TRGB	&	\cite{gorski11}	\\
1280	&	18	&		&	25.54	&	0.03	&		&	CEP	&	\cite{pietr06b}	\\
\multicolumn{8}{c}{{\bf NGC 300}}	\\
1870	&	17	&	103	&	26.36	&	0.02	&	0.12	&	CEP	&	\cite{rizzi06}	\\
1880	&	40	&	30	&	26.37	&	0.05	&	0.03	&	CEP	&	\cite{gieren05b}	\\
1930	&	40	&	40	&	26.43	&	0.04	&	0.05	&	CEP	&	\cite{gieren04}	\\
\multicolumn{8}{c}{{\bf NGC 55}}	\\
1940	&	30	&		&	26.434	&	0.037	&		&	CEP	&	\cite{Gieren08}	\\
1910	&	44	&	79	&	26.4	&	0.05	&	0.09	&	CEP	&	\cite{pietr06}	\\
\multicolumn{8}{c}{{\bf NGC 7793}}	\\
3440	&	80	&	130	&	27.68	&	0.05	&	0.08	&	CEP	&	\cite{pietrzynski10}	\\
\multicolumn{8}{c}{{\bf M81}}	\\
3500	&	200	&		&	27.7	&	0.1	&		&	FGLR	&	\cite{kudritski12}	\\
\multicolumn{8}{c}{{\bf NGC 247}}	\\
3380	&	60	&		&	27.644	&	0.036	&		&	CEP	&	\cite{gieren09}	\\
3630	&	150	&	150	&	27.8	&	0.09	&	0.09	&	CEP	&	\cite{garcia08}	\\
\hline
\end{tabular}
\end{center}
\normalsize
Notes: DEBS = Detached Eclipsing Binaries ; RCS = Red Clump Stars ; RRL = RR Lyrae stars ; CEP = Cepheids ; W VIR = W Vir stars (Type 2 Cepheids) ; RTGB = Red Tip of Giant Branch ; FGLR = Flux weighted Gravity--Luminosity Relationship ; 

\normalsize
\end{table*}

\chapter{Le cas particulier de Polaris}\label{Polaris}

Polaris est la Céphéide la plus proche de nous à environ $d=99\pm2$ pc \citep{turner13b}. Sa période de pulsation est d'environ 4 jours. \citet{Dinshaw89} indiquent que l'amplitude de la courbe de vitesse radiale, à l'origine autour de $2K \simeq 4.5-5$ \kms, a diminué réguliérement depuis 1956, avec une  forte décroissance dans les années 80 pour s'établir autour de $1.5 \pm 0.08$ \kms en 1989. Cette diminution indiquait pour les auteurs que l'étoile devait sortir de la bande d'instabilité du côté rouge (premier passage) pour finalement s'arrêter de pulser autour de 1995. En 1993, \cite{fernie93} publient un papier intitulé {\it ‘‘Goodbye to polaris the Cepheid''} et mesure $2K = 0.59 \pm 0.19$\kms. Mais l'année suivante,  \cite{krockenberger94} publient {\it ‘‘Polaris the Cepheid: still pulsating''}. Deux ans plus tard, \cite{kamper98} trouvent une ‘‘erreur sérieuse'' dans les mesures spectroscopiques des années précédentes du fait d'une problème lié à la correction des raies telluriques dans la méthode de la cross-corrélation. Ils trouvent que l'amplitude de la vitesse s'est en fait stabilisée autour de 1.6 \kms, ce qui est ensuite confirmé par une étude importante de \citet{hatzes00} qui trouvent  $1.517\pm0.047$ \kms , tandis que la période de l'étoile augmente. Les auteurs observent également des vitesses résiduelles dans les  données spectroscopiques, qu'ils attribuent à des spots ou des modes non-radiaux. \cite{Evans02} étudient les raies en émission chromosphérique MgII et montrent par ailleurs que le changement de période de l'étoile (3.2 secondes par an) est assez commun pour les Céphéides pulsant dans le premier harmonique. \cite{Evans04} montrent alors à partir des données photométriques WIRE que l'étoile pourrait avoir une sorte d'effet Blazhko, typique des étoiles de type RR Lyrae, même si cet effet n'a jamais été observé pour des Céphéides avant cela. \cite{Turner05} étudient à nouveau le statut évolutionnaire de l'étoile et confirment que l'étoile est dans son premier passage dans la bande d'instabilité, tandis que \citet{usenko05} fournissent des paramètres fondamentaux très précis pour cette étoile. \citet{merand06} trouvent alors à l'aide de données FLUOR/CHARA une enveloppe autour de l'étoile contribuant pour $1.5\pm0.4$\% en terme de flux en bande K et située à $2.4\pm0.1$ rayon stellaire, le diamètre angulaire de l'étoile étant mesuré à $\theta= 3.123 \pm 0.008$ mas. \citet{usenko08} étudient ensuite le compagnon optique Polaris B et pensent qu'il s'agit d'une F3V. \citet{bruntt08} font alors une découverte intéressante: l'amplitude de vitesse radiale a augmenté durant les années 2003 à 2006, indiquant que l'amplitude ne décroît pas continûment mais fait l'objet d'un cycle. Les auteurs mentionnent également la découverte d'un signal dans les données WIRE de l'ordre de 2-6 jours qui pourrait être dû à la granulation. \citet{Evans08} découvrent pour la première fois le compagnon de Polaris Ab grâce à des images du HST et obtiennent une séparation de 0.17'', tandis que le compagnon visuel Polaris B se trouve à 18''. La masse de l'étoile est alors établie à $4.5\pm2.2$ masses solaires, et celle du compagnon Polaris Ab à $1.26\pm0.14$ masse solaire. \citet{lee08} confirment la remontée de l'amplitude "2K" de l'étoile à partir de données du spectrographe BOES.   \cite{stothers09} étudie la variation cyclique de Polaris (mais aussi V473 Lyr) et propose une nouvelle classe de Céphéides Blazhko. \cite{evans10} découvrent que Polaris est une source de rayons X avec le satellite CHANDRA. Afin de réconcilier les modèles d'évolution avec le changement de période de l'étoile, \citet{neilson12b} soutiennent que l'étoile perd probablement de la masse à hauteur de 10$^{-6}$ Ms/an. \cite{turner13b} montrent que les paramètres fondamentaux de l'étoile sont cohérents avec une Céphéide pulsant dans le mode fondamental et par ailleurs que la parallaxe Hipparcos de l'étoile est probablement sous-estimée, ce qui est aussitôt démenti par \citet{vanleeuwen13}. Récemment, \citet{neilson14b} montra à partir de modèle d'évolution que Polaris se trouve probablement au niveau du troisième passage dans la bande d'instabilité et indique que sa distance est forcément supérieure à 118pc. Ceci fut repris et contredit par \citet{fadeyev15} qui montra à partir de modèles hydrodynamiques que Polaris doit passer la bande d'instabilité pour la première fois et qu'elle pulse probablement dans le mode fondamental.

\chapter{La modélisation des Céphéides}\label{Annexe_mod}

\begin{table*}
\caption{Les différents modèles hydrodynamiques dans le monde: (S) correspond à une modélisation de la structure interne de l'étoile basée sur un modèle d'évolution, (E) correspond à la modélisation de l'enveloppe à partir d'un modèle de pulsation, et (A) indique qu'au moins quelques couches dans l'atmosphère sont considérées. Un transfert de rayonnement pour calculer une raie spectrale nécessite un nombre suffisant de couches dans l'atmosphère. }
\label{Tab_modeles}
\small
\begin{tabular}{|l|c|c|c|c|c|c|}
\hline
   Reference           & Parties de l'étoile           & Linéaire ou & Transfert    de     &  Convection  & Nbr.    \\
              & modélisées      & Non Linéaire  & Rayonnement   &              &     Dimensions         \\
              
\hline
\cite{buchler84}       & E              &   NL                  & NON              &   OUI        &    1D          \\
\hline
\cite{fokin96}          & E+A        &   NL                  & OUI              &   NON        &  1D           \\
\hline
\cite{sasselov92}      & E+A       &   NL (Piston)         & OUI              &   OUI        &   1D          \\
\hline
\cite{dorfi91}        & E+A       &   NL                  & NON              &   OUI        &  1D          \\
\hline
\cite{bono99}         & S+E+A  &    L                  & NON              &   OUI        &   1D          \\
\hline
\cite{baraffe98b}       & S+E+A  &    L                  & NON              &   OUI        &   1D          \\
\hline
\cite{mundprecht13} & S+E+A  &   NL   & NON &  OUI & 2D  \\  
\hline
\cite{geroux15}    & S + E + A & NL & NON & OUI & 3D \\ 
\hline
\cite{vasilyev17}    & S + E + A & NL & OUI & OUI & 2D \\ 
\hline
\end{tabular}
\normalsize
\end{table*}

\chapter{La relation brillance de surface - couleur}\label{Annexe_SB}

\begin{table*}
\setlength{\doublerulesep}{\arrayrulewidth}
\caption{\label{Tab.SB}  Liste des relations brillance de surface - couleur. A ceci il faut rajouter l'idée de \cite{gould14} d'utiliser les micro-lentilles gravitationnelles pour déterminer le diamètre angulaire d'une étoile, ainsi que l'approche des pseudo-magnitudes développée par \cite{chelli16}.}
\begin{center}
\tiny
\begin{tabular}{lccccr}
\hline
\hline
Nb*      &  Notes & Type Sp. &   Couleur & Précision (mag) &  Ref   \\
    \hline  
  18 &  1 &B-M &  B-V & 0.12  & \cite{wesselink69}\\
  27 & 2 & G-M, S, C &  V-R, R-I  & 0.04, 0.08  & \cite{barnes76}\\
  25 & 3& O-G &  B-V, V-R, R-I & 0.018, 0.025, 0.033  & \cite{barnes76b,barnes78}\\
  11 & 4& K-M &  V-R & 0.002 & \cite{dibenedetto87}\\
   44 & 5& F-M, C &  V-K & 0.03-0.09 & \cite{dibenedetto93}\\
    11 & 6& Ceps &  V-K & 0.04 & \cite{welch94}\\ 
   27 &  7& F-M &  V-R, V-K, J-K & 0.004-0.034 & \cite{fouque97}\\
   22 & 8& F, G, K &  V-K & 0.03 & \cite{dibenedetto98}\\
   239 &9&  B-M &  V-K, B-K & 0.04-0.05 & \cite{vanbelle99}\\
 116  &10&  Géantes et Ceps &  V-R, V-K, J-K & 0.01-0.02 & \cite{nordgren02}\\  
   221  &11&  A-M &  V-K, V-R, J-K & 0.009-0.05 & \cite{gro04b}\\   
   9 &12& Ceps & B-V, V-K, B-H  & 0.0006 & \cite{kervella04c}\\
  45 &13& A, G, K, M & B-UVRIJHKL & 0.02& \cite{kervella04}\\
  44 & 14 & A, F, G, K & V-K & 0.03& \cite{dibenedetto05}\\
> 100 &15& B, A, F, G, K, M & V-K & 0.26& \cite{bonneau06}\\
42 & 16&A-M & BVRI & 0.09 - 0.17& \cite{kervella08}\\
124 & 17& A-M &  BUVRIJHKL... & 0.08 - 0.15& \cite{boyajian14}\\
132 & 18& A-M & V-K & 0.04-0.16 & \cite{challouf14}\\
40 & 19& EBs &B-K, V-K  & 0.05-0.11 & \cite{graczyk17}\\
\hline
\end{tabular}
\end{center}
\end{table*}

%barnes 78 ?
% first calibration based on interfeormetry Welch (1994) ? 

Les notes ci-dessous se rapportent à la Table~\ref{Tab.SB}:

\begin{enumerate}

\item \cite{wesselink69}. 18 mesures de diamètres d'étoiles: 14 avec le ‘‘Narrabri Intensity Interferometer'' (NII)  \citep{brown67, brown74b}, à cela s'ajoutent le Soleil, 1 binaire à éclipses dont on connaît la distance, ainsi que  2 autres étoiles observées avec ‘‘The Michelson Interferometer at Mount Wilson Observatory'' \citep{pease21b,michelson21,pease21b}. 

\item   \cite{barnes76}. 27 mesures de diamètres d'étoiles: 16 par occultations lunaires et 11 par interférométrie (voir la Table 1 et les références associées).

\item   \cite{barnes76b,barnes78}. 25 mesures de diamètres d'étoiles: il s'agit de la liste des 32 diamètres mesurés par le NII \citep{brown67, brown74b} à laquelle on a enlevé 7 binaires.  

\item   \cite{dibenedetto87}. 11 mesures de diamètres effectuées avec l'I2T sur le Plateau de Calern \citep{blazit77,faucherre83}. La relation est donnée sur deux intervalle (coupure vers $(V-R)_\mathrm{0} \simeq 1.2$ séparant les types K et M) avec sur les deux portions de droite une très faible dispersion (rms) de $0.002$ mag. 

\item   \cite{dibenedetto93}. 44 mesures de diamètres d'étoiles: 11 mesures effectuées avec l'I2T sur le Plateau de Calern \citep{dibenedetto87} + 10 autres obtenues quelques années plus tard \citep{dibenedetto90} + 23 par différentes techniques (voir table 3 et 4 de  \cite{dibenedetto93}). Des relations (sans prise en compte de l'extinction) sont données pour les géantes: $F_\mathrm{V}=3.927 - 0.122 (V-K)$ (1.4 < V-K < 3.7) (rms = 0.003) et $F_\mathrm{V}=3.833 - 0.101 (V-K)$ (rms=0.01) ( V-K > 3.7). Pour les supergéantes, on a $F_\mathrm{V}=3.954 - 0.133 (V-K)$ (rms = 0.007) ou encore $F_\mathrm{Vo}=3.958 - 0.139 (V-K)_\mathrm{0}$ (si l'extinction est prise en compte). Enfin, des relations différentes de la forme ($0.5 \log \theta = (4.2207-a)+b*(V-K)_\mathrm{0}-0.1V_\mathrm{0}$) sont données pour différents types spectraux (voir la table 5) avec une dispersion s'échelonnant entre 0.03 mag (G-K de classe II à V) et 0.08 mag (MIV-V, Mira et C). Pour les supergéantes (généralement plus éloignées et sensibles au rougissement), l'extinction considérée est basée sur les formules classiques $A_\mathrm{V}=3.6 E(B-V)$ et $E(V-K)=3.17 E(B-V)$ \citep{lee70}. 

\item   \cite{welch94}. Premier étalonnage de la relation brillance de surface - couleur sur la base des Céphéides. Une relation est donnée entre $F_\mathrm{K}$ et $(V-K)_\mathrm{0}$.

\item   \cite{fouque97}. 27 mesures de diamètres d'étoiles par interférométrie Michelson, toutes géantes ou supergéantes (voir la table 1 de \cite{fouque97}). La sélection des étoiles est faite sur des critères de non-variabilité, d'une bonne précision sur le diamètre (<6\%) et une absorption interstellaire inférieure à 0.2 mag pour les géantes (voir la partie 2.1 dans le papier). La plus faible dispersion (rms = 0.004 mag) est obtenue pour les géantes avec l'indice $(V-K)_\mathrm{0}$: $F_\mathrm{V}=3.930_\mathrm{\pm0.012} - 0.124_\mathrm{\pm0.004}  (V-K)_\mathrm{0}$ (2.22 < V-K < 4.11).  Les extinctions utilisées $A_\mathrm{V}$ sont déduites de la littérature et discutées dans la table 1, tandis que la loi de rougissement utilisée (voir la discussion détaillée dans le papier)  est de: $E(V-K)=0.88A_\mathrm{V}$. La relation obtenue pour les supergéantes est la suivante: $F_\mathrm{V}=3.914_\mathrm{\pm0.023} - 0.119_\mathrm{\pm0.007}  (V-K)_\mathrm{0}$ (rms = 0.032 et 0.52 < V-K < 5.53). Les auteurs déduisent également une relation spécifique pour les Céphéides: $F_\mathrm{V}=3.947 - 0.131 (V-K)_\mathrm{0}$

\item   \cite{dibenedetto98}. Amélioration de l'échantillon de  \cite{dibenedetto93} avec l'ajustement d'une relation non-linéaire: $S_\mathrm{V}=2.563  + 1.493  (V-K)_\mathrm{0} - 0.046 (V-K)_\mathrm{0}^2$ pour (rms = 0.030) pour $-0.1 < (V-K)_\mathrm{0} < 3.7$. 

\item  \cite{vanbelle99}. 239 étoiles issues de l'interférométrie essentiellement. \cite{vanbelle99} introduit le concept de diamètre angulaire à la magnitude zéro: $\theta_\mathrm{V=0} = \theta*10^{\frac{V}{5}}$. L'idée est que pour comparer le diamètre angulaire d'étoiles à des distances différentes, on peut utiliser  le diamètre angulaire des étoiles à la magnitude $V=0$. Ainsi, il obtient: $\theta_\mathrm{V=0} = 10^{0.789\pm0.119 + 0.218\pm0.014 (V-K)}$ pour les variables, $\theta_\mathrm{V=0} = 10^{0.50\pm0.023 + 0.264\pm0.012 (V-K)}$ pour les étoiles de la séquence principale et enfin $\theta_\mathrm{V=0} = 10^{0.669\pm0.052 + 0.223\pm0.010 (V-K)}$ pour les classes III, II et I. 

\item  \cite{nordgren02}. 57 géantes observées avec NPOI et 59 Céphéides. Les relations obtenues sont  $F_\mathrm{Vo}=3.934_\mathrm{\pm0.005} - 0.123_\mathrm{\pm0.002}  (V-K)_\mathrm{0}$ (N=57, rms=0.011) pour les géantes et $F_\mathrm{Vo}=3.956_\mathrm{\pm0.011} - 0.134_\mathrm{\pm0.005}  (V-K)_\mathrm{0}$ (N=59, rms=0.026). 

\item   \cite{gro04b}. Liste de 221 étoiles issues de l'interférométrie avec une précision sur le diamètre assombri meilleure que 3\%. L'extinction considérée est basée sur la distance des objets ainsi que leurs coordonnées galactiques (référence citée dans le papier mais non référencée Grenon \& Gomez (1992)). La formule obtenue est (selon le concept de \cite{vanbelle99}): $\log \theta_\mathrm{V=0} =  0.584 \pm 0.014 + 0.245 \pm 0.005 (V-K)_\mathrm{0}$ (N=74, rms=0.024) pour les géantes (III) et $\log \theta_\mathrm{V=0} =  0.519 \pm 0.012 + 0.274 \pm 0.004 (V-K)_\mathrm{0}$ (N=20, rms=0.022)  pour les étoiles de la séquence principale et $\log \theta_\mathrm{V=0} =  0.607 \pm 0.019 + 0.243 \pm 0.008 (V-K)_\mathrm{0}$ (N=21, rms=0.046) pour les supergéantes. 

\item  \cite{kervella04c}. 9 Céphéides observées avec VINCI/VLTI \citep{kervella04a} qui amènent à la relation suivante: $F_\mathrm{V}=-0.1336\pm0.0008 (V-K) +  3.9530\pm0.0006$. L'extinction considérée est $A_\lambda=R_\lambda E(B-V)$ tandis que le rougissement s'écrit $R_\mathrm{K}=\frac{R_\mathrm{V}}{11}$ avec $R_\mathrm{V}=3.07 + 0.28 (B-V) + 0.04 E(B-V)$. 

\item  \cite{kervella04}. 29 étoiles différentes mais 45 mesures de diamètres angulaires obtenues avec les instruments suivants: NII \citep{brown67, brown74b}, Mk III \citep{shao88}, PTI \citep{colavita99} and NPOI \citep{armstrong98}. Plusieurs étoiles ont été exclues de l'échantillon de base: les systèmes doubles (dont des binaires à éclipses), $\beta$~Pic qui présente un environnement, $\alpha$~Eri un rotateur rapide et deux étoiles de très faible masse (présence de molécules, variabilité, activité chromosphérique). Le passage d'un diamètre uniforme $\theta_\mathrm{UD}$ à  $\theta_\mathrm{LD}$ est traité de manière homogène en utilisant les tables de \citep{claret00}. Les étoiles de l'échantillon sont à moins de 15 pc et l'extinction est négligée. La relation ainsi obtenue est de la forme: $ \log \theta_\mathrm{LD} = 0.0755 \pm  0.0008 (V-K) + 0.5170 \pm 0.0017- 0.2 K$ avec une dispersion inférieure à 1\% sur le diamètre (rms = 0.022). 

\item \cite{dibenedetto05}. Amélioration de l'échantillon de  \cite{dibenedetto98} avec l'ajustement d'une relation non-linéaire: $S_\mathrm{V}=2.565\pm0.016 + 1.483\pm0.015 (V-K)_\mathrm{0} - 0.044 \pm 0.005 (V-K)_\mathrm{0}^2$ pour (rms = 0.040) pour $-0.1 < (V-K)_\mathrm{0} < 3.7$. 

\item  \cite{bonneau06}. Le nombre d'étoiles  n'est pas donné mais les diamètres sont issus de mesures interférométriques, d'occultation lunaires et de binaires à éclipses \citep{barnes78, andersen91, segransan03, mozurkewich03}. L'extinction est calculée avec les équations suivantes: mag[$\lambda_\mathrm{0}]=$mag[$\lambda]$ - $A_\mathrm{\lambda}$,  $R_\mathrm{\lambda} = \frac{A_\lambda}{E(B-V)}$ et $A_\lambda = \frac{A_\mathrm{V} R_\lambda}{R_\mathrm{V}}$ avec $R_\mathrm{V}=3.10$ \citep{fitzpatrick99}. Les résultats sont donnés sous la forme de l'équation suivante: $\phi_\mathrm{V}=\frac{\theta}{9.306.10^{\frac{-m_V}{5}}}=\sum_k a_k CI^k$ avec avec $C_{0}=0.32561925$,    $C_{1}=0.31467316$,    $C_{2}= 0.09401181$,   $C_{3}=-0.0187446$,   $C_{4}=0.00818989$ (rms = 0.15). 

\item  \cite{kervella08}. 42 étoiles naines ou sous-géantes. Ce papier fait suite à \citet{kervella04c} et se limite au domaine visible. Des relations non-linéaires sont ajustées avec une dispersion s'échelonnant entre 0.09 et 0.17 magnitude. 

\item   \cite{boyajian14}. 124 étoiles de la séquence principale avec des mesures de diamètres meilleures que 5\%. L'extinction n'est pas prise en compte. Des relations sont données dans de nombreuses bandes photométriques et en V, (V-K), les auteurs obtiennent: $\log \theta_\mathrm{V=0} =  0.53246 \pm 0.00057 + 0.26382 \pm 0.00028 (V-K)$ (N=97 ; rms = 0.1). 
 
\item \cite{challouf14}. 132 étoiles dont 8 étoiles de type précoce observées par VEGA/CHARA. L'extinction est traitée de 7 manières différentes et une relation valable sur un intervalle important $-0.931 \leq V-K \leq 3.69$ est donnée sous la forme: $S_{v}=\sum_{n=0}^{n=5}C_{n}(V-K)^{n}_{0}$ avec $C_{0}=2.568\pm0.005$,    $C_{1}=1.690\pm0.011$,    $C_{2}= -0.524\pm0.016$,   $C_{3}=0.351\pm0.017$,   $C_{4}=-0.104\pm0.008$,  $C_{5}=0.011\pm0.001$. 

\item   \cite{graczyk17}. 40 binaires à éclipses galactiques dont on a la distance ont permis de déduire une relation brillance de surface - couleur du type: $S_\mathrm{V}=2.625\pm 0.015 + 0.959 \pm 0.009 (V-K)$ (N=28, rms=0.047). Concernant l'extinction, la carte de \cite{schlegel98} est utilisée. Concernant le rougissement, les auteurs considèrent: $R_\mathrm{V}=3.1$ d'après \citep{fitzpatrick07}.  
\end{enumerate}

%%%%%%

\chapter{{\it Calibrating the Cepheid period-luminosity relation from the infrared surface brightness technique. I. The p-factor, the Milky Way relations, and a universal K-band relation}}\label{storm11a}
\includepdf[pages=-,pagecommand={\thispagestyle{fancy}},offset=0 0]{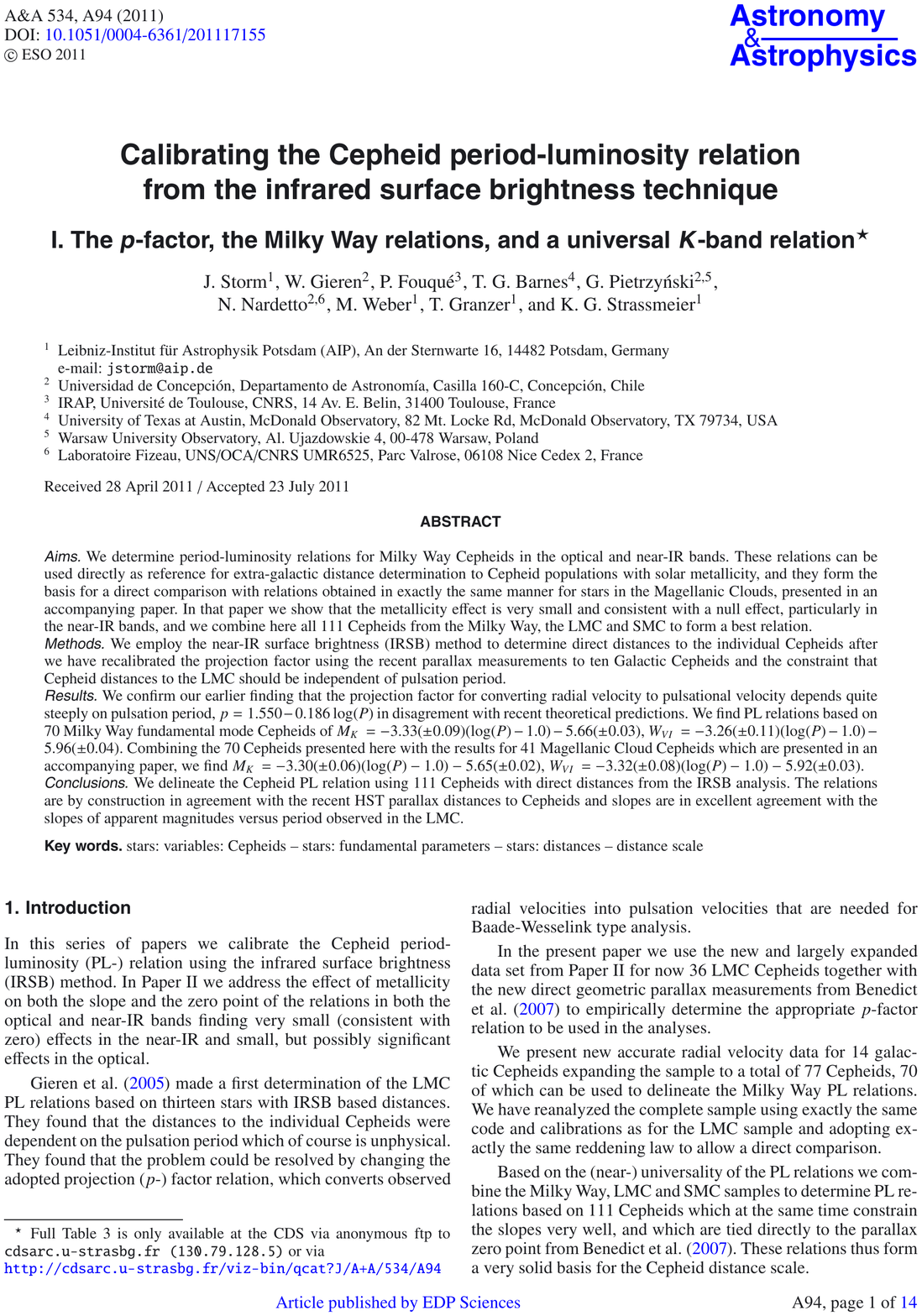}

\chapter{{\it Calibrating the Cepheid period-luminosity relation from the infrared surface brightness technique. II. The effect of metallicity and the distance to the LMC}}\label{storm11b}
\includepdf[pages=-,pagecommand={\thispagestyle{fancy}},offset=0 0]{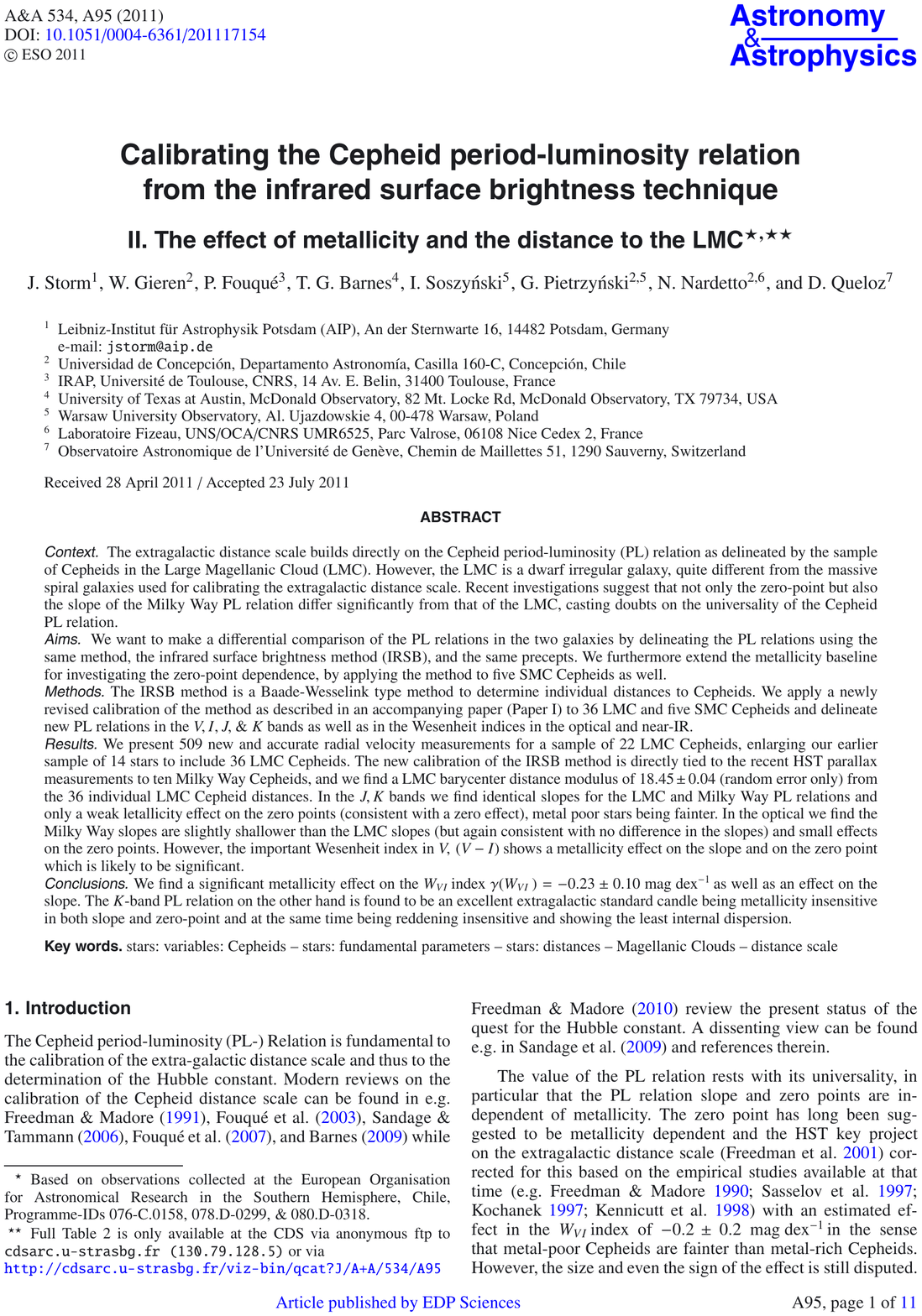}

\chapter{{\it An eclipsing-binary distance to the Large Magellanic Cloud accurate to two per cent }}\label{nature}
\includepdf[pages=-,pagecommand={\thispagestyle{fancy}},offset=0 0]{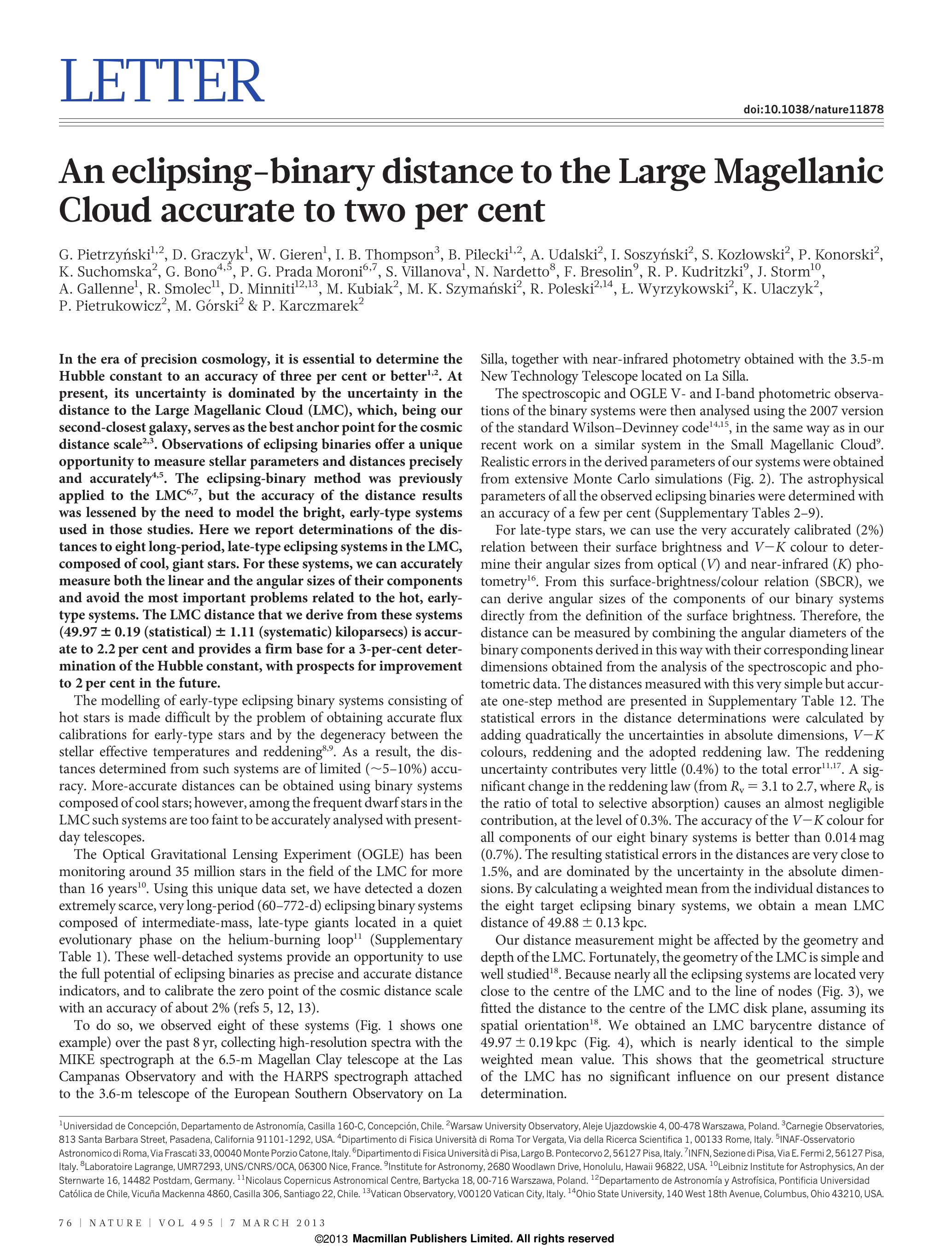}
\includepdf[pages=-,pagecommand={\thispagestyle{fancy}},offset=0 0]{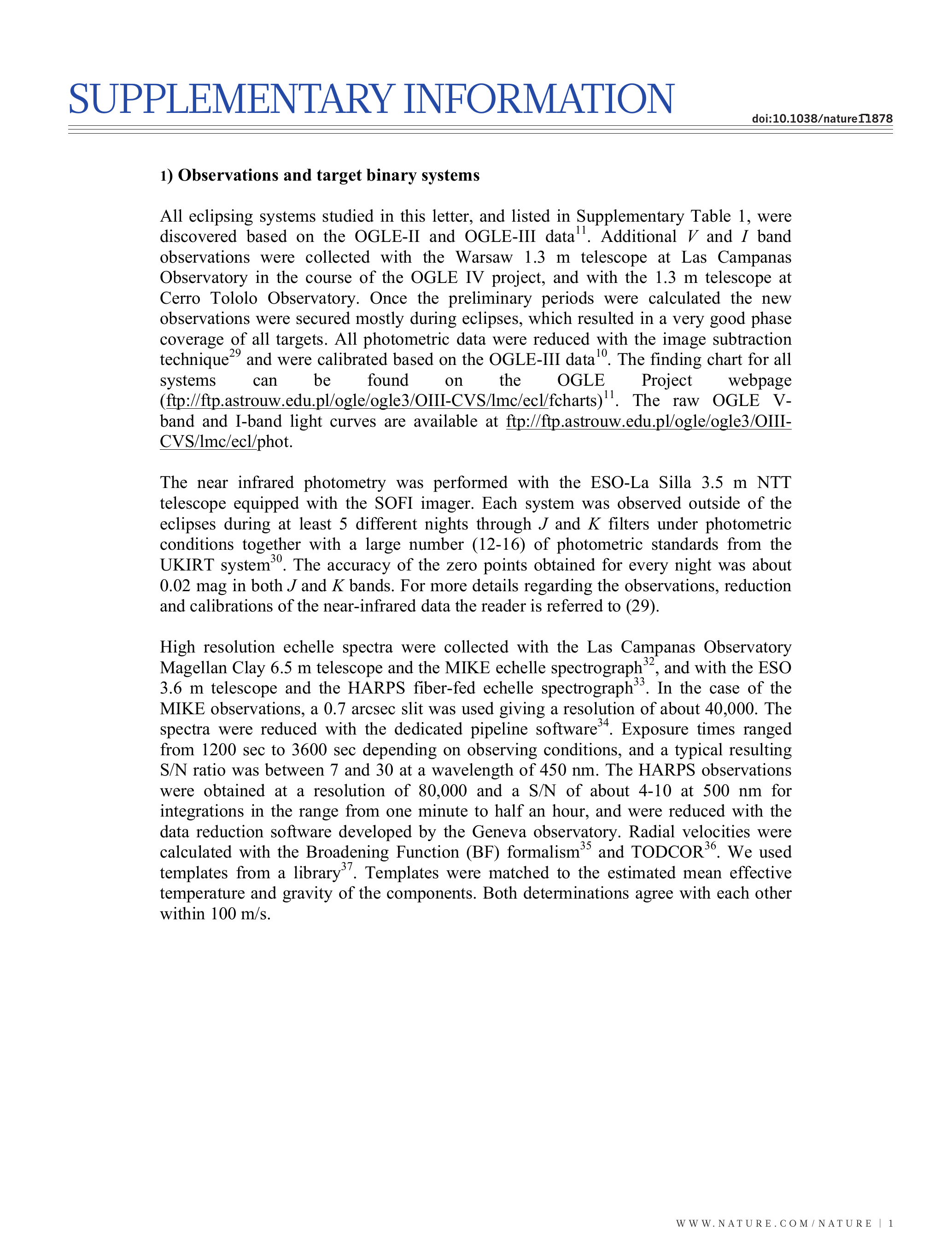}

%%%%%%

\chapter{{\it High-resolution spectroscopy for Cepheids distance determination. II. A period-projection factor relation}}\label{nardetto07}
\includepdf[pages=-,pagecommand={\thispagestyle{fancy}},offset=0 0]{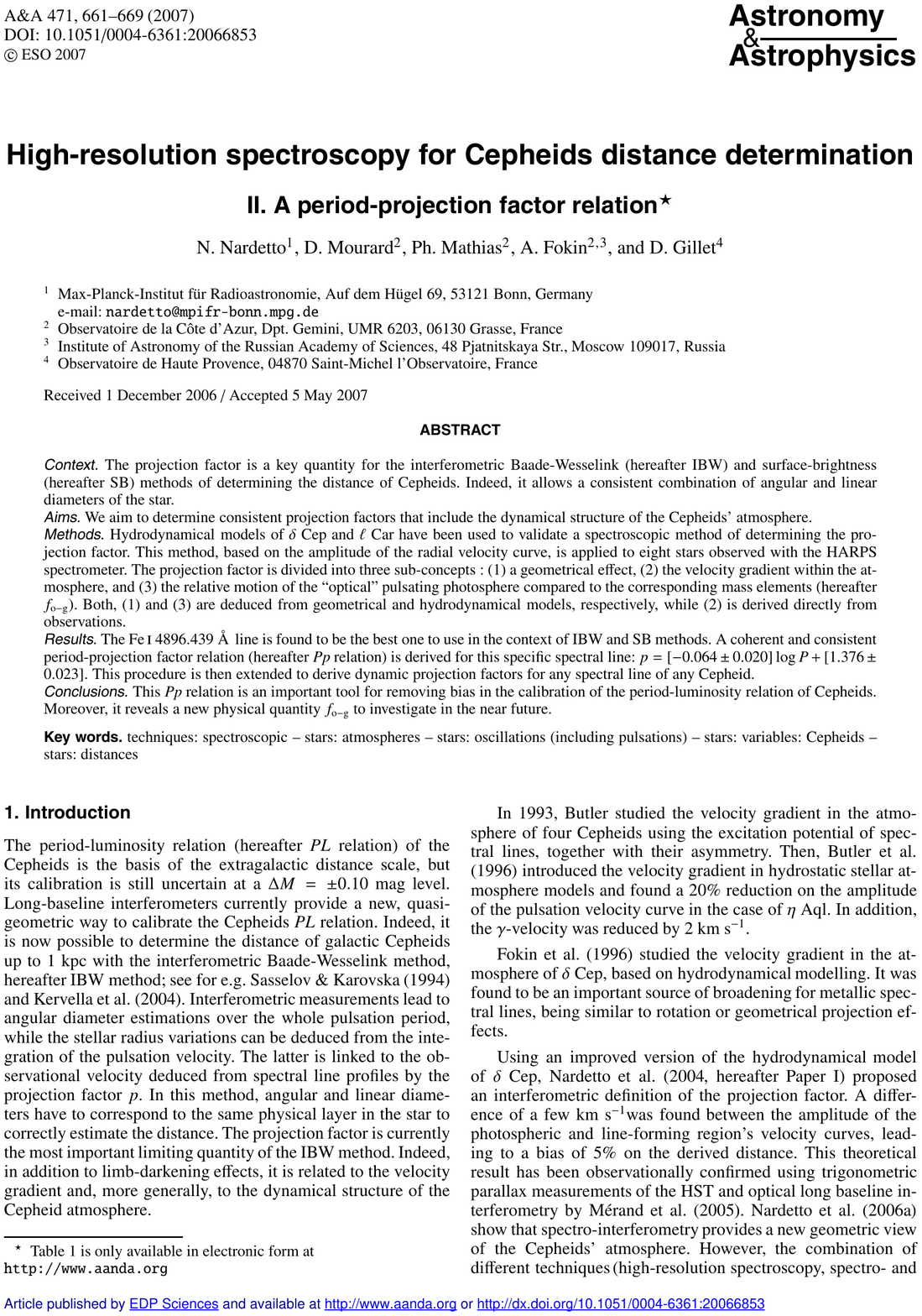}

\chapter{{\it High-resolution spectroscopy for Cepheids distance determination. V. Impact of the cross-correlation method on the p-factor and the {$\gamma$}-velocities}}\label{paperV}
\includepdf[pages=-,pagecommand={\thispagestyle{fancy}},offset=0 0]{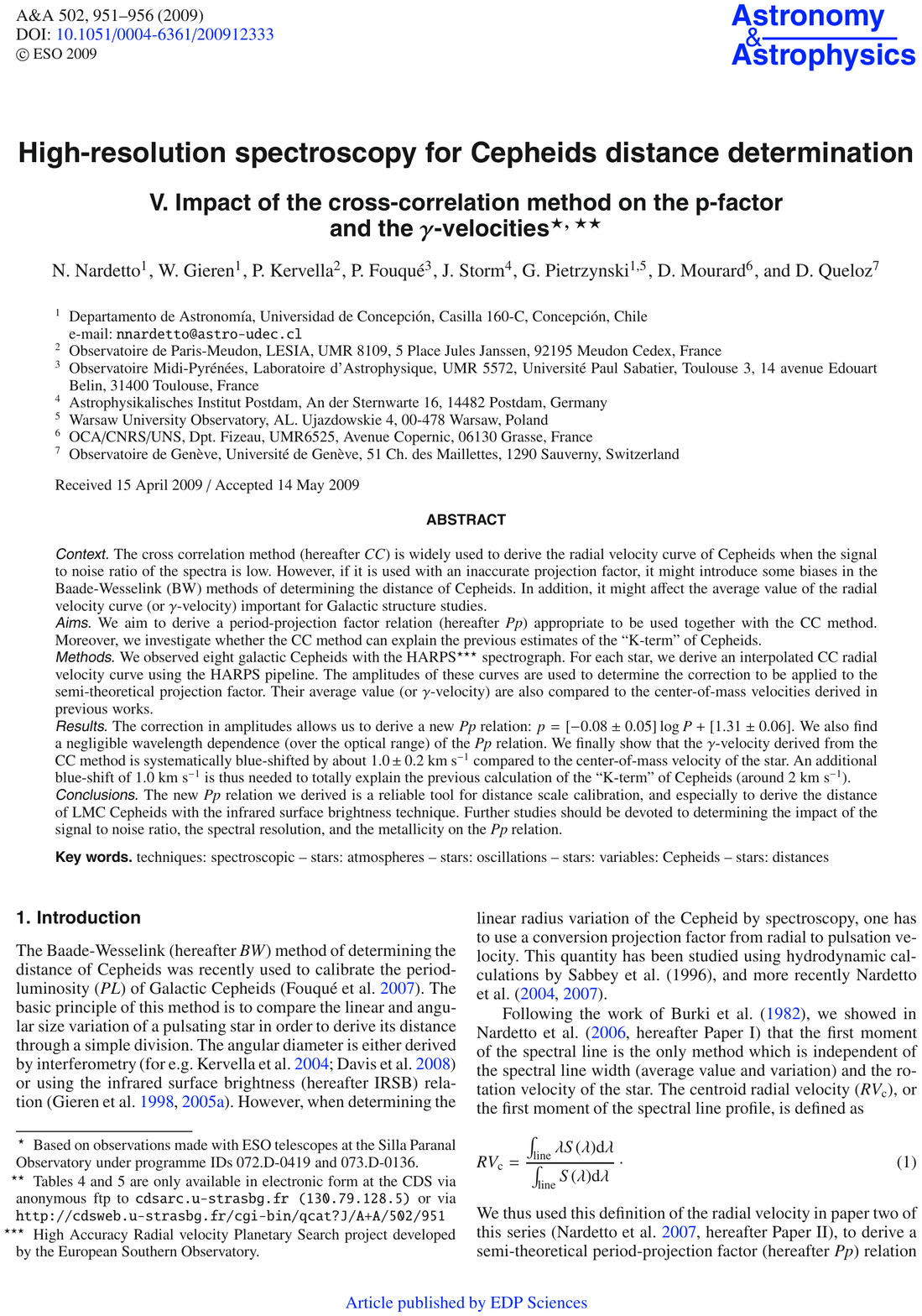}

\chapter{{\it The Baade-Wesselink p-factor applicable to LMC Cepheids}}\label{nardetto11b}
\includepdf[pages=-,pagecommand={\thispagestyle{fancy}},offset=0 0]{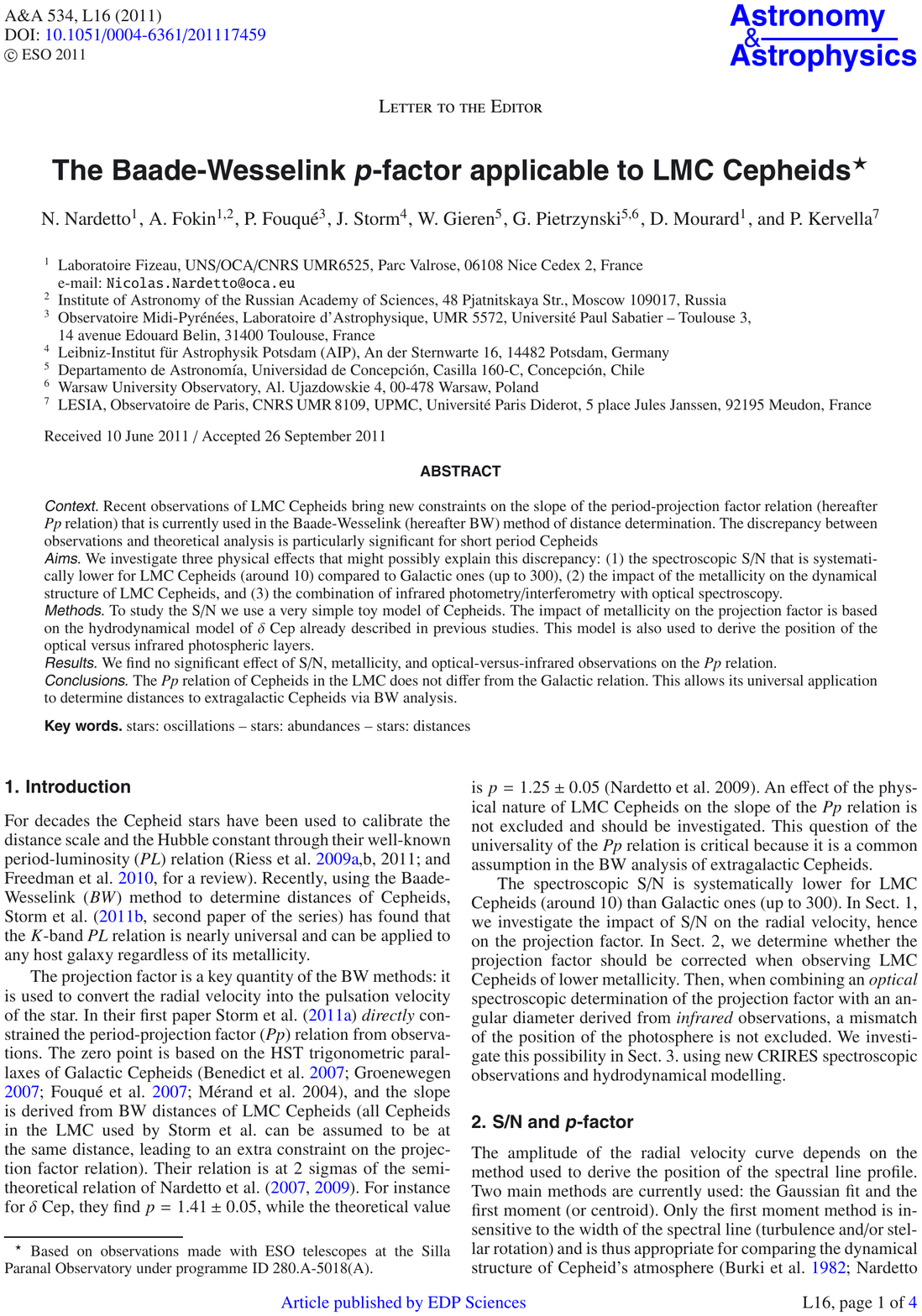}

\chapter{{\it HARPS-N high spectral resolution observations of Cepheids I. The Baade-Wesselink projection factor of {$\delta$} Cep revisited}}\label{nardetto17}
\includepdf[pages=-,pagecommand={\thispagestyle{fancy}},offset=0 0]{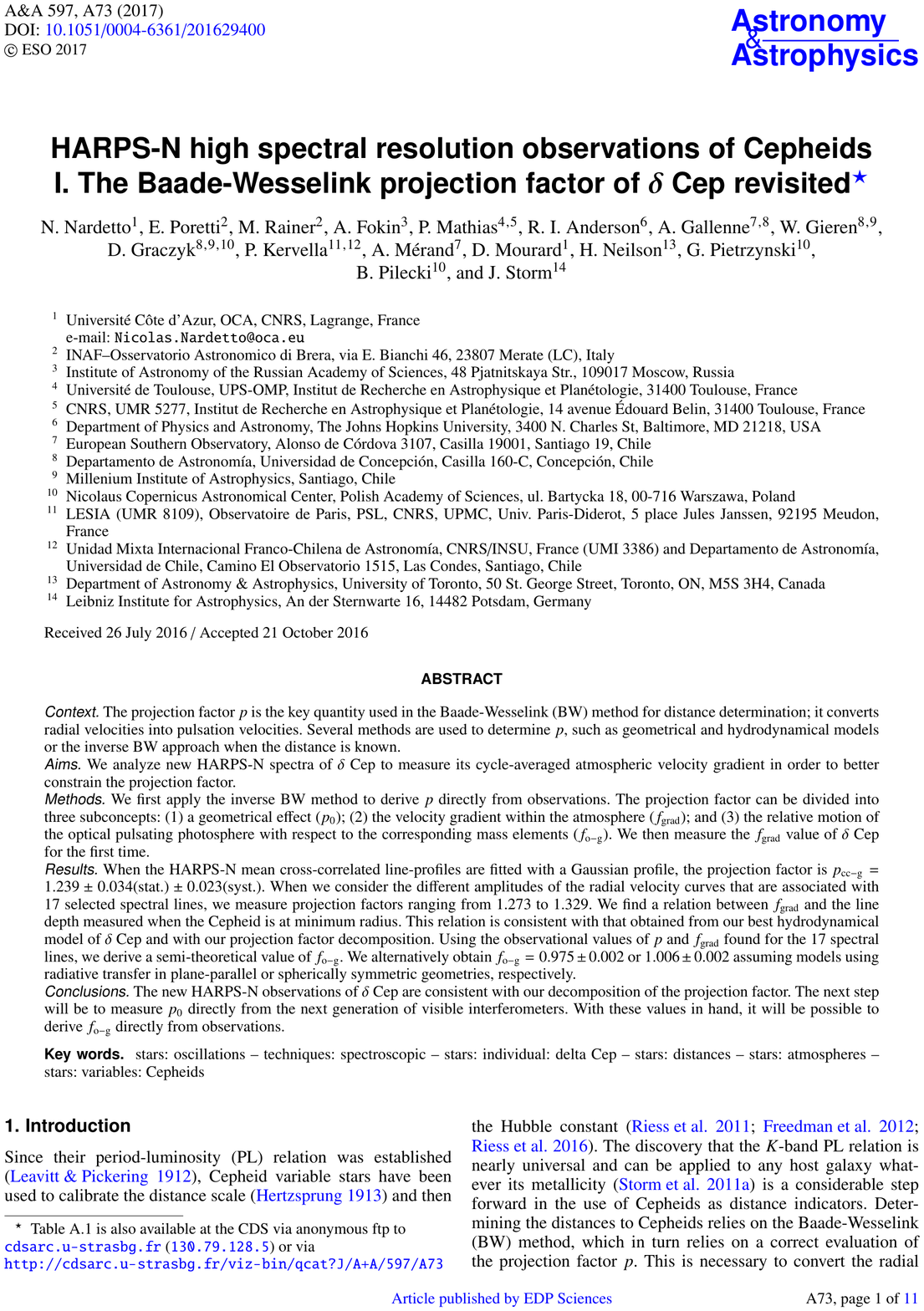} 

%%%%%%

\chapter{{\it High-resolution spectroscopy for Cepheids distance determination. III. A relation between {$\gamma$}-velocities and {$\gamma$}-asymmetries}}\label{nardetto08a}
\includepdf[pages=-,pagecommand={\thispagestyle{fancy}},offset=0 0]{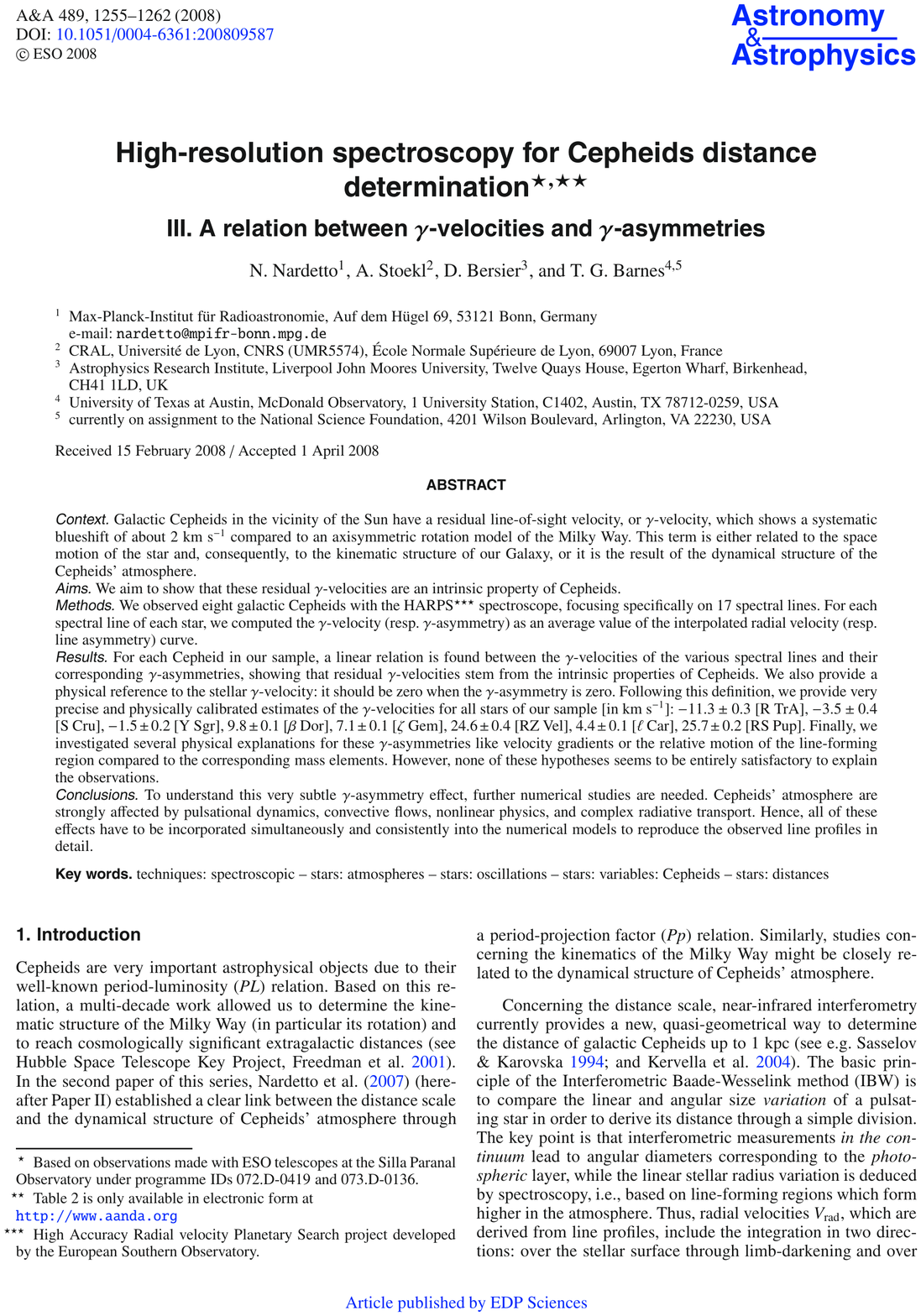}

%%%%%%

\chapter{{\it High-resolution spectroscopy for Cepheids distance determination. IV. Time series of H{$\alpha$} line profiles}}\label{nardetto08b}
\includepdf[pages=-,pagecommand={\thispagestyle{fancy}},offset=0 0]{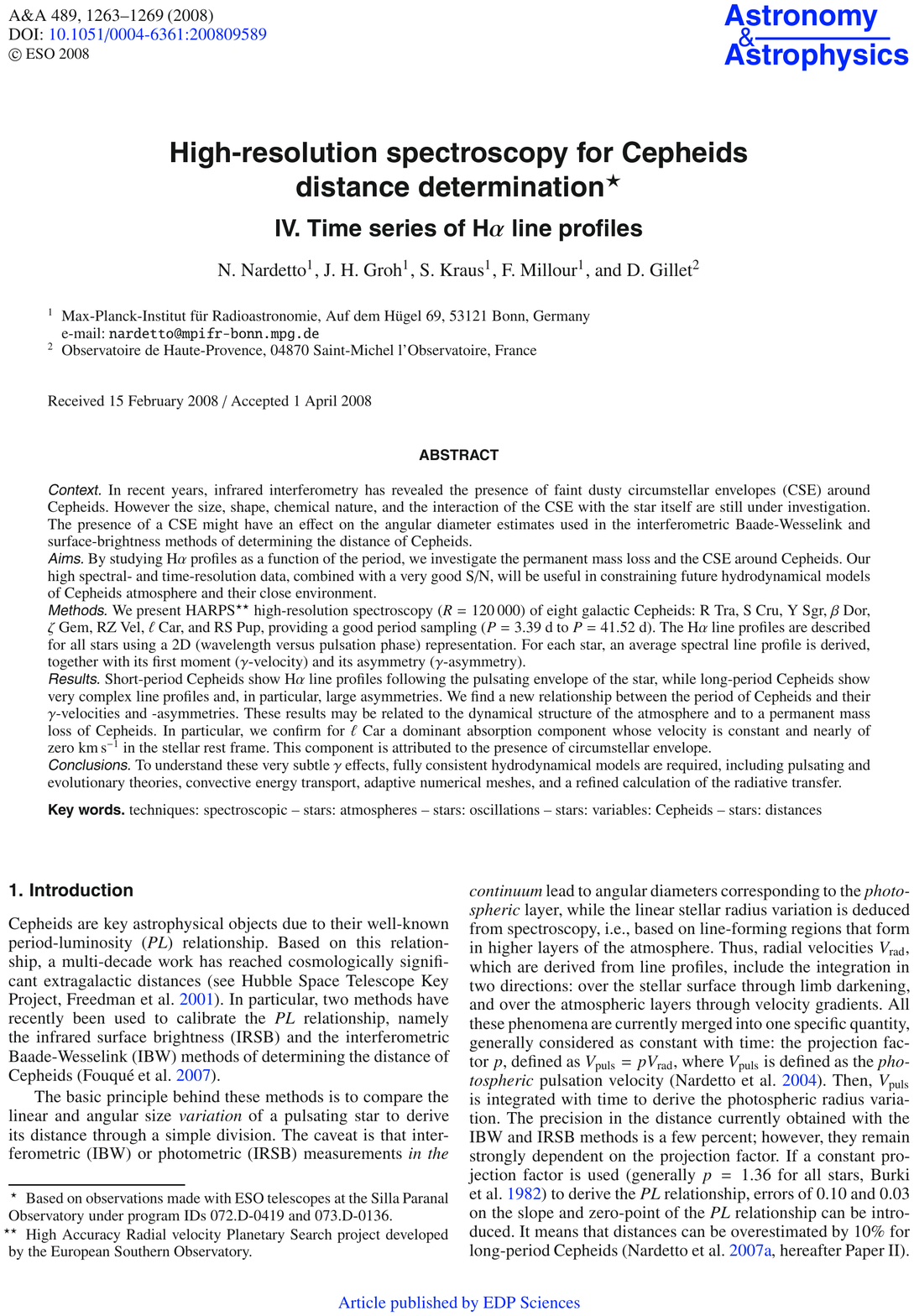}

\chapter{{\it VEGA/CHARA interferometric observations of Cepheids. I. A resolved structure around the prototype classical Cepheid {$\delta$} Cep in the visible spectral range}}\label{nardetto16a}
\includepdf[pages=-,pagecommand={\thispagestyle{fancy}},offset=0 0]{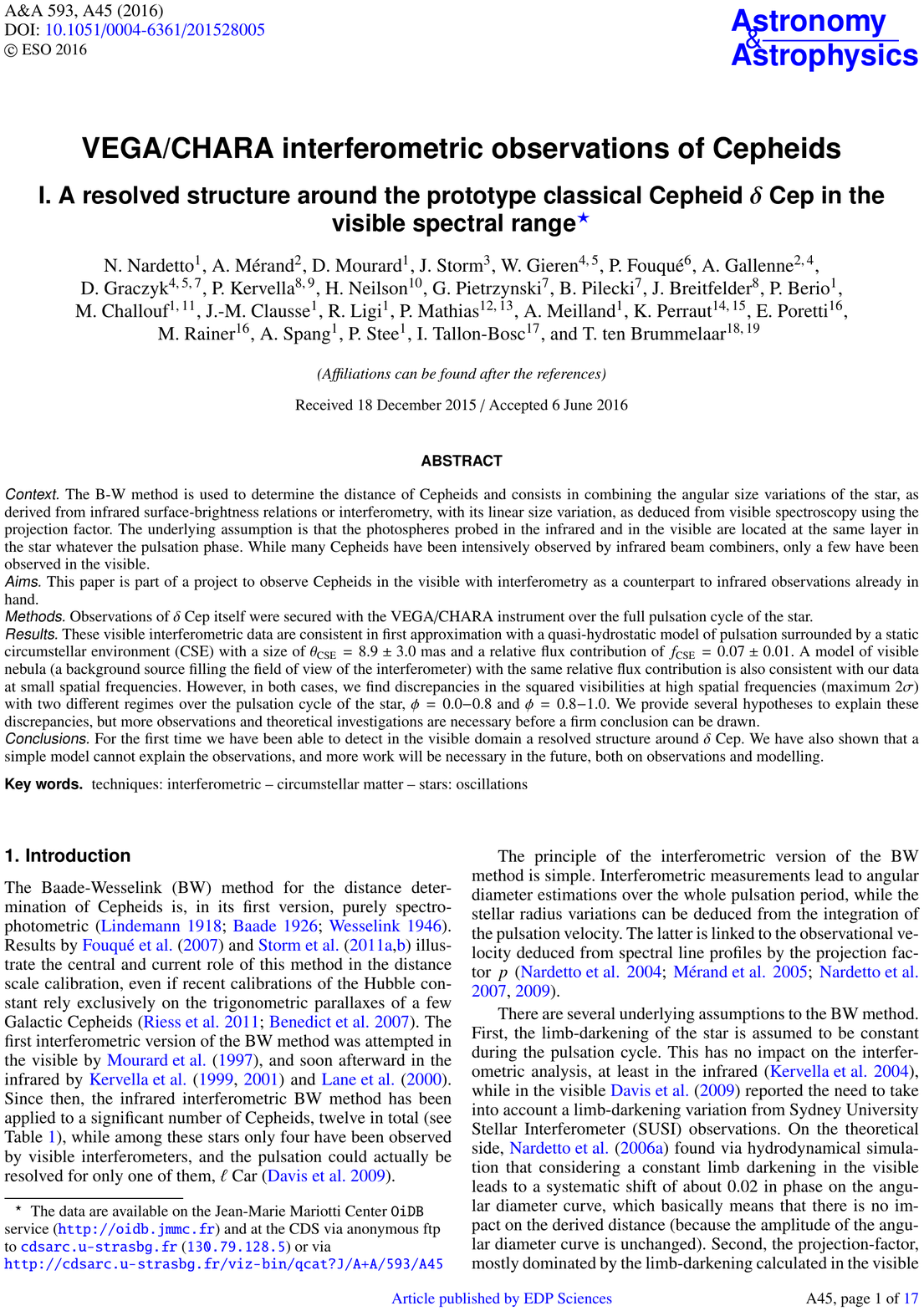} 

%%%%%%

\chapter{{\it Understanding the dynamical structure of pulsating stars: The center-of-mass velocity and the Baade-Wesselink projection factor of the {$\beta$} Cephei star {$\alpha$} Lupi}}\label{nardetto13}
\includepdf[pages=-,pagecommand={\thispagestyle{fancy}},offset=0 0]{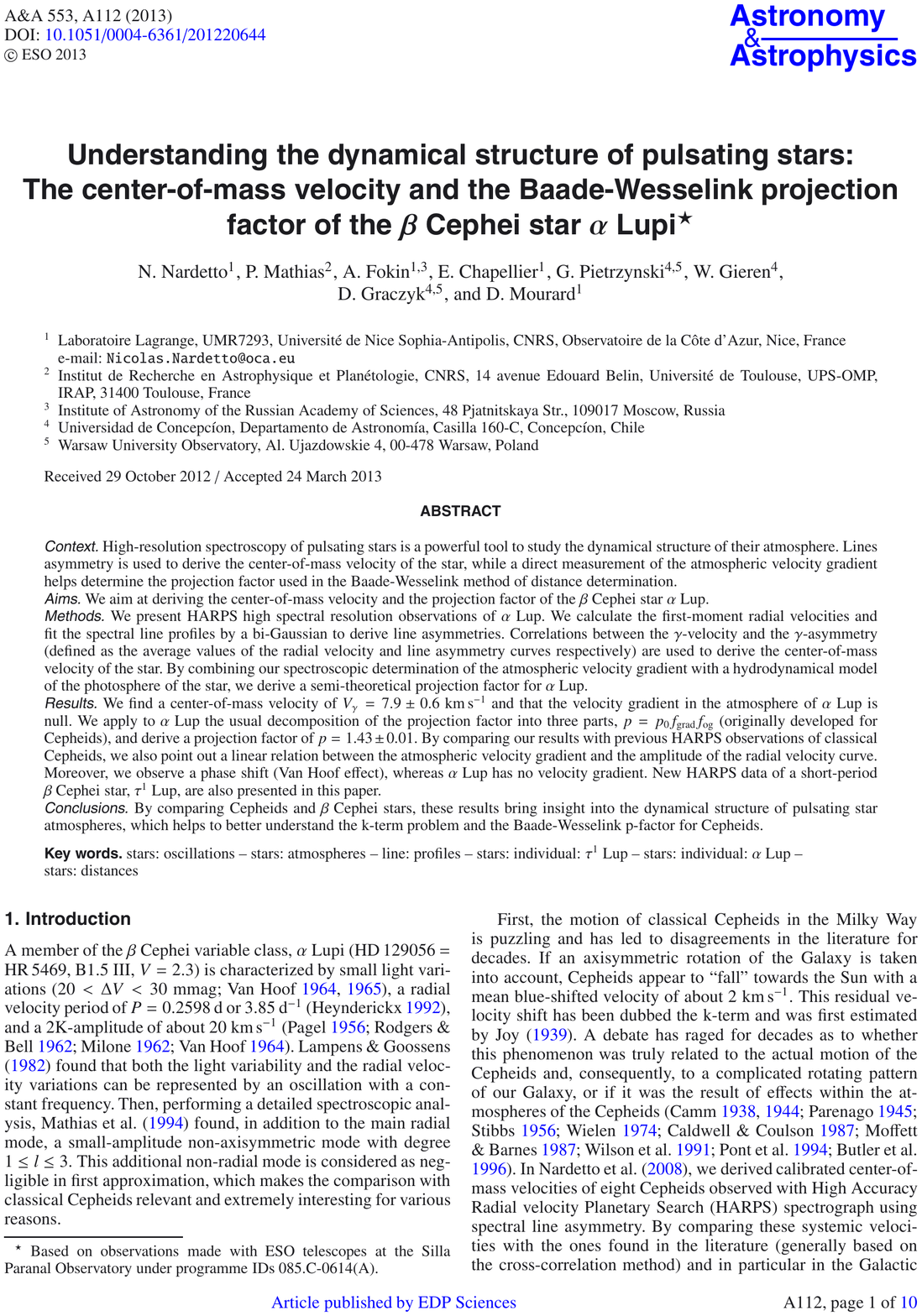}

\chapter{{\it Understanding the dynamical structure of pulsating stars: The Baade-Wesselink projection factor of the {$\delta$} Scuti stars AI Velorum and {$\beta$} Cassiopeiae}}\label{guiglion13}
\includepdf[pages=-,pagecommand={\thispagestyle{fancy}},offset=0 0]{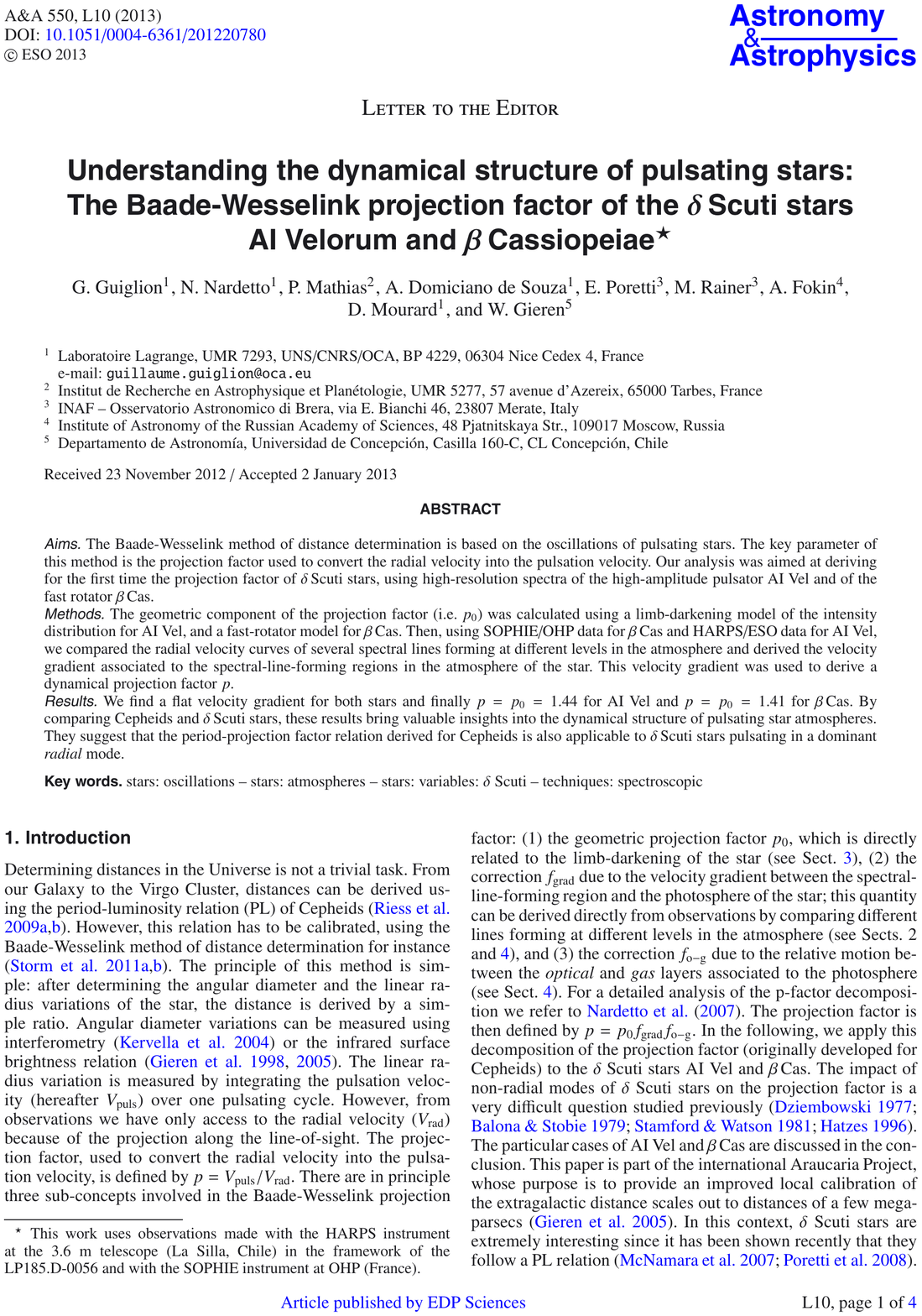}

\chapter{{\it Understanding the dynamical structure of pulsating stars. HARPS spectroscopy of the {$\delta$} Scuti stars {$\rho$} Puppis and DX Ceti}}\label{nardetto14}
\includepdf[pages=-,pagecommand={\thispagestyle{fancy}},offset=0 0]{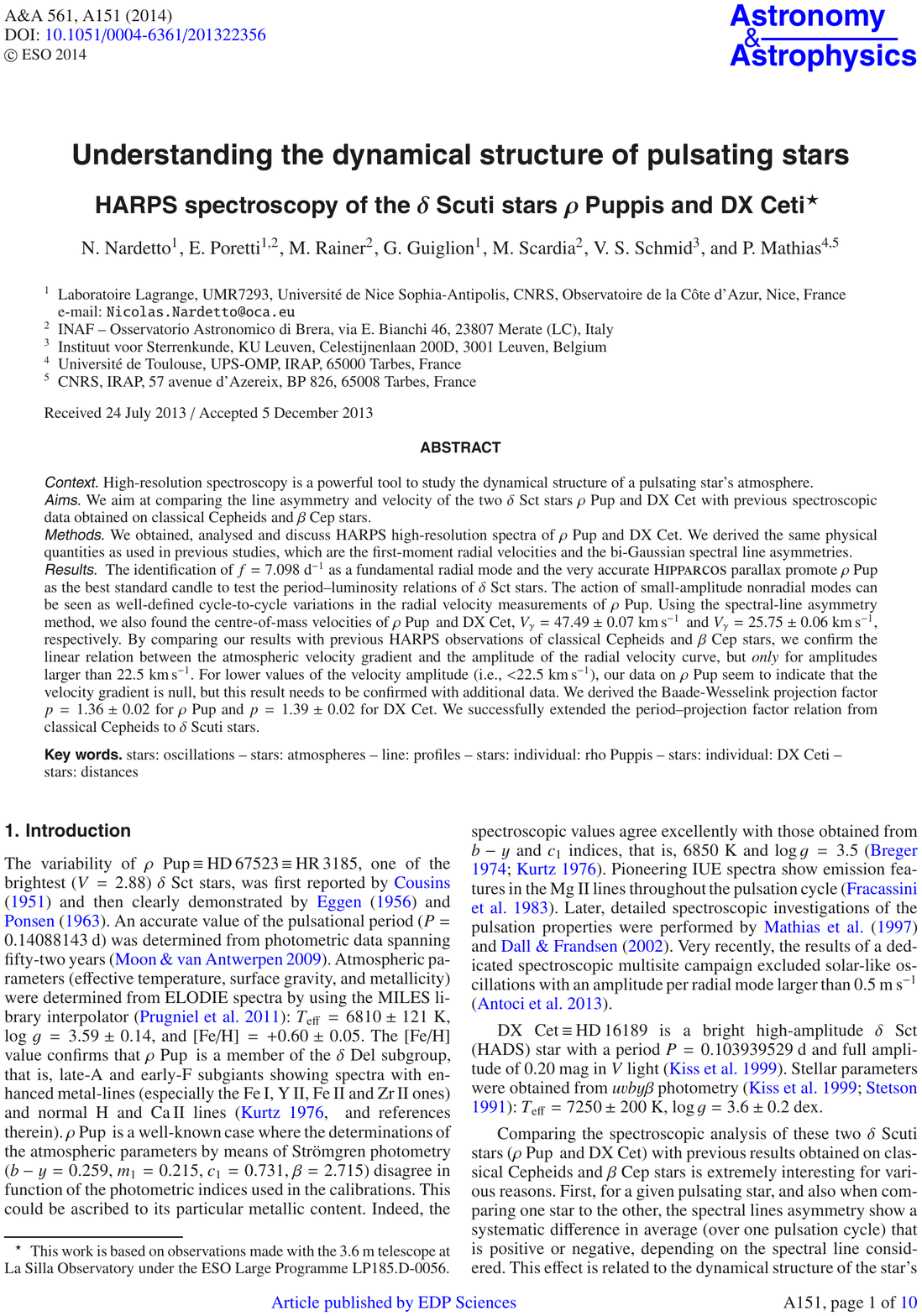}

\chapter{{\it An investigation of the close environment of {$\beta$} Cephei with the VEGA/CHARA interferometer}}\label{nardetto11a}
\includepdf[pages=-,pagecommand={\thispagestyle{fancy}},offset=0 0]{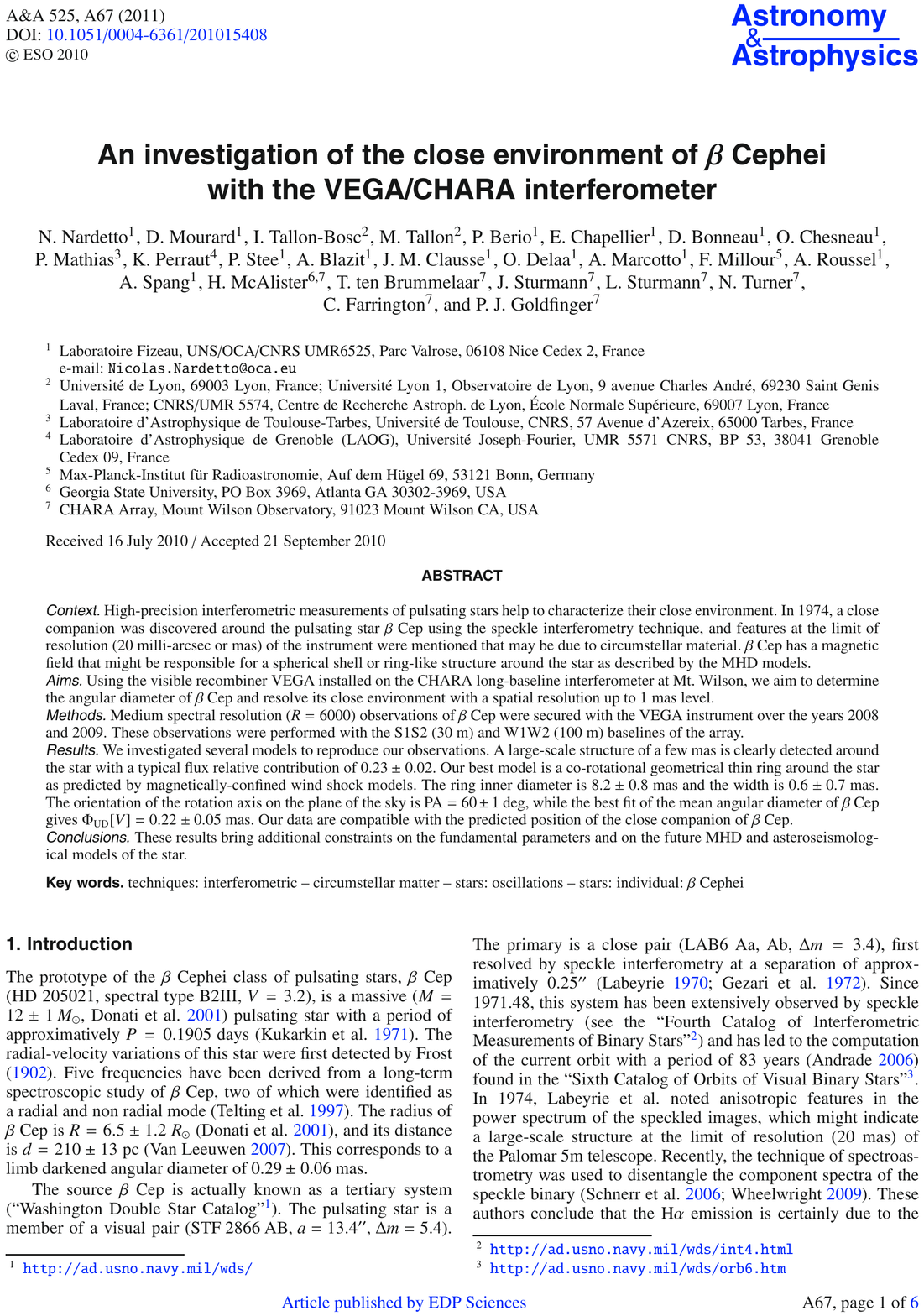} 

%%%%%%

\chapter{{\it Improving the surface brightness-color relation for early-type stars using optical interferometry}}\label{challouf14}
\includepdf[pages=-,pagecommand={\thispagestyle{fancy}},offset=0 0]{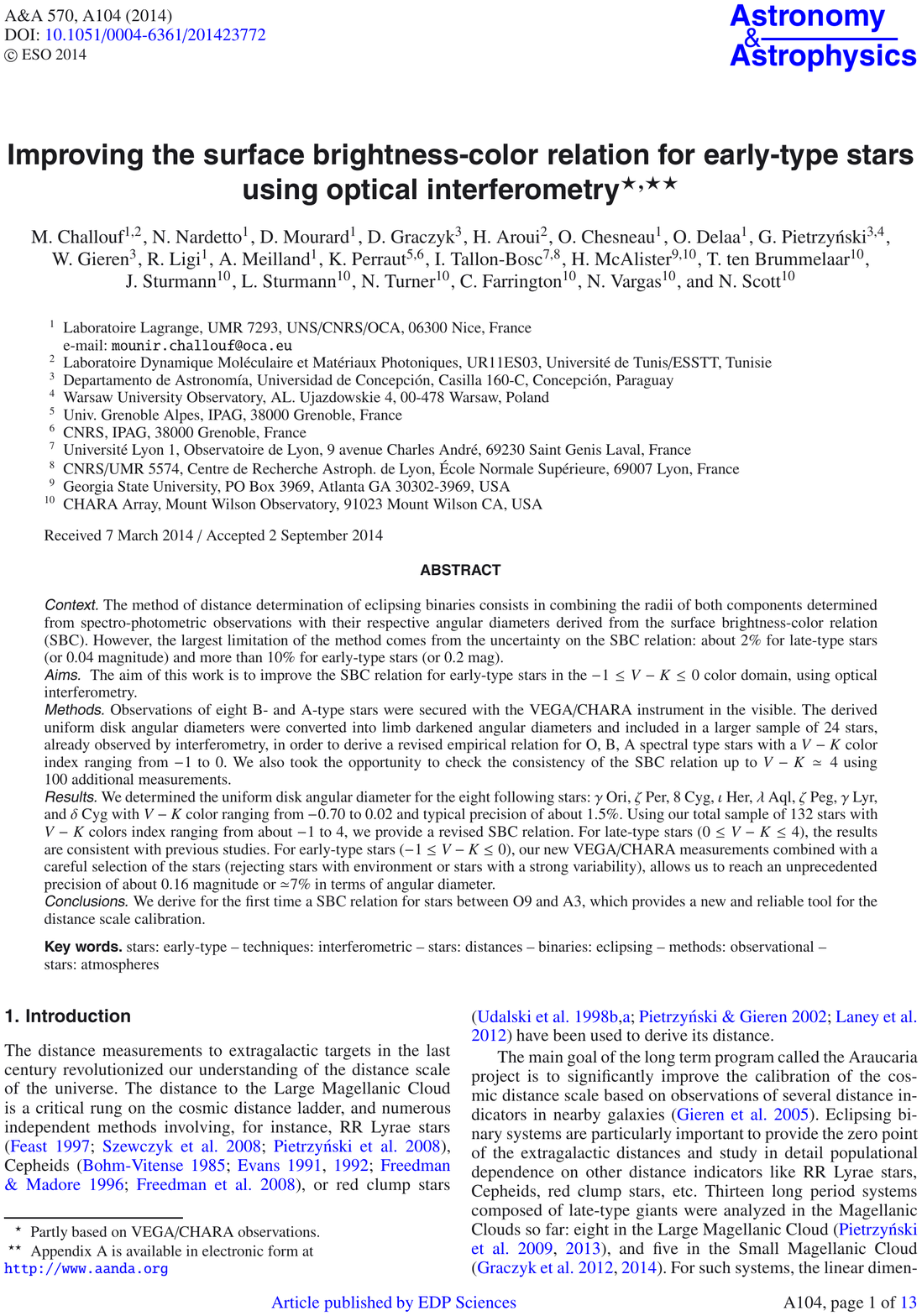} 

\chapter{{\it Theoretical impact of fast rotation on calibrating the surface brightness-color relation for early-type stars}}\label{challouf15}
\includepdf[pages=-,pagecommand={\thispagestyle{fancy}},offset=0 0]{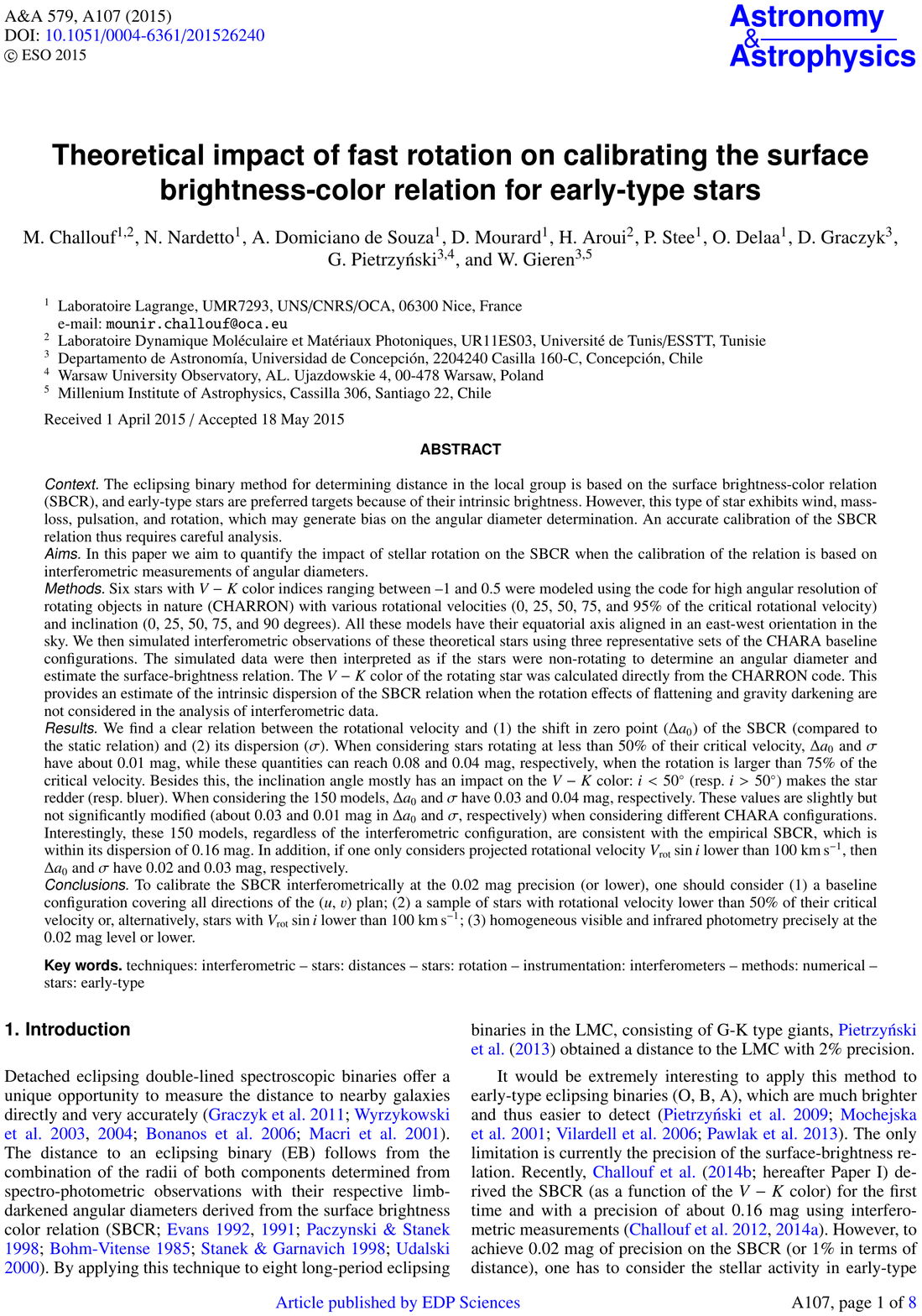} 

\chapter{{\it The Surface Brightness--Color Relations Based on Eclipsing Binary Stars: Toward Precision Better than 1\% in Angular Diameter Predictions}}\label{graczyk17}
\includepdf[pages=-,pagecommand={\thispagestyle{fancy}},offset=0 0]{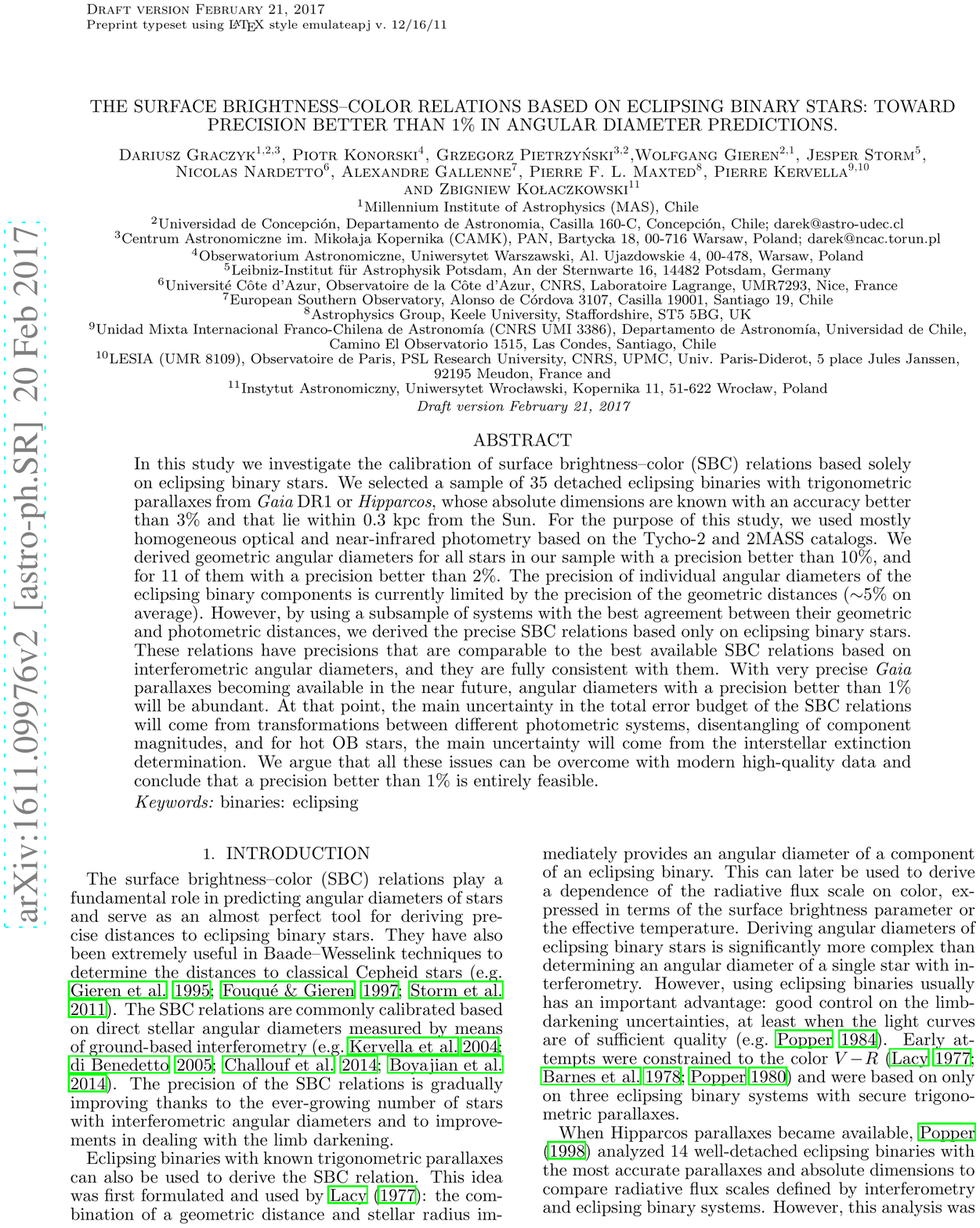}

%%%%%%
\chapter{{\it RR-Lyrae-type pulsations from a 0.26-solar-mass star in a binary system}}\label{pietrzynski12}
\includepdf[pages=-,pagecommand={\thispagestyle{fancy}},offset=0 0]{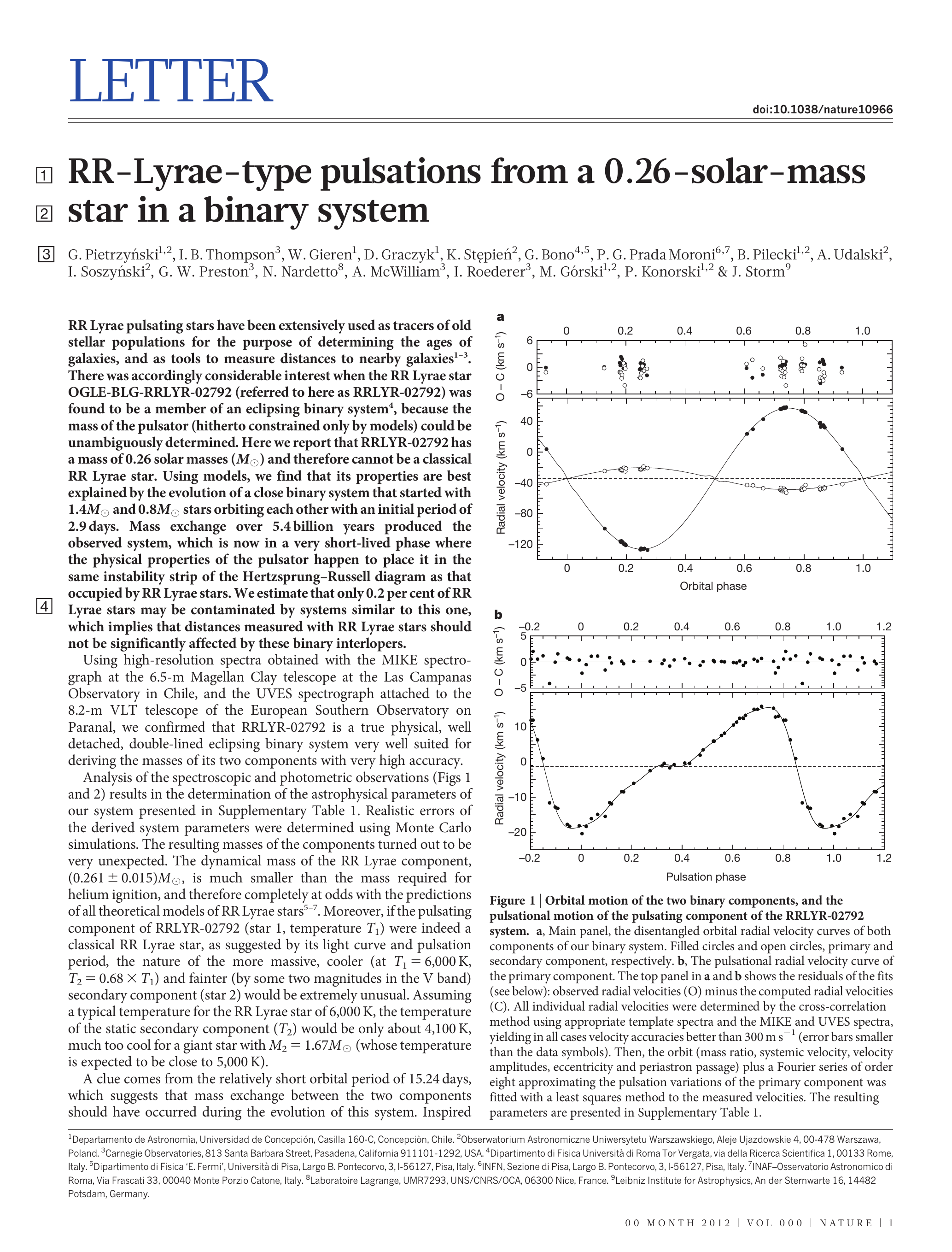} 

\chapter{{\it Physical parameters and the projection factor of the classical Cepheid in the binary system OGLE-LMC-CEP-0227}}\label{pilecki13}
\includepdf[pages=-,pagecommand={\thispagestyle{fancy}},offset=0 0]{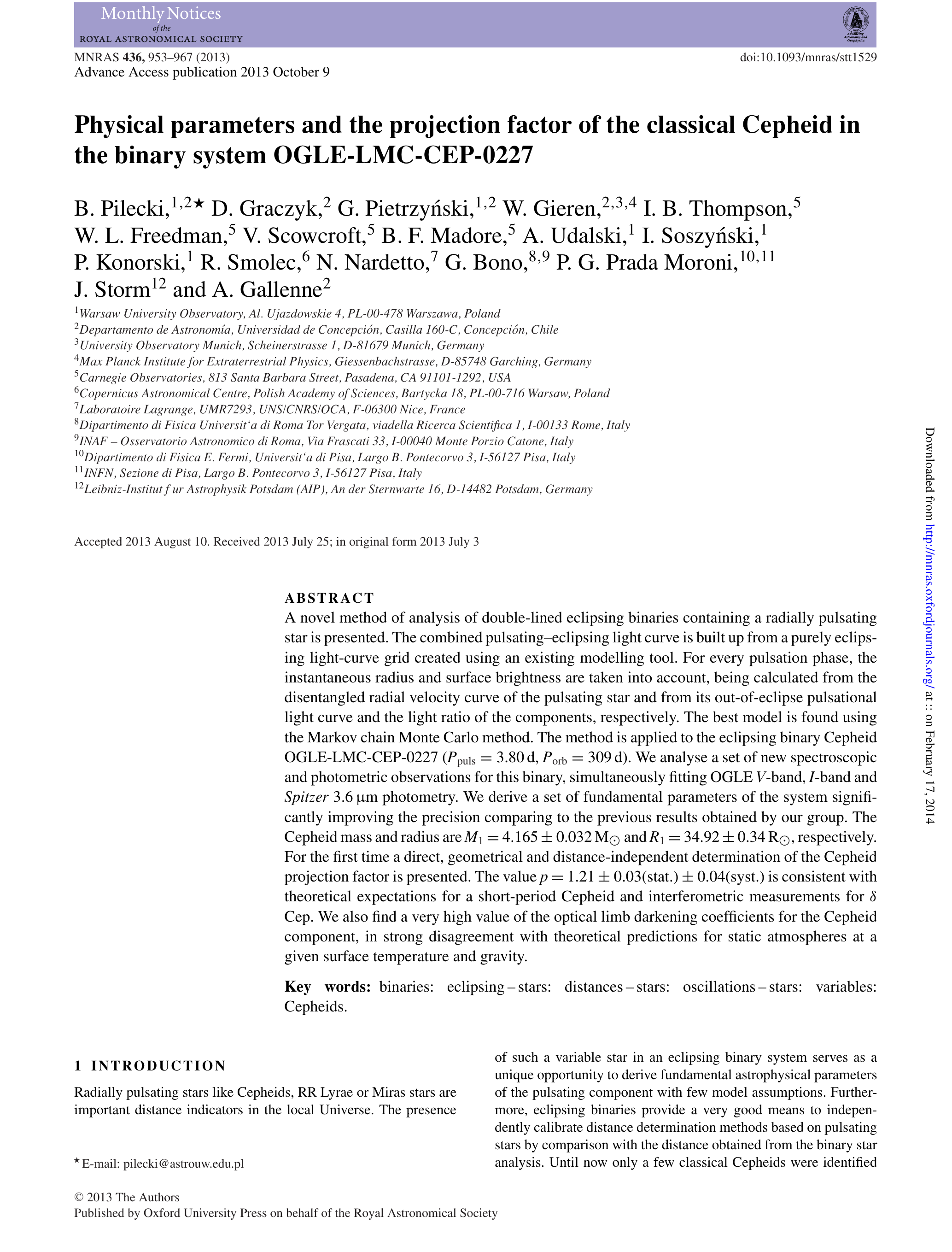}

\end{appendix}

%\newpage
%\strut \thispagestyle{empty} \vfill \pagebreak
\thispagestyle{empty}

{\Huge Résumé}

\vspace*{0.5cm}

Pour comprendre la nature de l'énergie noire, il faut mesurer le taux d'expansion de l'univers, c'est-à-dire la constante de Hubble ($H_\mathrm{0}$), avec une bonne précision, soit à mieux que 2\%. Les deux principales méthodes qui permettent de faire cela, le rayonnement de fond cosmologique et la détermination des distances dans l'univers présentent des désaccords significatifs, on parle de 'tension'. L'une des clefs pour résoudre cette tension se trouve très probablement dans l'étalonnage de la relation période-luminosité (PL) des Céphéides: quelle est son point-zéro ? Est-il sensible à la métallicité ? Peut-on réduire la dispersion de la relation ? La méthode des binaires à éclipses a permis récemment d'atteindre 2.2\% sur la distance du LMC, ce qui permet de contraindre le point-zéro de la relation PL et d'étudier l'effet de la métallicité. Mais cette méthode repose principalement sur l'hypothèse que les étoiles sont de bons corps noirs, ce qui n'est pas toujours le cas comme l'atteste les discordances dans les estimations des relations brillance de surface-couleur pour les étoiles de type O,A,B par exemple, utiles pour la détermination de distances de M31 ou M33. Par ailleurs, appliquer la méthode de Baade-Wesselink de détermination de distance aux Céphéides des nuages de Magellan est maintenant possible, mais la question clef de la valeur du facteur du projection et de sa dépendance avec la période des Céphéides demeure. Il est donc impossible de faire l'impasse sur la compréhension de la dynamique atmosphérique des Céphéides ($p$-facteur, $k$-facteur) et de leur environnement pour déterminer les distances dans l'univers. Ceci est d'autant plus vrai que les enveloppes des Céphéides pourraient avoir un effet sur la dispersion de la relation période-luminosité. L'arrivée des nouvelles générations d'interféromètres (MATISSE, SPICA), des parallaxes {\it Gaia}, dans le contexte du JWST, de PLATO et de l'ELT, permettront de répondre à ces questions fondamentales. 

\vspace*{0.5cm}

{\Huge Abstract}
\vspace*{0.5cm}

Determining the expansion of the universe, i.e. the Hubble constant ($H_\mathrm{0}$) to better than 2\% is required in order to understand the nature of dark energy. However, the two most accurate methods to do it, the cosmic microwave background and the distance scale ladder are inconsistent today, which is refereed as the 'tension'. One of the key to resolve this tension is related to the calibration of the period-luminosity (PL) of Cepheids: what is its zero-point ? Is it metallicity dependent ? Can we reduce the dispersion of the PL relation ? The eclipsing binaries method was recently used to determine the distance to LMC with a 2.2\% accuracy, which is crucial to constrain the zero-point of the PL relation and to study the impact of metallicity. However, this method is based on the hypothesis that stars are perfect blackbody, which is not always the case as shown by inconsistencies in the surface brightness - color relations of early type stars for instance, that are actually useful for the distance determination of eclipsing binaries in distant galaxies (i.e. M31, M33). On the other hand, it is now possible to apply the Baade-Wesselink method to Cepheids in the Magellanic Clouds, however the value of the projection factor and its dependence with the period of Cepheids remains a key issue. Understanding the dynamical structure of the atmosphere of Cepheids ($p$-factor, $k$-factor) and their environment cannot be circumvented in order to determine the distance in the universe. This is even more true that the environment of Cepheids could increase the dispersion of the PL relation. The next generation of interferometers (MATISSE, SPICA), the {\it Gaia} parallaxes, in the context of JWST, PLATO, and ELT will help to resolve these fundamental issues.

\end{document}